\documentclass[twoside,12pt]{article}
\usepackage{amssymb,amsmath}
\usepackage[dvipdfmx]{graphicx}
\usepackage{color}
\usepackage{bm}
\usepackage[breaklinks, colorlinks=true]{hyperref}

\graphicspath{{./Figs/}}

\newcommand{\feyn}[1]{
\setbox0=\hbox{\ensuremath{#1}}
\hbox to\wd0{\hbox to0pt{\hbox to\wd0{\hss/\hss}\hss}\box0}}

\newcommand{\comment}[1]{}  
\begin{document}
\title{ Heavy flavors under extreme conditions in high energy nuclear collisions }
\author{ Jiaxing Zhao$^1$, Kai Zhou$^2$, Shile Chen$^1$, and Pengfei Zhuang$^1$\\
\\
{\small $^1$Physics Department, Tsinghua University, Beijing 100084, China }\\
{\small $^2$Frankfurt Institute for Advanced Studies, Ruth-Moufang-Str. 1, 60438 Frankfurt am Main, Germany }}
\maketitle

\begin{abstract}
Heavy flavor hadrons have long been considered as a probe of the quark gluon plasma created in high energy nuclear collisions. In this paper we review the heavy flavor properties under extreme conditions and the realization in heavy ion experiments. After a short introduction on heavy flavor properties in vacuum, we emphasize the cold and hot nuclear matter effects on heavy flavors, including shadowing effect, Cronin effect and nuclear absorption for the former and Debye screening and regeneration for the latter. Then we discuss, in the frame of transport and coalescence models, these medium induced changes in open and closed heavy flavors in nuclear collisions and the comparison with nucleon-nucleon collisions. Considering the extremely strong electromagnetic and rotational fields generated in non-central nuclear collisions, which are widely studied in recent years, we finally investigate their effects on heavy flavor production and evolution in high energy nuclear collisions. 
\end{abstract}
\tableofcontents

\section{Introduction}
\label{introduction}
It is well-known that the symmetries of Quantum Chromodynamics (QCD) can be changed in multi-particle systems by vacuum excitation at finite temperature and vacuum condensation at finite density. When temperature and/or density is high enough, two QCD phase transitions happen, one is the deconfinement from hadron gas to quark gluon plasma (QGP), and the other is the chiral phase transition from chiral symmetry breaking to its restoration. In real case with nonzero pion mass in vacuum, the two phase transitions are expected to be of first order at high baryon density and become a crossover at high temperature, and there exists a critical point between the crossover and the first order phase transition. From the lattice QCD simulation at zero baryon density, the crossover temperature for chiral symmetry restoration is about $T_c = 155$ MeV~\cite{Bazavov:2011nk}. Such phase transitions are expected to be realized in the very beginning of our universe where the temperature is extremely high and in the core of compact stars where the baryon density is extremely high. In laboratories on the earth, the only way to realize the QCD phase transitions is through high energy nuclear collisions where a hot and dense fireball is formed in the central region of collisions. Such collisions happen at the Relativistic Heavy Ion Collider (RHIC) with colliding energy per pair of nucleons $\sqrt {s_{NN}}=200$ GeV and the Large Hadron Collider (LHC) with $\sqrt {s_{NN}}=2.76$ TeV.

The fireball formed in nuclear collisions is not a static system, it expands rapidly which leads to a continuous decrease of the temperature and density. When the temperature reaches the confinement value, the QGP, if it has formed in the early stage of the collisions, starts to hadronize into a gas of hadrons. Therefore, we cannot directly see the QGP in the final stage of nuclear collisions, and we need sensitive probes to signal the existence of the early QGP. Heavy flavor hadrons are such a probe due to the following reasons~\cite{Andronic:2015wma,Dainese:2016gch,Aarts,FCC}.

1. Since heavy quarks are so heavy that their masses are much larger than the temperature of the QGP created in nuclear collisions at RHIC and LHC energies, $m_Q\sim (1-5)$ GeV $\gg T\sim (0.3-0.5)$ GeV, thermal production in the QGP can be safely neglected and heavy quarks are almost entirely originated from the initial collisions and chemically decoupled from the medium (Note however, when the colliding energy is much higher than the LHC energy such as at the Future Circular Collider (FCC)~\cite{Dainese:2016gch,Aarts,FCC} with $\sqrt {s_{NN}}=39$ TeV, the plasma temperature could reach a value approaching to half charm quark mass, then the in-meidum production of charm quarks through gluon fusion will take place and would make a significant contribution to the total yield.). Considering that heavy quark masses are also much larger than the typical QCD scale, $m_Q\gg \Lambda_{QCD}$, their initial production is through hard QCD processes and can be solidly calculated through perturbative QCD (pQCD).

2. The initial heavy quarks are created with a time scale $\Delta \tau \sim 1/(2m_Q) \sim 0.07$ fm for charm quarks and $0.02$ fm for bottom quarks, which are much shorter than the QGP formation time $\tau_0 \sim 0.5$ fm/c at RHIC and LHC energies. Therefore, the cold and hot nuclear matter effects on heavy quarks can be simply factorized: The initial production process is modified by the cold nuclear matter effects like shadowing effect~\cite{Mueller:1985wy}, Cronin effect~\cite{Cronin:1974zm} and nuclear absorption~\cite{Gerschel:1988wn}, and then the created heavy quarks experience the whole space-time evolution of the fireball, lose part of their energy via interaction with the QGP constituents, and partially participate in the collective motion of the system. Considering the long and strong interaction with the QGP, heavy flavors are considered as a sensitive probe of the QGP.

3. Hadronization of partons is still an open question. In high energy nuclear collisions, one usually take quark coalescence on the hadronization hyper-surface of the fireball to calculate the final state hadron distributions, where the core quantity is the coalescence probability or the Wigner function which is normally treated as a Gaussian distribution with widths as free parameters. Considering the large mass of heavy quarks, one can neglect, as a first approximation, the heavy quark creation-annihilation fluctuations and calculate the wave function and then the Wigner function for heavy flavor hadrons in vacuum and at finite temperature in relativistic or even non-relativistic potential models~\cite{Satz:2005hx}. This provides a way to relate the heavy flavor production in nuclear collisions to understanding the QCD properties at finite temperature.

4. Considering the larger binding energy for heavy flavor hadrons in comparison with light hadrons, heavy flavors can in general case survive in the QGP phase. Since different heavy flavors have different binding energies, their surviving (or dissociation) temperatures should not be the same. For instance, there exists a sequential dissociation~\cite{Satz:2005hx} for charmonium states $J/\psi, \chi_c$ and $\psi'$ from the non-relativistic potential model with lattice simulated heavy quark potential at finite temperature~\cite{Petreczky:2010yn}. Therefore, unlike light hadrons which are all formed at the deconfinement phase transition, observed heavy flavors in the final state carry the information of the QGP at different stages and then can be used to probe the QGP structure. For instance, the bottom hadrons are more sensitive to the early stage of the QGP, while charm hadrons carry the information of the later stage of the QGP.

5. The electromagnetic field and rotational field generated in non-central nuclear collisions are extremely strong, and the quark spin interaction with the fields leads to many interesting quantum phenomena like chiral magnetic effect~\cite{Fukushima:2008xe} and chiral vortical effect~\cite{Kharzeev:2007tn,Kharzeev:2010gr} which are extensively studied in recent years. However, the lifetime of the electromagnetic field is very short and affects only those initially produced particles. Considering that heavy quarks are almost all produced in the very beginning of the collisions, they should be significantly affected by the fields. As a result, their modified properties will be inherited by heavy flavor hadrons in the final state.

The paper is organized as the follows. Since all the medium modifications are relative to the vacuum, we discuss, in the beginning of Section \ref{vacuum_and_medium}, heavy quark and heavy flavor hadron production in vacuum and their static properties in vacuum and medium in the frame of potential models. Then we focus, in the rest part of Section \ref{vacuum_and_medium}, on the cold and hot medium effects on heavy flavor hadrons. We will consider shadowing effect, Cronon effect, and nuclear absorption in cold nuclear matter and Debye screening and regeneration in hot nuclear matter. The heavy quark energy loss will be discussed together with open heavy flavor properties in hot medium in Subsection \ref{production}. We will also discuss shortly the heavy quark thermal production in hot medium in the end of Section \ref{vacuum_and_medium}. In Section \ref{open_collision}, we describe the above medium effects on open heavy flavors in high energy nuclear collisions. The space-time evolution of heavy quarks in nuclear collisions can be described by transport equations with collisional and radiative energy loss terms, and the production of open heavy flavors is usually controlled by coalescence mechanism for low momentum hadrons and fragmentation mechanism for high momentum hadrons. Heavy flavor baryons, especially multi-charmed baryons and their exotic states, are investigated in the frame of coalescence mechanism together with potential model. The hot medium effect on heavy flavor correlation in high energy nuclear collisions is investigated in the end of this section. The medium effects on closed heavy flavors in high energy nuclear collisions are described in Section \ref{closed}. To separate the novel QGP effect from the normal nuclear matter effect, we first consider the normal and anomalous charmonium suppression at SPS energy, and then take into account the recombination of those uncorrelated heavy quarks in the QGP at RHIC and LHC energies. Transport models are again used to describe the quarkonium motion in hot medium with both loss (dissociation) and gain (regeneration) terms. We show, in the end of this section, the calculated final state distributions like nuclear modification factor, elliptical flow and averaged transverse momentum and the comparison with the experimental data. The extreme conditions contain not only high temperature and high density but also strong electromagnetic and rotation fields. We discuss the behavior of heavy flavor hadrons in magnetized and rotational QGP created in non-central nuclear collisions in Section \ref{electro}. After a calculation of the external electromagnetic field and the feedback from the electrodynamics of the QGP, the motions of heavy quarks and heavy flavor hadrons in electromagnetic field are controlled respectively by transport and potential models with minimal coupling. Since the field breaks down the space symmetry, the heavy quark potential and collective flow for high momentum charmonia become anisotropic. The photoproduction of vector mesons in peripheral and especially ultra peripheral collisions is discussed in Subsection \ref{photon}. Finally, we discuss open and closed heavy flavors in a rotational field. We summarize the paper in Section \ref{conclusion}.

\section{Heavy flavors in vacuum and medium}
\label{vacuum_and_medium}
In this section, we first summarize heavy quark and heavy flavor hadron production mechanisms in vacuum, then focus on various cold and hot medium effects before and after the fireball formation in heavy ion collisions. Different from light quarks which are largely created in hot medium, heavy quarks are almost all produced through hard processes in the initial stage of the collisions and then pass through the fireball from the beginning to the end. Therefore, heavy quarks and heavy flavor hadrons in high energy nuclear collisions are sensitive to both the cold and hot mediums.

\subsection{Heavy quark and heavy flavor hadron production in vacuum}
\label{production}
The production of heavy quarks in elementary collisions can be treated by QCD factorization. The cross section at parton level can be calculated at leading order or next-to-leading order. The main uncertainty comes from the parton distribution function (PDF), especially in low-$x$ region. Therefore, studying the heavy flavor production can help to constrain the PDF in nucleons.

There are no free quarks in vacuum. Heavy quarks will finally undergo hadronization and become heavy flavor hadrons. Since hadronization is a non-perturbative process which happens at low energy scale with large coupling constant $\alpha_s$, we can't deal with it directly. Usually, people take fragmentation function to treat quark hadronization. For quakronia, one can describe it with effective field theory or potential model to project a heavy quark pair $Q\bar Q$ to a quarkonium.

\subsubsection{Heavy quark production}
\label{quark}
For heavy quark production in hadron-hadron collisions, the collinear factorization theorem~\cite{Collins:1985gm} is usually employed and has been proven to be valid. In standard QCD analysis, heavy quark production is dominated by the following factorization,
\begin{equation}
\sigma_{pp\to c\bar c}=\int_0^1dx_1 dx_2 \sum_{i,j} f_i(x_1,Q^2) f_j(x_2,Q^2)\sigma_{ij\to c\bar c}(x_1,x_2,Q^2),
\end{equation}
where $f_i$ is the parton distribution function in nucleons, $x_1$ and $x_2$ are the longitudinal momentum fractions of the parton $i$ and $j$, and $Q^2$ is the transform momentum square of the elementary process. Due to the large mass of heavy quarks in comparison with the typical QCD scale $\Lambda_{QCD}\sim 0.2$ GeV,  the cross section $\sigma_{ij\to c\bar c}$ can be perturbatively calculated as an expansion of the QCD coupling constant $\alpha_s$~\cite{Combridge:1978kx}.

At leading order (LO), there are mainly two sub-processes $g+g\to Q+\bar Q$ and $q+\bar q \to Q+\bar Q$. Considering the so-called unintegrated gluon distribution to account for the initial parton transverse momentum, the $k_T$-factorization approach~\cite{Catani:1990eg,Baranov:2000gv,Hagler:2000dda} is used to calculate the inclusive heavy quark production at leading order (LO), and the transverse momenta of the incident partons are incorporated by a random shift of these momenta ($k_T$ kick)~\cite{Frixione:1997ma}.
\begin{figure}[!htb]
{$$\includegraphics[width=0.6\textwidth]{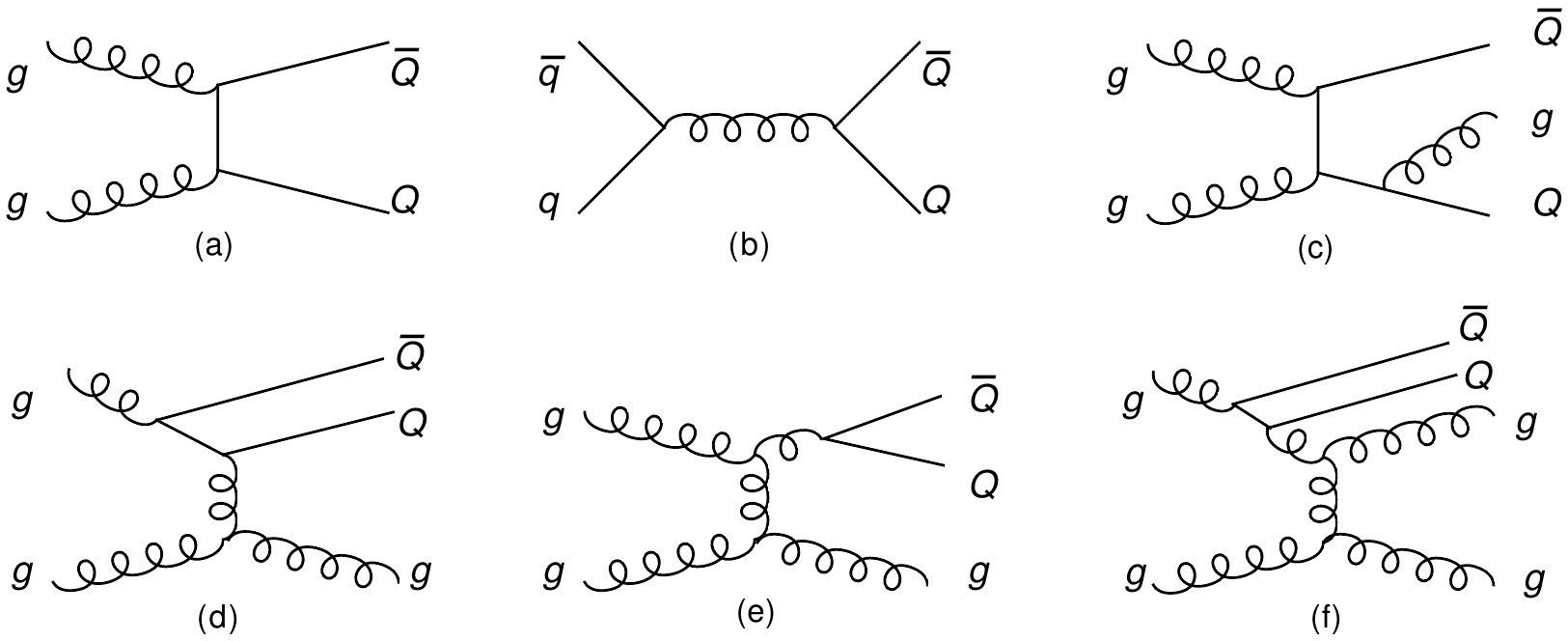}$$
\caption{Heavy quark production diagrams: (a) gluon fusion, (b)quark-antiquark annihilation, (c)pair creation with gluon emission, (d) flavor excitation, (e)gluon splitting, (f)together gluon splitting and flavor excitation.}
\label{fig1}}
\end{figure}

Recently, theoretical calculations on heavy quark production have been extended to next-to-leading order (NLO)~\cite{Nason:1989zy,Beenakker:1990maa,Mangano:1991jk}. There are generally two kinds of corrections: one is virtual one-loop correction to the subprocesses at leading order, and the other is the $2\to3$ process like $g+g\to Q+\bar Q+g$ shown in Fig.\ref{fig1}. All these processes have been calculated for heavy quark inclusive production cross section in Fixed-Flavour-Number Scheme (FFNS)~\cite{Nason:1989zy,Beenakker:1990maa}, Zero-Mass Variable-Flavor-Number Scheme (ZM-VFNS)~\cite{Cacciari:1993mq,Kniehl:1995em},  General-Mass Variable-Flavor-Number Scheme (GM-VFNS)~\cite{Kramer:2001gd,Kniehl:2005mk}, Fixed-Order plus Next-to-Leading-Log (FONLL)~\cite{Cacciari:2001td} and so on. Note that, these analytic treatments are applicable only for calculating infrared and collinear safe quantities due to the perturbative requirement, and therefore only the inclusive transverse momentum and rapidity distributions of heavy quarks can be investigated with these fixed-order calculations. In many cases and for some special purpose, people would like to access also the fully exclusive final state observables like angular correlations. To this end, the Monte-Carlo generators like PYTHIA~\cite{Sjostrand:2006za} and MC@NLO~\cite{Frixione:2002ik} can provide a more complete description for the final state hadrons through parton showering and modeling of further hadronization including decay and even detector response constraints.

\subsubsection{Open heavy flavor production}
\label{open}
Due to the non-perturbative property of hadronization process in QCD, people usually use a suitable fragmentation function $D_Q^H(z)$ to describe the transition of a heavy quark with momentum $p_Q$ into a heavy flavor hadron $H$ with momentum $p_H=zp_Q$,
\begin{equation}
d\sigma_H=d\sigma_{Q\bar Q} \otimes D_Q^H(z).
\end{equation}
The fragmentation function is universal, it can be measured in $e^+ + e^-$ annihilation experiments and then used to describe hadron production in hard QCD processes. One thing needs to mention is that, the NLO calculation for heavy quark production usually needs a harder fragmentation function than LO to compensate for the softening effects of the gluon emissions.

There are two main types of fragmentation functions. One is scale independent, like Lund String fragmentation function~\cite{Bowler:1981sb,Andersson:1983ia} used in PYTHIA and Peterson fragmentation function~\cite{Peterson:1982ak}. The Peterson fragmentation function can be expressed as 
\begin{equation}
D_Q^H(z)={1\over z \left (1-{1\over z}-{\epsilon \over 1-z}\right)^2}.
\end{equation}
For heavy flavors, the parameter $\epsilon$ is fixed by experimental data of $D$ mesons in $p+p$ and $e^++e^-$ collisions. The other is scale dependent, including the one based on Perturbative Fragmentation Function (PFF) approach~\cite{Cacciari:1993mq} used in FONLL~\cite{Cacciari:2001td} and the one based on Binnewies-Kniehl-Kramer (BKK) approach~\cite{Binnewies:1997xq,Binnewies:1998vm} used in GM-VFNS~\cite{Kramer:2001gd,Kniehl:2005mk}.

The fragmentation function based on PFF approach is given by a convolution of a perturbative fragmentation function of a parton $k$ into a heavy quark $Q$, $D_k^Q(z, \mu_F)$, with a scale-independent fragmentation function $D_Q^H(z)$ describing the hadronization of the heavy quark into a hadron $H$. The scale dependence is governed by the DGLAP evolution equation and the boundary condition which can be calculated perturbatively. The fragmentation function based on BKK approach cannot be split up into a perturbative and a non-perturbative part. The boundary condition at an initial scale $\mu_F=m_Q$ is determined by experimental data for the full non-perturbative fragmentation function $D_k^H(z,\mu_F)$, while the larger scale $\mu_F$ is controlled by the DGLAP equation.

\subsubsection{Closed heavy flavor production}
\label{close}
The theoretical study on closed heavy flavor production involves both perturbative and non-perturbative aspects of QCD. While the charm quark production cross section can be well calculated in the frame of pQCD, the subsequent soft interaction required to form a quarkonium is still theoretically not well understood. We need various mechanisms to describe the quarkonium production in $pp$ collisions.
In the following, we briefly discuss those non-perturbative models and their differences: the Colour-Evaporation Model (CEM), the Colour-Singlet Model (CSM) and the Colour-Octet Mechanism (COM), the latter two are encompassed in an effective theory named Non-Relativistic QCD (NRQCD).

$Color$-$evaporation$ $model$ (CEM)~\cite{Fritzsch:1977ay,Amundson:1995em,Bedjidian:2004gd}.
The quarkonium production cross section is expected to be directly connected to producing a $Q\bar Q$ pair in an invariant-mass region where its hadronization into a quarkonium is possible, that is between the kinematical threshold to produce a quark pair, $2m_Q$, and that to create the lightest open-heavy-flavor hadron pair $2m_H$,
\begin{equation}
\sigma_{CEM}^{pp\to \Psi}= f_{\psi} \int_{4m_Q^2}^{4m_H^2}d\mu^2\int_0^1dx_1 dx_2 \sum_{i,j} f_i(x_1,Q^2) f_j(x_2,Q^2)\sigma^{ij\to c\bar c}(x_1,x_2,Q^2; \mu^2),
\end{equation}
where the phenomenological factor for a given spin $J_Q$ of the quarkonium is $f_{\psi}=1/9(2J_Q+1)/\sum_i(2J_i+1)$. One assumes that a number of non-perturbative-gluon emissions occur once the $Q\bar Q$ pair is produced and that the quantum state of the pair at its hadronization is essentially decorrelated at least color-wise-with the state at its production.

$Color$-$singlet$ $model$ (CSM)~\cite{Chang:1979nn,Baier:1981uk,Berger:1980ni}. It assumes that a quarkonium is a bound state with a highly peaked wave function in the momentum space. Therefore, the cross section for quarkonium production should be expressed as the production of a heavy-quark pair with almost zero relative velocity,
 \begin{equation}
\sigma_{CSM}^{pp\to \Psi}=|\psi(0)|^2\int_0^1dx_1 dx_2 \sum_{i,j} f_i(x_1,Q^2) f_j(x_2,Q^2)\sigma^{ij\to [c\bar c]}(x_1,x_2,Q^2),
\end{equation}
where $|\psi(0)|^2$ is the square of the Schr\"odinger wave function at the origin in the position space.

$Color$-$octet$ $model$ (COM)~\cite{Bodwin:1992qr} and NRQCD~\cite{Bodwin:1994jh,Chao:2012iv,Butenschoen:2012px,Gong:2012ug}.
One can express more rigorously the hadronization probability of a heavy-quark pair into a quarkonium via long-distance matrix elements (LDMEs). Different from the usual expansion in powers of $\alpha_s$, NRQCD introduces an expansion in relative velocity $v$. The leading order contribution of NRQCD is the CSM, while the higher-Fock states (in $v$) contain the non-perturbative transitions between the colored states and the physical mesons,
 \begin{equation}
\sigma_{COM}^{pp\to \Psi}=\sum_n \langle O_n^\psi \rangle\int_0^1dx_1 dx_2 \sum_{i,j} f_i(x_1,Q^2) f_j(x_2,Q^2)\sigma^{ij\to [c\bar c]_n}(x_1,x_2,Q^2),
\end{equation}
where $ \langle O_n^\psi \rangle$ is the LDMEs, and $n$ denotes the additional quantum numbers (angular momentum, spin, and color).

\subsection{Heavy flavor properties in vacuum}
\label{property}
We have discussed the production mechanisms of open and closed heavy flavors. In this section, we summarize the theoretical studies on the properties of heavy flavors in vacuum. The non-perturbative QCD calculations, including Lattice QCD simulations~\cite{Gattringer:2010zz} and effective QCD sum rules~\cite{Reinders:1984sr,Shuryak:1981fza,Bagan:1992tp}, have been used to study heavy flavor hadrons for many years and give a good description for the hadron mass spectra~\cite{Bagan:1992tp,AliKhan:1999yb,Morita:2009qk}. Considering the large mass of heavy quarks, the creation and annihilation can be safely neglected, and we can use effective field theories of QCD, such as NRQCD and potential NRQCD~\cite{Brambilla:2004jw}, and even non-relativistic and relativistic potential models~\cite{Matsui:1986dk, Bali:2000gf} to comprehensively and simply describe heavy flavors in vacuum and medium. In potential models, a problem of quantum field theory becomes a problem of quantum mechanics, and the heavy flavor properties are clearly controlled by the Schr\"odinger or Dirac equations for the heavy quark pair.

\subsubsection{Non-relativistic potential model}
\label{potential}
The non-relativistic potential model, based on Schr\"odinger equation, has been successfully used to describe the properties of quarkonia for many years. We will show here the framework of $N$-body Schr\"odinger equation which can be used to treat $N$-body bound states. The Schr\"odinger equation to describe the wave function $\Psi({\bf r}_1,...,{\bf r}_N)$ and energy $E$ for a $N$-quark system is
\begin{equation}
\left( \sum_{i=1}^N{\hat {\bf p}_i^2 \over 2m_i} +V \right)\Psi({\bf r}_1,...,{\bf r}_N)=E\Psi({\bf r}_1,...,{\bf r}_N)
\end{equation}
with $V=\sum_{i<j}V({\bf r}_i, {\bf r}_j)$. As a usually used approximation, we have here neglected the three-body and other higher order potentials and expressed the total potential as a sum of pair interactions. From the quark model or leading order of perturbative QCD calculation, the diquark potential is only one half of the quark-antiquark potential, $V_{QQ}=V_{Q\bar Q}/2$. We assume that such a relation still holds in the case of strong coupling. The central part of the potential between a pair of quark and antiquark in vacuum is the Cornell potential, and the spin-spin interaction part can be taken from the lattice studies~\cite{Kawanai:2011jt},
\begin{equation}
V_{Q\bar Q}(|{\bf r}_{ij}|)=-{\alpha \over |{\bf r}_{ij}|}+\sigma |{\bf r}_{ij}|+ \beta e^{-\gamma |{\bf r}_{ij}|} {\bf s}_i \cdot {\bf s}_j,
\end{equation}
where ${\bf r}_{ij}={\bf r}_i-{\bf r}_j$ is the distance between the two quarks labeled with $i$ and $j$, and the parameters $\sigma$, $\alpha$, $\beta$ and $\gamma$ should be fixed by fitting experimental data.

We first factorize the $N$-body motion into a center-of-mass motion and a relative motion by introducing the Jacobi coordinates,
\begin{eqnarray}
&&\Psi({\bf r}_1,...,{\bf r}_N)=\Theta({\bf R})\Phi({\bf x}_1,...,{\bf x}_{N-1}),\nonumber\\
&&{\bf R}= {1\over M}\sum_{i=1}^Nm_i{\bf r}_i, \nonumber\\
&&{\bf x_j}=\sqrt{M_j m_{j+1}\over M_{j+1}\mu}\left( {\bf r}_{j+1}-{1\over M_j}\sum_{i=1}^j m_i {\bf r}_i\right)
\end{eqnarray}
with $M_j=\sum_{i=1}^j m_i$, $j=1,...,N-1$, the total mass $M=M_N$ and the reduced mass $\mu$. It is clear that, the bound state properties are only related to the relative motion of the system. There are many ways to solve the $3(N-1)$ dimensional relative equation, what we use here is the expansion method in terms of spherical harmonic functions~\cite{1998FBS,Barnea:1999be}.

By rewriting the relative coordinates ${\bf x}_1$,...,${\bf x}_{N-1}$ in terms of the hyperradius $\rho=\sqrt{x_1^2+...+x_{N-1}^2}$ and hyper angles $\Omega=\{\alpha_2,..,\alpha_{N-1}, \theta_1,\phi_1,...,\theta_{N-1},\phi_{N-1}\}$ with the definition of  $\alpha_i=x_i/\rho_i$ and $\rho_i=\sqrt{\sum_{j=1}^i x_j^2}$, the relative wave function is controlled by the Schr\"odinger equation,
\begin{eqnarray}
&& \left [ {1\over 2\mu}\left( -{1 \over \rho^{3N-4}}{d\over d\rho}\rho^{3N-4}{d \over d\rho}  + {\hat K_{N-1}^2\over \rho^2}\right) + V(\rho, \Omega) \right ]\Phi(\rho, \Omega) = E_r \Phi(\rho, \Omega),\\
&& \hat K_{N-1}^2 = -{\partial^2 \over \partial \alpha_{N-1}^2}+{(3N-9)-(3N-5)\cos(2\alpha_{N-1}) \over \sin(2\alpha_{N-1})}{\partial \over \partial \alpha_{N-1}}+{1\over \cos^2\alpha_{N-1}} \hat K_{N-2}^2+{1\over \sin^2\alpha_{N-1}} \hat l_{N-1}^2,\nonumber
\end{eqnarray}
where $\hat K_{N-1}^2$ is the hyper angular momentum operator with $\hat K_1^2=\hat l_1^2$ being exactly the particle angular momentum, and its eigenstate and eigenvalue are determined by
\begin{equation}
\hat K_{N-1}^2 {\mathcal Y}_\kappa (\Omega)= K(K+3N-5) {\mathcal Y}_\kappa (\Omega).
\end{equation}
Expanding the relative wave function in terms of the complete and orthogonal hyperspherical harmonic functions ${\mathcal Y}_\kappa (\Omega)$, $\Phi(\rho,\Omega)=\sum_\kappa R_\kappa(\rho){\mathcal Y}_\kappa (\Omega)$, the relative equation for $\Phi$ becomes a set of coupled radial equations for $R_\kappa$,
\begin{equation}
\left[{1\over 2\mu}\left({1\over \rho^{3N-4}}{d\over d\rho}\rho^{3N-4}{d\over d\rho}-{K(K+3N-5)\over \rho^2}\right)+E_r\right]R_\kappa = \sum_{\kappa'}V_{\kappa\kappa'}R_{\kappa'}
\end{equation}
with the potential matrix
\begin{equation}
V_{\kappa\kappa'}=\int {\mathcal Y}_\kappa^*(\Omega)V(\rho,\Omega){\mathcal Y}_{\kappa'}(\Omega)d\Omega.
\end{equation} 
\begin{figure}[!htb]
{$$\includegraphics[width=0.3\textwidth]{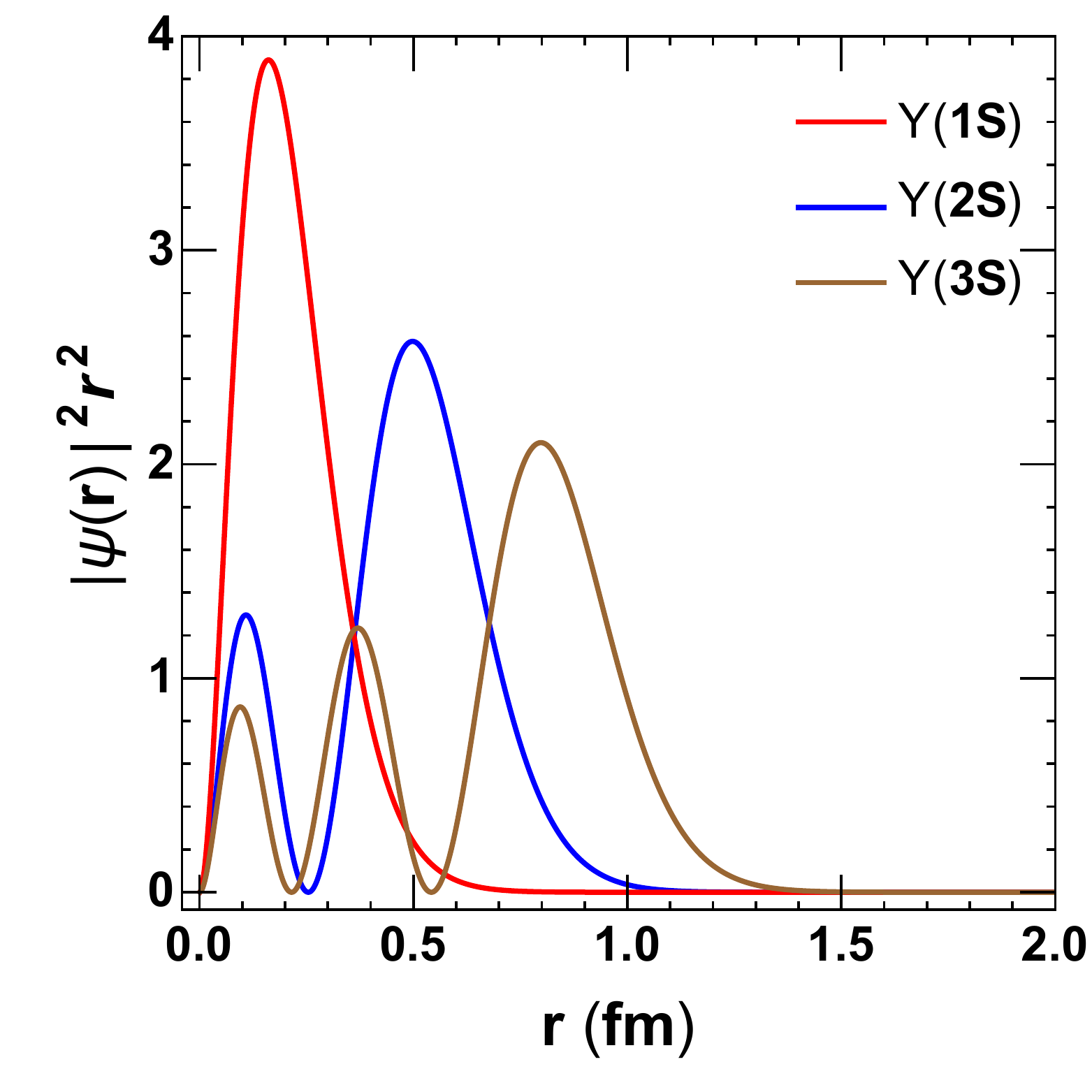} \ ~ \ ~ \includegraphics[width=0.3\textwidth]{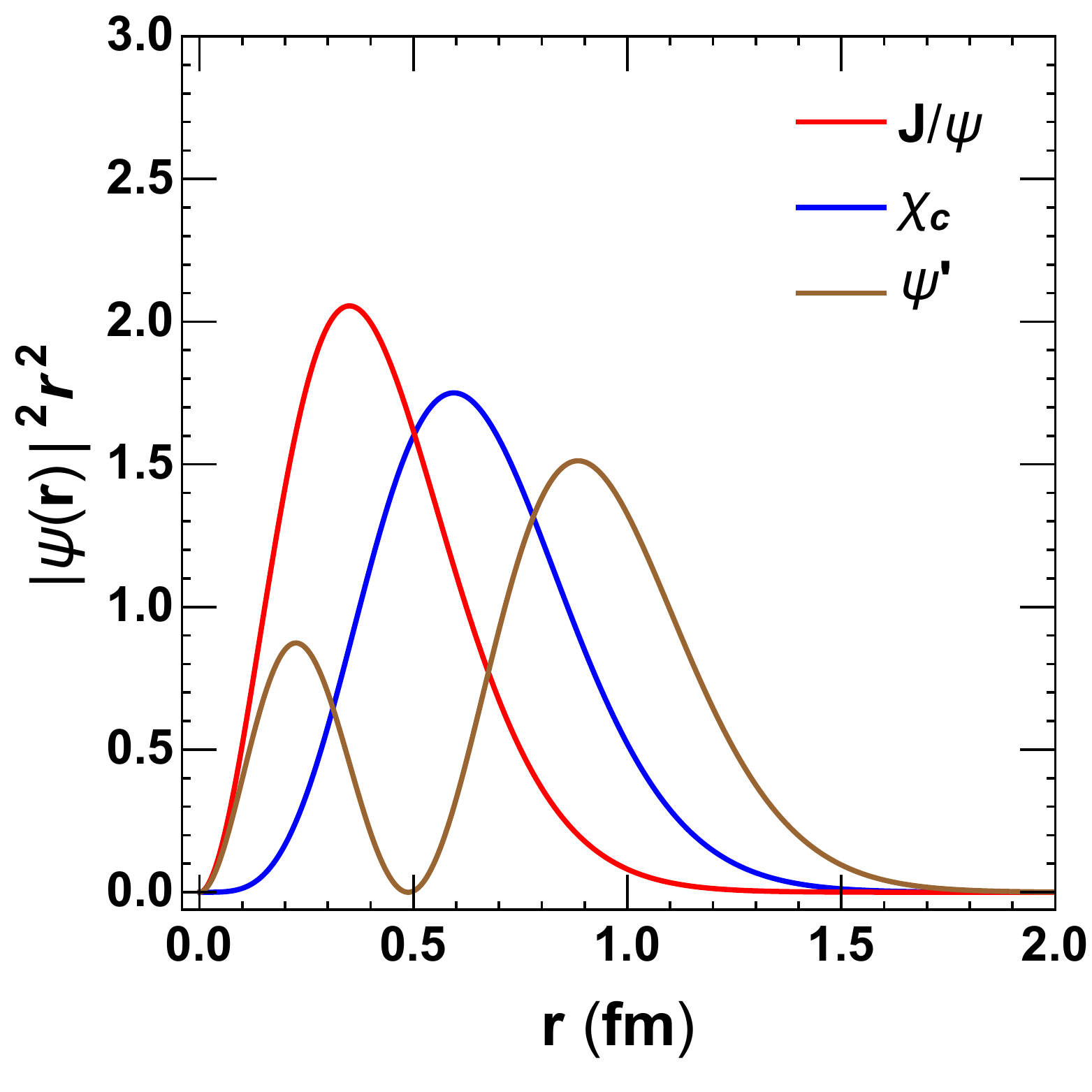}$$
\caption{The radial probability distribution for different charmonium (right panel) and bottomonium (left panel) states in vacuum. }
\label{fig2}}
\end{figure}

We now apply the $2$-body and $3$-body Schr\"odinger equations to quarknoia and heavy flavor baryons. For quarknoium systems with the global and relative coordinates ${\bf R} = 1/2({\bf r}_1+{\bf r}_2)$ and ${\bf r}={\bf r}_2-{\bf r}_1$, the relative motion is separated into a radial part
\begin{equation}
\left [ {1\over 2\mu}\left( -{d^2 \over d r^2}-{2\over r}{d \over dr}  + {l(l+1)\over r^2}\right) + V_{Q\bar Q}(r) \right ] R(r) = E_r R(r)
\end{equation}
and an angular part, the solution of the latter is the familiar spherical harmonic function $Y(\theta,\phi)$. With the quark mass $m_b=4.7$ GeV and $m_c=1.29$ GeV and the coupling constants $\alpha=0.4105$, $\sigma=0.2$ GeV$^2$, $\beta=0.318$ GeV for bottomonium and $2.06$ GeV for charmonium and $\gamma=1.982$ GeV, the calculated charmonium and quarkonium masses are shown in Tables \ref{table1} and \ref{table2}. The average radius is defined as $\langle r \rangle=\int |R(r)|^2r^3dr$. The wave functions are shown in Fig.\ref{fig2}. From the comparison with the experimental data, the non-relativistic potential model describes well all the quarkonium states, especially for the bottomonium states, since the heavier bottom quark leads to a better application of the non-relativistic Schr\"odinger equation.
\begin{table}[!hbt]
	\centering
	\begin{tabular*}{6.0in}{@{\extracolsep{\fill}}llcccccccc}
		\hline
		\hline
		State & $\eta_c(1S)$ & $J/\psi(1S)$& $ h_c(1P)$ & $\chi_c(1P)$& $\eta_c(2S)$ & $\psi(2S)$ & $h_c(2P)$  & $\chi_c(2P)$   \\
		\hline
		$M_{Exp} \text{(GeV)}$ & 2.981 & 3.097 & 3.525 & 3.556 & 3.639 & 3.686 & - & 3.927 \\
		\hline
		$M_{Th} \text{(GeV)}$ & 2.967 & 3.102 & 3.480 & 3.500 & 3.654 & 3.720 & 3.990 & 4.000 \\
		\hline
		$\langle r \rangle \text{(fm)}$  & 0.365 & 0.427 & 0.635 & 0.655 & 0.772 & 0.802 & 0.961 & 0.980 \\
		\hline
		\hline
	\end{tabular*}
	\caption{The experimentally measured~\cite{Tanabashi:2018oca} and model calculated masses $M_{Exp}$ and $M_{Th}$ and model calculated averaged radius $\langle r \rangle$ for charmonium states. }
	\label{table1}
\end{table}

\begin{table}[!hbt]
	\centering
	\begin{tabular*}{7.0in}{@{\extracolsep{\fill}}llccccccccc}
		\hline
		\hline
		State & $\eta_b(1S)$ & $\Upsilon(1S)$& $ h_b(1P)$ & $\chi_b(1P)$& $\eta_b(2S)$ & $\Upsilon(2S)$ & $h_b(2P)$  & $\chi_b(2P)$ &$\Upsilon(3S)$   \\
		\hline
		$M_{Exp} \text{(GeV)}$ & 9.398 & 9.460 & 9.898 & 9.912 & 9.999 & 10.023 & - & 10.269 & 10.355 \\
		\hline
		$M_{Th} \text{(GeV)}$ & 9.397 & 9.459 & 9.845 & 9.860 & 9.957 & 9.977 & 10.211 & 10.221 & 10.325\\
		\hline
		$\langle r \rangle \text {(fm)}$  & 0.200 & 0.214 & 0.377 & 0.387 & 0.465 & 0.474 & 0.597 & 0.603 & 0.680 \\
		\hline
		\hline
	\end{tabular*}
	\caption{The experimentally measured~\cite{Tanabashi:2018oca} and model calculated masses $M_{Exp}$ and $M_{Th}$ and model calculated averaged radius $\langle r \rangle$ for bottomonium states. }
	\label{table2}
\end{table}

For heavy flavor baryons with the global and relative coordinates
\begin{eqnarray}
&&{\bf R} = {1\over M}(m_1{\bf r}_1+m_2{\bf r}_2+m_3{\bf r}_3), \nonumber\\
&&{\bf x_1}=\sqrt{{m_1m_2\over (m_1+m_2) \mu}}\left({\bf r}_2-{\bf r}_1\right),\nonumber\\
&&{\bf x_2}=\sqrt{(m_1+m_2)m_3 \over (m_1+m_2+m_3)\mu}\left({\bf r}_3-{m_1{\bf r}_1+m_2{\bf r_2}\over m_1+m_2}\right),
\end{eqnarray}
the relative motion is controlled by
\begin{equation}
\left [ {1\over 2\mu}\left( -{d^2 \over d\rho^2}-{5\over \rho}{d \over d\rho}  + {\hat K_2^2\over \rho^2}\right) + V(\rho, \Omega) \right ]\Phi = E_r \Phi.
\end{equation}
Since the potential $V(\rho, \Omega)$ depends on both the hyperradius $\rho$ and the 5 angles $\Omega=\{\alpha,\theta_1,\phi_1,\theta_2,\phi_2\}$, the relative motion cannot be further factorized into a radial part and an angular part. When the three quarks are the same, for the ground states $\Omega_{ccc}$ and $\Omega_{bbb}$, the potential is reduced to~\cite{Zhao:2017gpq},
\begin{equation}
V(\rho, \Omega)=3V_{QQ}(|{\bf r}_{12}|)=3V_{QQ}(\sqrt{2}\rho\sin \alpha).
\end{equation}
To simplify the relative motion, we take the angle averaged potential
\begin{equation}
\bar V(\rho) = 3 {16\over \pi}\int_0^{\pi/2} V_{QQ}(\sqrt{2}\rho\sin \alpha)\cos^2\alpha \sin^2\alpha d\alpha
\end{equation}
as the effective potential in the relative motion. Under this approximation, the relative equation of motion can be factorized into the radial equation
\begin{equation}
\left [ {1\over 2\mu}\left( -{d^2 \over d \rho^2}-{5\over \rho}{d \over d\rho}  + {K(K+4)\over \rho^2}\right) + \bar V(\rho) \right ] R(\rho) = E_r R(\rho)
\end{equation}
and the angular equation with again the solution of the spherical harmonic function. The radial wave function $R(r)$ satisfies the normalization condition $\int |R(\rho)|^2\rho^5d\rho =1$, and the root-mean-squared radius is defined as
\begin{equation}
r_{rms}^2={m_1m_2{\bf r}_{12}^2+m_2m_3{\bf r}_{23}^2+m_3m_1{\bf r}_{31}^2 \over (m_1+m_2+m_3)^2}={\mu \over m_1+m_2+m_3}({\bf x}_1^2+{\bf x}_2^2).
\end{equation}

With the same parameters used for quarkonia, we solved the 3-quark Schr\"odinger euqation. The prediction on heavy flavor baryon mass and root-mean-squared radius is shown in Table \ref{table3}. 
\begin{table}[!hbt]
	\centering
	\begin{tabular*}{4.5in}{@{\extracolsep{\fill}}llcccccc}
		\hline
		\hline
		State & $\Omega_{ccc}$ & $\Omega_{ccb}$  & $\Omega^*_{ccb}$  & $\Omega_{bbc}$  & $\Omega^*_{bbc}$ & $\Omega_{bbb}$   \\ \hline
		$J^P$ & ${3\over 2}^+$ & ${1\over 2}^+$ & ${3\over 2}^+$ & ${1\over 2}^+$ & ${3\over 2}^+$ & ${3\over 2}^+$  \\
		\hline
		$M_{Th} \text{(GeV)}$ & 4.797 & 8.143 & 8.207 & 10.920 & 10.953 & 14.363  \\
		\hline
		$r_{rms} \text{(fm)}$  & 0.289 & 0.200 & 0.211 & 0.171 & 0.175 & 0.153  \\
		\hline
		\hline
	\end{tabular*}
	\caption{The model calculated mass $M_{Th}$ and root-mean-squared radius $r_{rms}$ for heavy flavor baryon states. }
	\label{table3}
\end{table}

\subsubsection{Relativistic potential model}
\label{potential2}
A nature question we ask ourselves is the relativistic correction to the dynamical evolution of a quarkonium. The correction to a bottomonium is expected to be neglected safely, but for a lighter charmonium state like $J/\psi$, the correction might be remarkable. Let us qualitatively estimate the relativistic effect on the quarkonium potential before a strict calculation. Neglecting the quark spin, the relative part of the Hamiltonian for a pair of heavy quarks can be approximately written as a non-relativistic form,
\begin{equation}
H=\sqrt{\mu^2+p^2}-\mu+V(r)\approx {p^2\over 2\mu}+V_{eff}.
\end{equation}
The effective potential $V_{eff}=V-p^4/(8\mu^3)<V$. Since the relativistic correction leads to a deeper potential well, the quarkonium becomes a more deeply bound state.

A direct way to perturbatively include relativistic corrections order by order is in the frame of non-relativistic quantum mechanics~\cite{Isgur:1979be,Capstick:1986bm,DeRujula:1975qlm}. A problem in this treatment is the spin interaction, it cannot be self-consistently included in non-relativistic systems. On the other hand, if we want to extend the application of the potential model from pure heavy quark hadrons to open heavy flavors including light quarks, the kinematics correction for light quarks cannot be treated as a perturbation. The first covariant treatment of a relativistic bound-state problem is the Bethe-Salpeter equation~\cite{Salpeter:1951sz,Logunov:1963yc,Gross:1969rv}. The covariant wave equation proposed by Sazdjian~\cite{Sazdjian:1986aw,Sazdjian:1988be} provides a way to obtain relativistic bound-state wave functions. It has proved that the interaction potential and the wave function of the bound state are related in a definite way to the kernel and the wave function of the Bethe-Salpeter equation. In the meantime, Crater and Alstine derived the two-body Dirac equation~\cite{Crater:1983ew,Crater:1987hm,Crater:2010fc} from Dirac's constraint mechanics and supersymmetry.

For two relativistic spin-one-half particles interacting through scalar and vector potentials, the Dirac equations for the wave function $\Psi=\{\psi_1,\psi_2,\psi_3,\psi_4\}^T$ of the two particles can be expressed as~\cite{Crater:1983ew}
\begin{eqnarray}
&& \mathcal{S}_1\Psi = \left[\gamma_5\left(\gamma^\mu (p_\mu-A_\mu)+m+S\right)\right]_1\Psi=0,\nonumber\\
&& \mathcal{S}_2\Psi = \left[\gamma_5\left(\gamma^\mu (p_\mu-A_\mu)+m+S\right)\right]_2\Psi=0.
\end{eqnarray}
Clearly,the operators $\mathcal{S}_1$ and $\mathcal{S}_2$ for the two particles are independent and they commute with each other $[\mathcal{S}_1,\mathcal{S}_2]=0$ which results in some restrictions on the relativistic four-vector potential $A_\mu$ and scalar potential $S$. Taking Pauli reduction and scale transformation in center-of-mass frame, the relative motion can be expressed as a four-component relativistic Schr\"odinger-like equation,
\begin{equation}
\left[p^2+\Phi_{12}\right]\psi=b^2\psi
\label{TBDE}
\end{equation}
with $\psi$ being the four-component spinor and $\Phi$ the interaction potential~\cite{Crater:2010fc}. The relativistic corrections to the non-relativistic potential, containing the Darwin term and many spin interaction terms, are self-consistently included in the total potential $\Phi_{ij}$ between two particles labeled by $i$ and $j$,
\begin{eqnarray}
{\Phi}_{ij} &=& 2m_{ij}S+S^2+2\epsilon_{ij}A-A^2+\Phi_D+{\boldsymbol \sigma}_i\cdot{\boldsymbol \sigma}_j\Phi_{SS}\nonumber\\
&&+{\bf L}_{ij}\cdot({\boldsymbol \sigma}_i+{\boldsymbol \sigma}_j)\Phi_{SO}+{\bf L}_{ij}\cdot({\boldsymbol \sigma}_i-{\boldsymbol \sigma}_j)\Phi_{SOD}+i{\bf L}_{ij}\cdot({\boldsymbol \sigma}_i\times{\boldsymbol \sigma}_j)\Phi_{SOX}\nonumber\\
&&+({\boldsymbol \sigma}_i\cdot\hat{\bf r}_{ij})({\boldsymbol \sigma}_j\cdot\hat{\bf r}_{ij}){\bf L}_{ij}\cdot({\boldsymbol \sigma}_i+{\boldsymbol \sigma}_j)\Phi_{SOT}+(3({\boldsymbol \sigma}_i\cdot\hat{\bf r}_{ij})({\boldsymbol \sigma}_j\cdot\hat{\bf r}_{ij})-{\boldsymbol \sigma}_i\cdot{\boldsymbol \sigma}_j)\Phi_T.
\label{rpotential}
\end{eqnarray}
The explicit expressions for the Dawin term $\Phi_D$, spin-spin interaction $\Phi_{SS}$, spin-orbital interactions $\Phi_{SO}, \Phi_{SOD}, \Phi_{SOX}$ and $\Phi_{SOT}$ and tensor interaction $\Phi_T$ can be found in Ref.~\cite{Crater:2010fc}. The non-relativistic central potential between a quark and its antiquark can be separated into two parts, $V(r)=A(r)+S(r)$, where $A$ and $S$ control, respectively, the behavior of the potential at short and long distances. In vacuum, one usually takes the Cornell potential, $A(r)=-\alpha/r$ and $S(r)=\sigma r$. The Coulomb part dominates the wave function around $r=0$, and the linear part leads to the quark confinement.

Separating the radial part of Eq.\ref{TBDE} from the angular part, the radial wave functions of the spin-singlet $u_0$ and one of the spin-triplet $u^0_1$ with quantum numbers $n^{2s+1}l_j = n^1l_l$ and $n^3l_l$ are controlled by two coupled equations, and the other two states $u_1^+$ and $u_1^-$ of the triplet with quantum numbers $n^3l_{l+1}$ and $n^3l_{l-1}$ are controlled by the other two coupled equations~\cite{Liu:2002cn,Shi:2013rga}. By solving the two-body Dirac equation, the light and heavy meson spectra can be successfully described~\cite{Crater:2010fc}.

For baryon systems, Sazdjian has deduced a relativistic and covariant wave equation for three-body bound states~\cite{Sazdjian:1988be}. The baryon wave function $\Psi({\bf r}_1,{\bf r}_2,{\bf r}_3)$ is controlled by the Schr\"odinger-like equation,
\begin{equation}
\left[\sum_{i=1}^{3} {{\bf p}_i^2\over 2\epsilon_i} + \sum_{i<j}^3{\epsilon_i + \epsilon_j\over 2\epsilon_i \epsilon_j}\Phi_{ij}\right]\Psi =E\Psi,
\label{dirac}
\end{equation}
where ${\bf r}_i$ and ${\bf p}_i$ are the quark coordinates and momenta, $E=1/6\sum_{ij}(\epsilon_j^2 - m_j^2)/\epsilon_i$ is the energy eigenvalue related to the effective quark mass $\epsilon_i$ and vacuum quark mass $m_i$, the baryon mass $M_B$ is determined by the coupled equations,
\begin{equation}
\epsilon_i = {M_B\over 3} + {1\over 3}\sum_{j\neq i}{m_i^2-m_j^2\over \epsilon_i+\epsilon_j}.
\label{mass}
\end{equation}
Note that, for the two quark interaction, we still take the short and long range potentials $A$ and $S$ as one half of the corresponding ones in quark-antiquark interaction~\cite{Shi:2019tji}. The coordinate transformation is similar to the one in non-relativistic model by replacing the vacuum mass $m_i$ by the effective mass $\epsilon_i$,
\begin{eqnarray}
{\bf R} &=& {\epsilon_1{\bf r}_1 + \epsilon_2{\bf r}_2 + \epsilon_3{\bf r}_3 \over \epsilon_1+\epsilon_2+\epsilon_3},\nonumber\\
{\bm \rho} &=& \sqrt{{\epsilon_1 \epsilon_2 \over (\epsilon_1+\epsilon_2)\mu}}({\bf r}_1 - {\bf r}_2),\nonumber\\
{\bm \lambda} &=& \sqrt{{\epsilon_3\over \mu (\epsilon_1+\epsilon_2)(\epsilon_1+\epsilon_2+\epsilon_3)}}\left[\epsilon_1({\bf r}_3-{\bf r}_1)+\epsilon_2({\bf r}_3-{\bf r}_2)\right].
\label{jacobi}
\end{eqnarray}
To solve this 3-body Dirac equation, we express the total wave function as the product of the ones in flavor, spin and coordinate spaces and expand the one in coordinate space in terms of two body spherical harmonic oscillators~\cite{Shi:2019tji,Whitney:2011aa},
\begin{equation}
\left| \Psi \right> = \sum_{FSC} C_{FSC} \left| \Psi_{FSC} \right>
\end{equation}
with
\begin{equation}
\left| \Psi_{FSC} \right> = \left| F \right> \times \left| S \right> \times \left| n_\rho l_\rho m_\rho n_\lambda l_\lambda m_\lambda \right>.
\label{hilbert}
\end{equation}

Taking into account the complete and orthogonal conditions for the states $\left| \Psi_{FSC}\right>$, the eigenstate problem of the three-body system, $\hat H\left| \Psi \right> = E \left| \Psi \right>$, becomes a matrix equation for the coefficients $C_{FSC}$,
\begin{equation}
\sum_{F',S',C'} \left< \Psi_{FSC} \right|\hat H \left| \Psi_{F'S'C'} \right> C_{F'S'C'} = E C_{FSC}.
\end{equation}

By solving the two- and three-body Dirac equations numerically, one can systematically study the heavy flavor mesons and baryons in vacuum. Unlike the previous calculations~\cite{Crater:2010fc,Whitney:2011aa} where the parameters in the model, including coupling constants and vacuum quark masses, are taken different values for meson sector and baryon sector, here the parameters are taken the same values for both mesons and baryons. The results are shown in Tables \ref{table4} and \ref{table5}.
\begin{table}[!hbt]
	\centering
	\begin{tabular*}{5.0in}{@{\extracolsep{\fill}}llcccccc}
		\hline
		\hline
		State & $D^0$ & $D^{*0}$& $ D^+ $ & $D^{*+}$ & $D_s $ & $ D_s^* $   \\
		\hline
		$M_{Exp} \text{(GeV)}$ & 1.865 & 2.007 & 1.870 & 2.010 & 1.968 & 2.112\\
		\hline
		$M_{Th}\text{(GeV)}$ & 1.908 & 2.057 & 1.908 & 2.057 & 2.006 & 2.165\\
		\hline
		$r_{rms}\text{(fm)}$  & 0.41 & 0.48 & 0.41 & 0.48 & 0.39 & 0.46 \\
		\hline
		\hline
		State & $B^0$ & $B^{*0}$& $ B^- $ & $B^{*-}$ & $B_s $ & $ B_s^* $   \\
		\hline
		$M_{Exp}\text{(GeV)}$ & 5.280 & 5.325 & 5.279 & 5.325 & 5.367 & 5.415\\
		\hline
		$M_{Th}\text{(GeV)}$ & 5.310 & 5.365 & 5.310 & 5.365 & 5.402 & 5.467\\
		\hline
		$r_{rms}\text{(fm)}$  & 0.44 & 0.47 & 0.44 & 0.47 & 0.41 & 0.44 \\
		\hline
		\hline
	\end{tabular*}
	\caption{The experimentally measured~\cite{Tanabashi:2018oca} and model calculated masses $M_{Exp}$ and $M_{Th}$ and model calculated root-mean-squared radius $r_{rms}$ for heavy flavor mesons.}
	\label{table4}
\end{table}

\begin{table}[!hbt]
	\centering
	\begin{tabular*}{5.0in}{@{\extracolsep{\fill}}llcccccc}
		\hline
		\hline
		State & $\Lambda_c^+$ & $\Sigma_c^+$& $ \Xi_c^+ $ & $\Omega_c^0$ & $\Xi_{cc}^+ $ & $ \Omega_{cc}^+ $   \\
		\hline
		$M_{Exp}\text{(GeV)}$ & 2.286 & 2.453 & 2.468 & 2.695 & 3.619 & -\\
		\hline
		$M_{Th}\text{(GeV)}$ & 2.383 & 2.356 & 2.517 & 2.660 & 3.616 & 3.746\\
		\hline
		$r_{rms}\text{(fm)}$  & 0.29 & 0.29 & 0.29 & 0.29 & 0.28 & 0.27 \\
		\hline
		\hline
		State & $\Lambda_b^+$ & $\Sigma_b^+$& $ \Xi_b^- $ & $\Omega_b^-$ & $\Xi_{bb}^- $ & $ \Omega_{bb}^- $  \\
		\hline
		$M_{Exp}\text{(GeV)}$ & 5.620 & 5.811 & 5.795 & 6.046 & - & - \\
		\hline
		$M_{Th}\text{(GeV)}$ & 5.744 & 5.720 & 5.871 & 6.007 & 10.195 & 10.318\\
		\hline
		\hline
	\end{tabular*}
	\caption{The experimentally measured~\cite{Tanabashi:2018oca} and model calculated masses $M_{Exp}$ and $M_{Th}$ and root-mean-squared radius $r_{rms}$ for heavy flavor baryons.}
	\label{table5}
\end{table}

\subsubsection{Effective field theory}
\label{eft}
While the potential models have made some success in explaining the properties of heavy flavor hadrons in vacuum, we should keep in mind the condition to apply the models. Their connection with the QCD parameters is not transparent, the scale at which they are defined is not clear, and they cannot be systematically improved. From perturbative QCD, the potential models are valid only up to $O(\alpha_s^2)$. The question is that to which extent the potential picture is applicable. To answer this question, it is necessary to develop a formalism where the uncertainty produced by using a Schr\"odinger equation with a potential obtained from QCD instead of doing the computation in full QCD can be made quantitatively. 

For heavy quarks with large mass, the velocity $v$ is believed to be a small quantity, $v \ll1$. Therefore, a non-relativistic picture holds. This produces a hierarchy of scales: $m_Q\gg m_Qv\gg m_Qv^2$ for a heavy flavor system~\cite{Bodwin:1994jh,Caswell:1985ui}. The inverse of the soft scale, $m_Qv$, gives the size of the bound state, and the inverse of the ultrasoft scale, $m_Qv^2$ (usually $v_c^2\approx 0.3$ and $v_b^2\approx 0.1$), gives the typical time scale. In QCD another physically relevant scale need to be considered is the scale $\Lambda_{QCD}$ at which non-perturbative effects become important. Note that, the heavy quark mass $m_Q$ is also much larger than $\Lambda_{QCD}$. Since the hard ($m_Q$), soft ($m_Qv$) and ultrasoft ($m_Qv^2$) scales are clearly separated, two effective field theories can be introduced by sequentially integrating out $m_Q$ and $m_Qv$. One is the non-relativistic QCD (NRQCD)~\cite{Caswell:1985ui} by integrating out the hard scale $m_Q$, and the other is the potential NRQCD (pNRQCD)~\cite{Brambilla:1999xf} by further integrating out the soft scale $m_Qv$. This sequence of effective field theories has the advantage that it allows disentangling of perturbative contributions from nonperturbative ones to a large extent. 
 
The idea of NRQCD is to separate the scale $m_Q$ from the scales $m_Qv$, $m_Qv^2$ and $\Lambda_{QCD}$ by integrating out the degrees of freedom of momenta $m_Q$. These degrees of freedom include relativistic heavy quarks, light quarks, and gluons with momenta larger than $m_Q$. These degrees of freedom can be compensated by a set of new local operators and their coefficients in the Lagrangian which can be perturbatively matched to QCD. In NRQCD, heavy quarks, instead of being represented by a bispinor field, are represented by two spinor fields, one for the heavy quark and the other for the heavy anti-quark. The Lagrangian density of NRQCD can be expressed as ${\mathcal L}_{NRQCD}={\mathcal L}_{g}+{\mathcal L}_l+{\mathcal L}_\psi+{\mathcal L}_\chi+{\mathcal L}_{\psi \chi}$, where ${\mathcal L}_g$ and ${\mathcal L}_l$ are respectively from gluons and light flavors, ${\mathcal L}_\psi$ and ${\mathcal L}_\chi$ are from heavy quarks and anti-quarks, and ${\mathcal L}_{\psi \chi}$ is from the additional color singlet and color octet four-fermion interaction terms~\cite{Brambilla:1999xf}. The NRQCD approach has been widely applied in the phenomenological studies of quarkonium spectrum and inclusive decay widths in proton-proton collisions~\cite{Chao:2012iv,Butenschoen:2011yh,Butenschoen:2012qr,Han:2014kxa,Bain:2017wvk}. 
 
 If we concern only the quarkonium binding properties, we can further integrate out the scale $m_Qv$ from NRQCD to get the potential NRQCD. In pNRQCD, aiming to establish a power counting, it is more convenient to represent the quark-antiquark pair by a wave-function field $\Psi({\bf x}_1, {\bf x}_2, t)_{\alpha\beta}\equiv \psi_\alpha({\bf x}_1,t) \chi^\dag_\beta({\bf x}_2, t)$. This wave-function field can be uniquely decomposed into the singlet-field and octet-field components $S$ and $O$. In this case, the degrees of freedom in pNRQCD are the singlet and octet fields composed of heavy quark and anti-quark interacting with ultra-soft gluons.  The Lagrangian density of pNRQCD up to order $p^3/m_Q^2$ can be expressed as~\cite{Brambilla:2004jw,Brambilla:1999xf}
\begin{eqnarray}
{\mathcal L}_{pNRQCD}&=&\int d^3r {\mathrm{Tr}} \Big [S^\dag (i\partial_0-H_S)S + O^\dag (i\partial_0 -H_O)O  \Big ]  \nonumber \\
&+& V_A(r){\mathrm{Tr}}[O^\dag {\bf r}\cdot g{\bf E}S+ S^\dag {\bf r}\cdot g{\bf E}O]+{V_B(r)\over 2}{\mathrm{Tr}}[O^\dag {\bf r}\cdot g{\bf E}O+ O^\dag O{\bf r}\cdot g{\bf E}]+{\mathcal L}'_g+{\mathcal L}'_l,
\label{pnrqcd}
\end{eqnarray}
with
\begin{eqnarray}
H_S&=& \{ c_1^s(r), {{\bf p}^2 \over 2\mu} \}+c_2^s(r){{\bf P}^2 \over 2M}+V_S^{(0)}+{V_S^{(1)}\over m_Q}+{V_S^{(2)}\over m_Q^2}, \nonumber \\
H_O&=& \{ c_1^o(r), {{\bf p}^2 \over 2\mu} \}+c_2^o(r){{\bf P}^2 \over 2M}+V_O^{(0)}+{V_O^{(1)}\over m_Q}+{V_O^{(2)}\over m_Q^2},
\label{pnrqcd}
\end{eqnarray}
where we have taken $m_1=m_2=m_Q$, $\mu=m_Q/2$ is the reduced mass, $M=2m_Q$ is the total mass, ${\bf p}$ is the relative momentum, ${\bf P}$ is the center of mass momentum, and ${\bf E}$ represents the chromoelectric field. ${\mathcal L}'_g$ and ${\mathcal L}'_l$ describe the contributions from gluons and light quarks with momenta $\lesssim m_Qv$. The singlet and octet potentials $V_S$ and $V_O$ appear as parameters of the effective field theory and can be defined at any order in perturbation theory (for $m_Qv\gg\Lambda_{QCD}$) by matching pNRQCD to NRQCD at the scale $m_Qv$. The ultrasoft gluons contribute to dipole-like transitions between the color singlet and octet states ($V_A$ term) and within the color octet states ($V_B$ term). The static and the $1/m_Q$ potentials $V_{S(O)}^{(1)}$ are real-valued functions depending only on $r$. The $1/m_Q^2$ potentials $V_{S(O)}^{(2)}$ have imaginary parts proportional to $\delta^{(3)}({\bf r})$ and real parts that can be decomposed into spin-independent and spin-dependent components~\cite{Brambilla:2004jw}. The imaginary parts come from the matching coefficients of the four-fermion operators in NRQCD. The high order potentials can be treated as relativistic corrections in potential model. At leading order, there are the matching coefficients $c_1^s$=$c_2^s$=$c_1^o$=$c_2^o$=$V_A$=$V_B$=1, $V_S^{(0)}=-C_F\alpha_s/r$ and $V_O^{(0)}=(1/2N_c)\alpha_s/r$. That's what we are  familiar with. 

From the first line of the Lagrangian ${\mathcal L}_{pNRQCD}$, the evolution of the singlet and octet wavefunctions are governed by potentials. However, the existence of dipole-like interactions makes the singlet and octet quarkonium states coupled to each other and can not be evolved separately with a simple Schr\"odinger equation. Since pNRQCD has potential terms, it embraces potential models. The pNRQCD provides a new interpretation of the potentials that appear in the Schr\"odinger equation in terms of a modern effective field theory Language.
 
According to the relative size of $\Lambda_{QCD}$ compared to the scales $p(\sim m_Qv)$ and $E(\sim m_Qv^2)$, the pNRQCD can be divided into weak coupling regime and strong coupling region. For $p\gg \Lambda_{QCD}$, the integration of degrees of freedom of energy scale $p$ can be done in perturbation theory. Hence we do not expect a qualitative change in the degrees of freedom but only a lowering of their energy cutoff. For $p\gg \Lambda_{QCD}\gg E$, it is better to think in terms of hadronic degrees of freedom below the scale of $\Lambda_{QCD}$. If one switches off the light fermions (Goldstone boson fields), the only degree of freedom left is the singlet field interacting with a potential, and the pNRQCD is reduced to a pure two-particle nonrelativistic quantum-mechanical system~\cite{Brambilla:2004jw,Brambilla:1999xf}. The pNRQCD can also be used to study the spectrum of heavy flavor hadrons and decay widths~\cite{Kniehl:2002br,Brambilla:2001qk,LlanesEstrada:2011kc,Kniehl:2002yv}.

The non-perturbative approach is needed in strong coupling region and for the study of mesons with one light quark. First principle calculations in lattice QCD give a good description of the nonperturbative behavior of heavy quarks and quarkonia. There has been tremendous progress in lattice QCD calculations of heavy flavor hadrons, including quarkonia at zero temperature~\cite{Donald:2012ga,Basak:2013oya} and finite temperature~\cite{Datta:2003ww,Asakawa:2003re,Umeda:2002vr,Ding:2012sp,Mocsy:2013syh,Ohno:2011zc}. The results from lattice QCD are in good agreement with experiment data and can explain the hyperfine splitting of baryon states at zero temperature. At finite temperature, the hot medium will change the quarkonium properties, not only a shift of the peak position but also an increase of the width. The lattice studies of heavy flavor hadrons include also open heavy flavor mesons and doubly and triply charmed baryons~\cite{Basak:2013oya,Bowler:1996ws,Namekawa:2013vu,Mathur:2018rwu,Padmanath:2019ybu,Mathur:2018epb}. For a pointlike meson, the meson operator can be expressed as ${\mathcal O}({\bf x},t)=\bar q({\bf x},t)\Gamma q({\bf x},t)$, while for extended meson, as used in~\cite{Larsen:2019bwy}, the operator becomes ${\mathcal O}_i({\bf x},t)=\sum_{\bf r}\psi_i({\bf r})\bar q({\bf x+r},t)\Gamma q({\bf x},t)$. The vertex operators $\Gamma=1, \gamma_5$, $\gamma_i$, $\gamma_5 \gamma_i\ (i = 1,2,3)$ correspond to scalar, pseudoscalar, vector, and axial-vector channels, respectively. This allows us to select the spin and angular momentum properties of the fluctuations contributing to the correlation function. The Euclidean meson correlation function is defined as
\begin{eqnarray}
C({\bf x}, t)=\langle {\mathcal O}({\bf x}, t){\mathcal O}^\dag(0,0) \rangle.
\end{eqnarray}
The correlation function in momentum space $C({\bf p}, t)$ can be obtained via Fourier transformation. The meson spectral function which can be extracted from the quarkonium correlators is expressed as 
\begin{eqnarray}
C({\bf p}, t)=\int_0^\infty d\omega \rho({\bf p}, \omega) K(\omega, t) 
\end{eqnarray}
with the integration kernel $K(\omega, t)$. A heavy flavor bound state appears as a peak in the spectrum function which allows us to read-off the mass, binding energy and lifetime. This framework can be extended to finite temperature to study the medium effect.

The effective field theory NRQCD provides an alternative way to study heavy quarkonia on the lattice~\cite{Lepage:1992tx}. Lattice NRQCD has been successfully used for precise spectroscopy at zero temperature~\cite{Mathur:2018epb,Meinel:2010pv,Dowdall:2011wh,Mathur:2016hsm} and finite temperature~\cite{Larsen:2019bwy,Aarts:2012ka,Aarts:2010ek,Kim:2018yhk}. In lattice NRQCD, the evolution of the heavy quarks is separated from the QCD medium, we just need to populate the spacetime grid with light degrees of freedom (gluons and light quarks). That reduces the computational cost of the Euclidean heavy quark propagator. In the meantime, the absence of a transport peak contribution simplifies the extraction of the spectra from the Euclidean correlation function~\cite{Rothkopf:2019ipj}. 

\subsection{Cold nuclear matter effects}
\label{cold}
The cold nuclear matter effects are intrinsic to heavy ion interactions. While people usually focus on the hot nuclear matter effects which are the necessary condition to produce QGP, the cold nuclear matter effects characterize the initial condition of the hot and dense fireball. The baseline for open and closed heavy flavor production and suppression in heavy ion collisions should be determined from the studies on cold nuclear matter effects. On the other hand, the experimental and theoretical studies on the cold nuclear matter effects present a way to understand the parton distributions in nuclei, especially at low momentum. Since the cold nuclear matter is a many-body system with strong interaction, there is at the moment no first principle way to include all the cold nuclear matter effects, and the current study depends on effective models. There are several cold nuclear matter effects on heavy flavor hadrons: modification of parton distribution functions in nuclear matter compared to that in a free nucleon (shadowing), parton multiple scattering in nuclear matter before the hadron formation (Cronin effect), and absorption of hadrons in nuclear matter after their formation (nuclear absorption).

\subsubsection{Shadowing effect}
\label{shadowing}
\begin{figure}[!htb]
	{$$\includegraphics[width=0.75\textwidth]{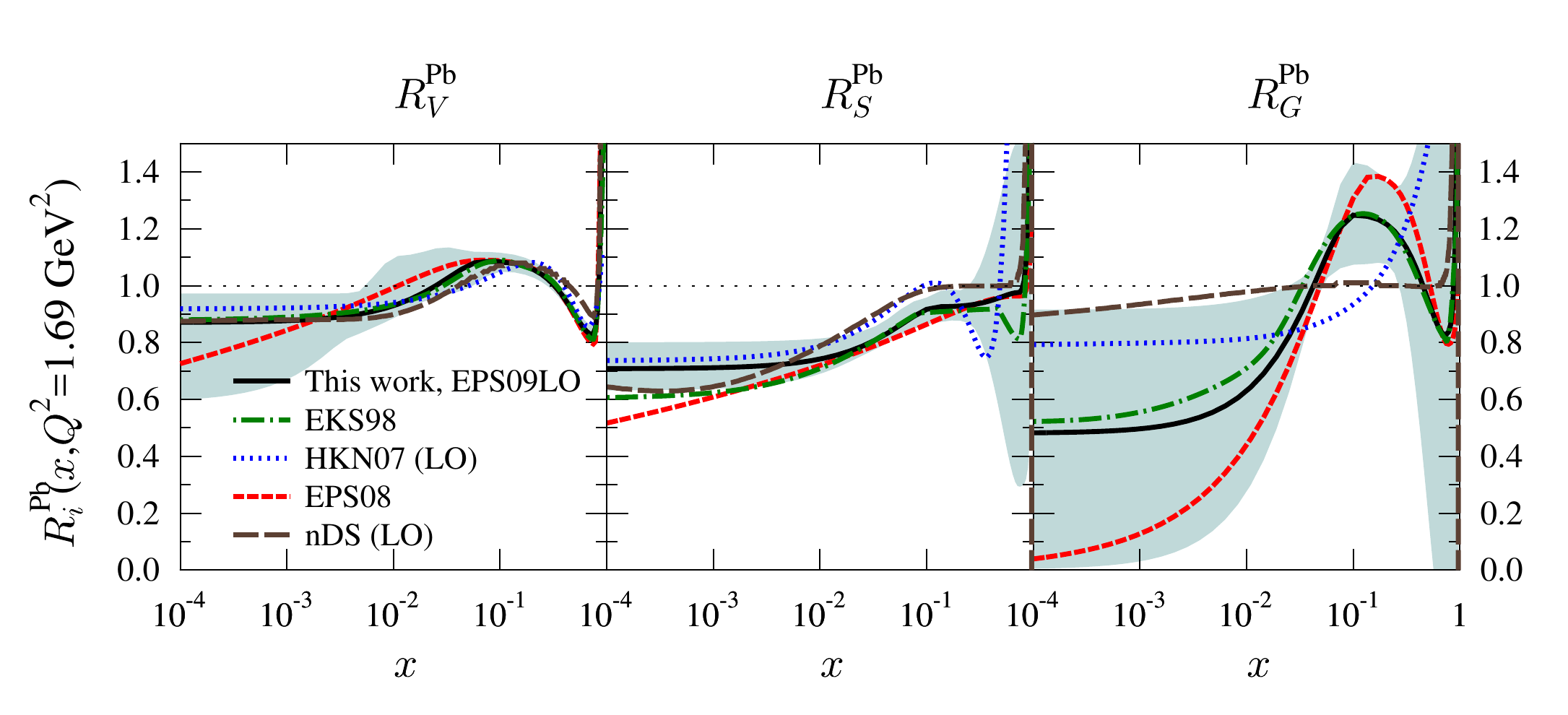}$$
		\caption{The shadowing modification factors for averaged valence quarks, sea quarks and gluons at $Q^2=1.69$ GeV$^2$, calculated with different models. The figure is taken from Ref.~\cite{Eskola:2009uj}.}
		\label{fig3}}
\end{figure}

The distribution function $f_i^A(x,Q^2)$ for parton $i$ in a nucleus differs from a simple superposition of the distribution function $f_i(x,\mu_F)$ in a free nucleon. The nuclear shadowing effect is described by the modification factor,
\begin{equation}
R_i^A(x,Q^2)={f_i^A(x,Q^2)\over Af_i(x,Q^2)},\ \ \ \ \ \ \  i= q,\bar q, g
\end{equation}
where $x$ and $Q^2$ are the parton longitudinal momentum fraction and transverse momentum scale. There are different models to parameterize the nuclear shadowing function $R_i^A$. The modification factors for valence quarks, sea quarks, and gluons calculated in different models are shown in Fig.\ref{fig3}. The KHN07~\cite{Hirai:2007sx} and nDC/nDSg~\cite{deFlorian:2003qf} indicate little shadowing, the EKS98~\cite{Eskola:1998df} and EKS09~\cite{Eskola:2009uj} suggest moderate shadowing effect, while the EPS08~\cite{Eskola:2008ca} gives large shadowing.

We take in the following numerical calculations EKS98 to describe the nuclear shadowing, it gives almost the average value of the other models. In the frame of EKS98, the shadowing function at the initial scale $Q_0^2=2.25$ GeV$^2$ is parameterized in accordance with the $e+p$ experimental data, and then by solving the DGLAP equation which characterizes the parton distribution equation, the $x$ and $\mu_F$ dependence of the shadowing is obtained. The nuclear effect depends strongly on the parton momentum fraction $x$. In small $x$ limit ($x<0.025$), $R_i^A <1$ means a shadowing effect, but at intermediate $x$ ($0.025<x<0.3$), $R_i^A$ indicates an anti-shadowing effect. In large $x$ limit, there is again $R_i^A<1$ indued by the EMC effect~\cite{Norton:2003cb} at $0.3<x<0.8$ and $R_i^A>1$ due to the Fermi motion at $x>0.8$.

We now discuss how the shadowing affects the quarkonium production in nuclear collisions. At RHIC and LHC energies, the gluon fusion $g+g\to (Q\bar Q)+g$ is the main source to create a $Q\bar Q$ pair. Assuming that the emitted gluon in the process is soft in comparison with the initial gluons and the produced quarkonium and can be neglected in kinematics, corresponding to the picture of color evaporation model at leading order, the longitudinal momentum fractions of the two initial gluons are calculated from momentum conservation,
\begin{equation}
x_{1,2}={\sqrt{m_\psi^2+p_T^2}\over \sqrt{s_{NN}}}e^{\pm y},
\end{equation}
where $y$ is the quarkonium rapidity. In central rapidity region around $y=0$, the two gluons have the same $x=x_1=x_2$. For charmonia in the transverse momentum region $0<p_T<5$ GeV/c, one has $0.18<x<0.34$ at SPS energy $\sqrt{s_{NN}}=17.3$ GeV, $0.016<x<0.029$ at RHIC energy $\sqrt{s_{NN}}=200$ GeV and $0.0011<x<0.0021$ at LHC energy $\sqrt{s_{NN}}=2.76$ TeV. This means that the cold nuclear matter effect is reflected as anti-shadowing at SPS, weak shadowing (anti-shadowing) at RHIC and strong shadowing at LHC.

To account for the spatial dependence of the shadowing in a finite nucleus, one assumes that the inhomogeneous shadowing is proportional to the parton path length through the nucleus~\cite{Klein:2003dj}, which amounts to considering the coherent interaction of the incident parton with all the target partons along its path length. Therefore, one replaces the homogeneous modification factor $R_i(x,Q^2)$ by an inhomogeneous one ${\mathcal R}_i(x,Q^2, x_T)$,
\begin{eqnarray}
{\mathcal R}=1+A(R_i-1)T_A({\bf x}_T)/T_{AB}(0),
\end{eqnarray}
where ${\bf x}_T$ is the transverse position of the colliding tube, $T_{AB}(b)=\int d^2{\bf x}_T T_A({\bf x}_T)T_B({\bf x}_T-{\bf b})$ is determined by the thickness functions $T_A({\bf x}_T)$ and $T_B({\bf x}_T-{\bf b})$ controlled by the nuclear geometry, and ${\bf b}$ is the impact parameter.

Replacing the free gluon distribution $f_g$ by the modified distribution $\bar f_g =Af_g {\mathcal R}_g$, we get the shadowing effect on the quarkonium distribution in $A+B$ collisions,
\begin{eqnarray}
f({\bf x}, {\bf p},\tau_0|{\bf b})&=&{(2\pi)^3\over E_T\tau_0}\int dz_Adz_B \rho_A({\bf x}_T, z_A)\rho_B({\bf x}_T, z_B)\nonumber\\
&&\times {\mathcal R}_g(x_1,\mu_F, {\bf x}_T){\mathcal R}_g(x_2,\mu_F, {\bf x}_T-{\bf b})\bar f_{pp}({\bf x}, {\bf p}, z_A, z_B|{\bf b}),
\label{shadowing}
\end{eqnarray}
where $E_T$ is the quarkonium transverse energy, $\tau_0$ is the formation time of the QGP, $z_A$ and $z_B$ are the longitudinal coordinates of the two colliding nucleons, and $\rho_A$ and $\rho_B$ are the nucleon distribution functions in nucleus A and B. The Cronin effect on the transverse momentum distribution is included in the effective quarkonium distribution $\bar f_{pp}$ in $p+p$ collisions.

\subsubsection{Cronin effect}
\label{croin}
\begin{figure}[!htb]
	{$$\includegraphics[width=0.3\textwidth]{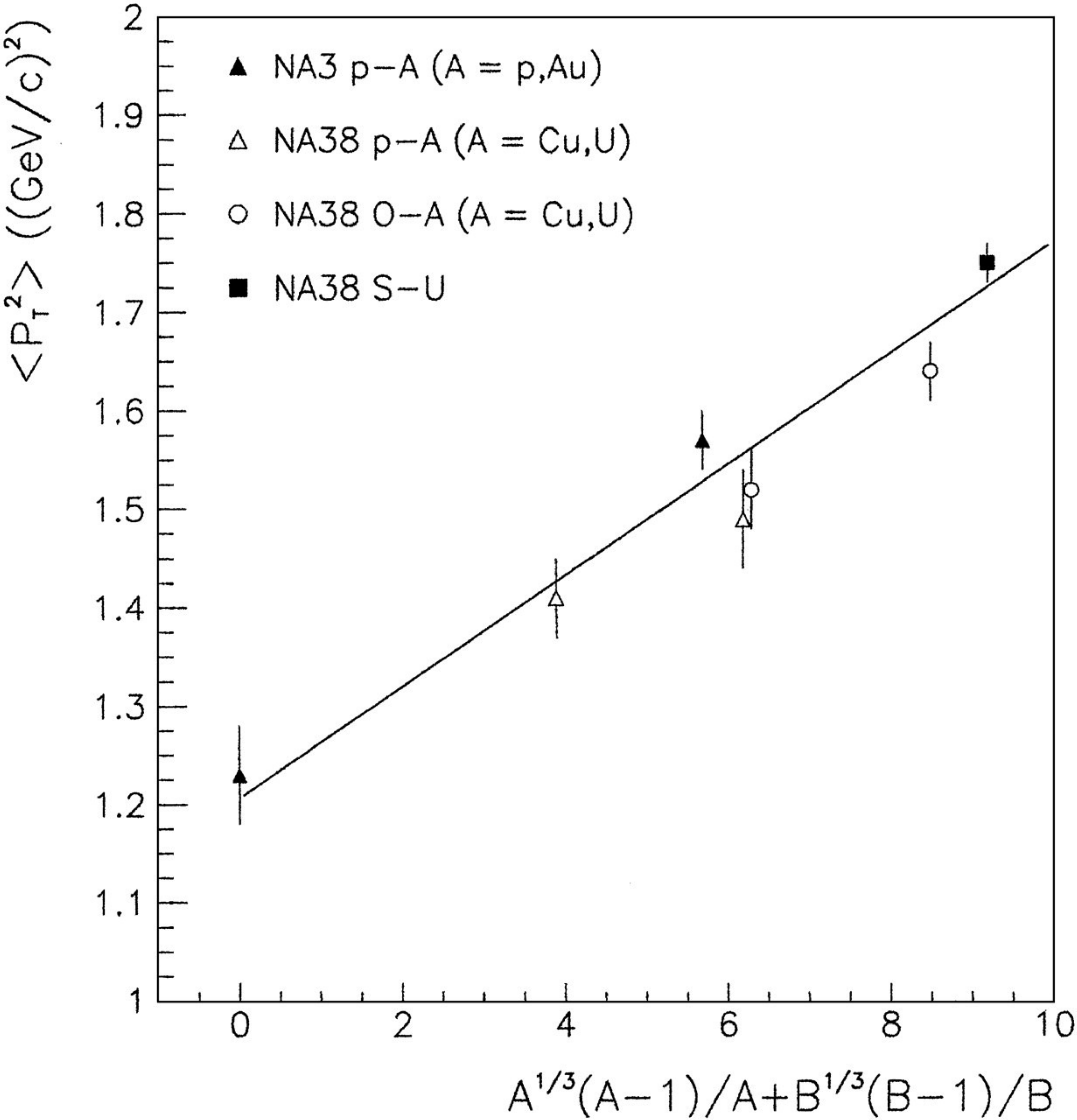}$$
		\caption{The $J/\psi$ averaged transverse momentum square as a function of the size of the colliding system in $A+B$ collisions at SPS energy. The factor $(A-1)/A$ and $(B-1)/B$ are used to remove the nuclear matter effect in the projectile and target when it is a proton. The figure is taken from Ref.~\cite{Gerschel:1998zi}.}
		\label{fig4}}
\end{figure}

We now discuss parton multiple scatterings before the $Q\bar Q$ formation. This is the Cronin effect and leads to a quarkonium transverse momentum broadening in nuclear collisions. Let us consider the main $Q\bar Q$ production channel, the gluon fusion, in $A+B$ collisions. Before two gluons fuse into a quarkonium, they acquire additional transverse momentum via multiple scattering with the surrounding nucleons in nuclei $A$ and $B$, and this extra momentum would be inherited by the produced quarkonium. Inspired from a random-walk picture, one obtains the averaged transverse momentum square of the produced quarkonia,
\begin{equation}
\langle p_T^2\rangle_{AB} = \langle p_T^2\rangle_{pp}+a_{gN} l,
\end{equation}
where the Cronin parameter $a_{gN}$ is the averaged quarkonium transverse momentum square obtained from the gluon scattering with a unit of length of nucleons, usually extracted from corresponding $p + A$ collisions where the produced quarkonia suffer from only the cold nuclear matter effect, and $l$ is the mean trajectory length of the two gluons in the projectile and target nuclei before $Q\bar Q$ formation, determined by the nuclear geometry ($l\approx r_0 (A^{1/3}+B^{1/3})$). The experimentally measured slope shown in Fig.\ref{fig4} for nuclear collisions at SPS energy can be well described by the Cronin effect with the parameter $a_{gN}=0.1$ GeV$^2$/fm.

With the averaged value $\langle p_T^2\rangle_{AB}$, we take a Gaussian smearing for the modified transverse momentum distribution in effective $p+p$ collisions,
\begin{equation}
\bar f_{pp}={1\over \pi a_{gN}l}\int d^2 {\bf p'}_T \exp(-{{\bf p'}_T^2 \over a_{gN}l})f_{pp}(|{\bf p}_T-{\bf p'}_T|,p_z),
\end{equation}
where $f_{pp}$ is the quarkonium momentum distribution in free $p+p$ collisions without Cronin effect. Considering the absence of $p + A$ data at LHC energy, we take $a_{gN}=0.15$ GeV$^2$/fm from empirical estimations~\cite{Vogt:2001nh}.

\subsubsection{Nuclear absorption}
\label{absorb}
Even without any QGP formation, the quarkonia produced inside a nuclear environment will show a suppression, due to the nuclear absorption. Suppose a quarkonium is produced inside the projectile or target nucleus. On its way out, the quarkonium has inelastic interaction with the surrounding nucleons and suffers from a suppression. The quarkonium surviving probability in $A+B$ collisions can be expressed as
\begin{eqnarray}
S_{AB}&=&{1\over AB}\int d^2{\bf s} dz_A dz_B\rho_A({\bf b},z_A) \rho_B({\bf s-b},z_B)\nonumber\\
&&\times \exp\left[-\sigma_{abs}\left(\int_{z_A}^\infty dz' \rho_A({\bf b},z')+\int_{-\infty}^{z_B} dz' \rho_B({\bf s-b},z')\right)\right],
\label{absAB}
\end{eqnarray}
where $\rho_A$ and $\rho_B$ are the nuclear density profiles. The value of the absorption cross section (the dissociation cross section of quarkonium with nucleons) $\sigma_{abs}$ is fixed by fitting experimental data.

\begin{figure}[!htb]
	{$$\includegraphics[width=0.33\textwidth]{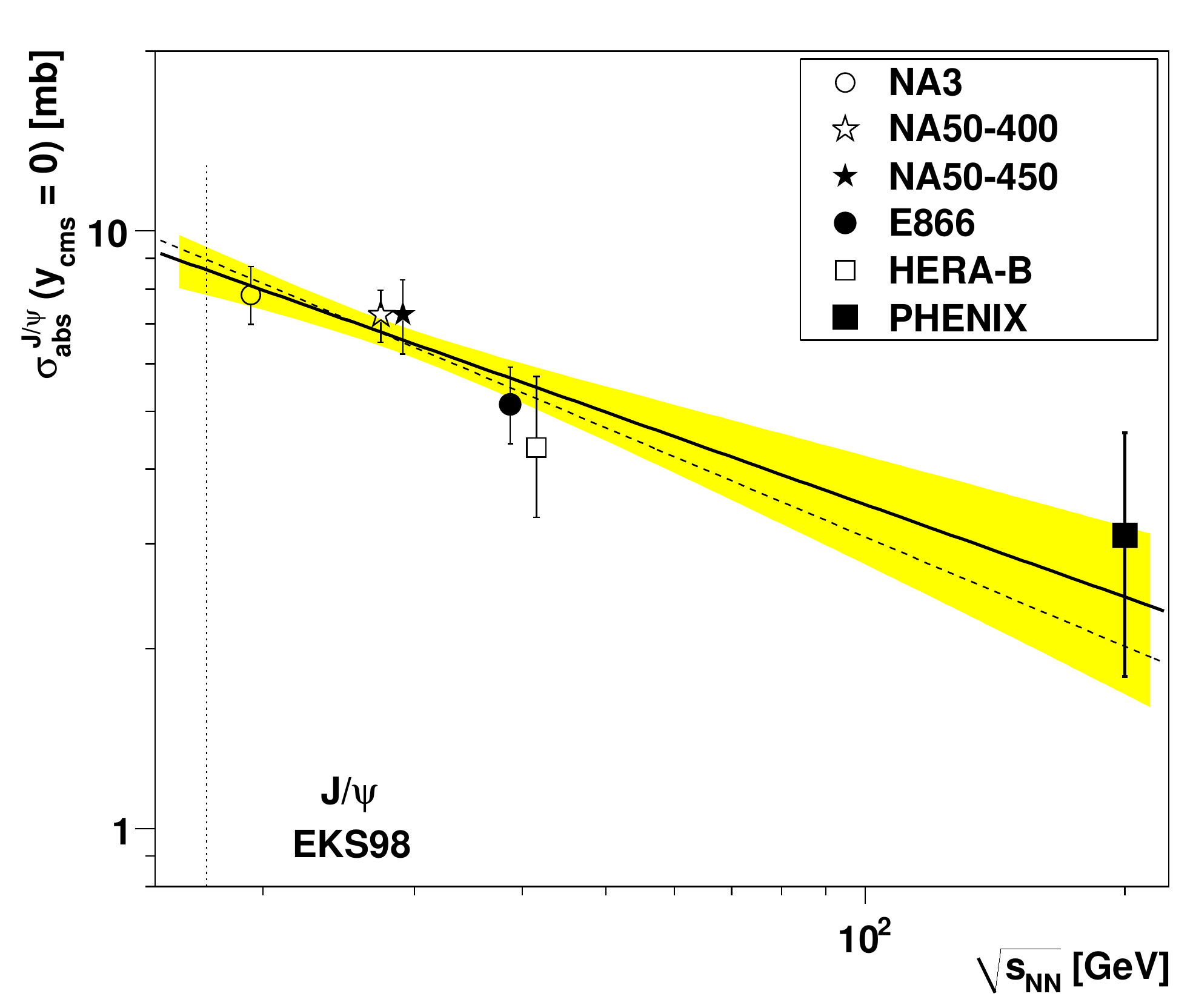} \ ~ \ ~ \ ~ \includegraphics[width=0.28\textwidth]{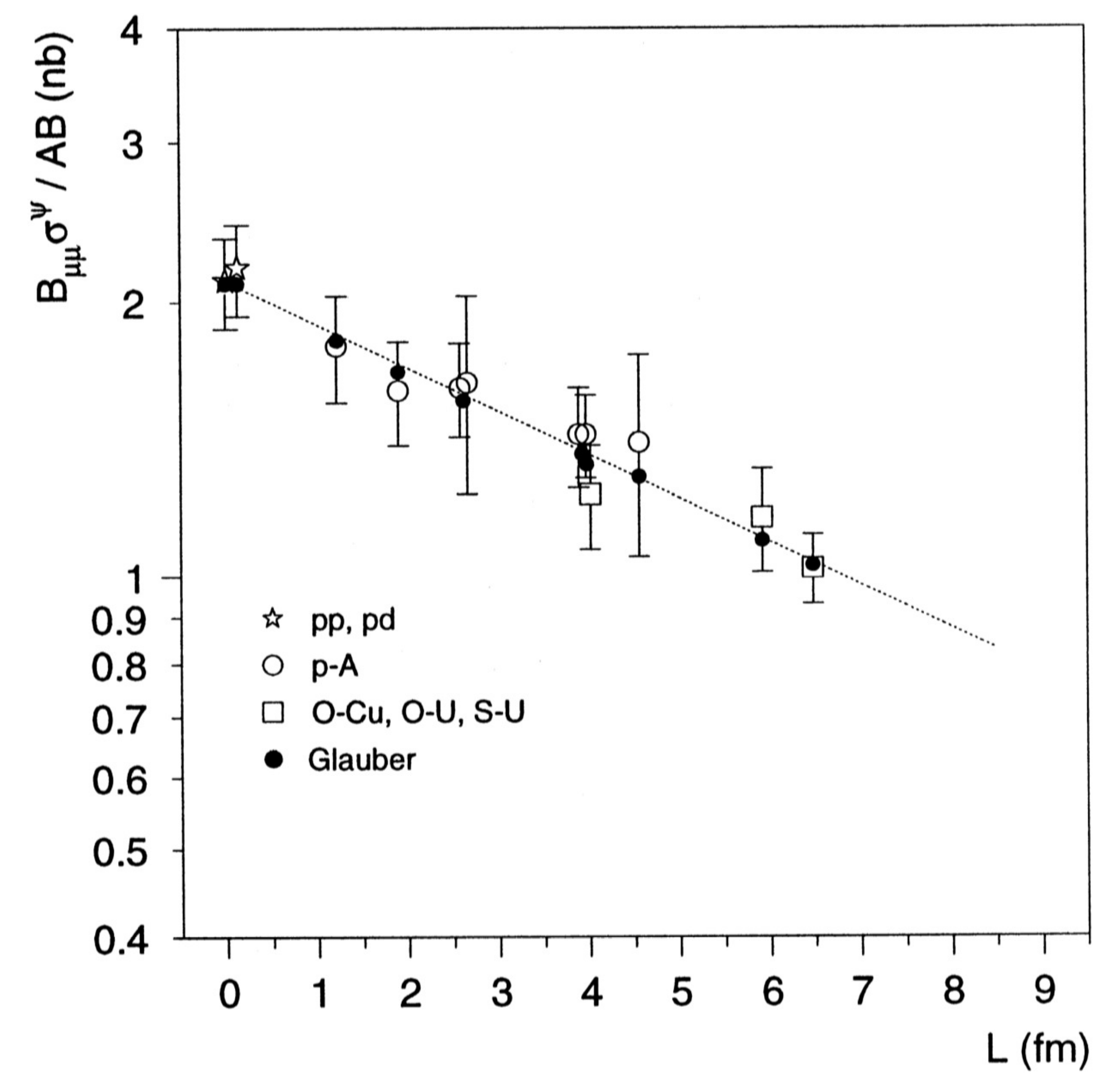}$$
		\caption{The energy dependence of $J/\psi$ nuclear absorption cross section $\sigma_{abs}$ at mid-rapidity (left panel) and the cross section per nucleon-nucleon collision as a function of $c\bar c$ traveling length in the nuclear matter (right panel). The figures are taken from Refs.~\cite{Brambilla:2010cs,Abreu:1999nn}.}
		\label{fig5}}
\end{figure}

The quarkonium formation time is neglected in Eq.\ref{absAB}. Considering a nonzero quarkonium formation time $\tau_f$, the absorption cross section $\sigma_{abs}$ is for a pre-meson, not a fully developed quarkonium, and should be time dependent. Since the time scale for a $Q\bar Q$ pair production, $1/(2m_Q)$, is very short, the quarkonium formation time can be considered as the time scale from a colored or color-neutral $Q\bar Q$ pair to a well developed quarkonium. If the pair is initially produced as a small color-singlet, the absorption cross section is small at the initial time, increases with the proper time, and finally reaches the maximum value at the formation time. From the kinematics, one can estimate the formation time $\tau_f\sim 0.5$ fm/c, and the time dependence of the cross section can be expressed as $\sigma_{abs}(\tau)\sim (\tau/ \tau_f)^2$. At high colliding energies, the collision time for the two colliding nuclei to pass through each other is very short, so the quarkonium states will experience negligible nuclear absorption. This can be seen clearly from the energy dependence of the $J/\psi$ absorption cross section at central rapidity~\cite{Lourenco:2008sk,Brambilla:2010cs} shown in Fig.\ref{fig5}. Considering the larger size of the excited states, the absorption cross section for the excited states should be larger than that for the ground state.

If the produced pair is in octet state, however, it will immediately interact with the nuclear matter with a large cross section, since it is a colored object. In this case, it has often been assumed that all precursor quarkonium states will interact with the same cross section. From the comparison with the SPS data~\cite{Abreu:1999nn}, see Fig.\ref{fig5}, the averaged nuclear absorption cross section at SPS energy is extracted to be $\sigma_{abs}=6.5\pm 1.0$ mb for both $J/\psi$ and $\psi'$. This, however, looks in contrast with the early photon production data where the inelastic $J/\psi$+nucleon cross section ($3.5\pm 0.8$ mb) is significantly smaller, and the inelastic cross section for $\psi'$ is almost four times the value for $J/\psi$. This indicates that, the $c\bar c$ states suffering from nuclear absorption have already (at least partially) evolved into their final states. The relation between the total surviving probability of $J/\psi$ and its averaged traveling length $L$ in the nuclear matter  can be fitted well by a simple form,
\begin{equation}
S_{AB}=e^{-L\rho_0\sigma_{abs}},
\end{equation}
where $\rho_0\approx 0.17fm^{-3}$ is the nucleon density in the center of a heavy nucleus. This relation can be derived from Eq.\ref{absAB} when $\sigma_{abs}$ is very small.

The attenuation of the $Q\bar Q$ dipole during the formation time may offer an alternative explanation to the observed nuclear absorption. The color exchange interaction of the $Q\bar Q$  dipole with the medium leads to a break-up of the colorless dipole~\cite{Kopeliovich:2012be}.

\subsection{Hot nuclear matter effects}
\label{hot}
We now turn to the hot nuclear matter effects which are directly related to the creation of the QGP in high energy nuclear collisions. We first discuss Debye screening and collisional dissociation which change the initially produced quarkonium surviving probability in the QGP, then consider the quarkonium regeneration in the QGP which offers the second source for the quarkonium production, and finally estimate the heavy quark regeneration in hot medium.

\subsubsection{Debye screening}
\label{debye}
Hadronic matter undergoes a transition to a deconfined plasma phase of quarks and gluons at temperature $T_c\sim160$ MeV and zero baryon chemical potential $\mu_B=0$. In the deconfined phase, the property of a quarkonium bound state differs significantly from that in vacuum. The static potential $V(r)=-\alpha/r+\sigma r$ between $Q$ and $\bar Q$ in vacuum consists of a one-gluon exchange part and a confined part. At finite temperature, the potential is screened in analogy to Debye screening in a colored electromagnetic plasma.

Suppose we put a pair of $Q\bar Q$ in a soup of light quarks and gluons, the string tension $\sigma(T)$ which controls the long-range force is strongly reduced by the hot medium and approaches zero $\sigma(T)\to0$ for $T>T_c$. On the other hand, the $Q\bar Q$ pair will change the original charge distribution. The charge rearrangement leads to the Debye screening: The charge density around $Q$ seen by $\bar Q$ decreases and the Coulomb potential becomes the Yukawa potential,
\begin{eqnarray}
-{\alpha \over r}\to -{\alpha \over r}e^{-r/r_D},
\end{eqnarray}
where $r_D$ is the Debye screening length. When the screening length is shorter than the distance between $Q$ and $\bar Q$, $\bar Q$ can not see $Q$, and the bound state disappears. This is the picture of Debye screening.
From the Abelian approximation and pQCD calculation with colored gluons to the lowest order, the screening length is inversely proportional to the temperature of the QGP,
\begin{equation}
r_D=\left\{
\begin{array}{rcl}
\sqrt{6\over g_qe_q^2}{1\over T},       &      & \text{Abelian approximation}\\
{1\over \sqrt{({N_c\over 3}+{N_f \over 6})g^2}}{1\over T},       &      & \text{pQCD}
\end{array} \right.
\end{equation}
\begin{figure}[!htb]
	{$$\includegraphics[width=0.4\textwidth]{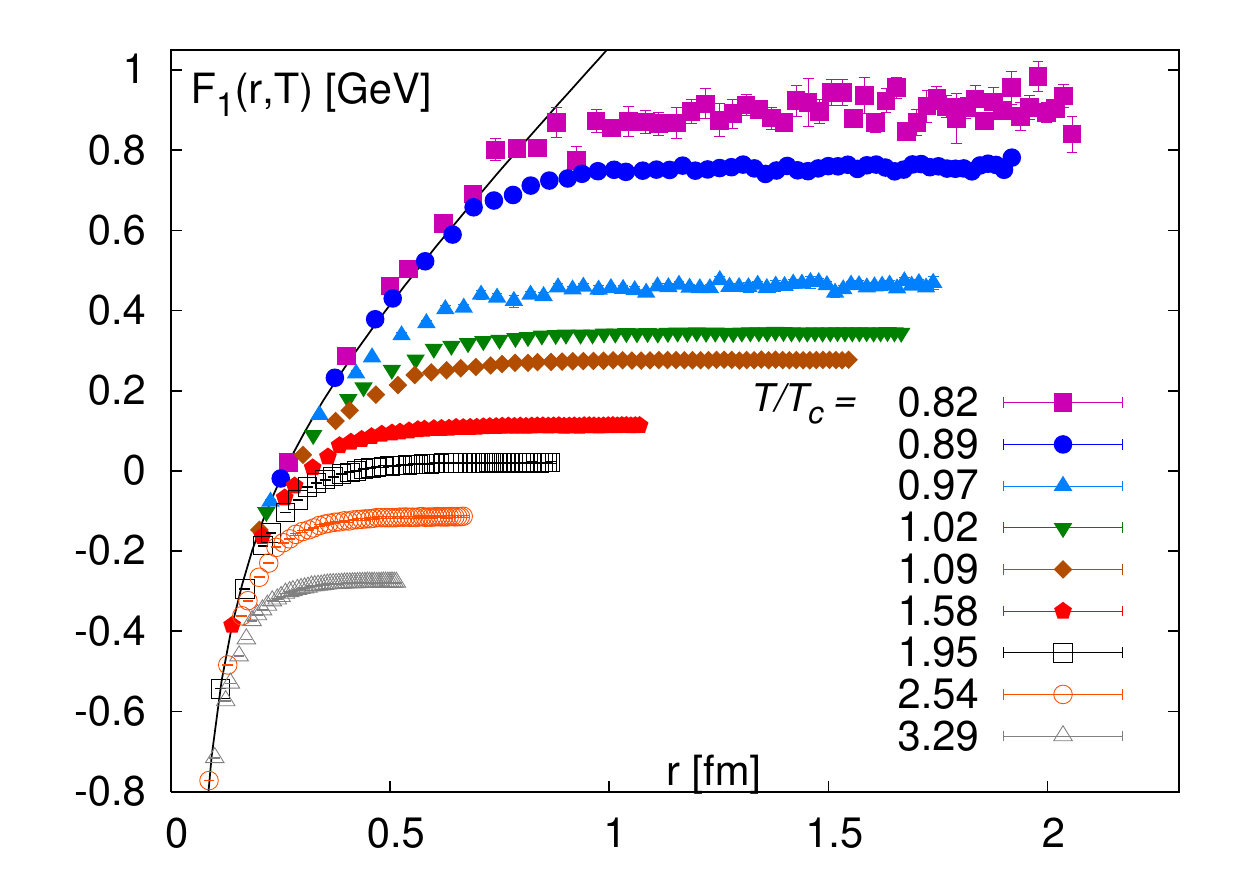}$$
		\caption{Static quark anti-quark singlet free energy calculated in 2+1 flavor lattice QCD at different temperatures. The figure is taken from Ref.~\cite{Petreczky:2010yn}.}
		\label{fig6}}
\end{figure}

The dissociation temperature $T_d$ can be calculated in potential model at finite temperature. The potential between the two quarks depends on the dissociation process in the medium. For a rapid dissociation where there is no heat exchange between the heavy quarks and the medium, the potential is just the internal energy $U$, while for a slow dissociation, there is enough time for the heavy quarks to exchange heat with the medium, the potential is then the free energy. The lattice simulated free energy between a pair of heavy quarks is shown in Fig.\ref{fig6}. As expected, in the zero-temperature limit the free energy coincides with the Cornell potential. At finite temperature, the free energy is temperature independent at sufficient short distance $r$ between the two quarks and becomes saturated at large $r$. Since a constant in free energy does not mean any interaction force, the range of interaction decreases with increasing temperature. The free energy can be parameterized as,
\begin{equation}
\label{v}
F_{Q\bar Q}(r,T)={\sigma \over m_D}\left[ {\Gamma(1/4) \over 2^{3/2}\Gamma(3/4)}-{\sqrt{m_D r} \over 2^{3/4}\Gamma(3/4)}K_{1/4}(m_D^2r^2) \right]-\alpha \left[m_D+{e^{-m_D r}\over r} \right],
\end{equation}
where $\Gamma$ is the Gamma function, $K$ is the modified Bessel function of the second kind, and the temperature dependent parameter $m_D(T)$, namely the screening mass or the inverse screening radius, can be extracted from fitting the lattice simulated free energy~\cite{Zhao:2017gpq,Digal:2005ht,Karsch:2003jg}.
From the thermodynamic relation $F=U-TS<U$ where $S$ is the entropy density, the surviving probability of a quarkonium state with potential $V=U$ is smaller than that with $V=F$. 
\begin{table}[!hbt]
	\centering
	\begin{tabular*}{6.5in}{@{\extracolsep{\fill}}llcccccccc}
		\hline
		\hline
		State & $J/\psi(1S)$& $\chi_c(1P)$ & $\psi(2S)$ & $\Upsilon(1S)$ & $\chi_b(1P)$& $\Upsilon(2S)$ & $\chi_b(2P)$ &$\Upsilon(3S)$   \\
		\hline
		$T_d/T_c(V=U)$  & 2.1 & 1.16 & 1.12 & $> $4.0 & 1.76 & 1.6 & 1.19 & 1.17 \\
		\hline
		$T_d/T_c(V=F)$  & 1.21 & $<$1.0 & $<$1.0 & 3.0 & 1.12 & 1.08 & 1.0 & $<$1.0 \\
		\hline
		\hline
	\end{tabular*}
	\caption{The quarkonium dissociation temperatures calculated by non-relativistic potential model with internal energy $U$~\cite{Satz:2005hx} and free energy $F$.}
	\label{table6}
\end{table}

The low and up limits of the quarkonium dissociation temperature can be calculated through the non-relativistic potential model with potential $V=F$ from the lattice simulation and $V=U=F+TS$. Substituting the potential $V(r,T)$ into the radial Schr\"odinger equation in the rest frame of the $Q\bar Q$ pair, we can obtain the binding energy $\epsilon(T)$ and the wave function $\psi(r,T)$ which determines the averaged size of the bound state $\langle r \rangle(T)$. The dissociation temperature $T_d$ can be determined by
\begin{equation}
\epsilon(T_d)=0\ \ \text{or}\ \ \langle r \rangle(T_d)=\infty.
\end{equation}
By solving the Schr\"odinger equation, the dissociation temperatures for different quarkonium states are listed in Table~\ref{table6}. For all $\Upsilon$ states, the dissociation temperatures are about 30\% lower in the case of $V = F$ compared to that with $V = U$. It is easy to understand that, a loosely bound quarkonium is easy to be dissociated and a tightly bound quarkonium is hard to be melted.

The dissociation is also investigated in the frame of relativistic Dirac equation~\cite{Shi:2013rga,Guo:2012hx}. In comparison with the non-relativistic calculation, the $J/\psi$ dissociation temperature for $V=F$ increases from $1.26 T_c$ to $1.35 T_c$, the relativistic correction is 7\%. For $V = U$, the dissociation temperature goes up from the non-relativistic value $2.1 T_c$ to $2.38 T_c$, and the relativistic correction becomes 13\%. The relativistic potential model can also be used to estimate the flavor dependence of the melting temperature for open heavy flavor mesons. The sequential melting temperature can explain the difference in meson elliptic flow observed in heavy ion collisions~\cite{Shi:2013rga}.

\subsubsection{Complex potential and spectrum}
\label{spectrum}
In above screening picture, medium effects are understood in terms of a temperature-dependent potential. With this potential, the quarkonium dissociation at finite temperature is reasonably discussed and successfully applied to quarkonium production in heavy ion collisions~\cite{Satz:2005hx,Digal:2001iu,Karsch:1987pv}. However, a derivation of the in-medium heavy quark potential from QCD shows that not all the medium effects can be incorporated into a screened potential~\cite{Burnier:2015tda}. 

The effective field theory approaches, such as NRQCD and pNRQCD, have given a good description of quarkonium spectra. However, when extending to finite temperature, things become much more complicated due to the presence of additional thermal scales $T$, $m_D\sim gT$ and $g^2T$ (with $g$ being the gauge coupling, $g^2=4\pi \alpha_s$). The maximum temperature of the QGP produced in heavy ion collisions is about $600$ MeV. When considering effective field theory descriptions for quarkonium in a thermal medium, one needs to deal with the relationship between different scales $1/r \sim m_Qv$, $T$ and $gT$. Let us first consider the weak-coupling regime. In a wide range of temperature, larger or smaller than the inverse distance between the heavy quark and antiquark (short distance means that $1/r$ is much larger than the typical hadronic scale $\Lambda_{QCD}$), people obtained the leading thermal effect on the potential in the frame of effective field theories~\cite{Brambilla:2008cx,Brambilla:2010vq}. The quark anti-quark potential at finite temperature becomes complex, and the explicit expression of the potential depends on the relationship between the quarkonium size and the temperature of the medium. The imaginary part of the potential comes from two main mechanisms: One is the imaginary part of the gluon self-energy induced by the Landau damping phenomenon, and the other is the quark-antiquark color singlet to color octet transition~\cite{Brambilla:2008cx}. At a very high-temperature, the Landau damping dominates and the potential is controlled by hard-thermal loop (HTL) resummed perturbation theory~\cite{Laine:2006ns}. The short-distance analysis is a valuable tool for studying the thermal dissociation of the lowest quarkonium resonances, the inclusion of the non-perturbative scale $\Lambda_{QCD}$ in the analysis may become necessary for studying excited states.
\begin{figure}[!htb]
{$$\includegraphics[width=0.3\textwidth]{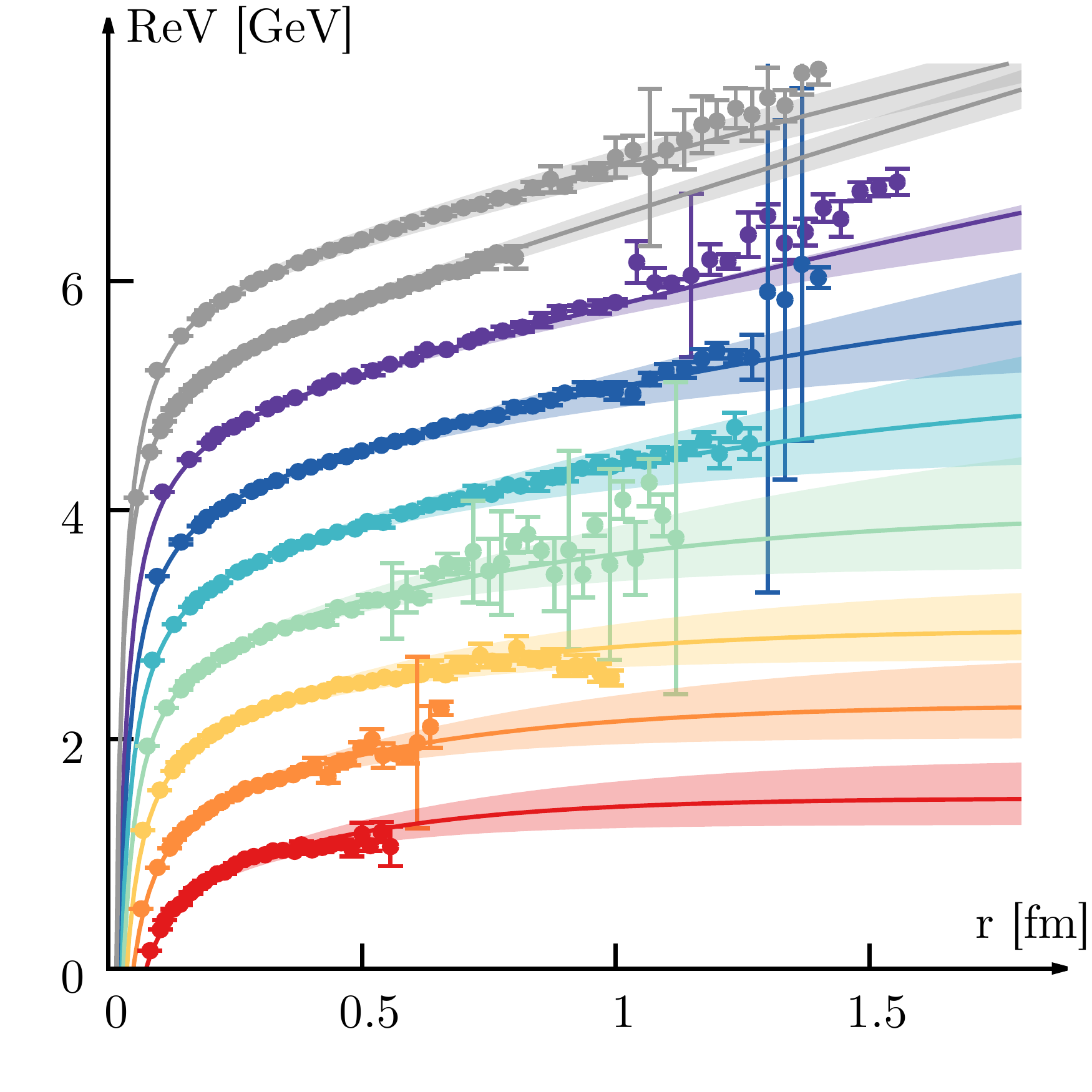}\ ~ \   ~ \  ~ \includegraphics[width=0.3\textwidth]{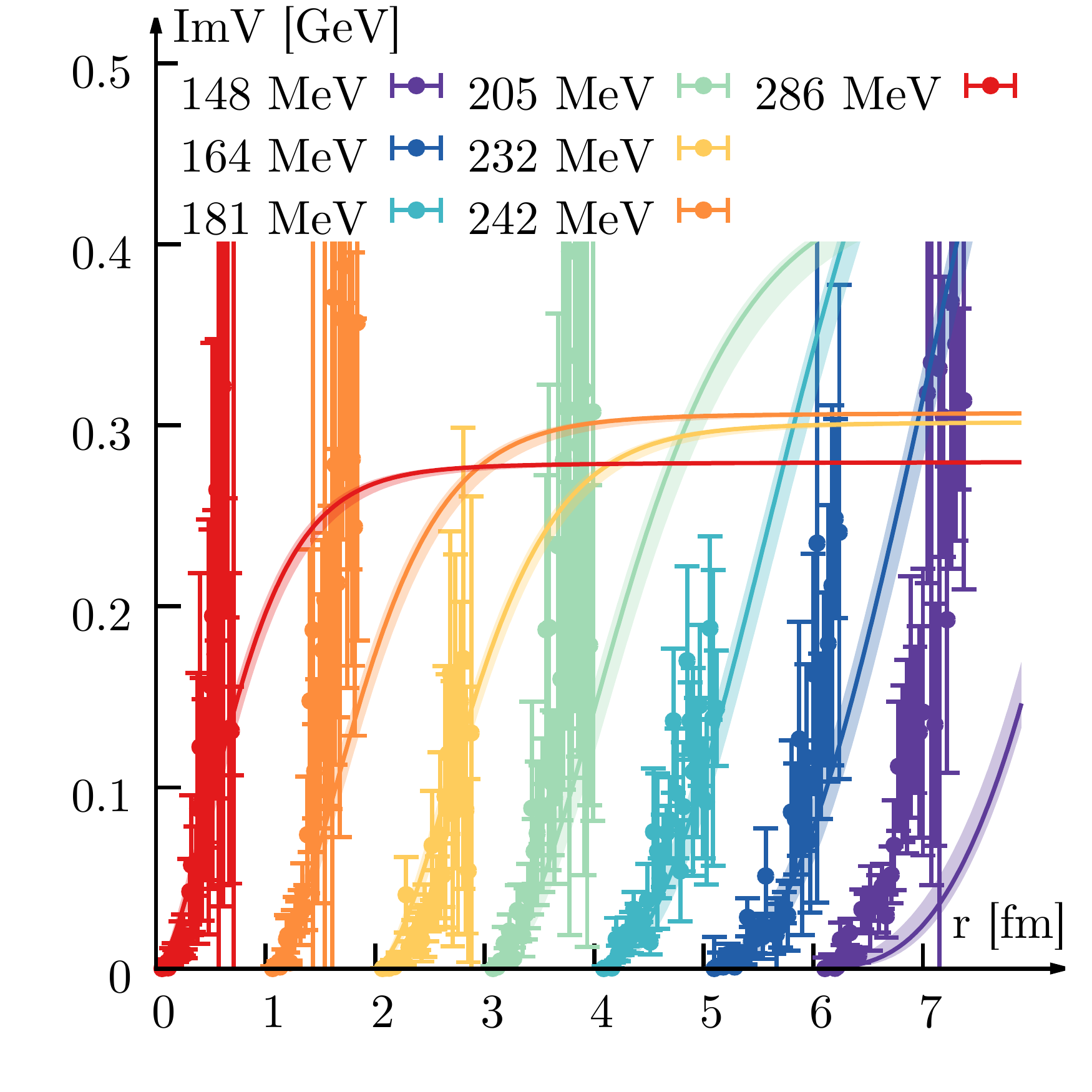}$$
\caption{Real (left panel) and imaginary (right panel) parts of the in-medium heavy quark potential from lattices QCD. The points are the lattice results, solid lines are the best fit, and shaded bands show the error bars. The figure is taken from Ref.~\cite{Lafferty:2019jpr}}
\label{fig7}}
\end{figure}

Attempts to study pNRQCD beyond weak coupling are presented in Refs.~\cite{Burnier:2015tda,Petreczky:2010tk}. Out of the weak-coupling regime, the potential can be extracted non-perturbatively from imaginary time simulations by inspecting the spectral function of the Wilson loop. The real-time heavy quark potential $V_S^{(0)}$, which is the leading order contribution to the color singlet potential in the heavy quark velocity expansion, can be represented as 
\begin{eqnarray}
&&V(r)=\lim_{t\to \infty}{\partial_t W(r,t) \over W(r,t)}, \nonumber\\
&&W(r, t)=  \left \langle \mathrm{Tr} \left (\exp \left [-ig\int dx^\mu A_\mu^a T^a \right] \right) \right \rangle
\end{eqnarray}
at finite temperature. The real-time definition of the static potential $V(r)$ is formulated in Minkowski space and thus not directly amenable to an evaluation in lattice QCD. The strategy~\cite{Rothkopf:2011db} is to evaluate the real-time definition using Euclidean lattice QCD simulations. It leads to a simple relation,
\begin{eqnarray}
W(r, \tau)=\int d\omega e^{-\omega \tau} \rho(r, \omega) \leftrightarrow \int d\omega e^{-i \omega t} \rho(r, \omega)=W(r, t).
\end{eqnarray}
This relates the potential to the spectral function which in principle can be obtained from lattice QCD,
\begin{eqnarray}
V(r)=\lim_{t\to \infty}{\int d\omega \omega e^{-i \omega t} \rho(r, \omega) \over \int d\omega e^{-i \omega t} \rho(r, \omega)}. 
\end{eqnarray}
However, extracting the spectrum from Euclidean time simulation data is an inherently ill-defined inverse problem, as one seeks to determine the form of a continuous function from a finite and noisy set of individual points. If a well defined spectral feature is present and in the shape of a skewed Breit-Wigner, its position and width encode the real and imaginary part of $V$ respectively. The first extraction of the potential from Wilson line correlators using both the novel Bayesian inference prescription method and the appropriate fitting strategy was presented in Refs.~\cite{Burnier:2014ssa,Burnier:2016mxc}. The real part of the potential $\mathrm{Re}$V in a gluonic medium and a realistic QCD with quarks is close to the color singlet free energy in Coulomb gauge, and shows Debye screening above the (pseudo-)critical temperature $T_c$. The imaginary part $\mathrm{Im}$V is estimated in the gluonic medium which is of the same order of the magnitude as in hard-thermal loop resummed perturbation theory in the deconfined phase. The real and imaginary parts of the real-time potential in full QCD~\cite{Burnier:2015tda} are shown in Fig.~\ref{fig7}. The in-medium behavior of $\mathrm{Re}$V in the presence of dynamical quarks shows differences compared to quenched QCD, especially at lower temperatures. The potential models can provide an analytic parametrization of lattice QCD results and an intuitive physical picture to explain the temperature dependence of the lattice potential. In the past two decades, there are several proposals on how to construct appropriate analytic parameterizations of the quark potential at finite temperature~\cite{Digal:2005ht,Karsch:1987pv,Lafferty:2019jpr,Thakur:2013nia,Burnier:2015nsa,Guo:2019bwa}. The Gauss-law parametrization provides an efficient prescription to summarize the in-medium behavior of the non-perturbative heavy quark potential based on two vacuum parameters ($\alpha$ and $\sigma$) as well as the temperature-dependent Debye mass $m_D$~\cite{Digal:2005ht,Lafferty:2019jpr,Thakur:2013nia}. The recent study based on Gauss-law approach by using the HTL permittivity to modify the non-perturbative vacuum potential gives the Coulomb part of the potential,    
\begin{eqnarray}
&&\mathrm{Re}V_C(r)=-\alpha\left[ m_D+{e^{-m_D r}\over r}\right], \nonumber\\
&&\mathrm{Im}V_C(r)=-\alpha[i T\phi(m_Dr)],\nonumber\\
&&\phi(x)=2\int_0^\infty dz{z\over (z^2+1)^2}\left( 1-{\sin(xz)\over xz} \right),
\end{eqnarray}
and the string part of the potential,
\begin{eqnarray}
&&\mathrm{Re}V_S(r)={2\sigma \over m_D}-{e^{-m_Dr}(2+m_Dr)\sigma \over m_D}, \nonumber\\
&&\mathrm{Im}V_S(r)= {\sqrt{\pi} \over 4}m_D T \sigma r^3 G_{2,4}^{2,2}\left(_{{1\over2},{1\over2},-{3\over2},-1 }^{-{1\over2},-{1\over2}} \Big| {1\over 4}m_D^2r^2 \right),
\label{imp}
\end{eqnarray}
where $G$ is the Meijer-G function. Putting the real and imaginary parts together, one obtains the Gauss-law expression for the complex in-medium potential, $\mathrm{Re}V=\mathrm{Re}V_C+\mathrm{Re}V_S$, $\mathrm{Im}V=\mathrm{Im}V_C+\mathrm{Im}V_S$. When going into high-temperature region, $\mathrm{Im}V(r)$ is consistent with the result from pure HTL. The Debye mass $m_D$, which controls both $\mathrm{Re}V$ and $\mathrm{Im}V$, can be fixed by the real-part of the potential. The imaginary part of the potential is in good agreement with the lattice data, as shown in Fig.~\ref{fig7}.

In the previous studies, color screening is studied on the lattice by calculating the spatial correlation function of a pair of static quark and anti-quark, which propagates in Euclidean time from $\tau=0$ to $\tau=1/T$. Two types of correlation functions are usually calculated on the lattice. One is the normalized Polyakov loop correlator (color-averaged), and the other is the color singlet correlator~\cite{Petreczky:2010tk,Bazavov:2018wmo,Borsanyi:2015yka}. One can define the subtracted free energy of a static $Q\bar Q$ pair $F_{Q\bar Q}$ and the singlet free energy $F_S$, by taking the logarithm of these correlators. $F_{Q\bar Q}$ contains the contribution from both singlet free energy $F_S$ and octet free energy $F_O$. The singlet free energy in 2+1 flavor QCD is shown in Fig.~\ref{fig6}. One interesting finding is that $\mathrm{Re}V$ is close to the color singlet free energies $F_S$.  

The quarkonium dissociation can also be determined by the quarkonium spectral functions at finite temperature. Spectral functions are defined as the imaginary part of the retarded Green function of quarkonium operators. Bound states appear as peaks in the spectral functions. The peaks broaden and eventually disappear with increasing temperature. The disappearance of a peak signals the melting of the quarkonium state. 
\begin{figure}[!htb]
{$$\includegraphics[width=0.4\textwidth]{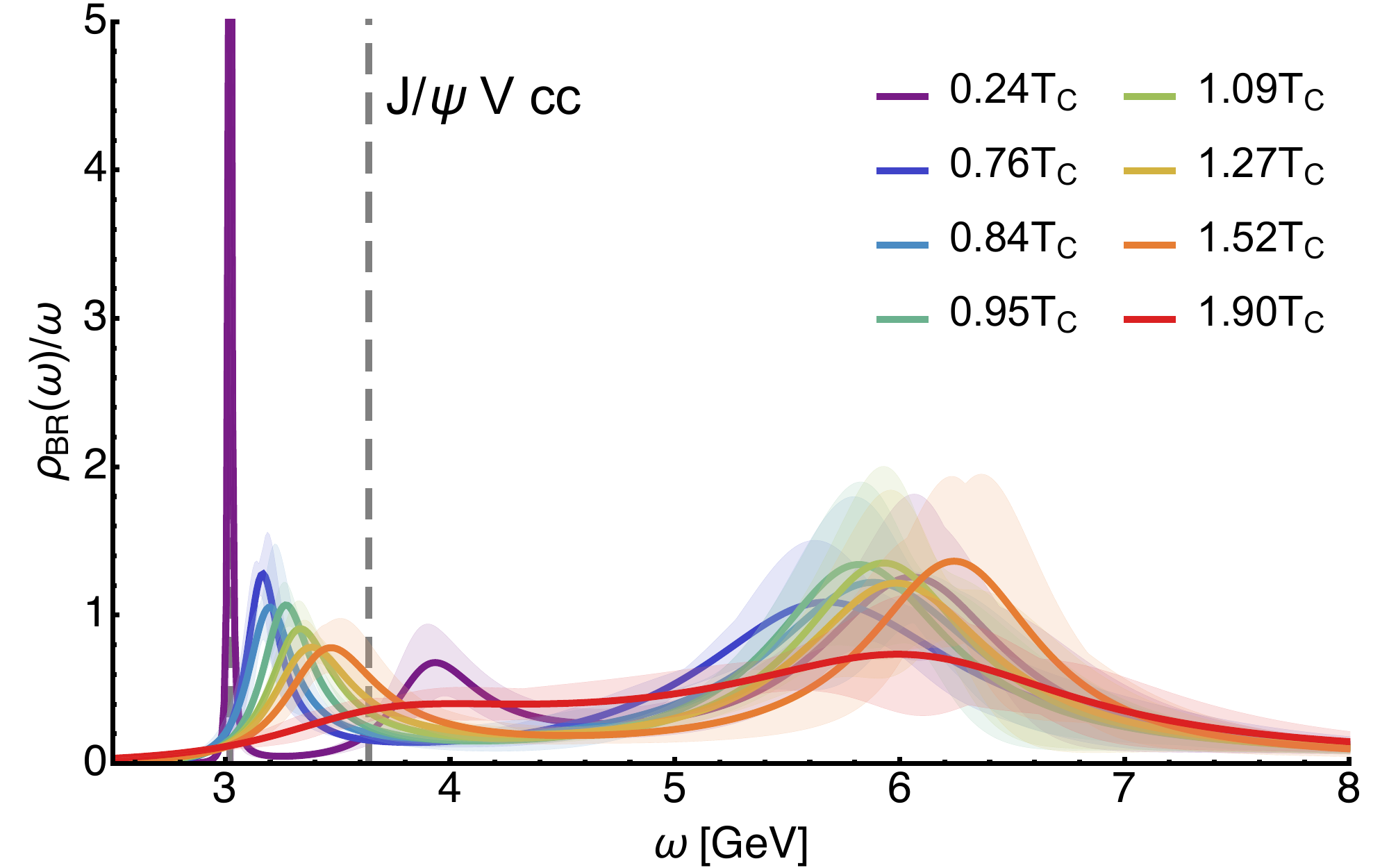}\ ~ \  ~  \ ~ \includegraphics[width=0.4\textwidth]{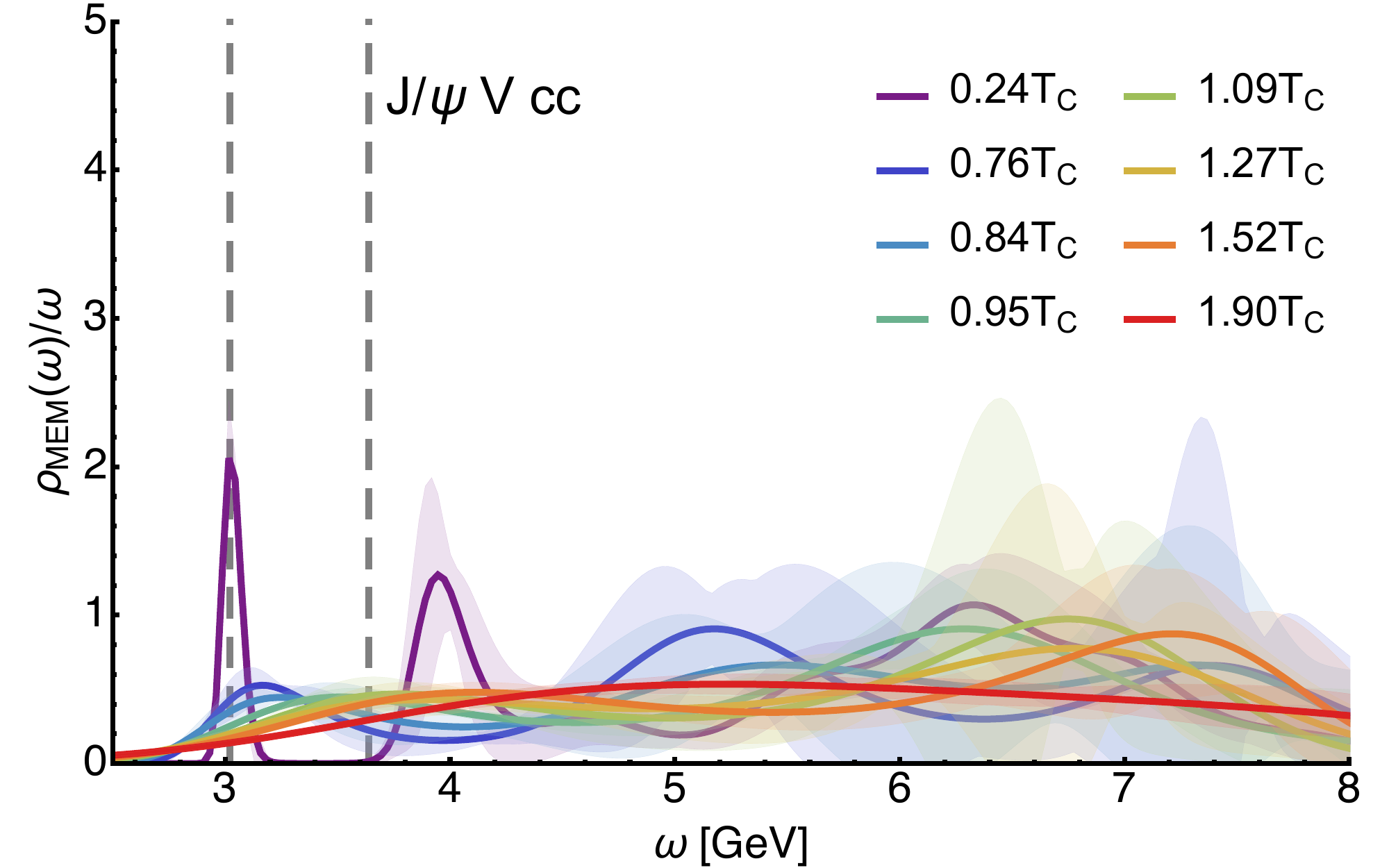}$$
\caption{ Spectral functions for the $c\bar c$ vector channels, obtained using the BR method(left panel) and MEM method(right panel). The figure is taken from Ref.~\cite{Kelly:2018hsi}.}
\label{fig8}}
\end{figure}

Spectral reconstructions have been carried out using many different methods in full lattice QCD~\cite{Datta:2003ww,Ding:2012sp,Laine:2008cf,Jakovac:2006sf,Aarts:2011sm,Aarts:2014cda,Ikeda:2016czj,Kelly:2018hsi,Burnier:2013nla}, such as Maximum Entropy Method (MEM) and BR method, the latter is based on Bayesian strategy and has significantly improved the uncertainties. The review paper~\cite{Rothkopf:2019ipj} gives a good description of the recent progress. The spectral functions for charmonia and open-charm-mesons reconstructed at finite temperature in a fully relativistic lattice QCD approach are shown in~\cite{Kelly:2018hsi}. The spectral functions for the $c\bar c$ vector channels are shown in Fig.~\ref{fig8} with different reconstruction methods. We can see that, with increasing temperature the ground state structure monotonically moves to higher frequency, and the strength of the peak is reduces continuously. The lattice NRQCD provides an alternative discretization of heavy quarks, which has been applied to the study of both bottomonia and charmonia at finite temperature~\cite{Aarts:2010ek,Kim:2018yhk,Aarts:2013kaa,Kim:2014iga}. The extraction of spectral functions from the meson correlation functions in lattice NRQCD is less demanding than in full QCD. Quantitatively robust determinations of in-medium ground state properties have been achieved. 

The lattice NRQCD has already led to a significantly improved understanding of the in-medium ground state properties for both bottomonia and charmonia. However, the excited states, as well as the continuum, are not well captured. To progress in this direction, one can turn to the effective field theory pNRQCD, which allows deriving the proper real-time in-medium potential systematically from QCD. The potential model that we used before is only the real part of the potential, and we calculated only the binding energy. The first computation of the in-medium spectral function using the perturbatively evaluated real-time potential was carried out in Ref.~\cite{Burnier:2007qm}. In Refs.~\cite{Burnier:2015tda,Lafferty:2019jpr,Burnier:2016kqm}, people use the lattice vetted potential, both the real and imaginary parts, to calculate the quarkonium in-medium spectral functions. In this process, one needs to first calculate the forward correlator, 
\begin{eqnarray}
&&i\partial_t D^>(t,{\bf r},{\bf r}')=({\mathcal H} -i |\mathrm{Im}V(r)|)D^>(t,{\bf r},{\bf r}'),\ \ \  t>0\ \nonumber \\
&&i\partial_t D^>(t,{\bf r},{\bf r}')=({\mathcal H} +i |\mathrm{Im}V(r)|)D^>(t,{\bf r},{\bf r}'), \ \ \ t<0
\end{eqnarray}
where the Hamiltonian is defined as ${\mathcal H}=2m_Q+{{\bf p}^2\over 2m_Q}+{l(l+1)\over m_Qr^2}+\mathrm{Re}V(r)$. The vector channel spectrum is obtained by taking the limit of the correlator in frequency space,
\begin{eqnarray}
\rho_V(\omega)=\lim_{{\bf r},{\bf r'}\to 0}{1\over 2}\tilde D(\omega, {\bf r}, {\bf r'}).
\end{eqnarray}

\begin{figure}[!htb]
{$$\includegraphics[width=0.42\textwidth]{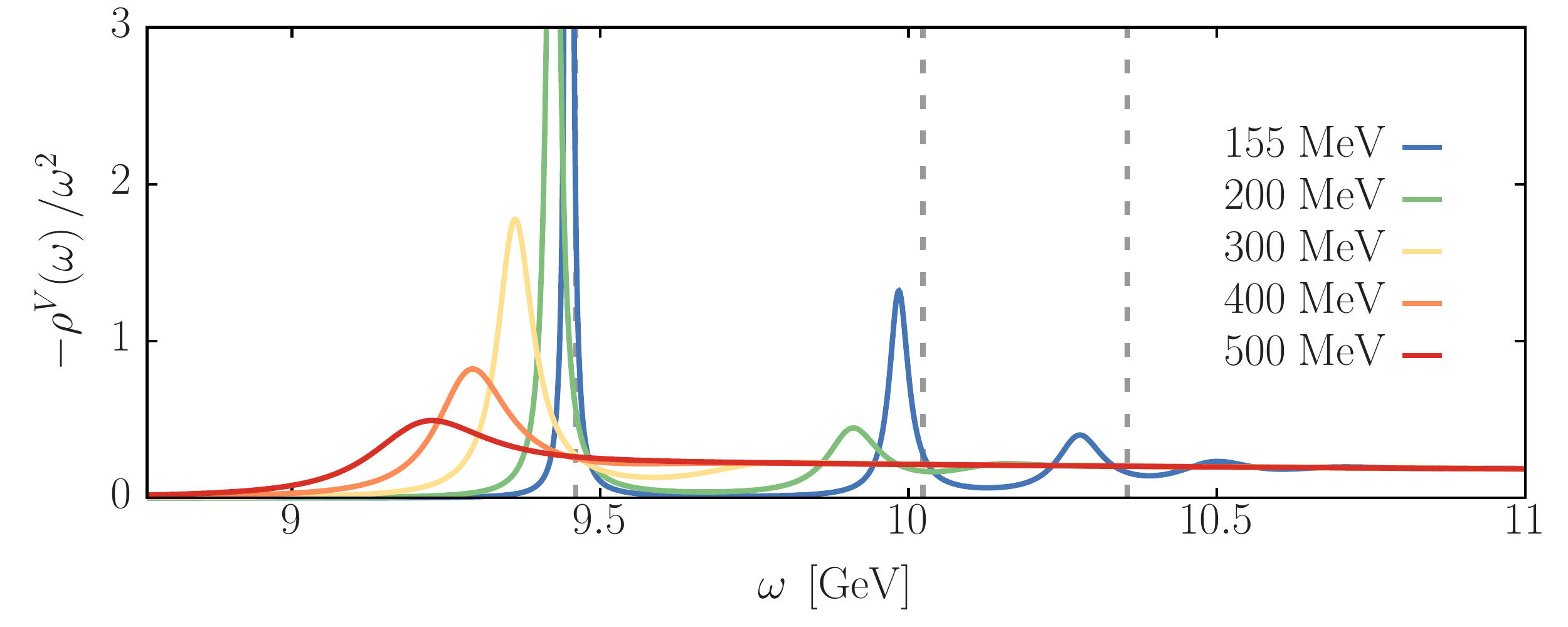}\ ~ \  ~
\includegraphics[width=0.42\textwidth]{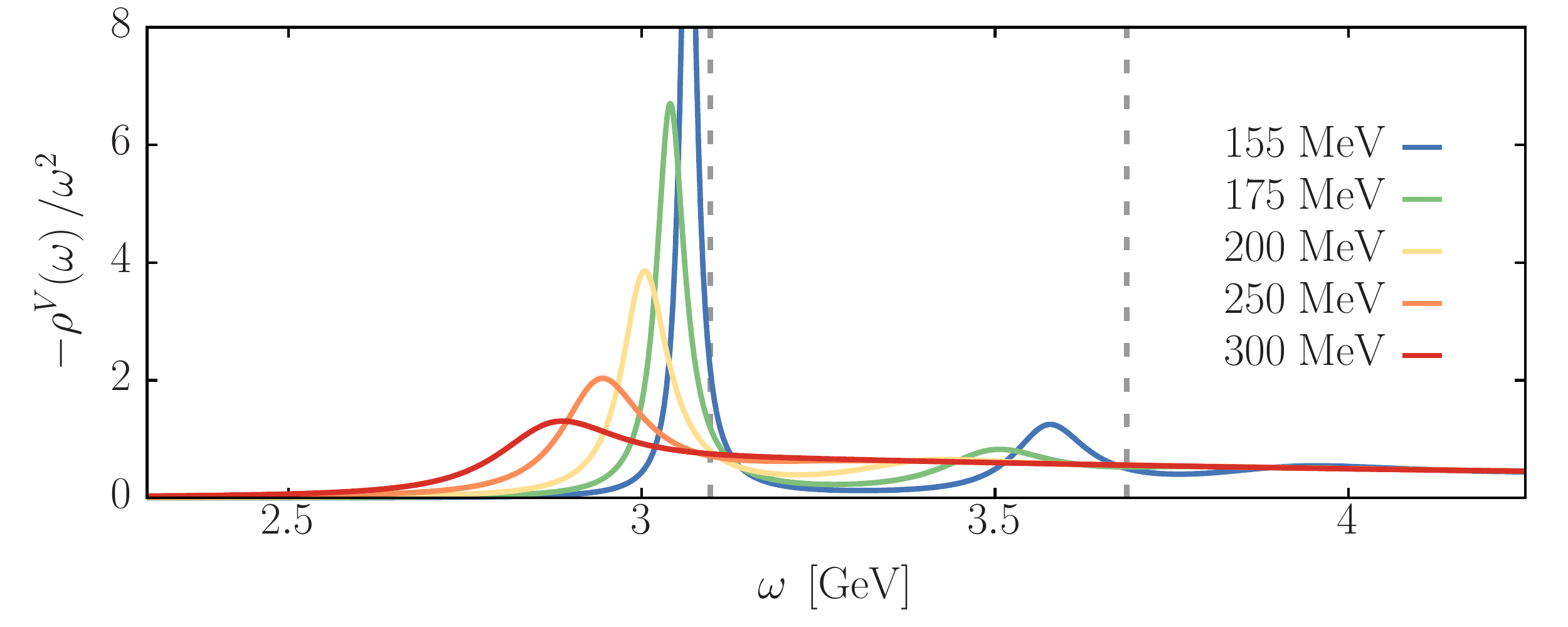}$$
\caption{ The S-wave in-medium spectral functions for bottomonia (left panel) and charmonia (right panel), based on the proper complex potential from lattice QCD. The figure is taken from Ref.~\cite{Lafferty:2019jpr}}
\label{fig9}}
\end{figure}

The temperature dependence of the energy $E$ and width $\Gamma$ can be derived by fitting the Breit-Wigner distribution. The spectral functions for $\Upsilon$ and $J/\psi$ are shown in Fig.~\ref{fig9}, and the P-wave states are showed in Ref.~\cite{Burnier:2015tda}. The position of the peak is shifted to a lower frequency as temperature increases, controlled by the real part of the potential (screening effect), and the main effect of the imaginary part is to broaden the peak, without changing its position.

As for the dissociation temperature, it is straightforwardly defined from the disappearance of the binding energy in the case of a real potential. For a complex potential, the situation becomes subtle, as the bound state broadens before it disappears. A popular choice is to define the melting of a state by the condition that its width equals to its binding energy, $\Gamma=E_{bind}$~\cite{Beraudo:2007ky}. With this condition, we can extract the dissociation temperature from the spectral function based on a proper complex potential. The results are shown in Tab.~\ref{table7}. We can see that the dissociation temperatures are different from the potential model result via solving the Schr\"odinger equation with a real potential, as showed in Tab.~\ref{table6}. 
\begin{table}[!hbt]
	\centering
	\begin{tabular*}{6.0in}{@{\extracolsep{\fill}}llcccccc}
		\hline
		\hline
		State & $J/\psi(1S)$ & $\psi(2S)$ & $\Upsilon(1S)$ & $\Upsilon(2S)$& $\Upsilon(3S)$ & $\Upsilon(4S)$   \\
		\hline
		$T_d$/ $T_c$  &$1.37^{+0.08}_{-0.07}$ & $<0.95$ & $2.66^{+0.49}_{-0.14}$ & $1.25^{+0.17}_{-0.05}$ & $1.01^{+0.03}_{-0.03}$ & $<0.95$  \\
		\hline
		\hline
	\end{tabular*}
	\caption{The quarkonium dissociation temperature extracted from the spectral function. The errors of $T_d$ are from the variation of the Debye screening mass $m_D$. The result is taken from Ref.~\cite{Burnier:2015tda}.}
	\label{table7}
\end{table}
For open charmed mesons ($D$ and $D^*$), their spectral functions are also reconstructed at finite temperature in lattice QCD~\cite{Kelly:2018hsi}. The results show that above $T_c$ $D$ and $D^*$ mesons become much different. The strength of the former continues to decease monotonically, while the latter shows a sudden rise, which hints that $D^*$ starts to be affected earlier than $D$. Moreover, smaller medium modifications for $D_s$ are observed in comparison with $D$, because $D_s$ is a much tighter bound state than $D$. These results may support the inference that $D_s$ and $D$ can survive above $T_c$. Charm fluctuations and correlations are also studied in the frame of lattice QCD. The results seem to imply that open heavy flavor mesons may exist in the quark-gluon plasma~\cite{Mukherjee:2015mxc}. The spatial correlation functions for open and closed charmed mesons have been studied in (2+1)-flavor lattice QCD~\cite{Bazavov:2014cta}. The significant in-medium modifications for $D_s$ meson is already at temperatures around the chiral crossover, while for $J/\psi$ and $\eta_c$ mesons the in-medium modifications remain relatively small around the crossover and become significant only above 1.3 times the crossover temperature. Including light quarks, the properties of open charmed mesons at finite temperature are studied via a two-body Dirac equation~\cite{Shi:2013rga} with the lattice simulated potential. It shows that the dissociation temperature of $D$ is larger than $T_c$ but smaller than the dissociation temperature for $J/\psi$. 

While we have obtained the quarkonium spectral functions and thermal width directly or indirectly from the lattice QCD, it is still insufficient to predict quarkonium production and decay rates in heavy-ion collisions, due to the following two reasons: 1) The above results are obtained in the equilibrium limit, but heavy quarks in heavy ion collisions are not fully thermalized with the medium; 2) Quarkonia produced in heavy ion collisions are not static in the medium, but move through the fireball with a finite velocity. These will be discussed in the next subsection. 

\subsubsection{Collisional dissociation and regeneration}
\label{regeneration}
The Debye screening is an explanation of quarkonium dissociation in a static and homogeneous medium. In a dynamical evolution of an inhomogeneous fireball, the quarkonium state will have probability to interact with particles in the medium, and this hit may induce a decay. There are two main collision processes, one is the gluon dissociation, and the other is inelastic parton scattering,
\begin{eqnarray}
 \text{Gluon dissociation:} \ && \psi+g \to Q +\bar Q; \nonumber\\
 \text{Inelastic scattering:} \ &&  \psi+ p \to Q +\bar Q + p, \  p=g,q(\bar q).
\end{eqnarray}
For the gluon dissociation, we can use a non-relativistic approximation to treat the quarkonium state. Taking into account the fact that the distance between the two heavy quarks in vacuum is short, the potential is mainly from the Coulomb part, the gluon dissociation cross-section can be approximately calculated by the Operator Production Expansion method (OPE)~\cite{Peskin:1979va,Bhanot:1979vb}. This is called Bhanot and Peskin approach. The leading order is a chromoelectric dipole interaction between the quarkonium and the gluon, and the dissociation cross-section can be analytically expressed as
\begin{eqnarray}
&&\sigma(1S)=A{(r-1)^{3/2}\over r^5}, \nonumber\\
&&\sigma(1P)=A{(r-1)^{1/2}(9r^2-20r+12)\over r^7}, \nonumber\\
&&\sigma(2S)=A{(r-1)^{3/2}(r-3)^2\over r^7}
\label{ope}
\end{eqnarray}
with $A=2^{11}\pi/\left(27\sqrt{m_Q^3\epsilon(1S)}\right)$, heavy quark mass $m_Q$, and $r=\omega/\epsilon$, where $\epsilon$ is the binding energy of the quarkonium, $\omega=p^\mu k_\mu/m_\psi=(s-m_\psi^2)/(2m_\psi)$ is the gluon energy in the quarkonium rest frame. Considering that the quarkonium mass $m_\psi$ is finite, the relativistic correction leads to a shift of the threshold energy for the dissociation process~\cite{Polleri:2003kn}. In contrast to the leading order counterparts which rapidly drop off with increasing incident gluon energy, the NLO cross sections exhibit finite value toward high energies, because of the new phase space opened up~\cite{Chen:2018dqg}.

The OPE method to derive the above dissociation cross-section is valid in the following cases: 1) the energy of the incoming gluon is much smaller than the binding energy of the quarkonium, $E_g\ll \epsilon$; 2) the quarkonium is a tightly bound state in vacuum and at low temperature; 3) the octet potential can be neglected, which means negligible final-state interactions. When the temperature of the fireball is higher than the in-medium binding energy $\epsilon<T$, the gluon-dissociation mechanism turns out to be inefficient in destroying quarkonium. At high temperature, especially when reaching the dissociation temperature $T_d$, the OPE fails to calculate the gluon dissociation. Aiming to describe the process in hot medium, we consider the geometric relation between the integrated cross-section and the size of the quarkonium,
\begin{equation}
\sigma(T)={ \langle r^2 \rangle_\psi(T) \over \langle r^2 \rangle_\psi(0)}\sigma(0),
\end{equation}
where the averaged size of the quarkonium at finite temperature can be derived from the potential model discussed above. The cross-section changes smoothly at low temperature, but increases rapidly at high temperature and finally approaches to infinity at $T\to T_d$. When the fireball temperature is above $T_d$, all the quarkonium states disappear due to the gluon dissociation. This behavior is in good agreement with the lattice QCD simulation in a static limit of the quarkonium spectra where the quarkonium peak disappears suddenly. 

When the energy of the incoming gluon is close or larger than the binding energy of the quarkonium, $E_g \gtrsim \epsilon$, another dissociation process, for instance, the inelastic parton scattering processes $\psi+p\to Q+\bar Q+p\ (p=g,q,\bar q)$ becomes important~\cite{Combridge:1978kx,Grandchamp:2001pf,Grandchamp:2002wp}. Since in the QGP charmonia are loosely bound states, the incoming parton could collide with the $c$ or $\bar c$ and leads to a dissociation of the bound state. In the limit of $E_g \gg \epsilon$, the interaction between $c$ and $\bar c$ inside the charmonium cannot interfere with the interaction between the incoming partons. The $p-\psi$ scattering cross-section should approach the value of $2\sigma_{pc}$, the sum of the probability of the scattering between the parton and one of the constituent charm quarks. Because this approximation neglects the bound-state effects, it is called quasi-free. The cross-section $\sigma_{pc}$ can be obtained via the perturbative QCD at leading-order~\cite{Combridge:1978kx}. At finite temperature, one needs to consider the thermal mass effect via for instance introducing a Debye mass $m_D=gT$ into the denominator of the $t$-channel gluon-exchange propagator, $1/t\to1/(t-m_D^2)$. Since the quasifree proce is not the only dissociation mechanism for $\psi$, for practical applications one needs to effectively parameterize other dissociation mechanisms into the quasifree processes by using the strong coupling constant $\alpha_s$ as an adjustable parameter~\cite{Grandchamp:2001pf,Grandchamp:2002wp}.
The dissociation rate can be calculated with the above-obtained cross sections $\sigma$,
\begin{equation}
\Gamma({\bf p}, T)= \sum_{p=g,q,\bar q}\int_{k_{min}}{d^3k \over (2\pi)^3}f_p(k, T)\sigma(s) v_{rel},
\label{decay}
\end{equation}
where $v_{rel}$ is the relative velocity between the partons and the quarkonium, and $f_p$ is the parton distribution function (Bose-Einstein distribution for gluons and Fermi-Dirac distribution for light quarks). The momentum $k_{min}$ is the minimum incoming momentum necessary to dissociate the bound state. The momentum dependence comes from the cross-section $\sigma$.

Recently, pNRQCD is used to study quarkonium dissociation rate in hot medium~\cite{Brambilla:2008cx,Brambilla:2010vq,Brambilla:2011sg,Brambilla:2013dpa}, which deepens our understanding of the previous knowledge. In Ref.~\cite{Brambilla:2008cx}, it is found that there are two mechanisms contributing at leading order to the quarkonium decay width: one is the Landau damping, and the other is the singlet-to-octet thermal breakup. These two mechanisms contribute to the thermal decay width in different temperature regions. The former dominates in the temperature region where the Debye mass is larger than the binding energy $m_D>\epsilon$, while the latter dominates at temperatures where the Debye mass is smaller than the binding energy $m_D<\epsilon$. In Ref.~\cite{Brambilla:2011sg}, the relationship between the singlet-to-octet thermal break-up and the so-called gluon-dissociation is investigated. The singlet-to-octet break-up width is given by the imaginary part of the leading heavy-quarkonium self-energy diagram in pNRQCD. From the width, one can extract the cross-section,
\begin{eqnarray}
\sigma(1S)={2^{10}\over 3}\alpha_sC_F\pi^2 {E_1^4 \over m \omega^5}\rho(\rho+2)^2\left(t^2(\omega)+\rho^2\right){\exp \left ({4\rho \over t(\omega)}\arctan\left(t(\omega)\right)\right) \over \exp \left({2\pi \rho \over t(\omega)} \right)-1}
\end{eqnarray}
with $t(\omega)=\sqrt{\omega/|E_1|-1}$, $\rho=1/(N_c^2-1)$ and the first energy level $E_1=-mC_F^2\alpha_s^2/4$. As we mentioned above, the final-state interactions (octet potential) is neglected in the previous calculation as shown in Eq.\ref{ope}. Theoretically, this assumption may be realized by taking the large-$N_c$ limit, because the octet potential at leading order equals to $V_o^{(0)}=\alpha_s/(2N_c r)$. Therefore, the limit $N_c\to \infty$ leads to Eq.\ref{ope} (multiplied by a factor 16 for polarization and colors factor). The singlet-to-octet thermal break-up expression allows us to improve the Bhanot–Peskin cross-section by including the contribution of the octet potential, which amounts to include final-state interactions between the heavy quark and antiquark.

In Ref.~\cite{Brambilla:2013dpa}, a similar analysis for the relation between the Landau-damping mechanism and the dissociation by inelastic parton scattering is performed. The Landau-damping mechanism corresponds to the dissociation by inelastic parton scattering. The dissociation cross-section and the corresponding thermal width in different temperature regimes are studied in an EFT framework where the bound state effects are systematically included, comparing with the previous work~\cite{Park:2007zza}. The result is consistent with the previous conclusion: Inelastic parton scattering is the dominant process in the temperature region where the Debye mass is larger than the binding energy, $m_D>E$. While the previously used quasi-free approximation is justified when the Debye mass $m_D$ is much larger than $mv$, the condition $m_D \gg m_Qv$ requires a temperature larger than the dissociation temperature $T_d$. For $m_Qv\gg m_D$, the quasi-free approximation is not justified, and its contribution is canceled by the bound-state effects. Therefore, the inelastic parton scattering cross-section for the ground state can be expressed as $\sigma^{1S}_p=\sigma_{pc}f(m_D, k)$, where $f(m_D, k)$ depends on the sequence of temperature $T$, the momentum of order $mv$, and Debye mass $m_D$. The cross section $\sigma_{pc}$ can be defined as 
\begin{eqnarray}
\sigma_{qc}=8\pi C_Fn_f \alpha_s^2 a_0^2, \ ~ \ ~  \sigma_{gc}=8\pi C_F N_c \alpha_s^2 a_0^2.
\end{eqnarray}
In the meantime, the convolution formula Eq.\ref{decay} can not be applied in the case of inelastic parton scattering, because there is a parton in the final state and one should take into account correctly the effect of Pauli blocking and Bose enhancement~\cite{Brambilla:2013dpa}.
 
\begin{figure}[!htb]
	{$$\includegraphics[width=0.35\textwidth]{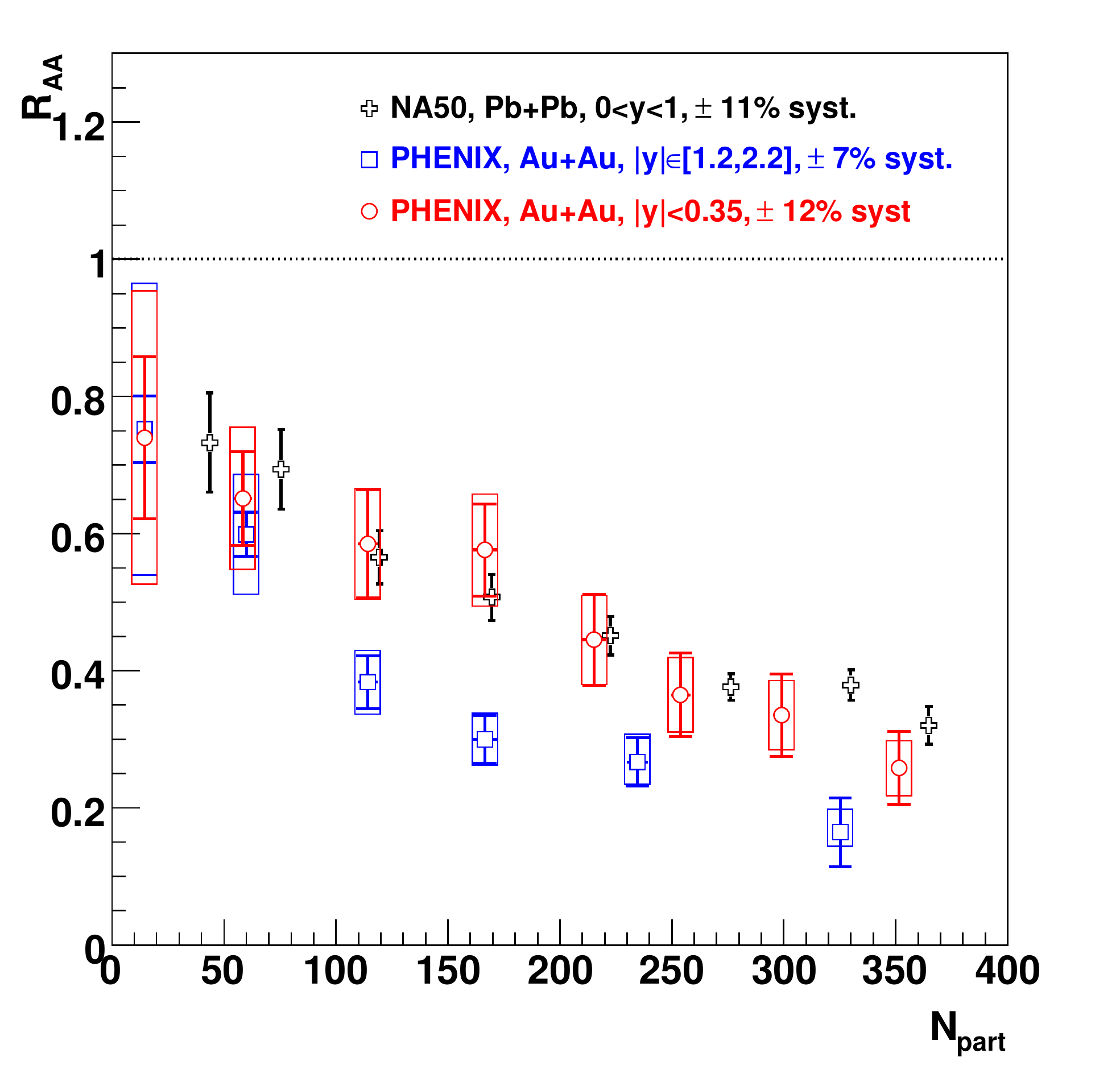}$$
		\caption{The $J/\psi$ nuclear modification factor as a function of the number of participants $N_{part}$ in nuclear collisions at SPS and RHIC energies. The figure is taken from Ref.~\cite{GranierdeCassagnac:2008ke}.}
		\label{fig10}}
\end{figure}

Now, let's consider the regeneration effect. While charm quark production at SPS is expected to be small, there are more than 10 $c\bar c$ pairs produced in a central Au+ Au collision at RHIC and probably more than 100 pairs at LHC~\cite{Gavai:1994in}. These uncorrelated charm quarks in the QGP can be recombined to form charmonium states. Therefore, there will be two production sources of quarkonia in heavy ion collisions in extremely high energies. Obviously, regeneration will enhance the quarkonium yield and alter its momentum spectra.
On the experimental side, the charmonium data at RHIC look difficult to be understood, if people consider only the Debye screening effect. Fig.\ref{fig10} shows the $J/\psi$ nuclear modification factor as a function of the number of participants in central and forward rapidity regions at SPS and RHIC energies~\cite{GranierdeCassagnac:2008ke}. There are two puzzles here. One is the almost same suppression at SPS and RHIC. Since the fireball at RHIC is much more hotter than at SPS, the Debye screening should be stronger, and therefore a stronger suppression is expected. The other puzzle is the rapidity dependence at RHIC, the suppression at forward rapidity is stronger than at central rapidity. Since the fireball temperature decreases with rapidity, the suppression should be stronger at central rapidity, if we take into account only the Debye screening. While the cold nuclear matter effects may play a role here~\cite{Ferreiro:2008wc,Kopeliovich:2010nw}, the puzzles are direct hints for the introduction of quarkonium regeneration: There are more heavy quarks and in turn stronger quarkonium regeneration at high energy and in central rapidity.

The dissociation rate is calculated in the frame of pNRQCD, and the regeneration rate can be modeled from detailed balance. The inverse of a dissociation process is the corresponding regeneration process,
 \begin{eqnarray}
  \text{Regeneration:} \  &&Q +\bar Q \to \psi+g, \nonumber\\
    &&Q +\bar Q + p \to \psi+ p. \  p=g,q(\bar q). 
\end{eqnarray}
From the detailed balance, namely the same transition probabilities for the dissociation and regeneration processes, we obtain the regeneration cross section corresponding to the gluon dissociation,
\begin{equation}
\sigma_{ Q +\bar Q \to J/\psi+g}(s)={4(s-m_{J/\psi}^2)^2\over 3s(s-4m_Q^2) } \sigma_{J/\psi+g\to Q +\bar Q}(s).
\end{equation}
The difference between the two cross sections is controlled by the degrees of freedom in the initial state and the flux factor.
Regeneration yield is a convolution of regeneration cross section and heavy quark distribution function. Quarkonium regeneration was studied in a pNRQCD-based Boltzmann equation~\cite{Yao:2017fuc}. Taking into accounting the coupling between two Boltzmann transport equations for heavy quarks and quarkonia, where the heavy quark distribution is not assumed as a parametrization but rather calculated from real-time dynamics, quarkonium dissociation and regeneration are calculated in a self-consistent way~\cite{Yao:2017fuc}. The regeneration process is also analyzed in the frame of perturbative QCD with parametrized non-thermal heavy quark distributions~\cite{Song:2012at}. We will discuss the detail in next Section.
\subsection{Heavy quark thermal production}
\label{thermal}
A much hotter medium will emerge at the Future Circular Collider (FCC) with colliding energy $\sqrt{s_{NN}}=39$ TeV. Gluons and light quarks inside the medium would be more energetic and denser. The thermal production of charm quarks via gluon fusion and quark and anti-quark annihilation may have a sizeable effect on charmonium regeneration.
\begin{figure}[!htb]
	{$$\includegraphics[width=0.3\textwidth]{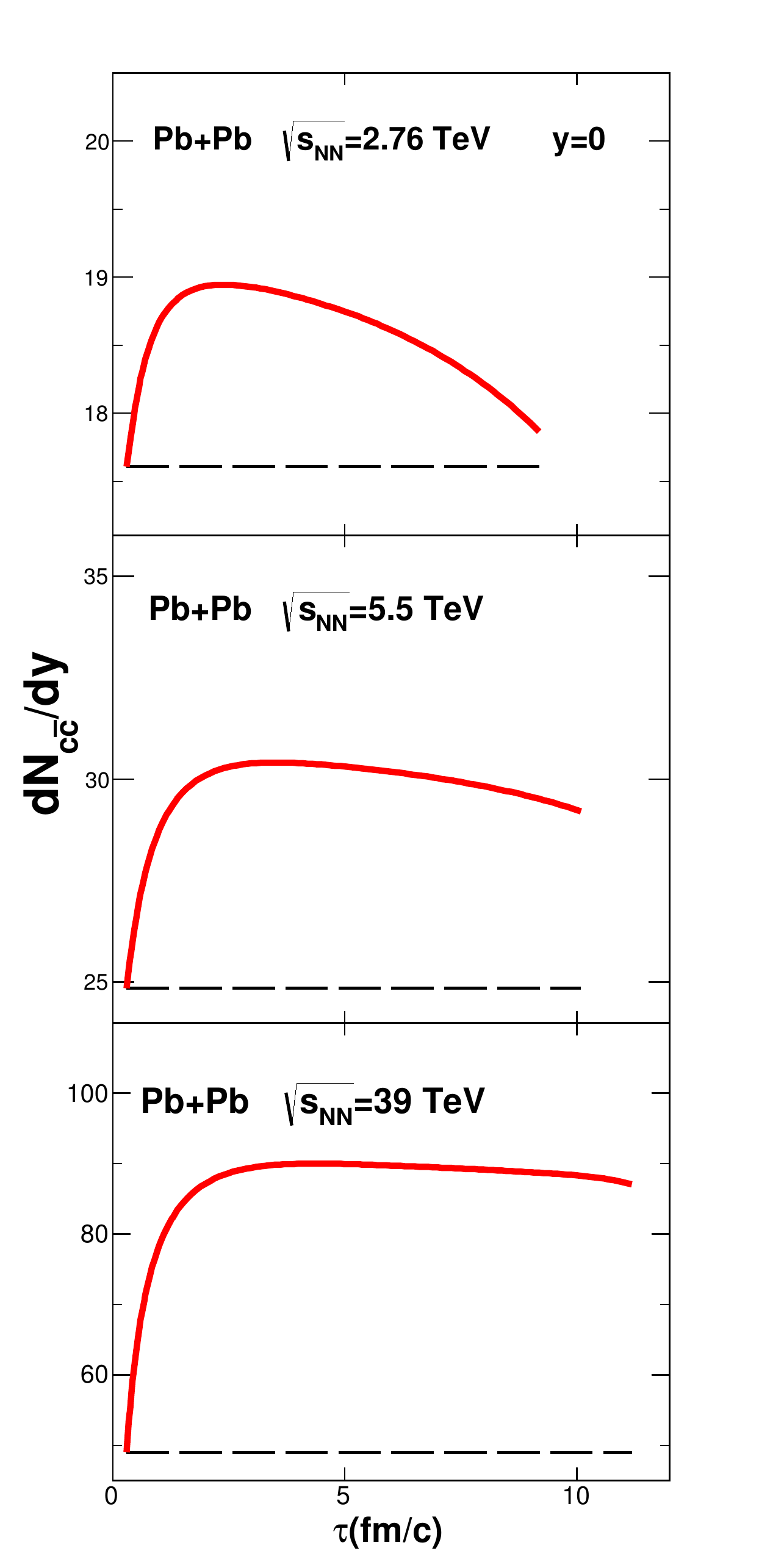}\includegraphics[width=0.3\textwidth]{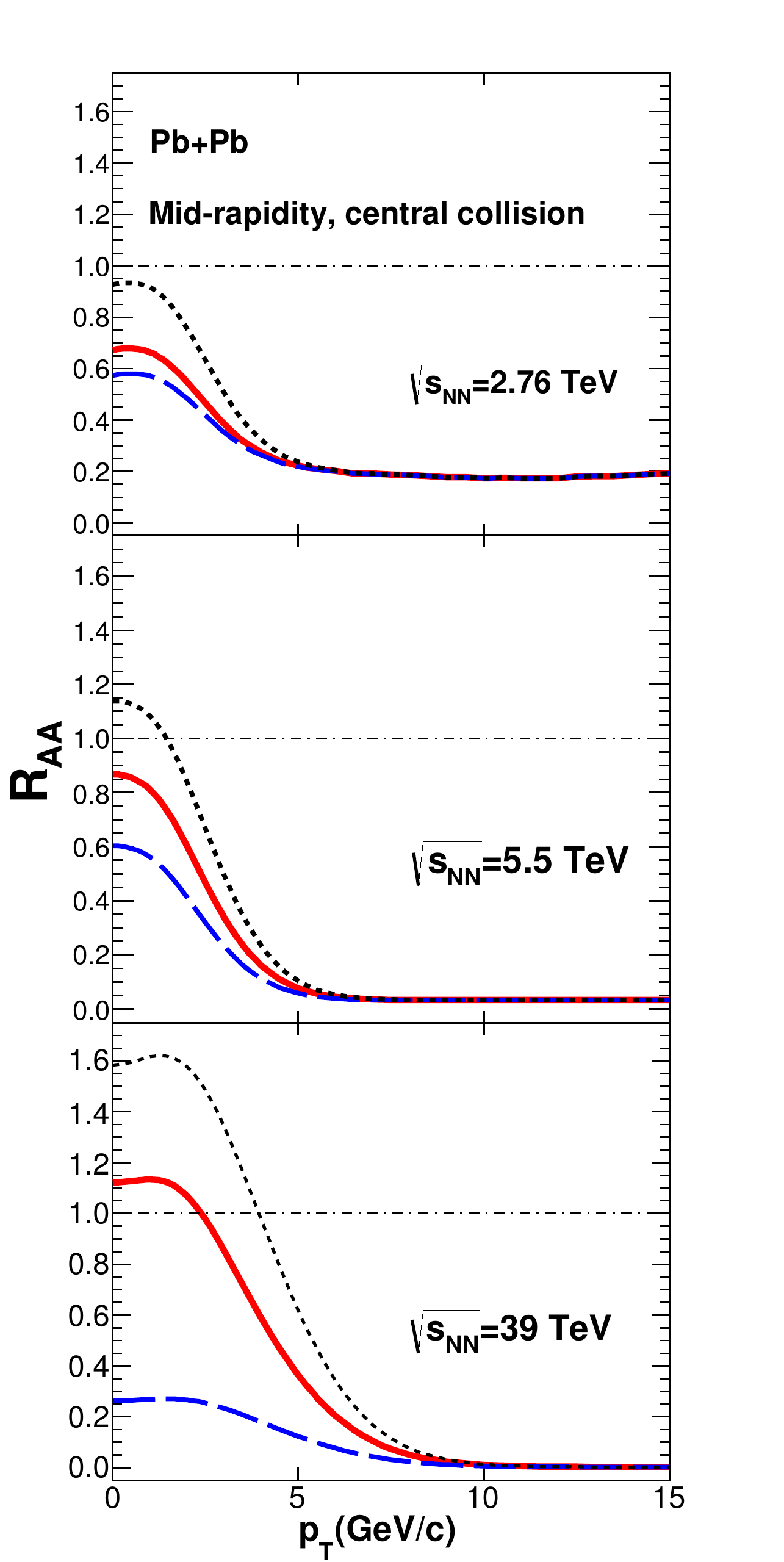}$$
		\caption{The time evolution of the charm quark pair number per unit of rapidity (left panel) and the $J/\psi$ differential nuclear modification factor $R_{AA}$ (right panel) in central Pb+Pb collisions at LHC and FCC energies. The solid and dashed lines are the calculations with and without charm quark thermal production, and the dotted lines represent the calculation without shadowing effect. The figures are taken from Ref.~\cite{Zhou:2016wbo}.}
		\label{fig11}}
\end{figure}

The cross section of heavy quark thermal production can be calculated via pQCD. To the next to leading order with QCD coupling constant $\alpha_s(m_c)$ and renormalization scale $\mu=m_c$, one obtains the thermal production cross section $\sigma_{gain}$~\cite{Zhang:2007dm}. By taking into account the detailed balance we then get the cross section $\sigma_{loss}$ for the inverse process. When the thermal production is included, the charm conservation in the medium is expressed as
\begin{equation}
\partial_\mu(n_{c\bar c}u^\mu)=r_{gain}-r_{loss},
\end{equation}
where $n_{c\bar c}$ is the charm quark pair density, and the loss and gain terms are expressed in terms of the cross sections $\sigma_{gain}$ and $\sigma_{loss}$ respectively. With Lorentz covariant variables $\eta$ and $\tau$, the conservation can be written as,
\begin{equation}
{1\over \cosh \eta}\partial_\tau n_{c\bar c}+\nabla \cdot(n_{c\bar c}{\bf v}_T)+{1\over \tau\cosh \eta}n_{c\bar c} = r_{gain}-r_{loss}.
\end{equation}
Assuming that the longitudinal motion of charm quarks satisfies the Bjorken expansion law in the mid-rapidity region, the charm quark pair density in the transverse plane $\rho_{c\bar c}$ defined by $n_{c\bar c}=\rho_{c\bar c}/\tau$ is controlled by the reduced rate equation,
\begin{equation}
{1\over \cosh \eta}\partial_\tau \rho_{c\bar c}+\nabla \cdot(\rho_{c\bar c}{\bf v}_T) = \tau(r_{gain}-r_{loss}).
\end{equation}

Combining this equation with the hydrodynamics to determine the fluid velocity $u_\mu$ and temperature $T$, the $c\bar c$ pairs produced in heavy ion collisions in central Pb+Pb collisions at $\sqrt{s_{NN}}=5.5$ TeV and $=39$ TeV are shown in Fig.~\ref{fig11}. We can see that at $5.5$ TeV, the number of thermally produced $c\bar c$ pairs, the difference between the solid and dashed lines, is much smaller compared with the initial production. However, the thermal production at FCC is almost as larger as the initial production~\cite{Zhang:2007dm,Zhou:2016wbo,Liu:2016zle}.

The extra charm quark pairs via the thermal production in the QGP will obviously enhance the charmonium yield at FCC~\cite{Zhou:2016wbo}. As shown in the right panel of Fig.~\ref{fig11}, the nuclear modification factor $R_{AA}$ at FCC is much larger than that at LHC energy and even larger than unit at low $p_T$.

\section{Open heavy flavors in high energy nuclear collisions}
\label{open_collision}
Heavy quarks have advantages to probe the QGP produced in high energy nuclear collisions. In this section, we discuss the energy loss mechanism of heavy quarks in hot medium and summarize various models on energy loss and transport approaches on heavy quark motion. The heavy quark energy loss is related to probing the QGP medium and the hadronization mechanism. We will compare different hadronization models and calculate the yield of multi-charmed baryons in heavy ion collisions. We will see that the production probability of the multi-charmed baryons is dramatically enhanced in A+A collisions in comparison with p+p collisions.

\subsection{Transport models with energy loss}
\label{transport}
The heavy quark interaction with hot QCD medium can be separated into perturbative and non-perturbative calculations. The perturbative processes, based on the assumption of weak interaction between heavy quarks and medium partons, can be divided into two parts according to the scattering diagram: the elastic collision and radiation. It is easy to understand that the perturbative calculation is not safe for a realistic application in studying a strongly coupled QGP. It is very hard to use a pure perturbative treatment to explain the experimentally measured nuclear modification factor $R_{AA}$ and elliptic flow $v_2$ for open heavy flavors. It is necessary for us to develop non-perturbative approaches. With the non-perturbatively calculated energy loss terms, one can use a transport approach to describe the evolution of heavy quarks in hot medium.

\subsubsection{Collisional energy loss}
\label{collision}
Elastic collisions would contribute to the lowest order to heavy quark scatterings off thermal partons in perturbative QCD. At leading order, heavy quarks interact with QGP medium partons through the following elastic scatterings: $Q+g\to Q+g$ and $Q+q(\bar q)\to Q+q(\bar q)$. Note that, the scattering off light quark or antiquark occurs only through $t$-channel process. While $Q+g$ scattering includes contribution from $s-$ and $u-$channels (like Compton scattering), the $t-$channel still dominates the scattering cross section. People usually dub the energy loss induced by these binary scatterings as collisional energy loss. It is often intuitive to calculate the energy loss per unit length $dE/dx$ to characterize the medium quenching effects on heavy quark motion,
\begin{equation}
{dE\over dx}={1\over 2E_Qv}\int_{p',k,k'} {1\over d_Q}\sum |\mathcal{M}|^2(2\pi)^4 \delta^4(p+k-p'-k')f_m(k)(1\pm f_m(k'))\omega,
\end{equation}
where $\omega=E_Q-E'_Q$ denotes the energy loss of the heavy quark in one collision, $v$ is the heavy quark velocity, and the tree-level amplitude square $|\mathcal{M}|^2$ is summed over the final spin states and averaged over the initial spin states~\cite{Combridge:1978kx}. The infrared divergence of the two $t-$channel processes are regulated by introducing a Debye screening mass $m_D=gT$ into the exchanged gluon propagator,
 \begin{equation}
|\mathcal{M}|^2\propto {\alpha_s^2 \over (t-m_D^2)^2}.
\end{equation}
Different from the above kinetic calculation, Thoma and Gyulassy (TG)~\cite{Thoma:1990fm} calculated the heavy quark collisional energy loss using techniques of classical plasma physics together with Hard-Thermal-Loop (HTL) corrected gluon propagator from thermal QCD. The screening effect on the infrared singularities from long-range Coulomb and magnetic interactions is automatically provided by the plasma effects in this formalism, but still, the choice of the upper limit $q_{max}$ suffers from ambiguity. Furthermore, the recoil effect which happens for large momentum transfer is not allowed in the calculation. Thus the calculation breaks down at some ultraviolet scale and becomes incomplete at hard momentum. This motivated Braaten and Thoma (BT) to take recourse of the resummation method~\cite{Braaten:1991jj,Braaten:1991we,Braaten:1989kk} to give a complete calculation on heavy quark collisional energy loss at leading order in $g$.

The momentum transfer in elastic scattering is separated into the hard and soft parts by an arbitrary intermediate scale $q^*$ which is chosen to satisfy $m_D\ll q^* \ll T$, with the Debye screening mass $m_D\sim gT$. Note that this condition requires a very small coupling constant to ensure $m_D\ll T$. The calculation is done within a microscopic kinetic based formalism, and the tree-level propagator for the exchanged gluon is used, where the infrared divergence in the $t-$channel is cut off by the lower limit $q^*$ in the integration. For the soft region $q <q^*$, which is related to frequent distant collisions, since the de Broglie wave length of the exchanged gluon is comparable with the plasma constituents' inter-distance, a microscopic kinetic description based on individual collisions is not applicable due to the absence of plasma effects. Inspired by TG's treatment~\cite{Thoma:1990fm}, the soft part in BT's calculation~\cite{Braaten:1991we} is evaluated via imposing the HTL resummed gluon propagator to self-consistently incorporate the screening of infrared divergence. After adding up the hard and soft contributions together, the dependence of the total result on the intermediate scale $q^*$ is cancelled automatically. The resulted energy loss for heavy quarks is~\cite{Braaten:1991we}
\begin{equation}
{dE\over dx} ={8\pi \alpha_s^2 T^2 \over 3}\left(1+{n_f \over 6}\right)\left[{1\over v}-{1-v^2\over 2v^2}\ln {1+v\over 1-v} \right]\ln\left(2^{n_f/(6+n_f)}B(v){ET\over m_gm_Q}\right)
\end{equation}
for $E\ll m_Q^2/T$ with $B(v)$ being a smooth function of heavy quark velocity $v$, and
\begin{equation}
{dE\over dx} ={8\pi \alpha_s^2 T^2 \over 3}\left(1+{n_f \over 6}\right)\ln\left(2^{n_f/2(6+n_f)}B(v)0.92{ET\over m_gm_Q}\right)
\end{equation}
for $E\gg m_Q^2/T$ with $m_g=(gT/\sqrt{3})\sqrt{1+n_f/6}$ being the thermal gluon mass, where $n_f$ is the flavor number.

The above calculation is at leading order. When extending to the next to leading order, Peigne and Peshier (PP) proposed to use the intermediate scale $t^*=\omega^2-q^2$ to correct the BT calculation for the hard part with $E\gg m^2_Q/T$ by going beyond the leading logarithmic accuracy~\cite{Peigne:2008nd}. They considered the running of the strong coupling to further improve the calculation of heavy quark collisional energy loss. Introducing running coupling in perturbative calculation can effectively account for higher-order corrections even in the non-perturbative regime in some cases~\cite{Peshier:2008zz}. The running coupling calculation gives
\begin{equation}
{dE\over dx} ={4\pi T^2 \over 3}\alpha_s(m_D^2)\alpha_s(ET) \left[(1+{n_f\over 6})\ln {ET\over m_D^2}+{2\over 9}{\alpha_s(m_Q^2)\over \alpha_s(m_D^2)}\ln {ET\over m_Q^2}+c(n_f)+... \right],
\end{equation}
where the Debye screening mass is self-consistently derived with the running coupling $m_D^2=(N_c/3)(1+n_f/6)4\pi \alpha(m_D^2)T^2$.

\subsubsection{Radiative energy loss}
\label{radiation}
It is widely believed that elastic scatterings dominate the energy loss for very slowly moving heavy quarks~\cite{Moore:2004tg}. For very fast heavy quarks, the radiative process---gluon bremsstrahlung---is commonly considered to win over the collisional energy loss mechanism. For the intermediate momentum region, both mechanisms would contribute to heavy quark energy loss. It is still an open question on the relative importance of the two mechanisms with respect to heavy quark momentum and medium temperature. In QGP, the heavy quark radiative energy loss proceeds through the following $2\to3$ scatterings: $Q+g\to Q+g+g$ and $Q+q(\bar q)\to Q+q(\bar q)+g$. The $Q+q$ scattering includes 5 Feynman diagrams, while the $Q+g$ scattering contains 16 diagrams. By a naive vertices counting, the matrix elements are proportional to $\alpha_s^3$ or $g^6$. It's argued in Ref.~\cite{Braaten:1991we} that the screening from the hot and dense medium at the scale $gT$ would in principle tame the divergence appeared in tree-level radiation and render the contribution being of the same order in coupling as the collisional one.

For the above $2\to3$ gluon bremsstrahlung processes, the exact matrix elements for light quark process $qq'\to qq'g$ has already been calculated~\cite{Berends:1981rb}, see also Ref.~\cite{Ellis:1985er} where the matrix elements for all $2\to3$ parton scattering processes are presented in $n$ dimensions. While the exact matrix elements for $Qq\to Qqg$ can be obtained from the calculation for $q\bar q\to Q\bar Qg$~\cite{Kunszt:1979iy} by crossing the two antiquarks, it is rather difficult to extend these expressions from vacuum to hot and dense medium. People thus usually used the Gunion-Bertsch (GB) formula~\cite{Gunion:1981qs} to approximate the matrix elements and calculate the cross section, which assumes that the induced gluon bremsstrahlung is associated with a single isolated collision and is factorized into elastic scattering multiplied by a gluon radiative probability. The spectrum of the soft gluon emission in pQCD in high energy limit is derived to be
\begin{equation}
{d\sigma^{2\to 3}\over d^2q_\bot dyd^2k_\bot}\approx{d\sigma^{2\to 2}\over d^2q_\bot}{dn_g \over dyd^2k_\bot}\approx {d\sigma^{2\to 2}\over d^2q_\bot}\cdot {C_A\alpha_s \over \pi^2}{q^2_\bot \over k_\bot^2({\bf k}_\bot-{\bf q}_\bot)},
\label{gemission}
\end{equation}
where $q_\bot$ and $k_\bot$ are transverse momenta of the exchanged and emitted gluon respectively, and $y$ is the longitudinal rapidity of the emitted gluon. The cross sections from both exact calculation and GB matrix elements are divergent for infrared and collinear configurations which can be seen from Eq.\ref{gemission}. They can be cured in thermal field theory by loop resummation.

At finite temperature, besides the Debye screening effect, a consideration about time scales brings it encountering the Landau-Pomeranchuk-Migdal effect (LPM)~\cite{Landau:1953um,Migdal:1956tc}, each radiated gluon needs some duration to be formed, named formation time $\tau_f$. If this formation time is larger than the scattering time or mean free path $\lambda$, multiple scattering would happen during one radiation, and the interference of scattering amplitudes will lead to a suppression of the gluon radiative process. The LPM effect is considered through multiplying the gluon radiative spectrum by a step function $\theta(\lambda-\tau_f)$. A lot of approaches have been developed to describe the radiative energy loss, like GLV (also called reaction operator approach)~\cite{Gyulassy:2000fs,Gyulassy:2000er}, Higher Twist (HT)~\cite{Guo:2000nz,Wang:2001ifa}, ASW~\cite{Wiedemann:2000za,Salgado:2003gb}, AMY~\cite{Arnold:2002ja,Arnold:2002zm} and so on.
\begin{figure}[!htb]
{$$\includegraphics[width=0.31\textwidth]{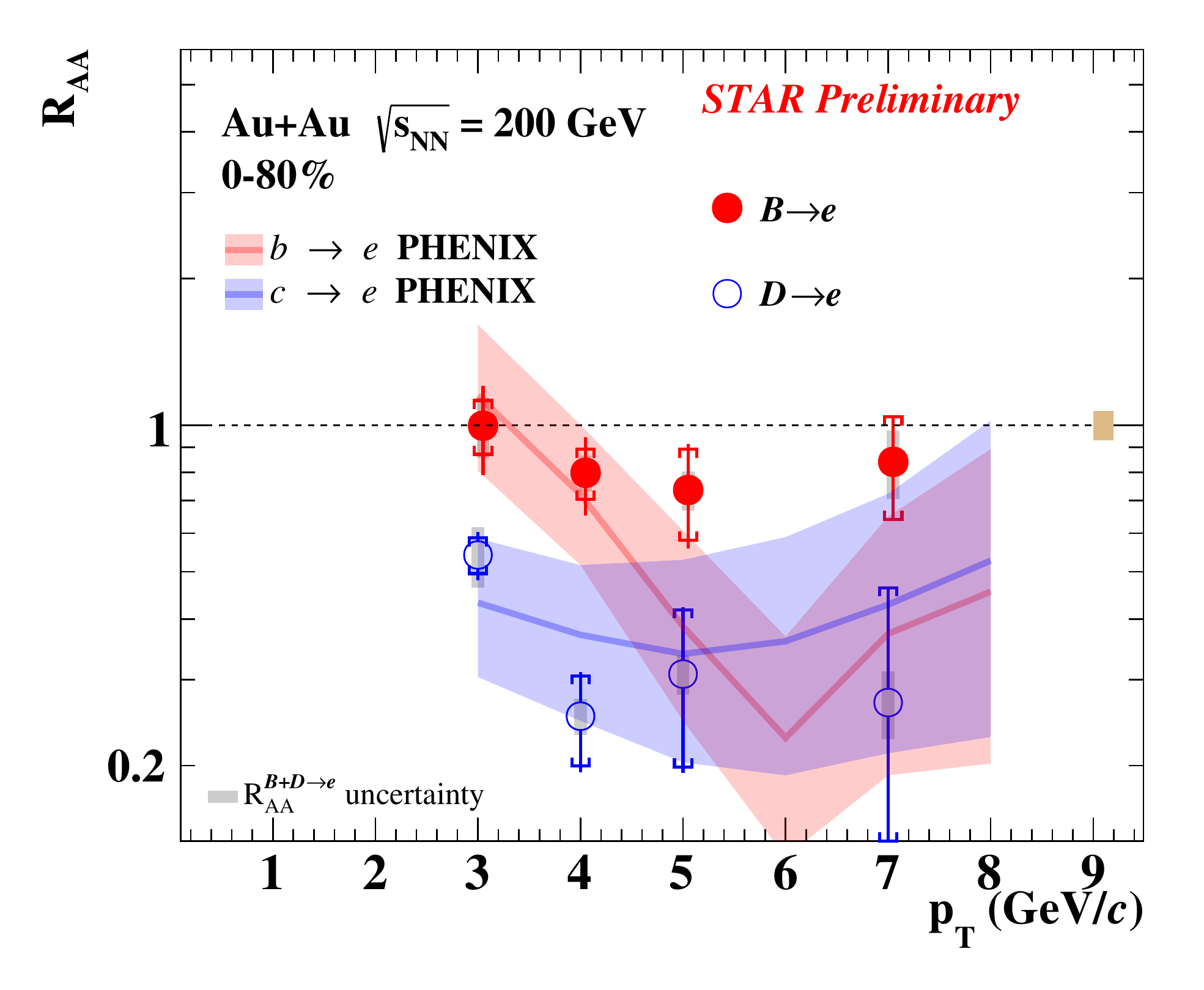} \ \ \ \ \includegraphics[width=0.26\textwidth]{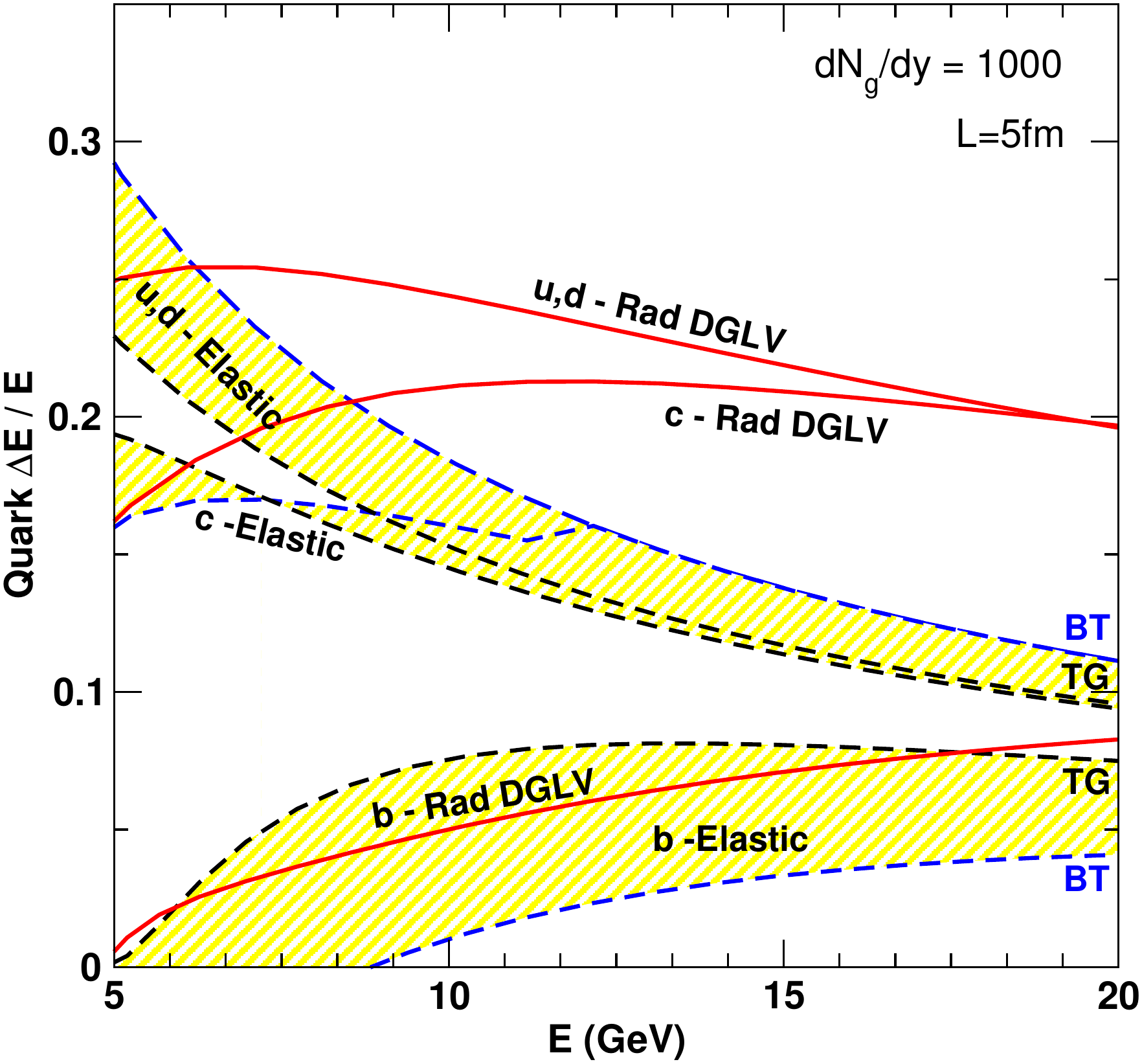}$$
\caption{ The experimentally measured $R_{AA}$ of charm and bottom separated single electrons in Au+Au collisions (left panel) and model calculated average energy loss $\Delta E/E$ for $u, d$ and $c, b$ quarks as functions of quark energy (right panel). The Bjorken expansion is used and the quark path length in QGP is assumed to be $L=5$ fm. The figures are taken from Refs.~\cite{Wicks:2005gt,Dong:2019byy}.}
\label{fig12}}
\end{figure}

In heavy quark sector, the first estimate of radiative energy loss in the QGP medium was made by Mustafa~\cite{Mustafa:1997pm}. For heavy quarks, there is an important effect named "dead cone" effect. It means that, the heavy quark mass suppresses the soft gluon radiation off heavy quarks with energy E in forward-direction within angle smaller than $\theta_D = m_Q/E$. Subsequently, the gluon bremsstrahlung of heavy quark energy loss is suppressed by a factor of $\mathcal{D}_{DK} = (1 + \theta_D^2 /\theta^2)^{-2}$ compared to light parton energy loss, and the enhancement of heavy-to-light hadron ratio is predicted in high transverse momentum regime~\cite{Dokshitzer:2001zm}. Based on the GLV opacity expansion, DGLV~\cite{Djordjevic:2003zk} derived the heavy quark energy loss due to medium-induced gluon radiation to all orders in opacity. To first order in opacity, the heavy quark total radiative energy loss including the plasmon asymptotic mass is
\begin{equation}
{\Delta E \over E}=\int dx d^2k_\bot x {dn_g \over dxd^2k_\bot},
\end{equation}
where $ xdn_g / (dxd^2k_\bot)$ is the radiative spectrum, given in~\cite{Djordjevic:2003zk}.

The improved DGLV calculation, including dynamical QCD medium effects in heavy quark radiative energy loss, was proposed by Djordjevic and Heinz (DH)~\cite{Djordjevic:2008iz}, it shows almost twice enhancement compared with previous calculation. An improved GB formalism with dead cone suppression factor is employed to estimate the heavy quark radiative energy loss, as shown in~\cite{Fochler:2013epa,Abir:2011jb} which gives a very compact and elegant expression for the gluon radiative spectrum of a heavy quark,
\begin{equation}
{dn_g \over d\eta d^2k_\bot}={C_A \alpha_s \over \pi}{1\over k_\bot^2}\left(1+{m_Q^2\over s\tan^2(\theta/2)}\right)^{-2}
\end{equation}
with the dead cone suppression factor $\mathcal{D}_{DK}=(1+{m_Q^2/s\tan^2(\theta/2)})^{-2}$ and $s=2E^2+2E\sqrt{E^2-m_Q^2}-m_Q^2$. The result is significantly different from the DGLV.

A comparison between radiative energy loss with collisional energy loss from pQCD calculation is made in Refs.~\cite{Mustafa:2004dr,Wicks:2005gt} and shown in Fig.\ref{fig12}. The radiative DGLV first-order energy loss is compared with the elastic collisional energy loss in TG or BT approximations. The yellow bands provide an indication of theoretical uncertainties in the leading log approximation to the elastic energy loss. We can see that, the elastic energy loss is comparable with radiative one up to $E\sim 10\ (20)$ GeV for charm and bottom quarks in a gluon plasma with initial density $dn_g/dy=1000$. Due to the mass-dependent energy loss effect~\cite{Zhang:2003wk,Armesto:2005iq}, both radiative and elastic energy losses for bottom quarks are significantly smaller than that for light and charm quarks, and the elastic and inelastic energy losses have almost the same magnitude in the whole energy region. Such a mass-dependent energy loss is observed at RHIC~\cite{Dong:2019byy} and LHC~\cite{Sirunyan:2018ktu}, as shown in Fig.\ref{fig12} with large uncertainties.

\subsubsection{Transport description}
\label{transport2}
We start with the Boltzmann equation for heavy quark distribution function $f_Q(t,{\bf x},{\bf p})$ in phase space,
\begin{equation}
\left[ {\partial \over \partial t} + {{\bf p}\over E}{\partial \over \partial {\bf x}}+{\bf F}\cdot {\partial \over \partial {\bf p}}\right] f_Q(t,{\bf x}, {\bf p}) =C[f_Q].
\end{equation}
On the left-hand side, the first two terms represent the space-time drifting of heavy quarks, and ${\bf F}$ denotes the external force executing on heavy quarks and its effect is relevant only when there are chromo electric and magnetic fields existed around heavy quarks. Usually the force term is neglected if not specified. On the right-hand side of the Boltzmann equation, $C[f_Q]$ is the collision term describing the interaction of heavy quarks with the medium degrees of freedom (thermal partons). It, in principle, can be written as a sum of different kinds of scattering processes,
\begin{equation}
C[f_Q]=C_{12}[f_Q]+C_{22}[f_Q]+C_{23}[f_Q]+...
\end{equation}
For the two-body scattering of a heavy quark off a thermal medium parton, the collision term is given by
\begin{eqnarray}
C_{22}[f_Q]&=&{1\over 2E_Q}\int_{k,k',p'}{1\over d_Q}\sum |\mathcal{M}_{22}|^2(2\pi)^4\delta^4(p+k-p'-k')\nonumber\\
&&\times\left[f_Q(p')f_q(k')(1-f_Q(p))(1\pm f_q(k))-f_Q(p)f_q(k)(1-f_Q(p'))(1\pm f_q(k'))\right]
\end{eqnarray}
where $\int_p$ is the shorthand for the standard invariant integration measure $\int {d^3{\bf p} \over (2\pi)^32E}$, $p, p', k, k'$ are the heavy quark and medium parton 4-momenta before and after the collision, $d_Q=6$ is the degeneracy of the heavy quark, $\mathcal{M}_{22}$ is the two-body scattering amplitude for the process $Q(p)+q(k)\to Q(p')+q(k')$, and $f_q$ is the phase space density of thermal partons. Quantum effects are included by associating the Fermi blocking or Bose enhancement factors $(1\pm f_q)$ with final states.

In principle, with the knowledge of medium information (phase space distribution of light partons in the medium) and also interaction matrix element $\mathcal{M}$, the space-time evolution for heavy quark distribution $f_Q$ can be fully determined by the Boltzmann equation for any given initial condition. From the structure of the collision integral, it is obvious that, in a static thermal bath with temperature $T$ the heavy quark distribution will reach thermal equilibrium in the long-time limit, $f_Q = d_Q\exp(-E_Q/T)$. The radiative processes can also be included in the collision term under local interaction assumption if the matrix element $\mathcal{M}_{23}$ can be given, but it's not easy compared to the elastic collisions due to the inherence induced interference. When the coupling is small, it's known that only for very fast particles the bremsstrahlung dominates the energy loss, while for moderately relativistic particles elastic collisions take over the energy loss~\cite{Moore:2004tg}.

When we focus on the modification of heavy quark distribution, it is convenient to make a momentum transfer ${\bf q=p-p'=k'-k}$ explicitly in the collision integral,
\begin{equation}
C_{22}[f_Q]=\int d^3{\bf q}\left[\omega({\bf p}+{\bf q},{\bf q})f_Q({\bf p}+{\bf q})-\omega({\bf p},{\bf q})f_Q({\bf p})\right]
\end{equation}
where $\omega({\bf p},{\bf q})$ is defined as the transition rate of the collisions. The transfer changes heavy quark momentum from ${\bf p}$ to ${\bf p-q}$, and the two terms in the bracket represent respectively the gain and loss probability. The transition rate $\omega({\bf p},{\bf q})$ can be related to the scattering matrix element,
\begin{equation}
\omega({\bf p},{\bf q})={1\over 2(2\pi)^2}\int_k{f_q(k)(1\pm f_q(k'))\over d_Q} { |\mathcal{M}_{22}|^2 \delta^0(E_Q+E_q-E'_Q-E'_q) \over E'_Q E'_q},
\end{equation}
where $E_q(k)$ is the energy of medium partons with momentum $k$.

For the above transport equation, it is very hard to solve it directly due to the integro-differential nature. One simple way to handle it is to use Relaxation Time Approximation (RTA) for the collision term and get a semi-analytical solution, see Ref.~\cite{Yan:2007zze}, from which one can get some feeling about the heavy quark relaxation rate by confronting to the
experimental data. An indirect way to solve the Boltzmann equation is through Monte Carlo cascade based on test particle ansatz for the distribution $f(t,{\bf x,p})\approx \sum_i \delta^3({\bf x-x_i}(t))\delta^3({\bf p-p_i}(t))$. The scattering processes can proceed using Monte Carlo sampling according to a scattering probability which can be deduced from the collision integral. For this kind of cascade simulation, see Refs.~\cite{Xu:2004mz,Molnar:2004ph,Zhang:2005ni}.

The heavy quark motion in hot medium can also be described by the Fokker-Planck Equation. The large mass $m_Q\gg T$ would get heavy quarks very likely to undergo a Brownian motion in the hot QCD medium. For thermal heavy quarks with typical thermal momentum $p\sim\sqrt{m_QT}$, the momentum transfer $q\sim T$ is softer with respect to heavy quark momentum ${\bf q^2\ll p^2}$, and the energy transfer is further suppressed with respect to the momentum transfer. Consequently, it would take many $(\sim p/T)$ collisions to change the heavy quark momentum by a factor of order one, which is just the characteristic of Brownian motion.

If the momentum transfer is small in one collision, we can expand the distribution $f_Q$ and transition rate $\omega$ in terms of ${\bf q}$, which leads to~\cite{Rapp:2009my}
\begin{equation}
\omega({\bf p}+{\bf q},{\bf q})f_Q({\bf p}+{\bf q})\approx \omega({\bf p},{\bf q})f_Q({\bf p})+{\bf q}\cdot {\partial \over \partial {\bf p}}\left[\omega({\bf p},{\bf q})f_Q({\bf p})\right]+{1\over2}q_iq_j{\partial^2 \over \partial p_i\partial p_j}\left[\omega({\bf p},{\bf q})f_Q({\bf p})\right]
\end{equation}
with $i,j=1,2,3$ denoting the three components of the momenta. Under this expansion, the collision term becomes
\begin{equation}
C[f_Q(t,{\bf p})]\approx \int d^3{\bf q} \left[ q_i{\partial \over \partial p_i} + {1\over2}q_iq_j{\partial^2 \over \partial p_i\partial p_j}\right]\omega({\bf p},{\bf q})f_Q({\bf p}),
\end{equation}
and the original Boltzmann equation is approximately expressed in the form of the Fokker-Planck equation,
\begin{equation}
{\partial \over \partial t}f_Q(t,{\bf p})={\partial \over \partial p_i} \left \{ A_i({\bf p})f_Q(t,{\bf p})+{\partial \over \partial p_j}\left[B_{ij}({\bf p})f_Q(t,{\bf p})\right] \right \}
\end{equation}
with the drag and diffusion coefficients
\begin{eqnarray}
A_i({\bf p})&=&\int d^3{\bf q} \omega({\bf p,q})q_i,\nonumber\\
B_{ij}({\bf p})&=&{1\over 2}\int d^3 {\bf q} \omega({\bf p,q})q_iq_j.
\end{eqnarray}
After many collisions with soft particles, the heavy quark will approach to the equilibrium distribution, given by the relativistic Boltzmann-Juttner distribution. This condition leads to the dissipation-fluctuation relation between drag and diffusion coefficients
 \begin{equation}
A_i({\bf p},T)=B_{ij}({\bf p},T){1\over T}{\partial E(p) \over \partial p_i}-{\partial B_{ij}({\bf p},T) \over \partial p_j}.
\end{equation}

Assuming that the background medium is in local equilibrium, $A_i$ and $B_{ij}$ depend only on the momentum ${\bf p}$ due to rotational symmetry in the local rest frame of the heat bath. Therefore, we can decompose them to
 \begin{eqnarray}
A_i({\bf p})&=&A(p)p_i,\nonumber\\
B_{ij}({\bf p})&=&B_0(p)P_{ij}^\bot({\bf p})+B_1(p)P_{ij}^\parallel({\bf p})
\end{eqnarray}
with the transverse and longitudinal projection operators $P_{ij}^\bot({\bf p})=\delta_{ij}-p_ip_j/{\bf p}^2$ and $P_{ij}^\parallel({\bf p})=p_ip_j/{\bf p}^2$.
The Fokker-Planck equation description of diffusion can be realized by the Langevin equation,
 \begin{eqnarray}
dx_j&=&{p_j\over E}dt,\nonumber\\
dp_j&=&-\Gamma(p,T)p_jdt+\sqrt{dt}C_{jk}({\bf p}+\xi d{\bf p},T)\rho_k,
\end{eqnarray}
where $\xi =0,1/2,1$ in stochastic process correspond to the pre-point Ito, the mid-point Stratonovic, and the post-point Ito realizations. From the relation between the Fokker-Planck equation and Langevin equation~\cite{Rapp:2009my}, we have
 \begin{eqnarray}
&&C_{jk}(p)=\sqrt{2B_0(p)}P_{jk}^\bot({\bf p})+\sqrt{2B_1(p)}P_{jk}^\parallel({\bf p}), \nonumber\\
&&\Gamma(p) p_j = A(p)p_j-\xi C_{lk}(p){\partial C_{jk}(p) \over \partial p_l}.
\label{eeq}
\end{eqnarray}
With a diagonal approximation of the diffusion coefficient, $B_0(p)=B_1(p)=D(p)$, and with the dissipation-fluctuation relation, Eq.\ref{eeq} is rewritten as  
\begin{eqnarray}
C_{jk}(p)&=&\sqrt{2D(p)}\delta_{jk}, \nonumber\\
\Gamma(p,T)&=&{1\over E}\left ({D(p)\over T}-(1-\xi){\partial D(p) \over \partial E} \right),
\end{eqnarray}
and the Fokker-Planck equation becomes
 \begin{eqnarray}
&&dx_j ={p_j\over E}dt,\nonumber\\
&&dp_j=-\Gamma(p,T)p_jdt+\sqrt{2dtD(|p+\xi dp|)}\rho_k.
\end{eqnarray}
In non-relativistic limit, both $D(p)=D$ and $\Gamma(p)=\gamma$ are momentum independent, and we have the limit $E\to m_Q$ and the Einstein's classical fluctuation-dissipation relation $D=\gamma m_Q T$, where $D$ is the momentum diffusion coefficient and related to the spatial diffusion coefficient $D_s$ via
\begin{equation}
D_s={T\over m_Q \gamma}={T^2\over D}
\end{equation}
or $D_s=T/(m_Q A(p=0,T))$.

The energy loss $dE/dx$ is related to the heavy quark drag coefficient. In last subsection we reviewed the perturbative calculations on heavy quark interaction in hot QCD medium. As a default setup, the background QGP medium is assumed to consist of weakly interacting quasiparticles (quarks and gluons). In phenomenological models, the information is treated to be in thermal equilibrium state from hydrodynamical simulation or non-equilibrium state from transport cascade simulation.

The NLO corrections~\cite{CaronHuot:2007gq} to heavy quark momentum diffusion in perturbative framework is found to be not convergent. This implies that the perturbative calculation about heavy quark motion is not safe for a realistic application in heavy ion collisions, and one should consider non-perturbative approaches. From the phenomenological investigation confronting experimental data, it is very hard to simultaneously explain the heavy quark quenching measured through $R_{AA}$ and the anisotropic flow $v_2$ observed at RHIC~\cite{Abelev:2006db,Adare:2006nq} and LHC~\cite{ALICE:2012ab} in the frame of perturbative treatment of heavy quark interaction with the medium. Going beyond the perturbative approaches, there are many effective models to calculate the drag coefficients, such as the T-matrix approach~\cite{Riek:2010fk,He:2014cla}, AdS/CFT~\cite{Berends:1981rb}, Dyson-Schwinger equation~\cite{Qin:2014dqa}, and quasi-particle model (QPM)~\cite{Berrehrah:2013mua,Das:2012ck}. The spectral function in low frequency regime may be related to the heavy quark diffusion coefficient via linear-response theory~\cite{Petreczky:2005nh}. The first principle calculation such as lattice QCD~\cite{Banerjee:2011ra,Ding:2015ona,Francis:2015daa} makes also some progress and gives some prediction.

\begin{figure}[!htb]
{$$\includegraphics[width=0.4\textwidth]{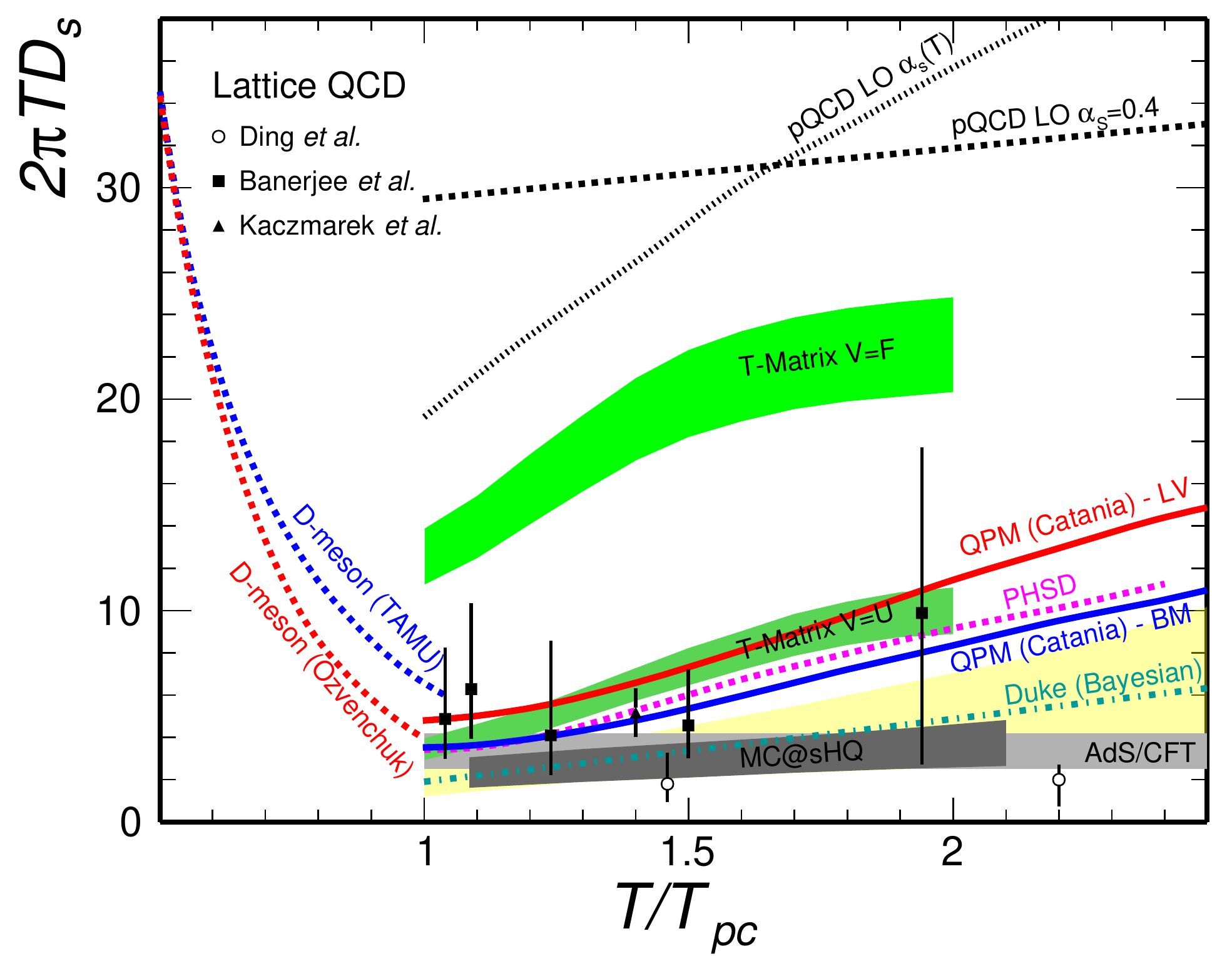} \ ~ \includegraphics[width=0.39\textwidth]{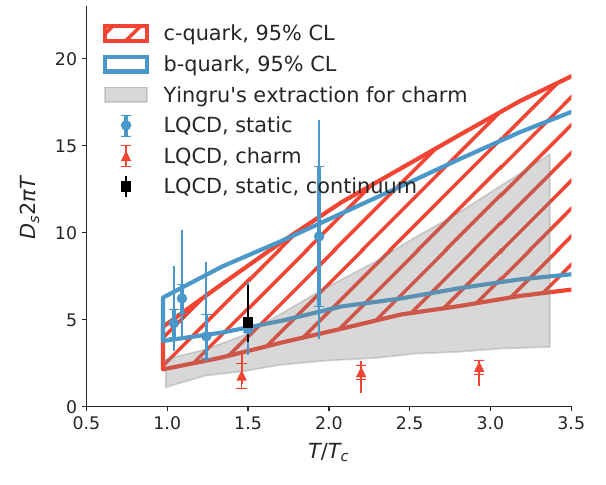}$$
\caption{ The spatial diffusion coefficient $D_s$ (in units of 2$\pi T$) as a function of temperature from different theoretical models and lattice QCD. The figures are taken from Refs.~\cite{Dong:2019byy,Ke:2018tsh}.}
\label{fig13}}
\end{figure}

The temperature-dependent spatial diffusion coefficient $D_s$ as shown in Fig.\ref{fig13}, including lattice QCD results, the leading order pQCD calculation with temperature dependent and independent coupling $\alpha_s$~\cite{Moore:2004tg,vanHees:2004gq}, the QPM calculation with Boltzmann and Langevin dynamics~\cite{Das:2015ana}, the PHSD transport calculation based on a dynamical QPM~\cite{Song:2015sfa} and MC@sHQ perturbative approach~\cite{Andronic:2015wma}. The recent development of Bayesian analysis fit based on the Duke Langevin model~\cite{Xu:2017obm} and Duke Linearized Boltzmann-Langevin mode~\cite{Ke:2018tsh} gives also the constraint on spatial diffusion coefficient $D_s$. The $D_s$ for D-mesons in hadronic matter~\cite{He:2011yi,Tolos:2013kva} is also calculated with the TAMU model and effective theory.

For solving the Langevin equation, one usually performs a Monte Carlo simulation by taking the test particle ansatz. There are many models to describe heavy quark transport and hadronization in heavy ion collisions, for instance the ones based on the Boltzmann equation like the Linear Boltzmann Transport (LBT) model~\cite{Cao:2016gvr,He:2015pra} and the Catania quasi-particle Boltzmann approach~\cite{Das:2015ana,Scardina:2017ipo}, the ones based on the Langevin equation like the Duke model with the Langevin approach~\cite{Cao:2013ita,Cao:2015hia} and the Texas A\&M University (TAMU) model~\cite{He:2011yi,He:2011qa}, and the Frankfurt Parton Hadron String Dynamics (PHSD) approach~\cite{Cassing:2009vt,Bratkovskaya:2011wp} based on the Kadanoff-Baym (KB) equation. The results and difference among these models were discussed in the review papers~\cite{Cao:2018ews,Rapp:2018qla}.

\subsection{Statistical hadronization model}
\label{statis}
The initial energy density in heavy ion collisions at RHIC and LHC, estimated with the Bjorken model, has exceeded the critical value calculated by lattice gauge theory~\cite{Bjorken:1982qr}. It is believed that the deconfinement phase transition from hadron gas to QGP happens during the evolution of the colliding system. It is also expected that the system reaches local equilibrium at $\tau<1$ fm after collisions. The quark hadronization is a fundamental process in QCD, and due to its non-perturbative nature, it is still an open question. Different from the hadronization process in elementary collision systems, the statistics plays an important role in parton hadronization in heavy ion collisions. If the colliding system keeps thermal equilibrium until the hadronization stage, the hadrons measured in the final stage will resemble a thermal equilibrium population, and the yield of hadrons can be described via a statistical approach.

The basic quantity required to compute the thermal composition of particle yields measured in heavy ion collisions is the partition function $Z$ in Grand Canonical ensemble,
 \begin{equation}
Z(T,V,\mu)=\text{Tr}e^{-\beta(H-\sum_i\mu_i n_i)},
\end{equation}
where $H$ is the Hamiltonian of the hadron resonance gas, including all mesons and baryons, see the details in Ref.~\cite{BraunMunzinger:2003zd}, $T=1/\beta$ is the fireball temperature, and $\mu_i=(\mu_B,\mu_S,\mu_Q,\mu_C)$ are the chemical potentials to keep charge conservation. For particle $i$ with baryon number $B_i$, electric charge $Q_i$, strangeness $S_i$, charm $C_i$, and spin-isospin degeneracy factor $g_i$, the partition function can be expressed as
\begin{equation}
\ln Z_i ={Vg_i \over 2\pi^2}\int_0^\infty \pm p^2dp \ln[1\pm \exp(-(E_i-\mu_i)/T)]
\end{equation}
with $\mu_i=\mu_BB_i+\mu_SS_i+\mu_QQ_i+\mu_CC_i$, $E_i=\sqrt{p^2+m_i^2}$ and the signs $\pm$ for bosons and fermions. The corresponding particle density can be calculated as
\begin{equation}
n_i={N_i\over V}=-{T\over V}{\partial \ln Z \over \partial \mu}={g_i\over 2\pi^2}\int_0^\infty {p^2dp \over \exp[(E_i-\mu_i)/T]\pm1]}.
\end{equation}

In order to describe the hadrons produced in heavy ion collisions, one needs to consider the contribution from resonance decay. All parameters, temperature $T$ and chemical potentials $\mu_i$, are determined by fitting the experimental data.
\begin{figure}[!htb]
{$$\includegraphics[width=0.37\textwidth]{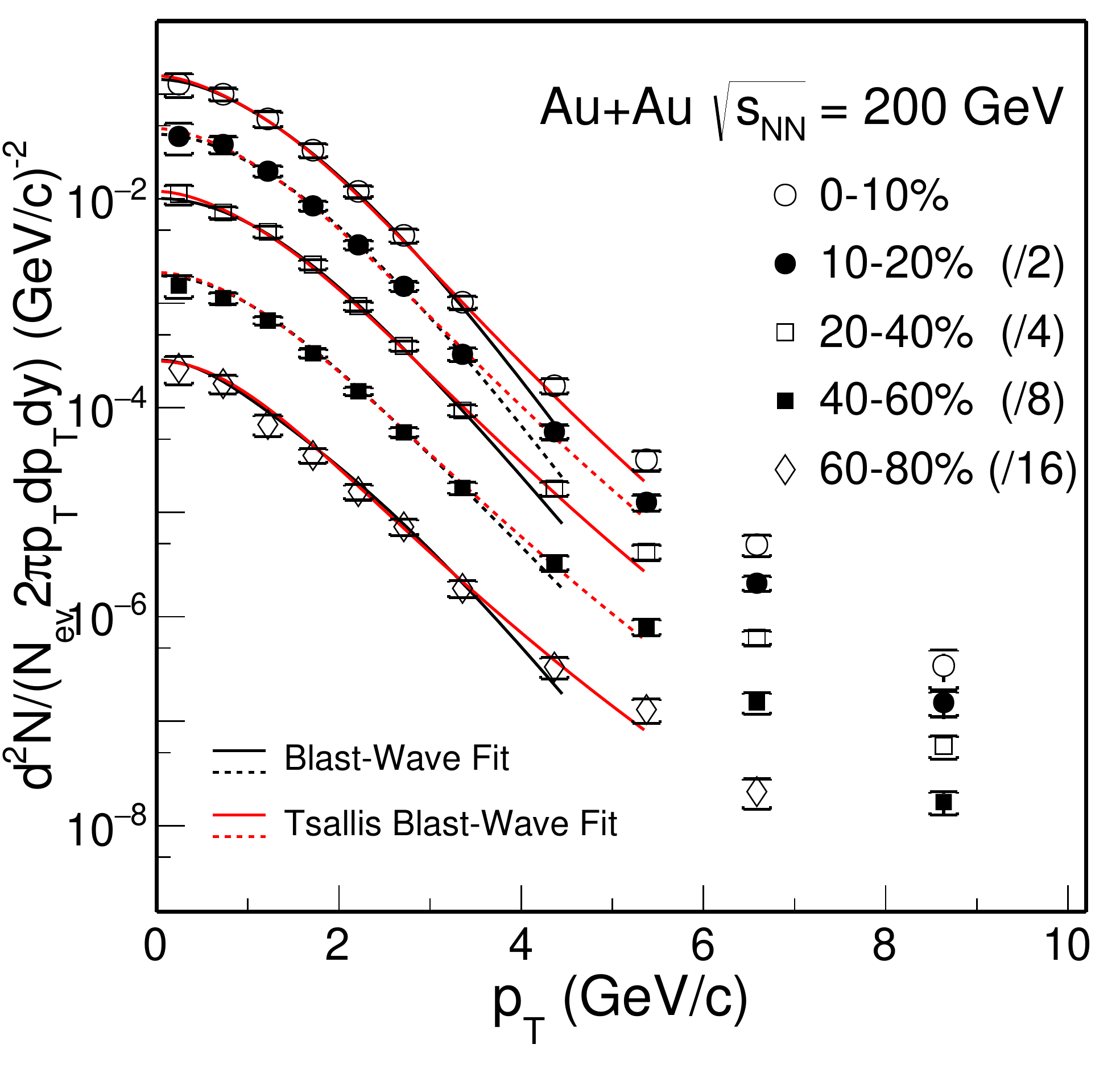}$$
\caption{ The experimentally measured $D^0$ yield at mid-rapidity as a function of transverse momentum for different centrality bins. The lines are from the Blast-wave model. The figure is taken from Ref.~\cite{Adam:2018inb}.}
\label{fig14}}
\end{figure}

For charm hadrons, the $c\bar c$ pairs are produced initially and the thermal production in QGP can be safely neglected at RHIC and LHC. In the statistical model, all hadrons freeze out at the phase boundary. Considering the charm quark number conservation during the hadronization, a factor $\gamma_c$ is introduced to guarantee the charm conservation~\cite{BraunMunzinger:2000px,Andronic:2003zv,BraunMunzinger:2007tn},
\begin{equation}
\label{cons}
N_{c\bar c}={1\over 2}\gamma_cV\sum_{single} n_i+\gamma_c^2V\sum_{double} n_i+...,
\end{equation}
where $\sum_{single} n_i$ means the summation over all singly charmed hadrons, and $\sum_{double}n_i$ means the summation over all doubly charmed hadrons. 

Aiming to study the transverse momentum spectrum of particles, the thermal statistical model was extended to the Blastwave model~\cite{Siemens:1978pb,Schnedermann:1993ws}. Different from the thermal model,  the medium is not static. The particle production can be described by the Cooper-Frye formula~\cite{Cooper:1974mv},
 \begin{equation}
E{dN\over d^3p}={g\over (2\pi)^3} \int d\Sigma_\mu(x)p^\mu f(x,p),
\end{equation}
where $g$ is the degeneracy factor. The particles produced on the freeze-out hypersurface $\Sigma_\mu$, which is fixed by the freeze-out temperature or the corresponding proper time $\tau_f$. The thermal distribution $f(x,p)=e^{-p^\mu u_\mu/T}$ is taken as the hadron distribution function with $p^\mu$ and $u_\mu$ being the hadron four-momentum and fluid cell velocity. For a boost invariant medium, instead of the variables $t, z$ and $p_z$, one can conveniently use the space-time rapidity $\eta=1/2\ln[(t+z)/(t-z)]$, proper time $\tau=\sqrt{t^2-r^2}$ and momentum rapidity $y=1/2\ln[(E+p_z)/(E-p_z)]$ to express
\begin{eqnarray}
 d\Sigma_\mu &=& \left(\cosh \eta, {\partial\tau_f\over \partial x}, {\partial \tau_f\over \partial y}, -\sinh \eta\right)\tau_f rdr d\eta d\phi, \nonumber\\
 p_\mu &=& \left(m_T\cosh y, p_T\cos\theta, p_T\sin\theta,m_T\sinh y\right),\nonumber\\ 
 u_\mu &=& \left(\cosh \rho \cosh \eta, \sinh \rho \cos \phi, \sinh \rho \sin \phi, \cosh \rho \sinh \eta\right).
\end{eqnarray}
In the Blastwave model where the hot medium with uniform temperature is considered as a cylinder along the collision direction and the transverse expansion is also taken to be uniform, integrating out $\eta$ and $\phi$ leads to the hadron transverse spectrum at freeze-out time $\tau_f$, 
\begin{equation}
E{dN\over d^3{\bf p}}={g \over 2\pi^2}\tau_f m_T \int_0^R rdr K_1({m_T\cosh \rho \over T})I_0({p_T\sinh \rho \over T}),
\end{equation}
where $m_T$ is the transverse energy, $K_1$ and $I_0$ are modified Bessel functions, and $R$ is the radius of the fireball at time $\tau_f$. In the frame of Blastwave model, the experimentally observed $D^0$ spectrum for different centrality classes can be explained quite well at low $p_T$~\cite{Adam:2018inb}, as shown in Fig~\ref{fig14}.

\subsection{Coalescence mechanism}
\label{coalescence}
Different from the statistical hadronization model which assumes thermalized hadrons and therefore does not care about the hadronization process from partons to hadrons, the coalescence model starts at parton level and describe the hadronization through parton coalescence at the deconfinement phase transition.

The experimental data at RHIC show that, baryons are less suppressed than mesons in the middle transverse momentum region $2$ GeV $<p_T<4$ GeV. The nuclear modification factor $R_{AA}$ reaches almost unity for protons and $\Lambda$ but is only one third for $\pi$ and $K$~\cite{Adler:2001bp,Adler:2002uv}. Meanwhile, people found that, one-third of the baryon elliptic flow and one-half of the meson flow are almost the same at RHIC and LHC~\cite{Sorensen:2003wi}. This enhancement of baryon to meson ratio~\cite{Hwa:2002tu} and the quark number scaling of the elliptic flow~\cite{Molnar:2003ff,Lin:2002rw} can be well described in the framework of coalescence models.

Inspired by the achievement for light hadrons, people apply the coalescence model or coalescence plus fragmentation model to study open and closed heavy flavor hadrons in heavy ion collisions~\cite{Greco:2003vf,Song:2015ykw,Lee:2007wr,Oh:2009zj,Plumari:2017ntm}. The models were also extended to the study of doubly and triply charmed hadrons~\cite{He:2014tga,Zhao:2016ccp,Zhao:2017yan}. In general coalescence models the produced hadron spectrum can be written as  
\begin{equation}
\label{coalescence}
\frac{dN_h}{d^2 {\bf p}_Td\eta} = C \int p^\mu d\sigma_\mu \prod_{i=1}^n{\frac{d^4 x_i d^4 p_i}{(2\pi)^3}}f_i(x_i,p_i) W_h( x_1,...,x_i, p_1,...,p_i),
\end{equation}
where the integration is on the coalescence hypersurface $\sigma_\mu(\tau_h,{\bf x})$ on the hadronization time $\tau_h$. For light hadrons the coalescence region is the hypersurface of the deconfinement phase transition, the coordinate $R_\mu=(t,R)$ on the hypersurface is constrained by the hadronization condition, $T(R_\mu)=T_c$, where $T_c$ is the critical temperature of the deconfinement phase transition, and the local temperature $T(R_\mu)$ and fluid velocity $u_\mu(R_\mu)$ are determined by the hydrodynamics for the QGP evolution. For heavy flavor hadrons, considering their larger binding energies compared with light hadrons, the hadronization hypersurface is in principle not the same with the phase transition hypersurface, they are hadronized earlier, see details in the next subsection.

The product in Eq.\ref{coalescence} is over the constituent quarks with $n=2$ for mesons and $n=3$ for baryons, and $p_\mu =(p_0,{\bf p})$ with $p_0=\sqrt{m_{h}^2+{\bf p}^2}$ and ${\bf p}=({\bf p}_T, p_z=p_0\sinh\eta)$ is the four-dimensional hadron momentum. The hadron coordinate $x$ and momentum $p$ are associated with the constituent quark coordinates $x_i$ and momenta $p_i$ via a Lorentz transformation~\cite{He:2014tga}. The constant $C$ is the statistical factor to take into account the inner quantum numbers in forming a colorless hadron.

$W_h$ in the hadron spectra Eq.\ref{coalescence} is the Wigner function or the coalescence probability for $n$ quarks to combine into a hadron. It is usually parameterized as a Gaussian distribution and the width is fixed by fitting the data in heavy ion collisions~\cite{Oh:2009zj}. For heavy flavor mesons and barons, we can derive their wave functions by solving the relativistic or non-relativistic potential models. The Wigner function in the center of mass frame for a charmed meson (similar treatment for charmed baryons) is directly from the wave function $\Psi(z)=\psi(r)Y(\Omega)$ where the radial part $\psi(r)$ is the solution of the potential model,
\begin{equation}
\label{wigner}
W_h(z,q)=\int d^4 y e^{-iqy}\Psi(z+y/2)\Psi^*(z-y/2).
\end{equation}

The distribution function $f_i$ in the spectra (\ref{coalescence}) is for the constituent quarks on the hadronization hypersurface. The light quarks $u$ and $d$ are thermalized and we take equilibrium distributions $f(x_i,p_i)=N_i/(e^{u_\mu  p_i^\mu/T}+1)$ with the degenerate factor $N_i=6$ and local velocity $u_\mu(x_i)$ and temperature $T(x_i)$ from the hydrodynamics. The strangeness enhancement, due to the thermal production in quark matter via for instance gluon fusion process $gg\to s\bar s$, was observed in heavy ion collisions and has long been considered as a signal of the quark matter formation~\cite{Rafelski:1982pu}. Considering that strange quarks may not reach a full chemical equilibrium at RHIC energy, their thermal distribution is multiplied by a fugacity factor $\gamma_s$. 

Different from light quarks and strange quarks,  the initially created charm quarks would interact with the hot medium and loss energy continuously. The charm quark distribution is determined by the transport model as shown in last subsection. Roughly speaking, it should be between two limits: the perturbative QCD limit without any energy loss and the equilibrium limit with full charm quark thermalization at the hadronization time. The experimental data of large elliptic flow of $D$ mesons~\cite{Adamczyk:2017xur,Abelev:2013lca} indicate that charm quarks, especially at low $p_T$, may have reached thermalization at the final stage. Therefore, one can reasonably take, as a good approximation, a kinetically thermalized phase-space distribution for charm quarks. 

\subsubsection{Coalescence versus fragmentation}
\label{fragmentation}
\begin{figure}[!htb]
	{$$\includegraphics[width=0.32\textwidth]{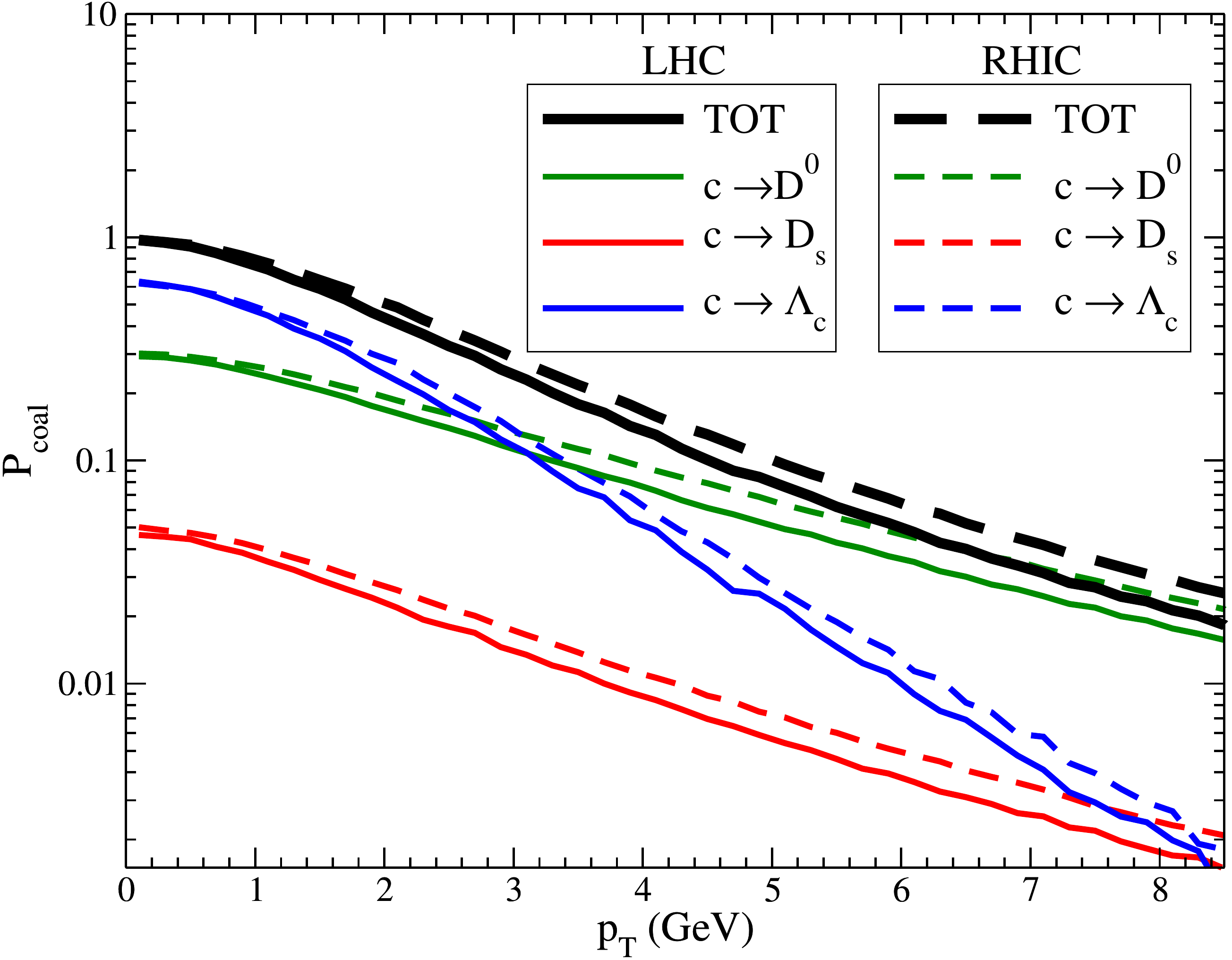}\ ~ \ ~ \ ~ \includegraphics[width=0.34\textwidth]{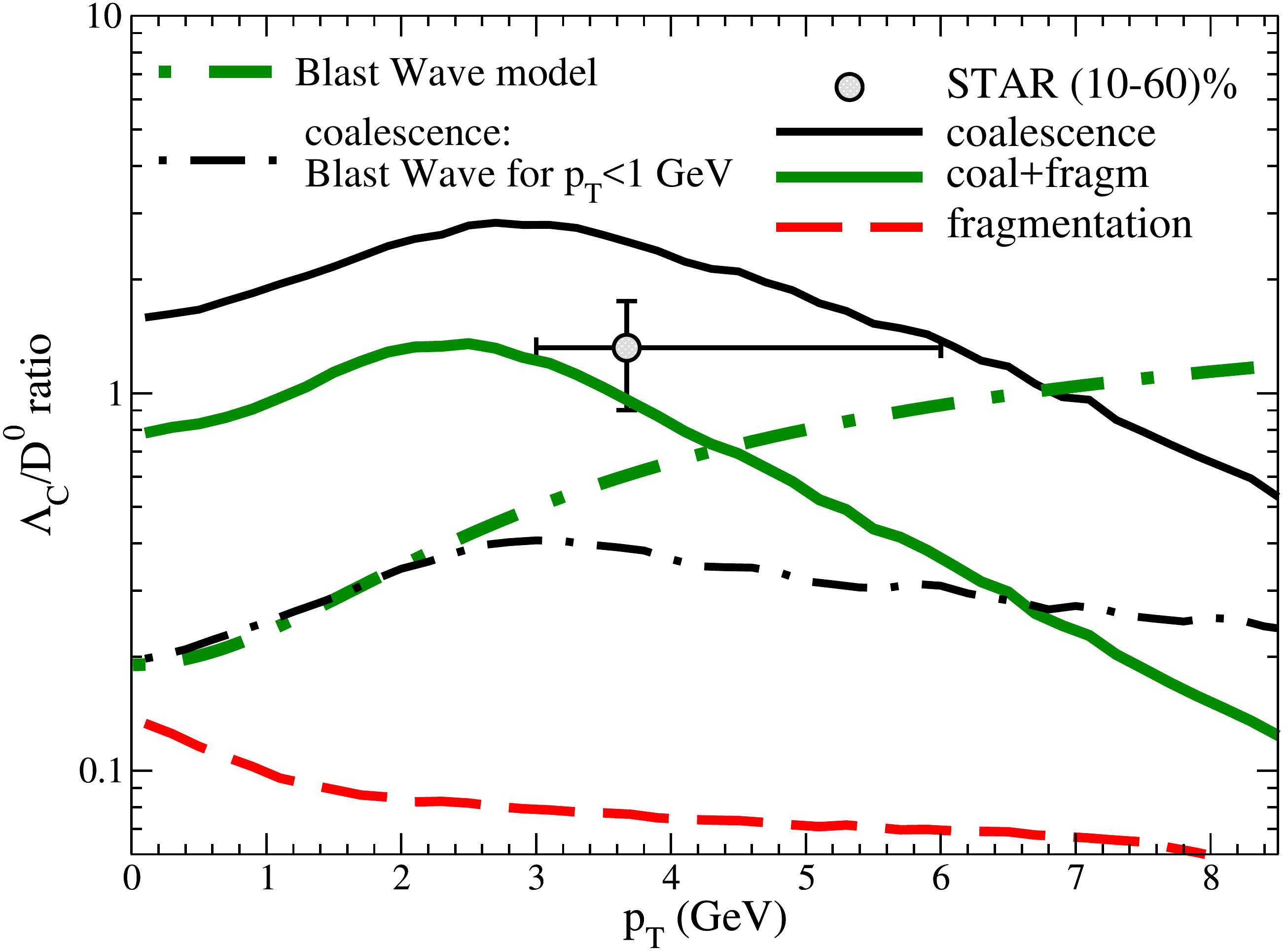}$$
		\caption{ The coalescence probabilities for charm quarks (left panel) and the yield ratio $\Lambda_c/D^0$ (right panel) as functions of transverse momentum $p_T$, calculated with different models. The figures are taken from Ref.~\cite{Plumari:2017ntm}.}
		\label{fig15}}
\end{figure}

There are two main hadronization mechanisms, one is the coalescence or recombination hadronization, the other is the fragmentation hadronization. The fragmentation hadronization has been successfully applied to describe a wide range of measured hadron spectra in elementary colliding system like $e^++e^-$. The coalescence hadronization has shown the power to describe the hadrons in heavy ion collisions, especially in the intermediate $p_T$ region. With increasing $p_T$ of heavy quarks, the coalescence probability decreases, and eventually,  the fragmentation mechanism takes over~\cite{Plumari:2017ntm,Fries:2003kq}. As shown in Fig.\ref{fig15}, the coalescence probability $P_{coal}$ for a charm quark with transverse momentum $p_T$ to hadronize into a charm hadron ($D$, $D_s$, $\Lambda_c$,...) according to the coalescence mechanism decreases very fast. This coalescence probability depends on both the Wigner function and light quark distribution. Considering the normalized total probability, the fragmentation probability is $P_{frag}=1-P_{coal}$.

Charm quarks with low transverse momentum $p_T$ are more likely to hadronize through coalescence with thermal partons from the QGP medium, while for charm quarks with high $p_T$ the fragmentation becomes the dominant mechanism. In some study of heavy quark hadronization, people firstly take coalescence mechanism and then consider fragmentation with probability $P_{frag}(p_T)$, when the coalescence probability $P_{coal}(p_T)$ is too small. 

\subsubsection{Energy conservation}
\label{econservation}
The main issue involved in the conventional coalescence model based on instantaneous projection is the energy conservation. The kinematics of the projection is $2\to1$ or $3\to1$, which makes it impossible to conserve 4-momentum. At intermediate transverse momentum ($p_T>m$), the kinematics is essentially collinear and the violation of energy conservation is suppressed by a factor of $m/p_T$ or $k_T/p_T$, where $k_T$ is the intrinsic transverse momentum of a parton inside the hadron. However, when extending to low $p_T$ where the collinearity disappears, the energy conservation is violated largely. Although the results from the conventional coalescence model are consistent with the experimental data at low $p_T$, the energy conservation should be considered in coalescence models.   

Energy conservation can be achieved through interaction with the surrounding medium. By introducing an effective mass distribution for quarks which satisfies both momentum and energy conservation, the obtained result agrees well with experimental data~\cite{Zimanyi:2005nn}. The coalescence model has also been extended to including finite width, which takes into account off-shell effects and allows to include the constraint of energy conservation~\cite{Ravagli:2007xx,Ravagli:2008rt}. The hadronization of constituent quarks is treated via resonance scattering within a Boltzmann transport equation. In the equilibrium limit, the asymptotic solution of the Boltzmann equation can be expressed as
\begin{equation}
 f_M^{eq}({\bf x,p})={E\over \Gamma m}\int{d^3{\bf p}_1d^3{\bf p}_2\over (2\pi)^6}f_q({\bf x, p_1})f_{\bar q}({\bf x, p_2})\sigma(s) v_{rel}({\bf p_1,p_2})\delta^{(3)}({\bf p-p_1-p_2}),
\end{equation}
where $v_{rel}$ is the relative velocity between $q$ and $\bar q$. The cross section can be taken as the relativistic Breit-Wigner distribution,
\begin{equation}
\sigma(s)=C{4\pi\over k^2}{(\Gamma m)^2\over (s-m^2)^2+(\Gamma m)^2},
\end{equation}
where $C$ is the degenerate factor, $m$ the meson mass, $k$ the quark 3-momentum in the CM frame, and $\Gamma$ the width of the meson resonance. This formulation conserves the 4-momentum and is applied to all resonances with mass close to or above the $q\bar q$ threshold. Finally, the yield can be calculated by using the Cooper-Frye formula, the result shows a good agreement with the experimental data and explains well the quark-number scaling of the elliptic flow $v_2$.

\subsubsection{Sequential coalescence}
\label{sequence}
What is the effect of charm conservation on charm quark hadronization? If all the charmed hadrons are produced at the same time, the charm conservation will contribute only a factor $\gamma_c$ to all the singly charmed hadrons and a factor $\gamma_c^2$ to all the doubly charmed hadrons, see Eq.\ref{cons}. In this case, the yield ratios of any two singly (doubly) charmed hadrons will not be affected by the conservation, since the effect is canceled. However, if the charmed hadrons are sequentially produced, such ratio will be changed by the charm conservation, since more charm quarks are involved in the early coalescence process and less charm quarks in the later coalescence process.   
\begin{figure}[!htb]
{$$\includegraphics[width=0.33\textwidth]{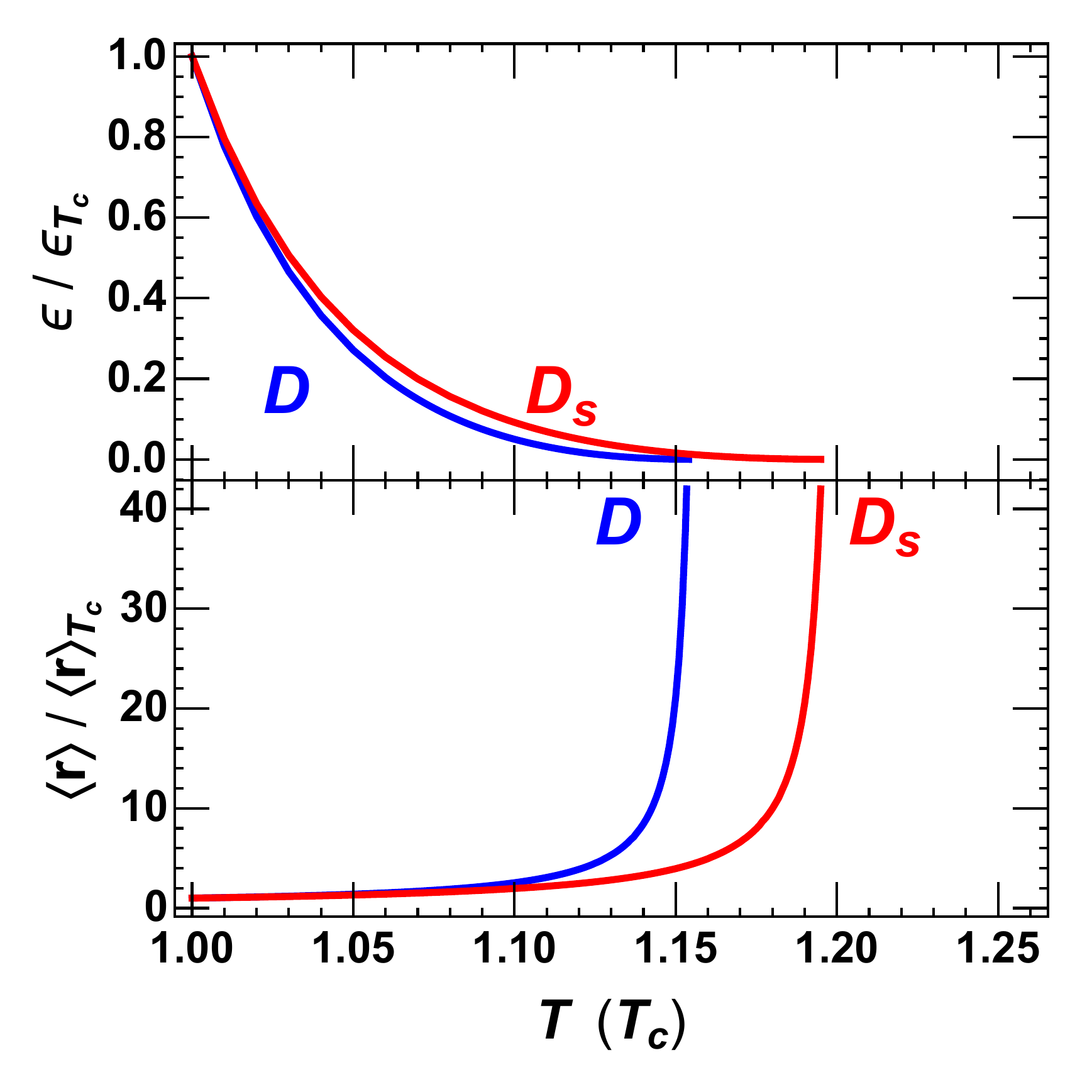}$$
\caption{ The binding energy and averaged radius of charmed mesons $D^0$ and $D_s$ as functions of temperature. }
\label{fig16}}
\end{figure}

By solving the two-body Dirac equation, which was successfully applied to the relativistic description of hidden and open charmed mesons at finite temperature, one can obtain the charmed meson binding energy $\epsilon_m(T)$ and the radial wave function $\psi(r,T)$. From the definition of meson melting $\epsilon_m(T_m)=0$ or $\langle r_m \rangle(T_m)\to \infty$, we extract the meson dissociation temperatures $T_{D_s^+}=1.2 T_c$ and $T_{D_s^{*+}}\simeq T_{D^0}\simeq T_{D^{*0}}\simeq T_{D^{*+}}=1.15 T_c$, as shown in Fig.\ref{fig16}. Considering the relation $V_{qq}\simeq V_{q\bar q}/2$ between quark-quark and quark-anti-quark potentials, the dissociation temperature of charmed baryons like $\Lambda_c$, $\Sigma_c$, $\Xi_c$ and $\Omega_c$ should be lower than that for charmed mesons~\cite{Zhao:2017gpq}. Since in heavy ion collisions, the fireball temperature decreases with time, the above sequential dissociation means that, $D_s^+$ should be produced first, then $D^0$, and finally the singly charmed baryons. In this case, the charm quark conservation will change the yield ratios of charmed hadrons. Note that, the effect of charm conservation on hadron production depends mainly on the relative order of production, one can take simply the coalescence temperature for charmed baryons as $T_c$ in the following calculation.
\begin{figure}[!htb]
{$$\includegraphics[width=0.42\textwidth]{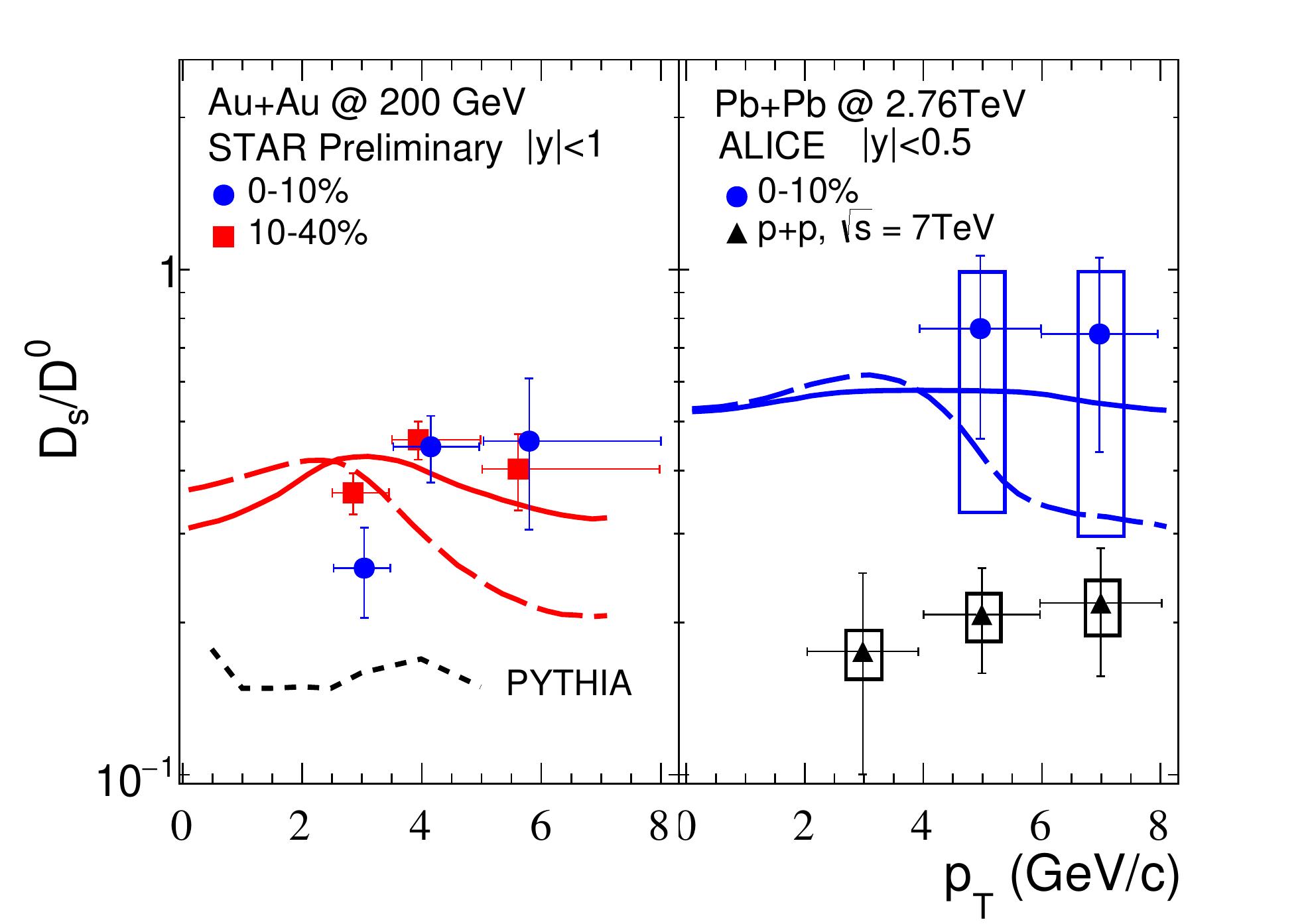}\includegraphics[width=0.42\textwidth]{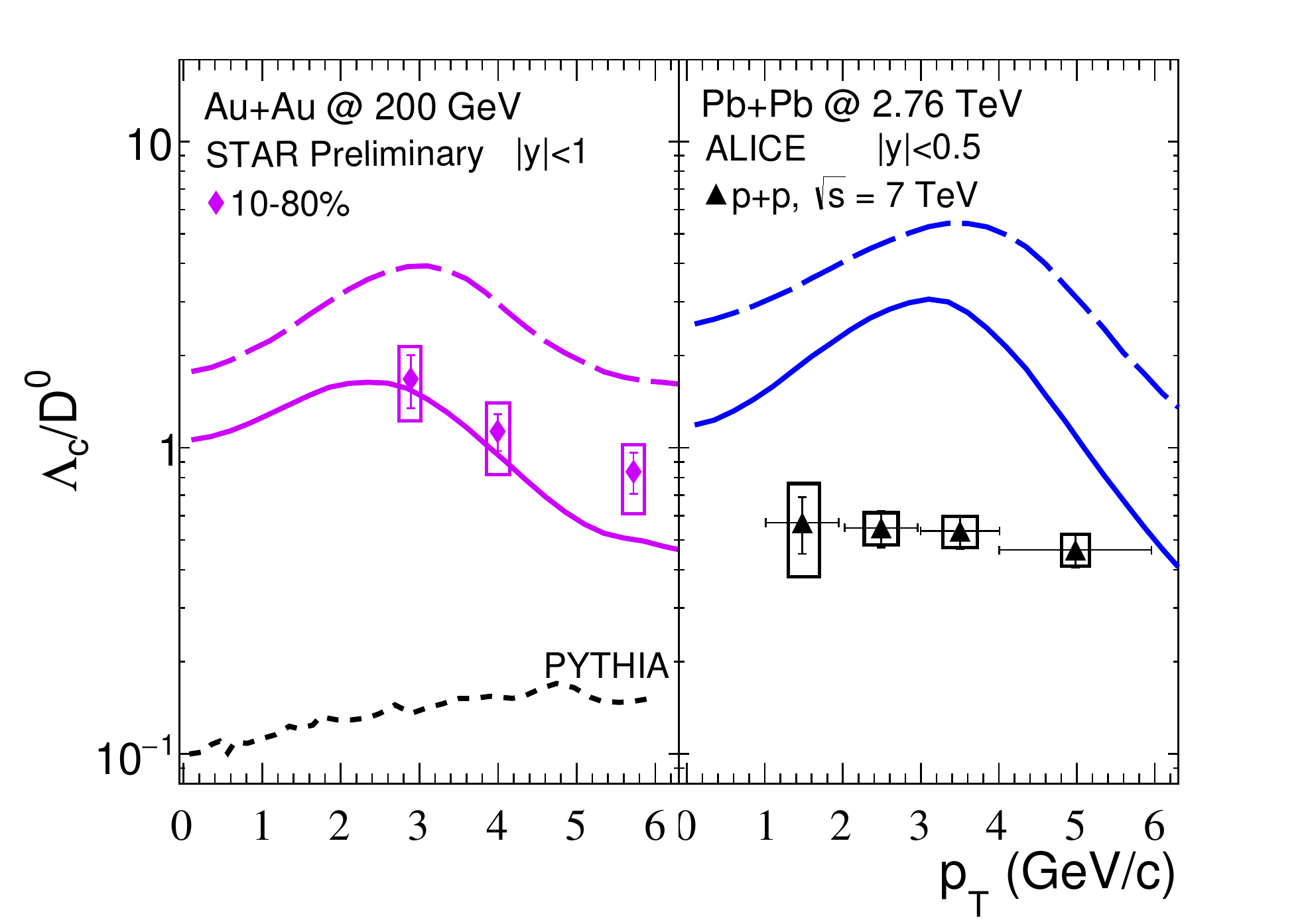}$$
\caption{ The yield ratios $D_s^+/D^0$ (left panel) and $\Lambda_c^+/D^0$ (right panel) as functions of transverse momentum in nuclear collisions at RHIC and LHC energies. The solid and dashed lines are from the sequential coalescence model with and without charm conservation. The figures are taken from Ref.~\cite{Zhao:2018jlw}. }
\label{fig17}}
\end{figure}

The calculated ratios $D_s^+/D^0$ and $\Lambda_c/D^0$ with normal and sequential coalescence models are shown in Fig.\ref{fig17}. Both the strangeness enhancement and charm conservation are responsible for the ratio enhancement. Since $D_s^+$ is produced earlier than $D^0$, more charm quarks are involved in the coalescence for $D_s^+$ in comparison with $D^0$. Therefore, the strangeness enhancement inducing $D_s^+$ enhancement will lead to a $D^0$ suppression by the charm conservation. This enhancement for $D_s^+$ and suppression for $D^0$ result in the strong $D_s^+/D^0$ enhancement. The other effect of the sequential coalescence is that the later produced $D^0$s are more thermalized in comparison with $D_s$s. Therefore, the $D^0$ distribution is shifted to the lower $p_T$ region in comparison with the $D_s$ distribution. As a competition of these two effects of sequential coalescence, the ratio $D^+_s/D^0$ will be significantly enhanced at high $p_T$ and weakly suppressed at low $p_T$.

Considering the difference in statistics for two- and three-quark states, the baryon to meson ratio in A+A collisions is dramatically enhanced in comparison with p+p collisions in coalescence models, see the yield ratio $\Lambda_c/D^0$ shown in Fig.\ref{fig17}. Since $\Lambda_c$ is produced later than $D^0$, the charm conservation reduces the ratio $\Lambda_c/D^0$.

\subsection{Multi-charmed baryons}
\label{multi}
The SU(4) quark model with four kinds of quarks u, d, s, and c predicts two 20-plet of baryons~\cite{Tanabashi:2018oca}. Among them, there are one triply charmed baryon $\Omega_{ccc}$ and six doubly charmed baryons $\Xi^+(ccd)$, $\Xi^{++}(ccu)$ and $\Omega^+(ccs)$ with spin 1/2 and 3/2. While the ground states of the baryons with one charm quark are all discovered, the doubly and triply charmed baryons are not yet observed. The experimental search for doubly charmed baryon $\Xi^+(ccd)$ lasts for decades. The
SELEX collaboration claimed the observation of $\Xi^+$~\cite{Mattson:2002vu}, but the Belle~\cite{Chistov:2006zj}, BaBar~\cite{Aubert:2006qw} and LHCb~\cite{Aaij:2013voa} collaborations failed to reproduce the results. The LHCb collaboration didn't claim that they have found doubly charmed baryon $\Xi_{cc}^{++}$ until 2017~\cite{Aaij:2017ueg}.

The reason why it is difficult to find multi-charmed baryons in elementary collisions at high energies is due to the fact that the production requires at least two or three pairs of charm quarks with small relative momenta in an event. This situation is however changed in high energy nuclear collisions. In a central heavy ion collision, there are plenty of off-diagonal charm quarks created in the early stage, and this leads to a much easier statistical production of multi-charmed baryons among all the charm quarks. The production of particles with two and three charm quarks in heavy ion collisions was systematically studied in the framework of a statistical coalescence model~\cite{He:2014tga,Zhao:2016ccp}.

Note that, if multi-charmed baryons are observed in heavy ion collisions, it is not only the discovery of new particles but also a clean signature of the QGP, since the coalescence of charm quarks is based on the assumption of QGP formation. In the coalescence model~\cite{He:2014tga,Zhao:2016ccp}, one can first solve the Schr\"odinger equation to get the multi-charmed baryon wave functions and then calculate the Wigner functions which serve as the coalescence probabilities. The multi-charmed baryon spectra in heavy ion collisions at RHIC and LHC energies can be computed on the hadronization hypersurface via coalescence mechanism. 

Considering the lower luminosity in nuclear collisions than in p+p collisions, one can introduce the effective cross section per binary collision $\sigma^{eff}_{AA}=\sigma_{AA}/N_{coll}$, where $\sigma_{AA}$ is the total production cross section of the charmed baryon in A+A collisions and $N_{coll}$ the number of binary collisions. From the calculation of $\Omega_{ccc}$, $\Xi_{cc}$ and $J/\psi$ in the frame of coalescence mechanism, the ratio of the effective cross sections $\sigma_{AA}^{eff}$ in A+A and $\sigma_{pp}$ in $p+p$ can be roughly expressed as~\cite{Zhao:2017gpq},
\begin{equation}
{\sigma^{eff}_{AA}(\Omega_{ccc})\over \sigma_{pp}(\Omega_{ccc})}:{\sigma^{eff}_{AA}(\Xi_{cc})\over \sigma_{pp}(\Xi_{cc})}:{\sigma^{eff}_{AA}(J/\psi)\over \sigma_{pp}(J/\psi)} \approx 10^2 : 10^1 : 10^0.
\end{equation}
It is clear that, the multicharmed baryon production is significantly enhanced in A+A collisions.
 
Heavy ion collisions provide not only the large probability to find multi charmed baryons but also a chance to search for exotic quantum states, such as Borromean state and Efimov state, at quark level.
Since the interaction between two heavy quarks is significantly reduced from a confinement potential in vacuum to a short range one above the critical temperature of deconfinement, it becomes possible to search for exotic states of triply charmed baryons at finite temperature. By solving the Schr\"odinger equation for a three charm quark system at finite temperature, it is found that, there exists a temperature region where the three charm quarks are still in bound state when the attractive interaction becomes too weak to bind any two charm quarks~\cite{Zhao:2017znk}. That is the so called Borromean state and might be realized in heavy ion collisions. The binding energies of the ground and excited states of the triply charmed baryon are also calculated near the resonance limit where the scattering length goes to infinity, and they satisfy precisely the scaling law for Efimov states~\cite{Zhao:2017znk}.

\subsection{Heavy flavor correlation}
\label{correlation}
We now consider charmed hadron-hadron correlation induced by charm quark energy loss in hot medium. In high energy hadron-hadron collisions, heavy quark pairs are produced back to back at leading order. A typical example of the production processes is the gluon fusion $gg\to c \bar c$. While the initial $k_T$ kick, high order processes, and fragmentation will smear the angular correlation, the approximate back-to-back production has been observed in the final state $D\bar D$ correlation~\cite{Lourenco:2006vw}. In high energy nuclear collisions, a hot and dense partonic medium is expected in the early stage. The interaction between charm quarks and the medium will modify the angular correlation and may lead to the absence of this back-to-back correlation in the final state~\cite{Zhu:2006er}.

As we emphasized above, the QGP medium formed in high energy nuclear collisions is not a static medium but a fast expanding fireball. The heavy quarks, which strongly interact with the hot medium, will carry the collective flow (or partonic wind) of the medium. Suppose a $c \bar c$ pair moves radially in the QGP,  when the velocity of $c (\bar c)$ which propagates towards the center of the fireball in the rest frame of the fluid element is less than the radial flow of the fluid element with respect to the laboratory frame, $c$ and $\bar c$ will both move towards outside, and the initial back-to-back correlation is turned to the same side correlation by the partonic wind. As a result, the $D\bar D$ correlation in the final state may be changed from the initial back-to-back correlation to the final near side correlation by the collective flow.  
\begin{figure}[!htb]
	{$$\includegraphics[width=0.34\textwidth]{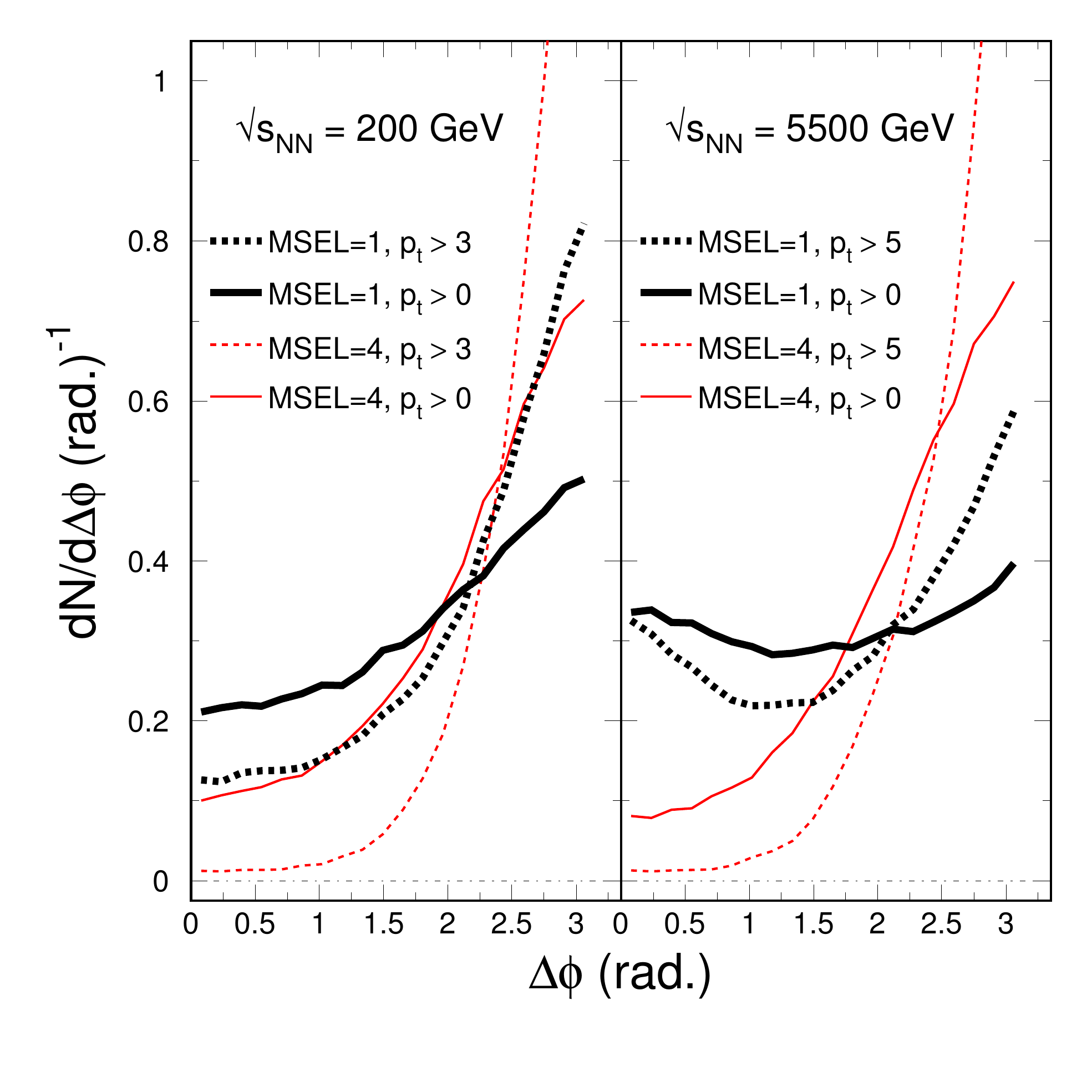} \ ~ \ ~  \includegraphics[width=0.34\textwidth]{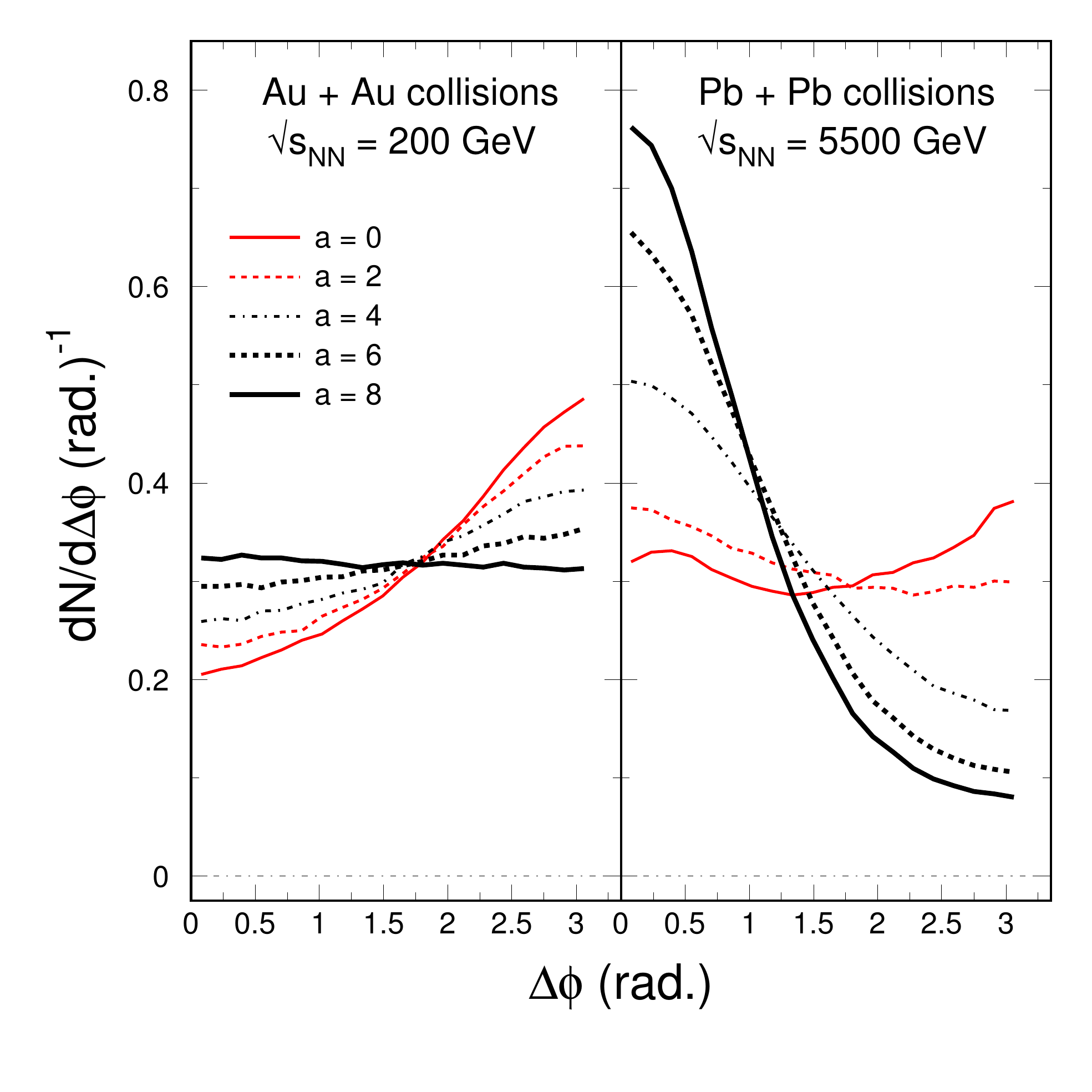}$$
		\caption{The $D\bar D$ angular correlations in p+p (left panel) and heavy ion collisions (right panel) at RHIC and LHC energies. The figures are taken from \cite{Zhu:2007ne}.}
		\label{fig18}}
\end{figure}

The comparison among the $D\bar D$ angular correlations in p+p collisions, Au+Au collisions at RHIC energy $\sqrt{s_{NN}}=200$ GeV and Pb+Pb collisions at LHC energy $\sqrt{s_{NN}}=5.5$ TeV is shown in Fig.\ref{fig18}~\cite{Zhu:2007ne}. The initial $c\bar c$ pair production in p+p collisions is simulated with PYTHIA~\cite{Sjostrand:2006za}, the motion of charm quarks in the medium is treated as a random walk by the Langevin equation, and the medium itself is described by hydrodynamic calculation. The effect of the partonic wind on the final $D\bar D$ correlation is obtained by solving the coupled nonrelativistic Langevin equation with the hydrodynamic evolutions. In p+p collisions, MSEL=4 and MSEL=1 indicate, respectively, the PYTHIA calculation to leading order and to next-to-leading order. In A+A collisions, the p+p collisions as input are taken to be the next-to-leading order. At RHIC, with increasing drag coefficient which describes the degree of the interaction between heavy quarks and the medium, the initial back to back $c\bar c$ ($D\bar D$) correlation gradually becomes flatter and flatter. For $a=6\times 10^{-6}(\text{fm})^{-1}(\text{MeV})^{-2}$, the correlation is totally washed out.
At LHC, the near side $c\bar c$ correlation becomes more and more visible with increasing drag coefficient. For $a=4\times 10^{-6}(\text{fm})^{-1}(\text{MeV})^{-2}$, the near side $c\bar c$ correlation is already very strong. This result indicates that, $c\bar c$ pairs are thermalized with the medium quickly and they move in the same direction as the partonic wind.

The study in Ref.~\cite{Cao:2015cba} shows that, the angular correlation of heavy flavor pairs, especially in the low and intermediate transverse momentum regime, is sensitive to the detailed energy loss mechanism of heavy quarks inside the QGP. The angular correlation seems less affected by the radiative energy loss, but the collisional energy loss affects the angular correlation much more and leads to a near side peak at $\Delta \phi=0$. This behavior is similar to the previous study~\cite{Zhu:2007ne}.

\section{Closed heavy flavors in high energy nuclear collisions}
\label{closed}
The ultimate goal of relativistic heavy ion collisions is to search for the new state of matter (QGP) at quark level and study its properties. A nuclear collision is not a simple superposition of nucleon collisions, and there should be cold and hot nuclear matter effects on partons and produced hadrons. However, not all the observed quarkonium suppression in nucleus-nucleus collisions relative to scaled proton-proton collisions is due to the QGP formation. For instance, the quarkonium suppression is already observed in proton-nucleus collisions where the hot medium effect on quarkonia is weak. It is necessary to find more sensitive signals to disentangle the hot and cold medium effects. In this section, we will discuss the regeneration idea and discuss two transport models for quarkonium motion in heavy ion collisions.

\subsection{Normal and anomalous suppression at SPS}
\label{sec:suppress}
In 1986, Matsui and Satz proposed $J/\psi$ suppression as a signature for the deconfinement phase transition in relativistic heavy ion collisions. $J/\psi$ suppression in nuclear collisions is defined as a lower $J/\psi$ yield per binary nucleon-nucleon collision in a heavy ion reaction, relative to elementary p+p reactions at the same energy.
To distinguish from the anomalous suppression induced by the hot nuclear matter effects, nuclear absorption is also called normal suppression. Note that the effect of nuclear absorption depends strongly on the passing time $\tau_d=2R_A/\sinh Y$ of the two colliding nuclei, where $R_A$ is the radius of the nuclei and $Y$ is their rapidity in the center of mass frame. While at SPS energy the collision time is about 1 fm/c and the normal suppression is large, the nuclear absorption in extremely high energy nuclear collisions should be small, due to the small collision time, e.g. 0.1 fm/c at RHIC and 1/200 fm/c at LHC. This can be seen from the energy dependence of the effective absorption cross section, shown in Fig.\ref{fig5}.

\begin{figure}[!htb]
	{$$\includegraphics[width=0.33\textwidth]{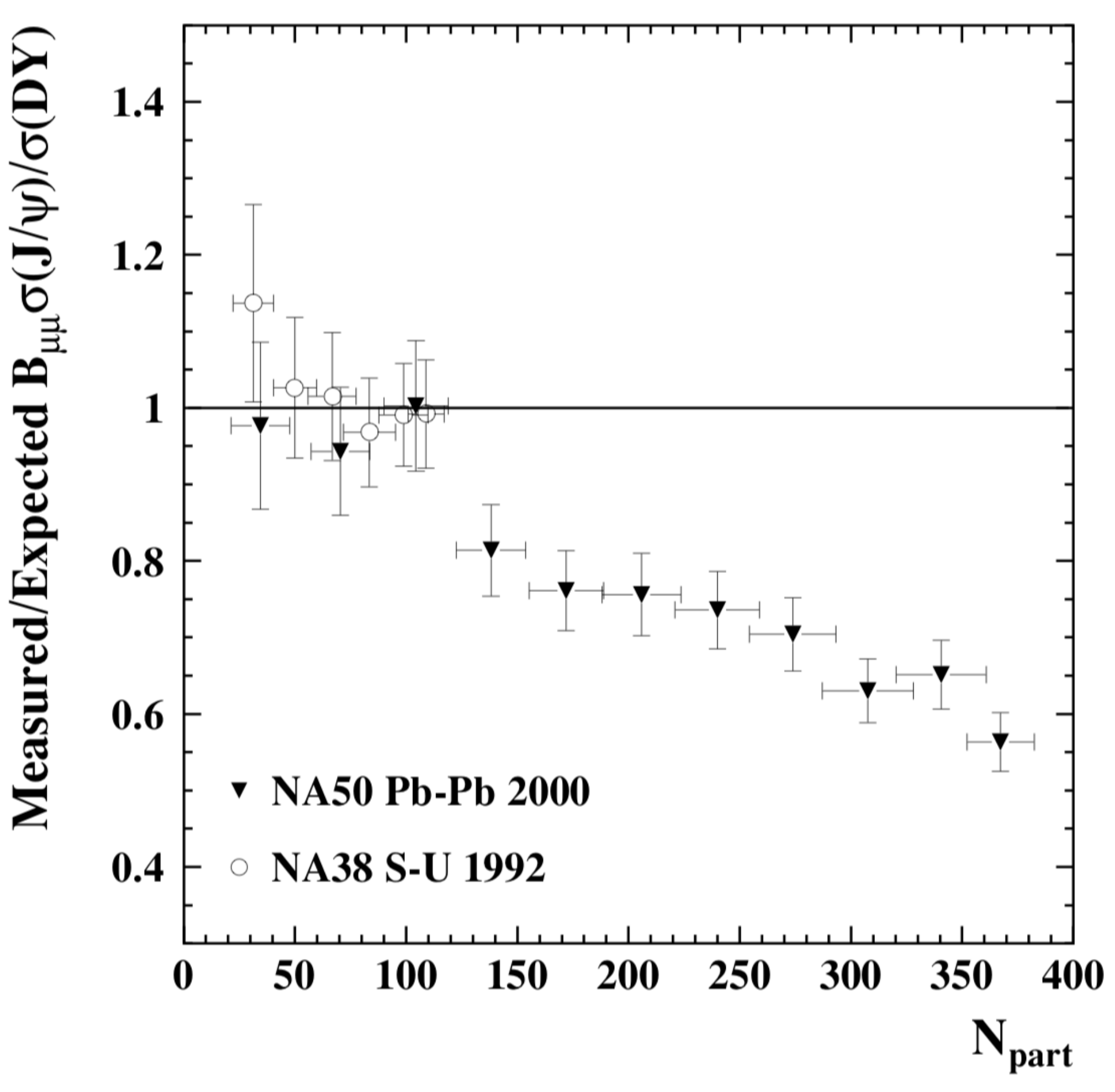} \ ~ \ ~ \ ~  \includegraphics[width=0.33\textwidth]{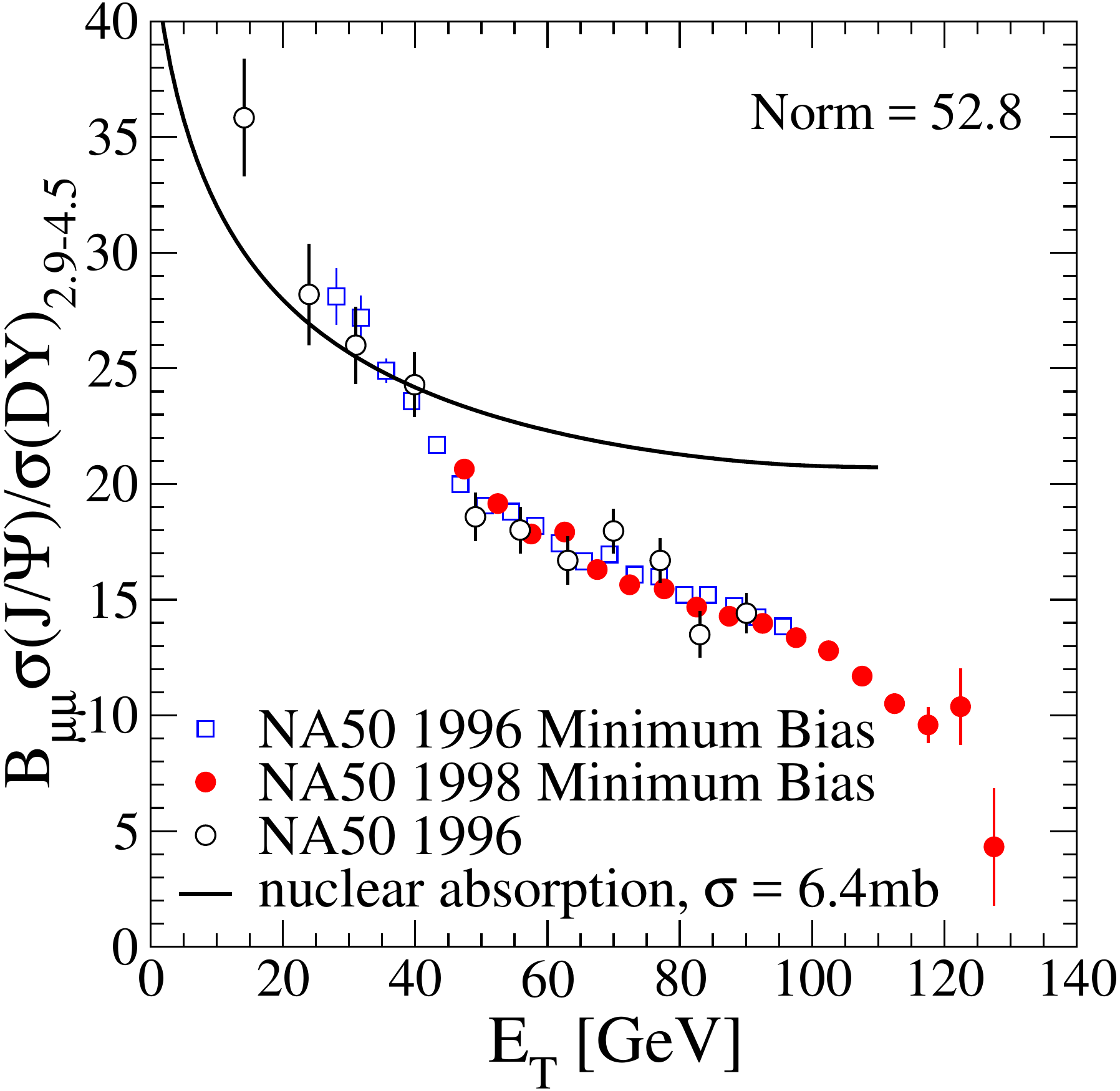}$$
		\caption{ The experimentally measured $J/\psi$ surviving probability relative to the normal suppression (left panel) and the $J/\psi$ yield (right panel) in Pb+Pb collisions at SPS energy. $N_{part}$ and $E_T$ are number of participants and transverse energy, and the line in the right panel is the model calculation with only normal suppression. The figures are taken from Refs.~\cite{Grandchamp:2002wp,Cortese:2005ns}.}
		\label{fig19}}
\end{figure}

While the nuclear absorption mechanism can well account for the experimental data in p+A and light nuclear collision systems at SPS energy, the experiments with heavy nuclear projectile and target (Pb+Pb) show that the suppression of $J/\psi$ (and $\psi'$) in semi-central and central collisions goes beyond the normal nuclear absorption, see Fig.\ref{fig19}~\cite{Cortese:2005ns}. This phenomenon, called anomalous $J/\psi$ suppression, is considered as one of the most important experimental results in relativistic heavy ion collisions at SPS. 

Various theoretical approaches have been put forward to explain the anomalous suppression. One mechanism is based on the original idea of Matsui and Satz: The Debye screening effect in the QCD medium created in the early stage of nuclear collisions leads to $J/\psi$ melting. In the picture of sequential suppression, with continuously increasing temperature of the fireball, $\psi'$ will melt first, then $\chi_c$ dissociates, and finally $J/\psi$ disappears. Considering the fact that about 40\% of the final state $J/\psi's$ originate from the decay of $\psi'$ and $\chi_c$, the anomalous $J/\psi$ suppression in Pb+Pb collisions at SPS is associated with the dissociation of $\psi'$ and $\chi_c$ in the produced fireball.

Along similar lines, $J/\psi$ suppression in the hot and dense medium has been described in a general threshold model without considering microscopic dynamics~\cite{Blaizot:1996nq}. In this model, the $J/\psi$ surviving probability is written as
\begin{equation}
S_{J/\psi}(b)=\int d^2s S_{J/\psi}^{nucl}(b,s) \Theta(n_c-n_p(b,s)),
\end{equation}
where $S_{J/\psi}^{nucl}(b,s)$ is the $J/\psi$ survival probability after the nuclear absorption, $b$ is the impact parameter, and $s$ is the transverse coordinate of $J/\psi$. The density $n_p(b,s)$ in the step function is proportional to the energy density of the matter at position $(b,s)$. In the hot and dense part of the fireball where $n_p$ is larger than a critical value $n_c$, all the $J/\psi$s are absorbed by the matter, and those $J/\psi$s outside the region suffer only normal suppression. The threshold density $n_c$ in this model is a parameter, it can be taken as the maximum $n_p$ in S+U collisions at SPS where no anomalous suppression is observed. If the matter with $n_p>n_c$ is QGP, the critical density $n_c$ can be considered as the threshold value to create the QGP. Despite its simplicity, the threshold model explains well the anomalous suppression in Pb+Pb collisions at SPS~\cite{Grandchamp:2001pf,Blaizot:2000ev}.

The above analyses based on the Debye screening effect depend typically on the assumption of a sharp transition of the inelastic charmonium widths from zero (stable below $T_D$) to infinity (dissolved above $T_D$). However, the volume of the produced fireball in relativistic heavy ion collisions is small and expands rapidly, implying rather fast temperature change and short fireball lifetime. In this case, the conclusion from the static Debye screening effect may deviate from the real system, and it becomes essential to include the concrete interactions between partons and charmonia, leading to sizable inelastic reaction rates comparable to the fireball expansion (or cooling) rate. In particular, charmonia can be destroyed below the dissociation temperature and survive above the dissociation temperature. Such an important process in the QGP is the gluon dissociation process, $J/\psi +g \to c+\bar c$, an analogy to the photon dissociation process of electromagnetic bound states~\cite{Kharzeev:1994pz}. For small binding energies when approaching the dissociation temperature, the phase space for gluon dissociation shrinks and the next-to-leading order processes take over. The most notable inelastic parton scatterings are $Q\bar Q+p\to Q+\bar Q+p$ with $p$ being gluons and light quarks. 

Not only partons in the deconfined phase can induce anomalous suppression, but also the secondary particles like $\pi, \rho, \omega$ mesons (so-called comovers) in a hot and dense hadron gas can interact with charmonia inelastically and cause $J/\psi$ suppression~\cite{Gavin:1988hs,Vogt:1988fj}. The suppression due to the comover effect can be schematically expressed as,
\begin{equation}
S_{J/\psi}^{co}=e^{-\int d\tau \langle v\sigma_{co}\rangle \rho_{co}(\tau)},
\end{equation}
where $\rho_{co}(\tau)$ is the comover density at proper time $\tau$, and the inelastic cross section (multiplied by the relative velocity) is averaged over different kinds of comovers and the interaction energy. The comover density $\rho_{co}(\tau)$ is normally obtained through some kinds of evolution mechanism of the matter (generally assumed to be of the Bjorken-type) and is fixed by fitting the measured final state hadron yield $dN/dy$. The cross section $\sigma_{co}$ is an adjustable parameter in the calculation.

A more detailed description of the matter evolution together with a dynamical treatment of the interactions between charmonia and comovers has been carried out in hadronic transport models UrQMD~\cite{Spieles:1998pa} and HSD~\cite{Bratkovskaya:2003ux} where the $J/\psi$ motion is traced microscopically throughout the medium. By adjusting the comover cross sections (and possibly other parameters such as formation times), interactions at hadron level can reproduce the SPS data of $J/\psi$ suppression~\cite{Capella:2000zp}.

Motivated by the lattice QCD results of surviving $J/\psi$ bound state above $T_c$, the formation and evolution of $c$ and $\bar c$ correlations are treated more microscopically. In a weakly coupled QGP, charm quarks would fly away from each other as soon as enough energy is available, while in a strongly coupled QGP the strong attraction between quarks opens the possibility of returning to the $J/\psi$ ground state, leading to a substantial increase in survival probability~\cite{Young:2008he}.

\subsection{Regeneration at RHIC and LHC}
\label{regeneration}
The normal and anomalous suppressions discussed above describe the nuclear matter effects on the initially produced charmonia before and after the fireball formation. In nuclear collisions at SPS energy and below, there is at most one pair of $c\bar c$ produced in a central Pb+Pb collision ($\langle N_{c\bar c}\rangle\sim 0.2$ at $E_{lab}=158$ A GeV). If the $c$ and $\bar c$ can not form a (pre-resonant) charmonium bound state around their creation point, the probability to recombine into a resonant state in the medium is small and can be neglected. However, for nuclear collisions at RHIC and LHC energies, the situation becomes quite different. In a central Au+Au collision at the maximum RHIC energy, about 10-20 $c\bar c$ pairs are produced, and the uncorrelated $c$ and $\bar c$ from different pairs have a significant probability (proportional to the square of the number of $c\bar c$ pairs) to form a charmonium bound state in the medium. The $J/\psi$ regeneration in the partonic phase arises as a new mechanism for charmonium production in heavy ion collisions at RHIC and LHC.

\begin{figure}[!htb]
	{$$\includegraphics[width=0.317\textwidth]{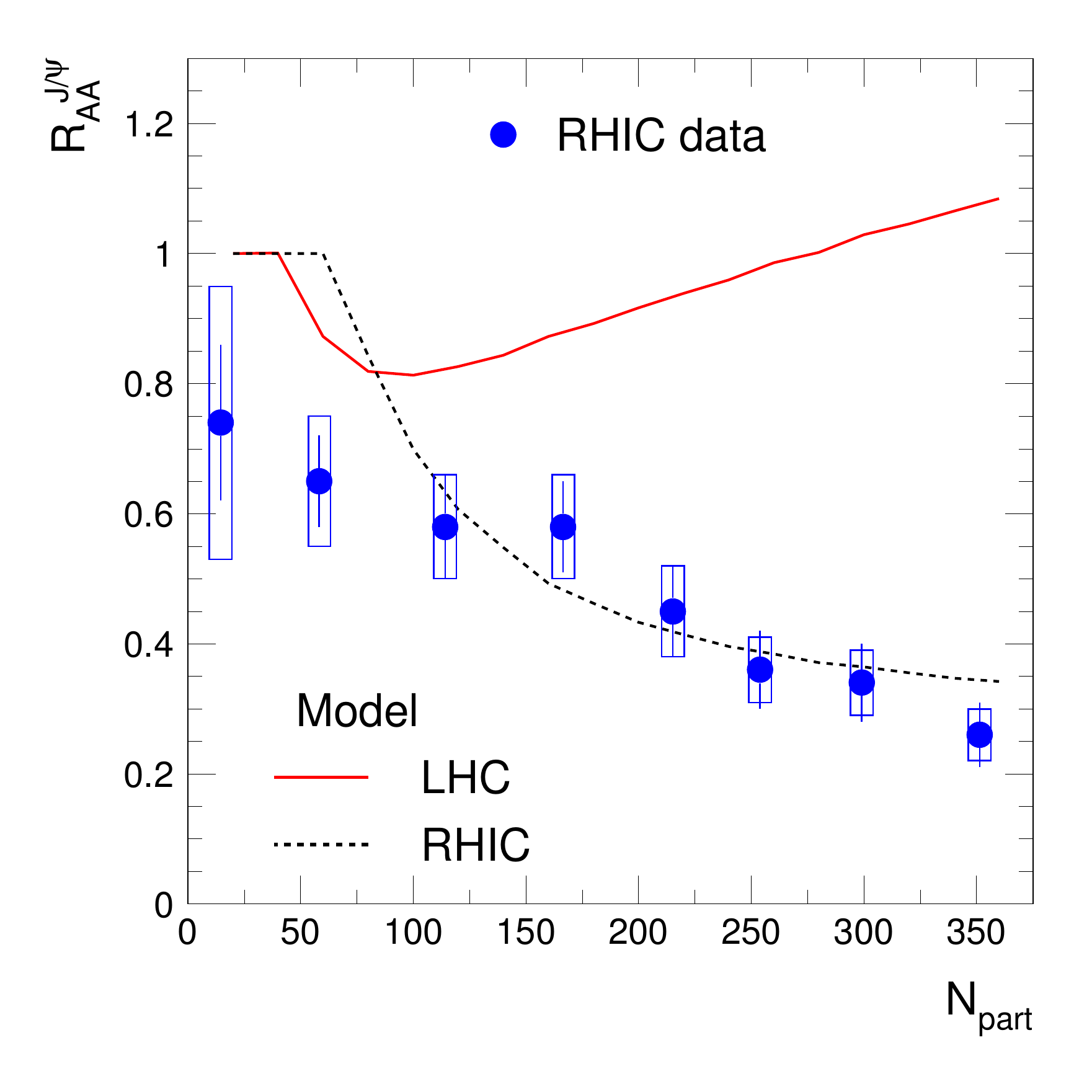} \ ~ \ ~  \includegraphics[width=0.4\textwidth]{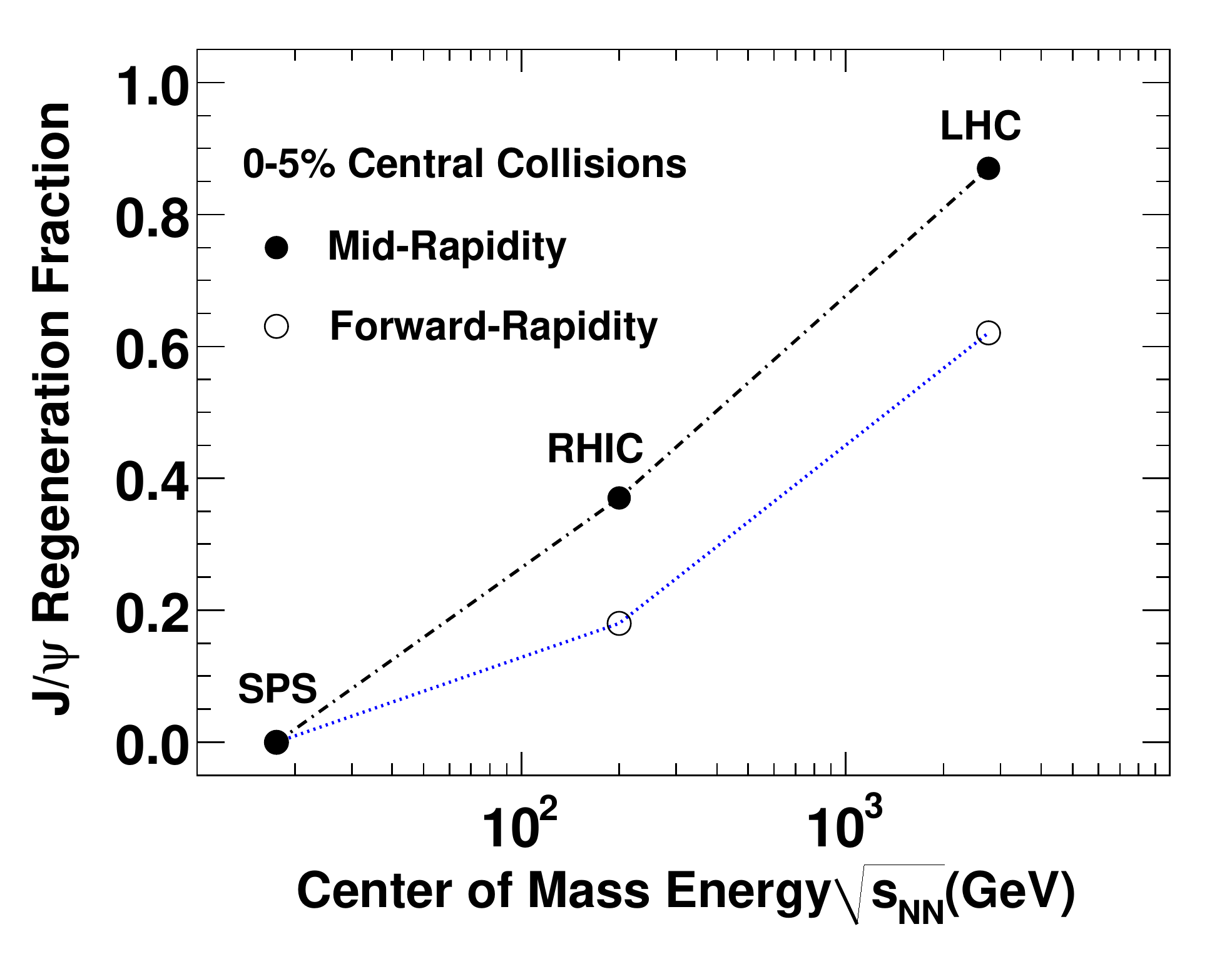}$$
		\caption{ The $J/\psi$ nuclear modification factor as a function of number of participants calculated in statistical model (left panel) and the $J/\psi$ regeneration fraction as a function of colliding energy calculated in transport model (right panel). The figures are taken from ~\cite{BraunMunzinger:2007tn,Zhou:2013aea}.}
		\label{fig20}}
\end{figure}

In the statistical hadronization model~\cite{BraunMunzinger:2000px}, the entire creation of charmonia occurs statistically at the hadronization of the QCD matter, i.e., at the critical temperature $T_c$. The model is a direct extension of the thermal model which describes the ratio of light hadron yields well in relativistic heavy ion collisions. Although charm quarks are far from (absolute) chemical equilibrium, they are assumed to be in thermal and relative chemical equilibrium in the statistical model and the mechanism for charmonium production is similar to that for light hadrons. The deviations of the nuclear modification factor
\begin{equation}
R_{AA}^{J/\psi}={dN_{J/\psi}^{AA}/dy \over N_{coll}\cdot dN_{J/\psi}^{pp}/dy}
\end{equation}
from unit characterizes the difference of $J/\psi$ production in A+A and p+p collisions. With initial charm quark distribution taken from perturbative QCD calculation for nucleon-nucleon collisions, the centrality dependence of the nuclear modification factor calculated in the statistical hadronization model is shown in Fig.\ref{fig20}.~\cite{BraunMunzinger:2007tn}. At top RHIC energy, it reproduces well the $J/\psi$ suppression in central Au+Au collisions at mid rapidity. At the LHC energy withy $\sqrt{s_{NN}}=5.5$ TeV, the charm production cross section grows by about an order of magnitude, resulting in a qualitatively new centrality dependence which predicts an enhancement of $J/\psi$ production for semi-central and central collisions. 

Including both initial production and regeneration in a transport model~\cite{Zhou:2013aea}, the regeneration fraction of $J/\psi$ in central Au+Au collisions at RHIC and Pb+Pb collisions at SPS and LHC is shown in the right panel of Fig.\ref{fig20}. As expected, the fraction depends strongly on the heavy quark number or the collision energy. At SPS, the initial production is almost the only production mechanism. At RHIC, both initial production and regeneration of charmonia play important roles. At LHC, the $J/\psi$ production is dominated by regeneration. The lower fraction in the forward rapidity is due to the rapidity distribution of heavy quarks.

\subsection{Transport models}
\label{transport}
From the lattice calculation of charmonium spectral functions, $J/\psi$ can exist in a thermal environment at temperatures above the deconfinement phase transition. Unlike the statistical model, charmonia in kinetic models~\cite{Thews:2000rj,Thews:2001hy} can be regenerated continuously throughout the QGP region, and the formed $J/\psi s$ reflect the initially produced charm quark spectra and the modification from the interaction with the medium. In kinetic models, $J/\psi$ production during the entire lifetime of the deconfined phase is dynamically calculated through production and dissociation processes at finite temperature and density. The simplest dissociation reaction utilizes the absorption of individual (deconfined) gluons in the medium to ionize the color singlet $g+J/\psi \to c+\bar c$, resulting in a $c\bar c$ pair in a color octet state. The inverse of this process serves as the corresponding production reaction, in which a $c\bar c$ pair in a color octet state emits a color octet gluon and falls into the color singlet $J/\psi$ bound state. The production cross section is obtained through the detailed balance. The competition between the $J/\psi$ production and suppression can be characterized by a transport equation with loss and gain terms.

Starting from the covariant Boltzmann transport equation
\begin{equation}
p^\mu \partial_{\mu} f_\psi= -\alpha E f_\psi +\beta E,
\end{equation}
where the left hand side is the diffusion term, the right hand side is the collision term which can be separated into the loss and gain terms, and $f_\psi({\bf p, x}, t)$ $(\psi=J/\psi, \psi', \chi_c)$ is the charmonium distribution function in phase space.

By integrating over the whole phase space on both sides of the equation, the left hand side becomes
\begin{equation}
\int {d^3xd^3p \over (2\pi)^3E}p^\mu \partial_{\mu} f_\psi= {d N_\psi(t) \over dt},
\end{equation}
where $N_\psi(t)= \int f_\psi d^3xd^3p/((2\pi)^3E)$ is the number of charmonia. If the spatial space is homogeneous, or the loss and gain terms $\alpha$ and $\beta$ are space independent ($\alpha({\bf p},{\bf x},t) \to \alpha({\bf p},t)$, $\beta({\bf p},{\bf x},t) \to \beta({\bf p},t)$), the right hand side is approximately
\begin{equation}
\int {d^3xd^3p \over (2\pi)^3E}\alpha({\bf p},{\bf x},t)E f_\psi \approx \Gamma_\psi(t) N_\psi(t),
\end{equation}
where $\Gamma_\psi(t)=\alpha(\bar p,t)$ is the charmonium decay rate at averaged momentum $\bar p$. Considering the equilibrium limit of the charmonium distribution in momentum space, the gain term is related to the loss term, $\beta({\bf p})= \alpha({\bf p})f_\psi^{eq}({\bf p})$, which leads to 
\begin{equation}
\int {d^3xd^3p \over (2\pi)^3E}\beta({\bf p},{\bf x},t)E = V\int {d^3p \over (2\pi)^3}\beta({\bf p},t) \approx \Gamma_\psi(t) N_\psi^{eq}(t).
\end{equation}
Putting all the terms together, the Boltzmann transport equation becomes the Rate equation~\cite{Grandchamp:2003uw, Zhao:2007hh, Zhao:2011cv, Du:2015wha},
\begin{equation}
{d N_\psi(t) \over dt}=-\Gamma_\psi(t) [N_\psi(t)-N_\psi^{eq}(t)],
\end{equation}
where $N_\psi^{eq}=\int f_\psi^{eq} d^3xd^3p/((2\pi)^3E)$ is the number of quarkonia in thermal equilibrium and can be calculated from the statistical model, as expressed in Eq.\ref{cons}. The charmonium dissociation rate $\Gamma_\psi$ which contains the microscopic dynamics, governs both suppression and regeneration processes,
\begin{equation}
\Gamma_\psi({\bf p},T)=\sum_i \int{d^3k \over (2\pi)^3}f_i(\omega_k, T)\sigma_{\psi i}^{diss}v_{rel},
\end{equation}
where $f_i$ is the parton or meson distribution, usually taken as thermal distribution, $\sigma_{\psi i}^{diss}$ is the charmonium dissociation cross section by the parton or meson, and $v_{rel}=\sqrt{(p^\mu k_\mu)^2-m_\psi^2m_k^2}/(E_\psi E_i)$ is the relative velocity with $k_\mu$ being the four-momentum of the parton or meson. There are two main mechanisms for charmonium dissociation in quark matter: the gluon dissociation and inelastic parton scattering.

The evolution of the boost invariant hot medium which is suitable for the medium in the mid-rapidity at RHIC and LHC is described by the fireball model approximately. The fireball model assumes that, the medium expands cylindrically and all thermodynamic quantities depend only on the proper time $\tau$. The fireball volume can be expressed as~\cite{Zhao:2007hh, Zhao:2011cv},
\begin{equation}
V_{FB}(\tau)=\left (z_0+v_z\tau+{1\over 2}a_z\tau^2\right)\pi \left(r_0+{1\over 2}a_\perp \tau^2 \right)^2,
\end{equation}
where $z_0\approx\tau_0 \Delta y$ is the initial longitudinal length related to the local thermalization time $\tau_0$, $r_0$ is the initial transverse radius depending on the nuclear parameters, and $v_z$, $a_z$ and $a_\perp$ are the longitudinal velocity, and longitudinal and transverse acceleration. The parameters can be fixed by experimental data of light hadron flow and spectrum. The temperature of the medium can be determined by the fireball expansion profile and the equation of state. The maximum of the transverse radius $R$ is given by the fireball expansion at freeze-out proper time $\tau_f$.

Almost all the models for quarkonium production in heavy ion collisions, with and without the assumption of a QGP and with and without regeneration mechanism, can describe the observed quarkonium yield after at least one parameter is adjusted. Since any transverse motion is created through interactions, transverse momentum distribution depends more directly on the production and regeneration mechanisms and therefore contains additional information about the nature of the medium and quarkonia. 

Anomalous suppression is not an instantaneous process but takes a certain time. During this time the produced quarkonia with high transverse momentum may leak out the parton plasma and escape the suppression. As a consequence, low $p_T$ quarkonia are more likely to be absorbed, and the average transverse momentum of the observed quarkonia will show an increase. A self-consistent way to incorporate the effect of leakage into the quarkonium production and suppression is through a transport equation in phase space~\cite{Zhu:2004nw,Yan:2006ve,Liu:2009wza,Zhou:2014kka}.

The medium created in high energy nuclear collisions evolves dynamically. In order to extract information about the medium by analyzing the quarkonium distributions, the hot medium and the quarkonium production processes must be treated dynamically. Due to its large mass, quarkonia are not fully thermalized with the medium and their phase space distribution should be governed by a transport equation including both initial production as well as regeneration. The charmonium distribution function in a nuclear collision with fixed impact parameter $b$, $f_\psi(p_T,x_T,y,\tau|b)$ $(\psi=J/\psi, \psi', \chi_c)$, is controlled by the classical Boltzmann transport equation,
\begin{equation}
\left[ \cosh(y-\eta){\partial \over \partial \tau} +{\sinh(y-\eta)\over \tau}{\partial \over \partial \eta}+{\bf v}_T\cdot \nabla_T \right] f_\psi=-\alpha f_\psi +\beta,
\end{equation}
where $\eta=1/2\ln{t+z\over t-z}$, $y=1/2\ln{E+p_z\over E-p_z}$ and $v_T=p_T/E_T$ are the quarkonium space-time rapidity, momentum rapidity and transverse velocity. The second and third terms on the left hand side arise from the free streaming of $\psi$ which leads to the leakage effect in the longitudinal and transverse direction. The anomalous suppression and regeneration mechanisms are reflected in the loss term $\alpha$ and gain term $\beta$. It is assumed that, the nuclear absorption of the initially produced $\psi$s cease before the medium is locally equilibrated at time $\tau_o$. Considering only the gluon dissociation process for the loss term and its inverse process for the gain term, $\alpha$ and $\beta$ are expressed as~\cite{Yan:2006ve,Liu:2009wza}
\begin{eqnarray}
\alpha({\bf p}_T,{\bf x}_T,\tau|{\bf b})&=&{1\over 2E_T}\int{d^3{\bf p}_g \over(2\pi)^32E_g}W_{g\psi}^{c\bar c}(s)f_g({\bf p}_g,{\bf x}_T,\tau)\Theta(T({\bf x}_T,\tau|{\bf b})-T_c), \\
\beta({\bf p}_T,{\bf x}_T,\tau|{\bf b})&=&{1\over 2E_T}\int {d^3{\bf p}_g \over(2\pi)^32E_g}{d^3{\bf p}_c \over(2\pi)^32E_c}{d^3{\bf p}_{\bar c} \over(2\pi)^32E_{\bar c}}  \nonumber\\
&\times& W_{c\bar c}^{g\psi}(s)f_c({\bf p}_c,{\bf x}_T,\tau|{\bf b})f_{\bar c}({\bf p}_{\bar c},{\bf x}_T,\tau|{\bf b})(2\pi)^4\delta^{(4)}(p+p_g-p_c-p_{\bar c}) \Theta(T({\bf x}_T,\tau|{\bf b})-T_c), \nonumber
\end{eqnarray}
where the transition probabilities $W_{g\psi}^{c\bar c}(T)$ and $W_{c\bar c}^{g\psi}(T)$ are determined by the cross sections $\sigma_{g\psi}^{c\bar c}(T)$ and $\sigma_{c\bar c}^{g\psi}(T)$, and the step function $\Theta(T-T_c)$ indicates that, only the regeneration and suppression in the QGP is taken into account and the hadron phase is neglected. With the known loss and gain terms, the transport equation can be solved analytically, and the result is shown as
\begin{eqnarray}
f({\bf p}_T,y,{\bf x}_T,\eta,\tau)&=&f({\bf p}_T,y,{\bf X}(\tau_0),H(\tau_0),\tau_0)e^{-\int_{\tau_0}^\tau d\tau' \alpha({\bf p}_T,y,{\bf X}(\tau'),H(\tau'),\tau')/\Delta(\tau') }\\
&+&\int_{\tau_0}^\tau d\tau' \beta({\bf p}_T,y,{\bf X}(\tau'),H(\tau'),\tau')/\Delta(\tau') e^{-\int_{\tau'}^\tau d\tau'' \alpha({\bf p}_T,y,{\bf X}(\tau''),H(\tau''),\tau'')/\Delta(\tau'')}\nonumber 
\end{eqnarray}
with
\begin{eqnarray}
&&{\bf X}(\tau')={\bf x}_T-{\bf v}_T[\tau\cosh(y-\eta)-\tau'\Delta(\tau')], \nonumber\\
&&H(\tau')=y-arcsinh(\tau/\tau' \sinh(y-\eta)), \nonumber\\
&&\Delta(\tau')=\sqrt{1+(\tau/\tau')^2\sinh^2(y-\eta)}.
\end{eqnarray}
The first and second terms on the right-hand side of the solution indicate the contributions from the initial production and continuous regeneration, respectively, and both suffer from anomalous suppression in the medium.

Suppose the medium is thermalized, the gluon distribution $f_g$ can then be taken as a thermal distribution. How is about the charm quark distribution $f_c (f_{\bar c})$ in the medium? From the experimental data at RHIC and LHC, the observed large quench factor for charmed mesons indicates that charm quarks interact strongly with the medium~\cite{Adamczyk:2017xur,Abelev:2013lca}. Therefore, one can reasonably take, as a good approximation, a kinetically thermalized phase-space distribution $f_c\sim 1/(e^{p\cdot u/T}+1)$ for charm quarks. Neglecting the creation and annihilation for $c\bar c$ pairs inside the medium, the spatial density of charm (anticharm) quark number $\rho_c=\int f_cd^3p/(2\pi)^3$ satisfies the conservation law $\partial_\mu(\rho_c u^\mu)=0$ with the initial density determined by the nuclear geometry,
\begin{equation}
\rho_c({\bf x},\tau_0|{\bf b})={T_A({\bf x}_T)T_B({\bf x}_T-{\bf b})\cosh \eta \over \tau_0}{d\sigma_{pp}^{c\bar c}\over d\eta},
\end{equation}
where $T_A$ and $T_B$ are the thickness functions at transverse coordinate ${\bf x}_T$ defined in the Glauber model, and $d\sigma_{pp}^{c\bar c}/d\eta$ is the rapidity distribution of charm quark production cross section in $p+p$ collisions~\cite{Abelev:2012vra}.

Where is the initial production and where is the cold nuclear matter effect? The initial production and the cold nuclear matter effects (shadowing, Cronin effect, and nuclear absorption) are reflected in the initial condition of the transport equation, $f({\bf p}_T,y,{\bf x}_T, \tau_0)$, which is controlled by Eq.\ref{shadowing}. Note that the shadowing effect changes not only the initial charmonium distribution but also the in-medium charmonium regeneration and the non-prompt contribution from the $B$ decay, by reducing the number of charm and bottom quarks. In principle, the shadowing should be centrality dependent. To simplify the numerical calculations, it is assumed that, a reduction of 20\% of the charm and bottom quark production cross sections is taken into account in the calculation, estimated from the centrality averaged EKS98 evolution~\cite{Eskola:1998df}.

The local temperature $T({\bf x}, t)$ and fluid velocity $u_{\mu}({\bf x}, t)$ used in the gluon and charm quark distribution functions and the loss and gain cross sections, are determined by the medium evolution. We employ the well tested 2+1 dimensional version of the ideal hydrodynamic equations,
\begin{eqnarray}
\partial_{\mu} T^{\mu v}=0, \ ~ \  \partial_{\mu} n^\mu=0
\end{eqnarray}
to simulate the evolution of the almost baryon-free medium created at RHIC and LHC, where $T^{\mu v}$ is the energy-momentum tensor of the medium, $n^\mu$ is the baryon number flow. The solution of the hydrodynamic equations provides the local temperature and fluid velocity of the medium. To close the hydrodynamical equations one needs to know the equation of state of the medium. The deconfined phase at high temperature is taken as an ideal gas of gluons and massless $u$ and $d$ quarks plus $s$ quarks with mass $150$ MeV, and the hadron phase at low temperature is considered as an ideal gas of all known hadrons and resonances with mass up to $2$ GeV~\cite{Sollfrank:1996hd}. There is a first-order phase transition between these two phases. The critical temperature is taken as $T_c=165$ MeV at vanishing baryon number density.

\subsection{Open Quantum System}
\label{oqs}
In order to describe both the dissociation and recombination processes in heavy ion collisions, the static description of heavy quarks via either lattice techniques in imaginary time or potential models is not appropriate, one needs a dynamical treatment of the heavy quark system. A real-time dynamics which goes beyond the equilibrium to understand the physics of heavy quark/quarkonium in hot medium is required. One development in the past decade is using the method of open quantum system to describe the real-time evolution of heavy quarkonia. In this framework, a quarkonium is treated as an open quantum system (also called reduced system or subsystem) that can dissipate or gain energy from the environment (for instance quark-gluon plasma). The environment and the heavy quark/quarkonium form a closed quantum system. This treatment allows one to follow the quantum dynamics of the subsystem and compute survival and formation probabilities of the heavy-quark bound states, as well as quantum decoherence. Here we just follow the standard deduction as shown in~\cite{oqs2002}. Assume the Hamiltonian of the closed system is given by
 \begin{eqnarray}
\hat H_{tot}=\hat H_s\otimes I_e+ I_s \otimes \hat H_e+\hat H_{int},
\end{eqnarray}
where $\hat H_s$ is the Hamiltonian of the subsystem, $\hat H_e$ the Hamiltonian of the environment, and $\hat H_{int}$ the interaction between the subsystem and the environment. Of particular interest is the probability $P(\psi(x,t)|\psi(x_0,t_0))$ to find the reduced system in a quantum state $\psi(x,t)$ at time $t$, given that it was in a state $\psi(x_0,t_0)$ at time $t_0$. This quantity can be written in terms of the density operator of the subsystem. Generally, the dynamics of an open quantum system is characterized by the total density matrix $\hat \rho_{tot}(t)$, which describes both the system and the environment. The total density operator of the closed system obeys the Liouville-von Neumann equation, 
\begin{equation}
i\hbar \dot {\hat{\rho}}_{tot}(t) = [\hat H_{tot}, \hat \rho_{tot}].
\end{equation}
If we concentrate on the open system, we can trace over the environment degrees of freedom and define the reduced density matrix $\hat \rho_s(t) = {\mathrm{Tr}}_e  \hat \rho_{tot}(t)$. After tracing over the environmental degrees of freedom, one gets a more complicated equation for $\hat \rho_s$, known as quantum master equation,
 \begin{equation}
i\hbar \dot {\hat{\rho}}_{s}(t) = {\mathrm{Tr}} _e[\hat H_{tot}, \hat \rho_{tot}]=[\hat H_s, \hat \rho_s]+{\mathrm{Tr}}_e [I_s \otimes \hat H_e + \hat H_{int}, \hat \rho_{tot}] \equiv \mathcal{L}{\hat{\rho}}_{s}(t),
\end{equation}
where $\mathcal{L}$ is a super-operator that describes the irreversible part of the dynamics of the subsystem propagating through the medium. In most cases, this equation cannot be solved analytically and even numerically. There are three main timescales in the system: the environment relaxation time scale $\tau_e$, the intrinsic timescale of the evolution of the subsystem $\tau_s$($\approx 1/|\omega'-\omega|$ with $\omega$ and $\omega'$ being the typical frequencies or energy levels), and the relaxation time scale of the subsystem $\tau_r$. If the environment relaxation time is much smaller than the relaxation time of the subsystem $\tau_e \ll \tau_r$ (Markovian approximation), the quantum master equation is simplified as a Lindblad equation (with $\hbar$=1)~\cite{Lindblad},
\begin{equation}
\dot {\hat{\rho}}_s(t) =-i [\hat H_s, \hat \rho_s]+\sum_{i=1}^N \gamma_i \left(L_i \hat \rho_s L_i^\dag -{1\over 2}L_i^\dag L_i\hat \rho_s -{1\over 2}\hat \rho_s L_i^\dag L_i \right),
\label{Lindblad}
\end{equation}
where the explicit form of the subsystem Hamiltonian $\hat H_s$, the Lindblad operators $L_i$ which describes the effect of the environment, and the coefficients $\gamma_i$ are derived from the master equation. Markovian approximation means that the memory effect of the medium can be largely neglected, and the time evolution implemented by the Lindblad equation is irreversible. If the intrinsic timescale of the evolution of the subsystem is much smaller than the relaxation time of the subsystem $\tau_s \ll \tau_r$, the evolution of the open quantum system moves to the quantum optical limit. If the environment relaxation time scale is much smaller than the intrinsic timescale of the evolution of the subsystem $\tau_e \ll \tau_s$, the evolution of the open quantum system tends to the quantum Brownian motion limit~\cite{oqs2002}. A well-known model of quantum Brownian motion is the Caldeira-Leggett model at high temperature~\cite{Caldeira:1982iu}. In this model, a (heavy) point-particle is coupled to a bath consisting of a large number of light harmonic oscillators. The condition of $\tau_e \ll \tau_s$ requires that the extension of the heavy quark is always smaller than the correlation length of the medium. The Caldeira-Leggett model can be used to describe the evolution of a single heavy quark. However, it fails to describe the quarkonium evolution when the radius of the quarkonium is with the same order or even larger than the medium correlation length. Feynman and Vernon developed a systematic treatment of how to derive the system-medium interaction~\cite{Feynman:1963fq}. The Lindblad master equation for in-medium quarkonium has been derived based on the Feynman-Vernon influence functional. The partition function of the total system is defined as (the closed-time path formalism)~\cite{Akamatsu:2012vt,Akamatsu:2014qsa}
\begin{eqnarray}
Z[\eta_1,\eta_2]&=&\int \mathcal{D}[\phi]_{1,2}\langle \phi_1|\hat \rho_{tot}|\phi_2 \rangle \nonumber \\
&\times& \exp\left[i\int_{t_0} d^4x\{ \mathcal{L}_{tot}(\phi_1)-\phi_1\eta_1 \}\right] \exp\left[-i\int_{t_0} d^4x\{ \mathcal{L}_{tot}(\phi_2)-\phi_2\eta_2 \}\right].
\end{eqnarray}
where $\phi=(A, q,\psi)$ represent the gluon, light quark, and heavy quark fields, $\mathcal{L}_{tot}$ is the total Lagrangian density, and $\eta_{1,2}$ are the sources. Assume that the initial density matrix can be factorized as $\hat \rho_{tot}=\hat \rho_s \otimes \hat \rho_e$ and that the environment reaches equilibrium, one can get
\begin{eqnarray}
Z[0,0]=\int \mathcal{D}[\psi]_{1,2}\langle \psi_1^\dag|\hat \rho_{s}|\psi_2 \rangle \exp\left[i S_{kin}[\psi_1]-iS_{kin}[\psi_2]+iS_{FV}[j_1,j_2]\right],
\end{eqnarray}
where $S_{kin}$ is the kinetic term of heavy quarks. The influence functional $S_{FV}$ defined as a functional of the heavy quark color current can be expressed as
\begin{eqnarray}
e^{iS_{FV}[j_1,j_2]}&=&\int \mathcal{D}[A,q]_{1,2}\langle A_1,q_1|\hat \rho_{e}|A_2,q_2 \rangle \nonumber \\
&\times& \exp\left[i\int_{t_0} d^4x\{ \mathcal{L}_{g+q}-gj_1^{a\mu}A_{1\mu}^a \}\right] \exp \left [-i\int_{t_0} d^4x\{ \mathcal{L}_{g+q}-gj_2^{a\mu}A_{2\mu}^a \} \right],
\end{eqnarray}
where $\mathcal{L}_{g+q}$ is the Lagrangian density for gluons and quarks. Considering the following assumptions: 1) one can take the non-relativistic limit for the heavy quark Lagrangian due to the large mass, 2) the medium temperature is high but much lower than the heavy quark mass, which allows a perturbative expansion of the influence functional, and 3) the intrinsic time scale of the heavy quark is long enough compared to the environment relaxation time $\tau_e \ll \tau_s$, the influence functional in the Markov limit can be divided into four terms with different physical meanings,
\begin{equation}
S_{FV}=S_{pot}+S_{fluct}+S_{diss}+S_{L},
\end{equation}
where $S_{pot}$ gives a potential between the two heavy quarks, $S_{fluct}$ accounts for the thermal fluctuations, $S_{diss}$ gives rise to the dissipative dynamics such as the drag force, and $S_{L}$ is a term which is useful to obtain Lindblad-form master equations. The explicit forms can be found in~\cite{Akamatsu:2014qsa}. We are interested in the evolution equation for the density matrix operator $\hat \rho_s(t)$ in the position basis. Aiming to do this, one can firstly project $\hat \rho_s(t)$ to the coherent states as generating functionals for heavy quarkonium. The density matrix in position space can be obtained via functional differentiation, and the time evolution equation for $\hat \rho_s(t)$ is obtained by the functional Schr\"odinger equation (where the Hamiltonian can be obtained by a Legendre transformation of the Lagrangian). This gives the master equation which is similar to the Lindblad equation. Comparing with the Lindblad form, this approach can tell us the explicit form of the Lindblad operators $L_i$ and coefficients $\gamma_i$ for the single heavy quark or quarkonium subsystem in various conditions~\cite{Akamatsu:2014qsa}. Besides, a non-trivial Hamiltonian for the subsystem can also been obtained. The master equations are equivalent to the stochastic Schr\"odinger equations in the recoilless limit,
\begin{eqnarray}
&i{\partial \over \partial t}\psi(t, {\bf r})= \hat H \psi(t, {\bf r}) \nonumber \\
&\hat H = -{\nabla_{\bf r}^2\over m_Q}+iC_FD(0)-(V({\bf r})+iD({\bf r}))(t^a \otimes t^{a*})+\theta^a(t, {\bf r}/2)(t^a\otimes 1)-\theta^a(t, -{\bf r}/2)(1 \otimes t^{a*}),
\end{eqnarray}
where $V({\bf r})$ is the real-part screened potential, $D({\bf r})={\mathrm{Im}}V(r)- {\mathrm{Im}}V(r=\infty)$ is the shifted imaginary part of the real-time potential, $C_F=(Nc^2-1)/2N_c$ is controlled by the color number $N_c$, $t^a$ is the color matrix, $\theta^a$ is the white noise with ensemble averages $\langle \theta^a(t, {\bf x})\rangle=0$ and $\langle \theta^a(t, {\bf x})\theta^b(s, {\bf y})\rangle=-D({\bf x-y})\delta(t-s)\delta^{ab}$, and $V({\bf r})$ and $D({\bf r})$ are defined by the retarded propagator and spectral function of gluons, which can be calculated via hard thermal loop resummed perturbation theory at high temperature. The stochastic Schr\"odinger equation describes the effect of thermal fluctuation on the quantum state of heavy quarks. Because of the thermal fluctuation, the wave function at distant points becomes decoherent. The decoherence induced by medium fluctuations is found to play an important role, in addition to the static screening. The decoherence depends on the interplay of two length scales~\cite{Akamatsu:2014qsa,DeBoni:2017ocl}, the correlation length of the noise $l_{fluct}\sim 1/gT$ and the coherence length of the bound states $l_{coh}$ (the range of the screened potential). If the correlation length is much larger than the coherence length $l_{fluct}\gg l_{coh}$, the noises for a heavy quark and a heavy antiquark are nearly canceled, the decoherence is inefficient, and the quarkonium dissociation requires a longer time. For $l_{fluct}\lesssim l_{coh}$, the decoherence is so efficient that the quarkonium dissociates quickly. At very high temperature, the correlation length will become smaller than all of the quarkonium lengths, and thus the medium is able to resolve even the most deeply bound states. The numerical simulations of the stochastic potential model can be found in Refs~\cite{Rothkopf:2013kya,Kajimoto:2017rel}. 

The master equation can be obtained with the definition of  $\hat \rho_s(t)=\langle \psi(t, {\bf r}) \psi^*(t, {\bf r'}) \rangle$. However, the stochastic Schr\"odinger equations can't be used to describe the irreversible processes such as momentum dissipation due to the time-dependent random potential. From the numerical side, solving the stochastic Schr\"odinger equation has a big advantage over solving the master equation because the dimension of the former is the square root of the latter. In the classical limit, the classical Langevin equation or Fokker-Planck equation can also be obtained in this framework~\cite{Akamatsu:2012vt}.

Recently several research groups are exploring open quantum systems to describe the quarkonium motion, by directly starting from effective field theories such as pNRQCD. The quarkonium system is described by the pNRQCD while the environment is a weakly coupled~\cite{Yao:2017fuc,Yao:2018nmy} or strongly coupled~\cite{Brambilla:2016wgg,Brambilla:2017zei} quark-gluon plasma. Under the Markovian approximation, it is shown that the Lindblad equation leads to a Boltzmann transport equation if a Wigner transformation is applied to the system density matrix. Now we follow the work in Ref.~\cite{Yao:2018nmy} to show the logic and approximation in the approach. As we discussed in the previous section, if integrating out the degrees of freedom of momenta $m_Q$ and $m_Qv$ from the full QCD, one can get the pNRQCD, the Lagrangian density is shown in Eq.\ref{pnrqcd}. When going to finite temperature, two extra scales emerge: the temperature $T$ and the Debye screening mass $m_D$. Here we focus on the low temperature region where the temperature of the medium is below the melting temperature ($T_d\sim m_Qv^2$) of the quarkonium, 
\begin{equation}
m_Q\gg m_Qv \gg m_Qv^2 \gtrsim T \gtrsim m_D.
\end{equation}
The evolution of the bound states can be achieved by taking the Wigner transformation of the corresponding matrix elements of the subsystem density matrix $\hat \rho_s$ in the singlet basis $|{\bf k}, nl, 1 \rangle$ (the in-medium dynamical evolution of open heavy quarks can also be described by Boltzmann equations via this approach),
\begin{equation}
f_{nl}({\bf x,k},t)\equiv \int {d^3 {\bf k'}\over (2\pi)^3}e^{i{\bf k'}\cdot {\bf x}} \langle {\bf k+{k'\over 2}}, nl, 1 |\hat\rho_s(t)|{\bf k-{k'\over 2}}, nl, 1 \rangle.
\end{equation}
Therefore, the evolution of the quarkonium system follows the Lindblad equation. We need to calculate all the terms on the right-hand side of (\ref{Lindblad}). 
\begin{eqnarray}
f_{nl}({\bf x,k},t+\Delta t) &= &f_{nl}({\bf x,k},t) -i\Delta t \int {d^3 {\bf k'}\over (2\pi)^3}e^{i{\bf k'}\cdot {\bf x}} \langle {\bf k+{k'\over 2}}, nl, 1 |[\hat H_s, \hat \rho_s(t)]|{\bf k-{k'\over 2}}, nl, 1 \rangle + ... \nonumber \\
&=& f_{nl}({\bf x,k},t) -i\Delta t \int {d^3 {\bf k'}\over (2\pi)^3}e^{i{\bf k'}\cdot {\bf x}} (E_{\bf k+{k'\over 2}}-E_{\bf k-{k'\over 2}})\langle {\bf k+{k'\over 2}}, nl, 1 |\hat \rho_s(t)|{\bf k-{k'\over 2}}, nl, 1 \rangle + ... \nonumber \\
&=& f_{nl}({\bf x,k},t) - \Delta t {\bf v} \cdot  \nabla_{\bf x} f_{nl}({\bf x,k},t)+ ...,
\end{eqnarray}
where ${\bf v=k}/(2m_Q)$ is the quarkonium velocity. The last step comes from the defintion of kinetic energy $E_{\bf k}=-|E_{nl}|+{\bf k}^2/(4m_{Q})$. The terms involving Lindblad operators can be divided into two sets. The one including $\hat \rho_s L_i^\dag L_i$ and $L_i^\dag L_i\hat \rho_s$ is directly related to the perturbative dissociation rate, as it is defined from the dipole transition from the singlet to octet. The other is $L_i \hat \rho_s L_i^\dag$ which is shown to be related to the quarkonium regeneration. Finally, one obtains the Boltzmann equation,
\begin{eqnarray}
{\partial \over \partial t}f_{nl}({\bf x,k},t) + {\bf v} \cdot  \nabla_{\bf x} f_{nl}({\bf x,k},t)= \mathcal{C}_{nl}^{(+)}({\bf x,k},t)-\mathcal{C}_{nl}^{(-)}({\bf x,k},t),
\end{eqnarray}
where $\mathcal{C}_{nl}^{(+)}$ and $\mathcal{C}_{nl}^{(-)}$ are regeneration and dissociation terms, respectively.

Another different approach to consider the hot bath effect on quarkonium motion is the Schr\"odinger-Langevin equation~\cite{Katz:2015qja}, which is originally from the Heisenberg-Langevin equation (SLE).
The Schr\"odinger-Langevin equation is considered as an effective open quantum system formalism suitable for phenomenological application to a quantum system interacting with a thermal bath.
The Schr\"odinger-Langevin equation can be expressed as
\begin{equation}
i\hbar{\partial \psi \over \partial t}= \left[\hat H_0+\hbar A(S(x,t)-\int \psi^*S(x,t)\psi dx)-xF_R(t) \right],
\end{equation}
where $\hat H_0$ is the isolated Hamiltonian of the system, $A$ the friction (drag) coefficient, $S$ the phase of the wavefunction, and $F_R(t)$ the noise term which can be taken as a Gaussian white noise. The Schr\"odinger-Langevin equation includes a thermal fluctuation term $xF_R(t)$ and a dissipative term $\hbar A(S(x,t)-\langle S \rangle)$. The relationship between the intensity of the fluctuation and drag coefficient satisfies the fluctuation-dissipation relation. The Schr\"odinger-Langevin equation~\cite{Katz:2015qja} includes dissipation due to the friction, but it is not associated with a master equation (hence to an SSE). This means that it is not clear how to derive an SLE from the underlying theory. 
  
Recent studies show that the stochastic Schr\"odinger equation can be derived from a Lindblad equation via the quantum state diffusion method, and that the Boltzmann transport equations can also be obtained from the Lindblad equation based on pNRQCD via the weak coupling. These studies show a deep connection between the approaches of QCD based open quantum systems and phenomenological models. So far, these deductions are with perturbative approximation. How to build a framework to consider the non-perturbative effect is still an open question~\cite{Brambilla:2017zei,Brambilla:2019tpt}. 

\subsection{Observables}
\label{obs}
In this section, we discuss those quarkonium observables which are sensitive to the properties of the created QGP phase in high energy nuclear collisions. They are nuclear modification factor $R_{AA}$, elliptic flow $v_2$, and averaged transverse momentum square $\langle p_T^2\rangle$.

\begin{figure}[!htb]
	{$$\includegraphics[width=0.32\textwidth]{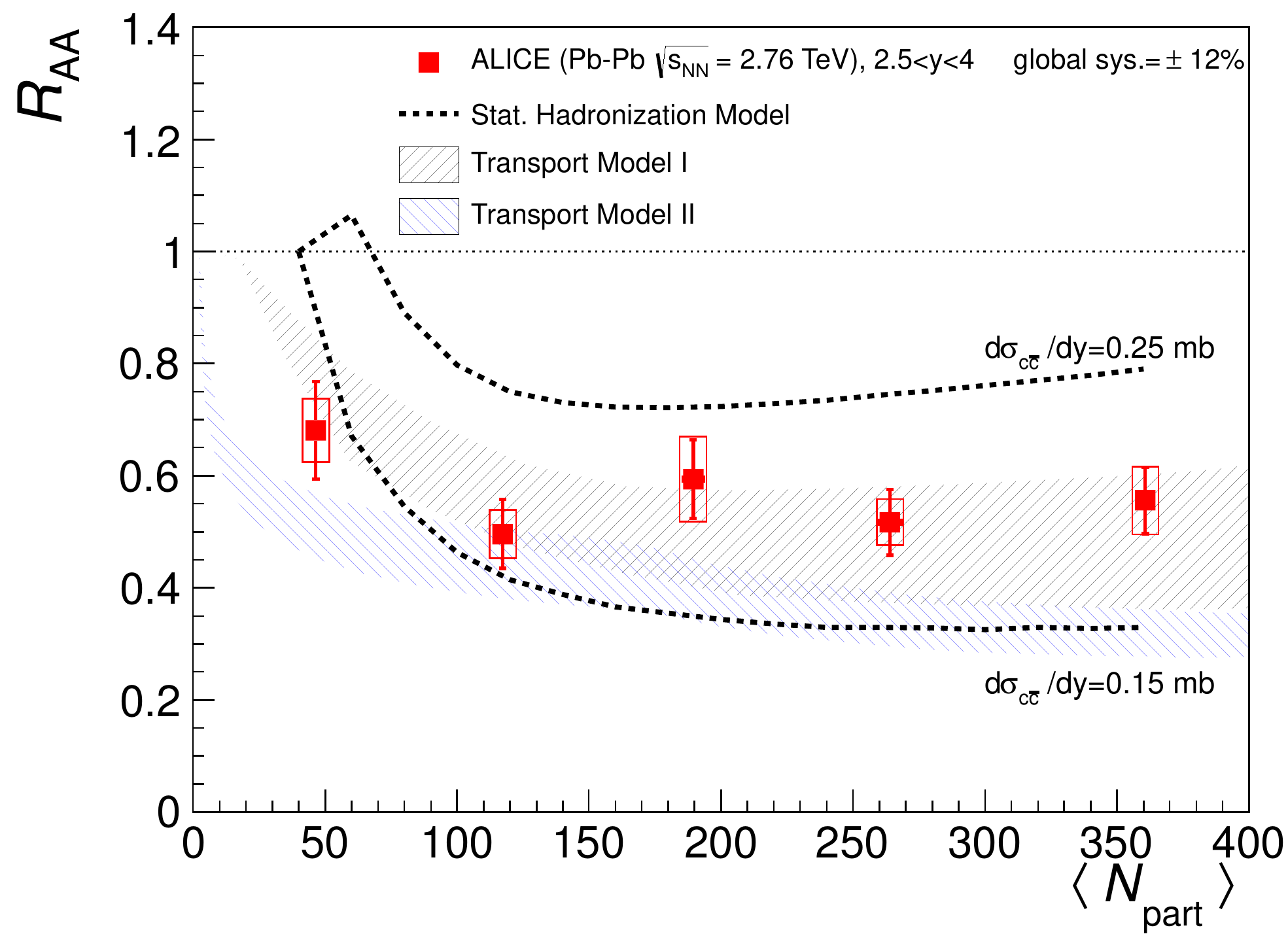} \ ~ \ ~ \ ~ \includegraphics[width=0.324\textwidth]{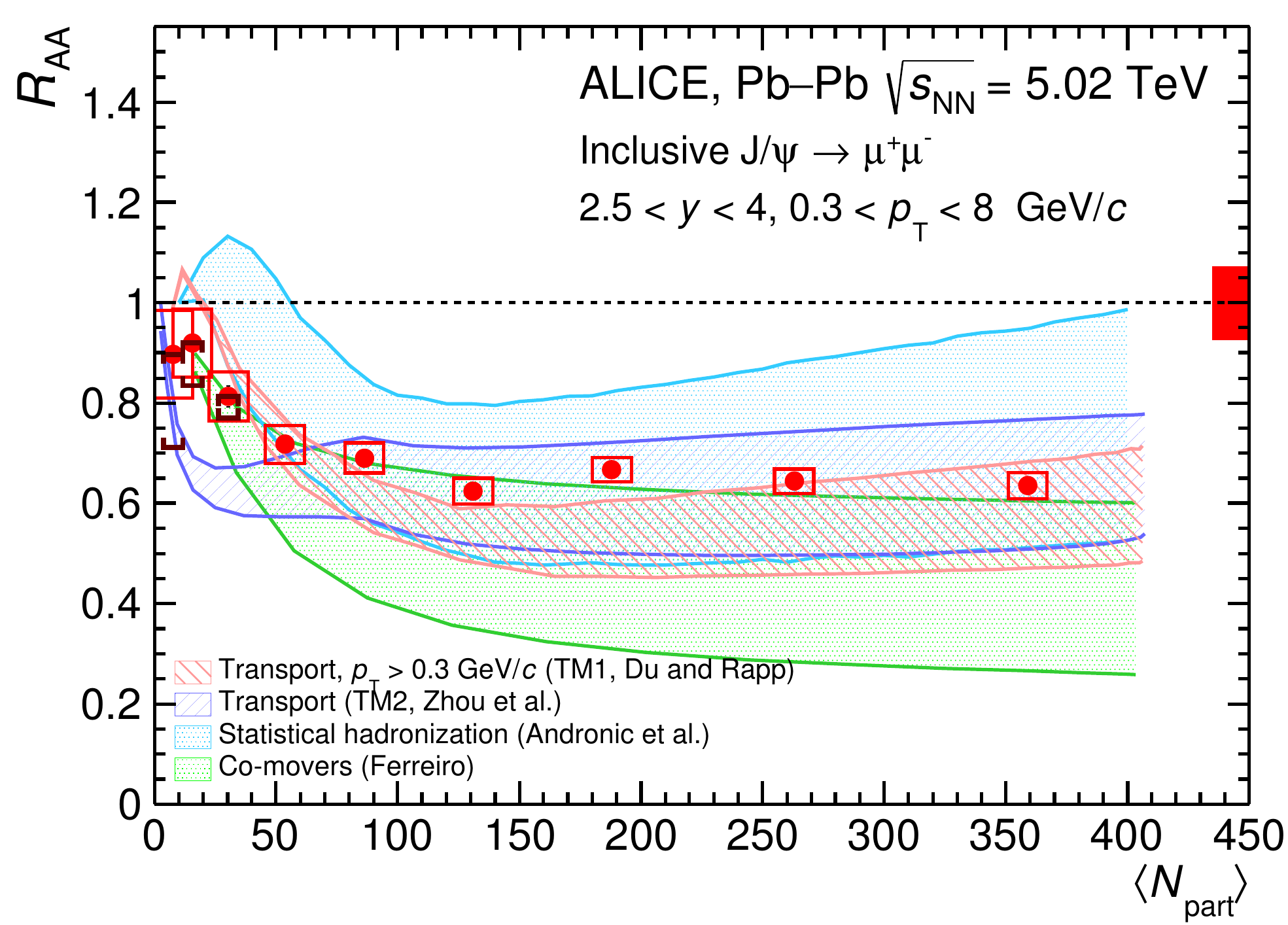}$$
		\caption{The experimentally measured and theoretically calculated nuclear modification factor $R_{AA}$ for inclusive $J/\psi$ at $\sqrt{s_{NN}}=2.76$ TeV (left panel) and $5.02$ TeV (right panel). The figures are taken from Ref.~\cite{Abelev:2012rv,Adam:2016rdg}.}
		\label{fig21}}
\end{figure}

The cold and hot nuclear matter effects on quarkonium yield can be described by the integrated or differential nuclear modification factor $R_{AA}^{\Psi}$ which describes the difference between a nuclear collision A+A and the simple superposition of p+p collisions. Any nuclear matter effect, cold or hot, will lead to a deviation of $R_{AA}$ from unit. The integrated $R_{AA}$ as a function of centrality for inclusive $J/\psi$ in Pb+Pb collisions at LHC energy~\cite{Abelev:2012rv,Adam:2016rdg} and the comparison with transport models and statistical approaches are shown in Fig.\ref{fig21}. With increasing centrality, the contribution from initial production drops down and the regeneration in the QGP goes up monotonically. The uncertainty in the charm quark production cross section in p+p collisions leads to a band for the regeneration. Different from the collisions at SPS energy where the regeneration can be neglected and at RHIC energy where the initial production is still a dominant component and the regeneration becomes equivalently important only in very central collisions, the regeneration at LHC energy becomes the dominant source of charmonium production in a wide centrality bin. The competition between the strong dissociation and regeneration leads to a flat structure for the total charmonium production at both forward rapidity and mid-rapidity.

One major feature of the hot and dense QCD medium created in heavy ion collisions is its perfect fluidity and strongly coupled nature, which are manifested as the large collective flow of the medium measured in A+A collisions and described well by hydrodynamic models. For non-central collisions, the azimuthal anisotropic flow is relevant for characterizing the flow feature of the medium, which is defined as the Fourier expansion coefficients on the final particle momentum spectra,
\begin{equation}
E{d^3N \over d^3{\bf p}}={d^2N \over 2\pi p_Tdp_T dy} \left (1+\sum_{n=1}^\infty 2v_n\cos [n(\phi-\Phi_{rp})]  \right ),
\end{equation}
where $\Phi_{rp}$ is the azimuthal angle of the reaction plane. The particular interest is the second coefficient, the elliptic flow $v_2=\langle p_x^2-p_y^2\rangle/\langle p_x^2+p_y^2\rangle$. While the anisotropic flows do not directly take the meaning of hydrodynamic flow (in this sense the radial flow is more relevant), it is known that they are closely related to the hydrodynamic evolution in the early stage and reflect the evolution of the flow features of the medium. Considering the large mass, heavy quarks are usually not expected to be easily involved in the medium flow unless they are thermalized to some extent. Therefore, the study of collective flow on heavy flavor particles can provide a direct measure of the degree of their interaction with the medium.
\begin{figure}[!htb]
	{$$\includegraphics[width=0.32\textwidth]{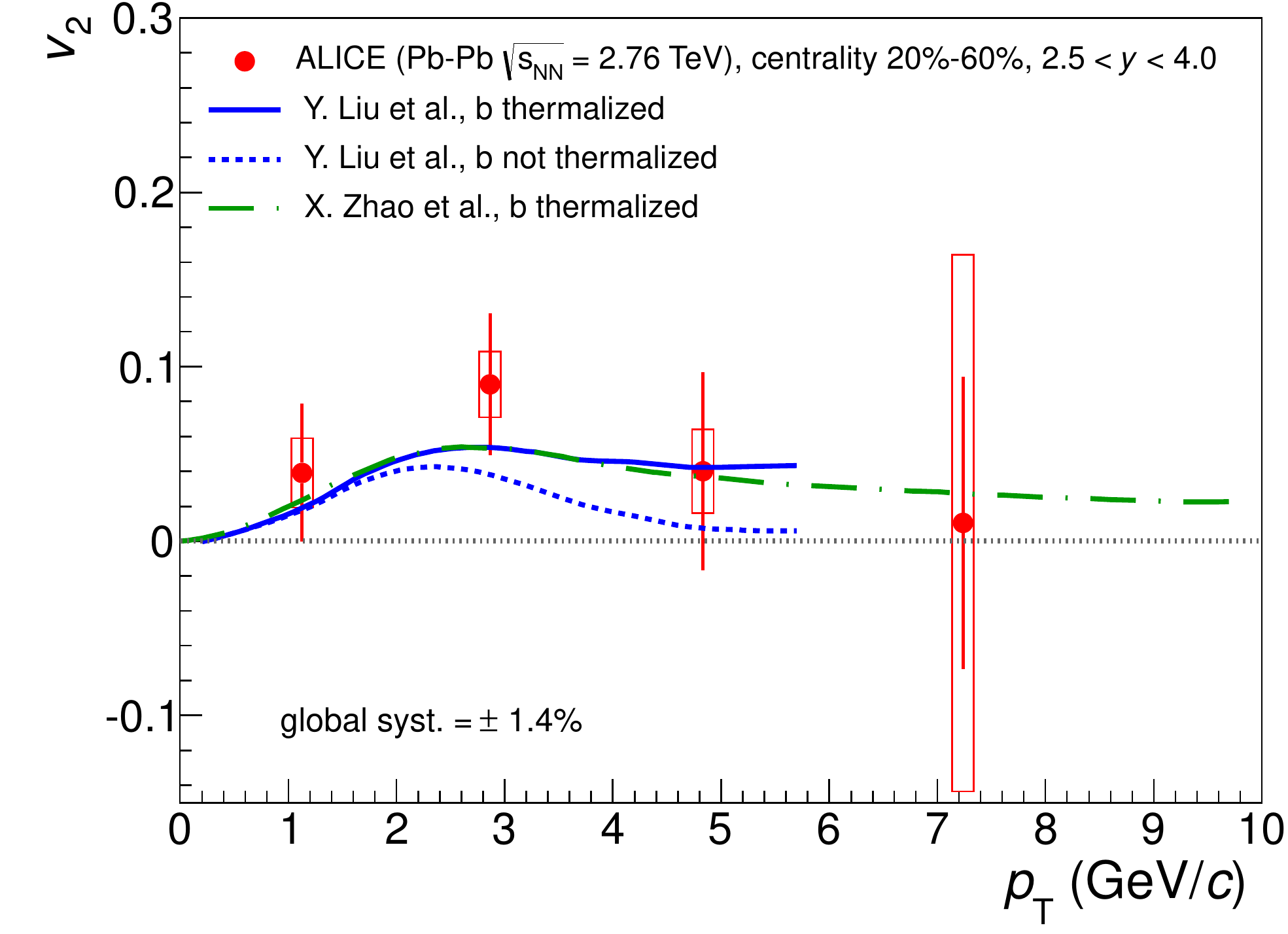} \ ~ \ ~ \ ~ \includegraphics[width=0.32\textwidth]{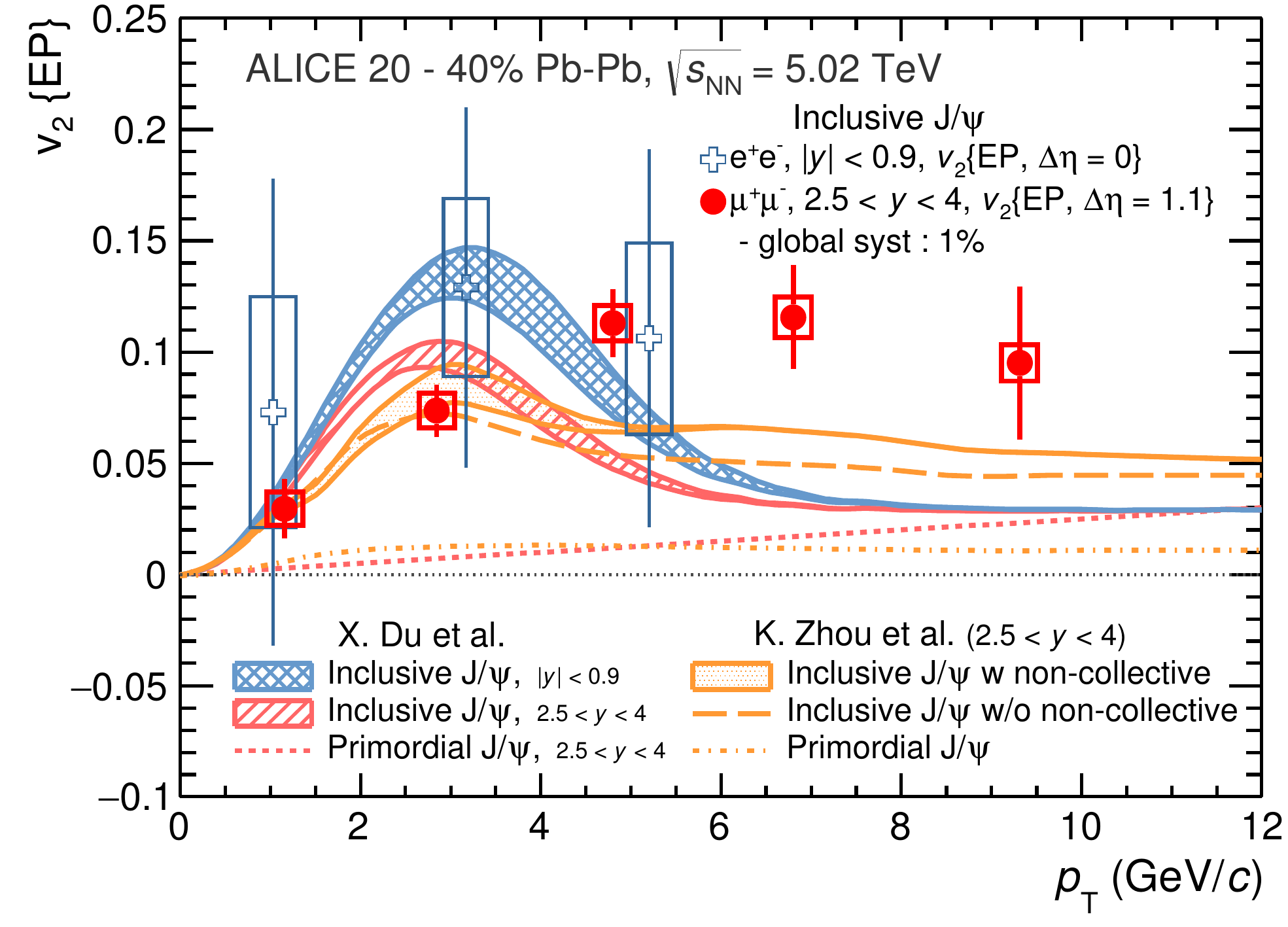}$$
		\caption{The inclusive $J/\psi$ elliptic flow $v_2$ as a function of transverse momentum at $\sqrt{s_{NN}}=2.76$ TeV (left panel) and $5.02$ TeV (right panel) and the comparison with the model calculations. The figures are taken from Ref.~\cite{ALICE:2013xna,Acharya:2017tgv}.}
		\label{fig22}}
\end{figure}

The $v_2$ of inclusive $J/\psi$ at LHC energy~\cite{ALICE:2013xna,Acharya:2017tgv} is shown in Fig.\ref{fig22}. The $J/\psi$s produced in initial collisions prior to the formation of the hot medium will not present any significant elliptic flow, due to the weak interaction between colorless charmonia and colored medium. The regenerated $J/\psi$s, on the other hand, may inherit flow from the partially or fully thermalized charm quarks. When we include the $J/\psi$ from bottom quark decay, the thermalized bottom quarks will contribute also to the charmonium flow. As one can see in the plot, when the bottom quarks are not thermalized with the medium, the $J/\psi$ elliptic flow quickly drops down and reaches zero at $p_T\sim 5$ GeV. However, the thermalized bottom quarks contribute a lot to the $J/\psi$ $v_2$ at high $p_T$. At this moment, the error bar of the experimental data is still too large to draw any conclusion about the bottom quark thermalization. 

While the differential nuclear modification factor $R_{AA}$ and elliptic flow $v_2$ tell us the importance of the regeneration mechanism, the quantity which is more sensitive to the thermalization of charm quarks and can be used to distinguish between the cold and hot medium effects is the ratio of averaged charmonium transverse momentum square in A+A and p+p collisions~\cite{Zhou:2014kka}, 
\begin{equation}
r_{AA}={\langle p_T^2\rangle _{AA} \over \langle p_T^2\rangle _{pp}}.
\end{equation}

\begin{figure}[!htb]
	{$$\includegraphics[width=0.33\textwidth]{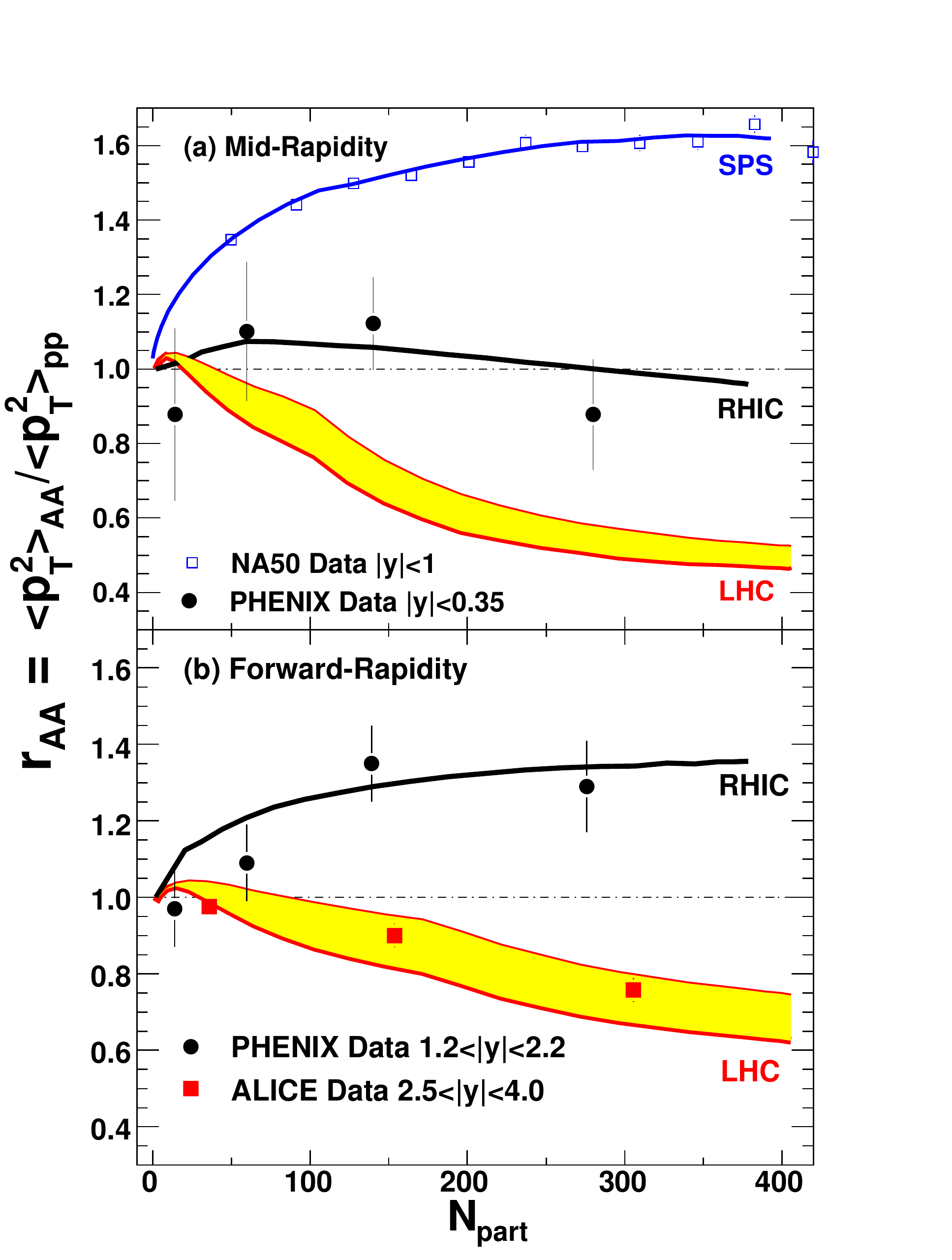}\includegraphics[width=0.33\textwidth]{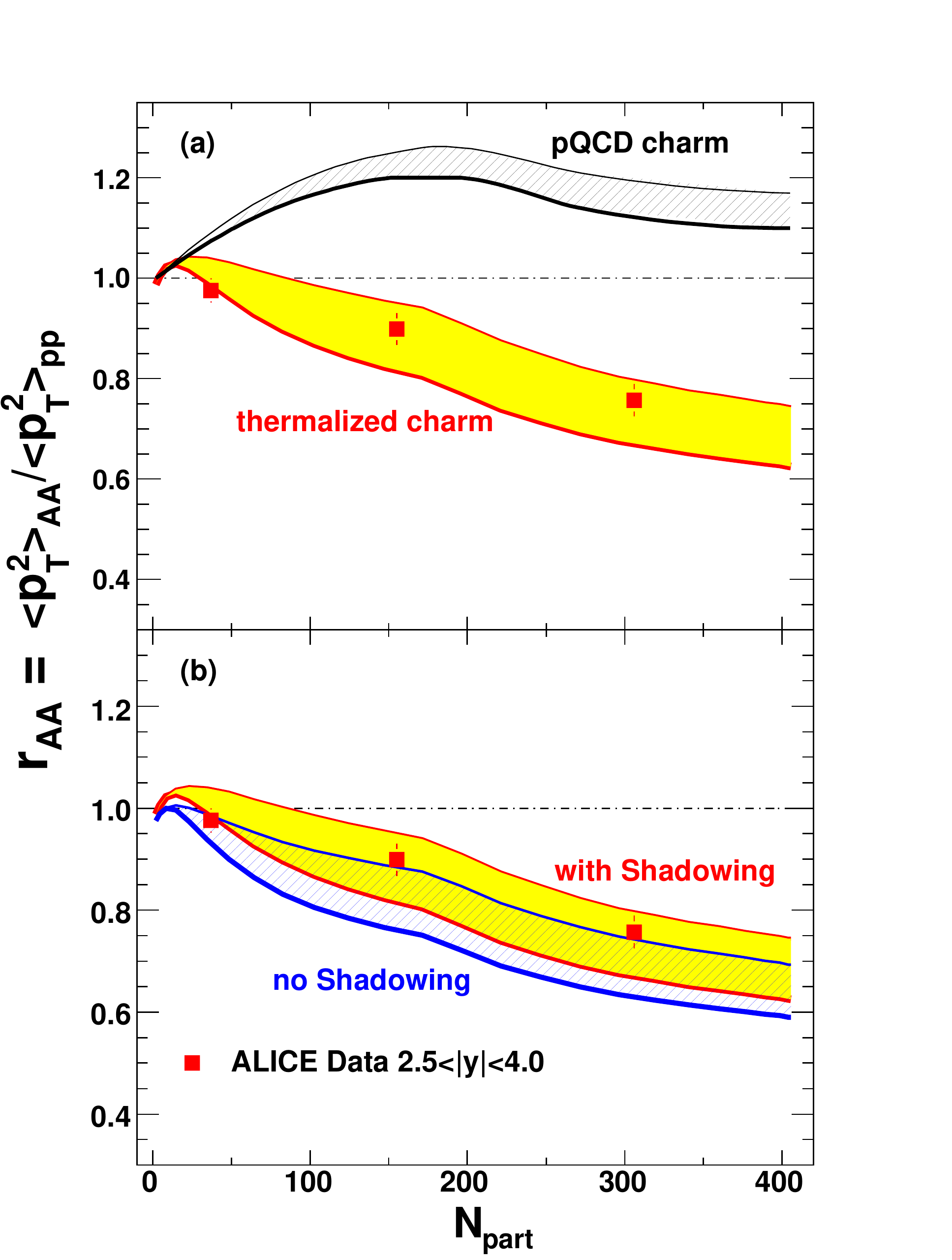}$$
		\caption{The $J/\psi$ nuclear modification factor for averaged transverse momentum $r_{AA}=\langle p_T^2\rangle_{AA}/\langle p_T^2\rangle_{pp}$ as a function of number of participants in mid and forward rapidity regions at SPS, RHIC and LHC energies (left panel) and its thermalization and shadowing dependence (right panel). The figures are taken from \cite{Zhou:2014kka}. }
		\label{fig23}}
\end{figure}

The calculated $r_{AA}$ in transport models and the comparison with experimental data at SPS, RHIC and LHC energies are shown in Fig.\ref{fig23} in mid and forward rapidity bins.
The energy dependence of $r_{AA}$ clearly reflects the underlying $J/\psi$ production and suppression mechanisms in high energy nuclear collisions. At lower collision energy where almost all the observed $J/\psi$s are from the initial production. In this case, the Cronin effect tends to increase the transverse momentum of the finally observed $J/\psi$s. Since the Cronin effect is proportional to the gluon traveling length in the nuclei, $r_{AA}$ increases monotonically versus collision centrality. In extremely high energy nuclear collisions, on the other hand, the regeneration for charmonia is significant. Although the initially produced heavy quarks carry high transverse momentum, they lose energy when passing through the medium. The heavy quark distribution in the medium should be between the pQCD distribution which is the limit without energy loss and can be simulated by some Monte Carlo generators like PYTHIA and the thermal distribution which is the other limit with full energy loss and determined by the medium temperature and flow. As a consequence of the increasing regeneration fraction with colliding energy, the competition between the initial production which controls high $p_T$ charmonium production and the regeneration which dominates the low momentum charmonium production leads to the decrease of the values of the ratio $r_{AA}$ from SPS to LHC. The predicted $r_{AA}$ at mid rapidity for heavy ion collisions at LHC is below unity and decreases toward more central collisions. The prediction has been confirmed by the experimental data in the forward rapidity window, as shown in the lower panel. At RHIC, the competition between the initial gluon scattering and the final stage regeneration leads to a weak centrality dependence for the mid-rapidity $r_{AA}$. At the forward rapidity, due to the smaller heavy quark production cross section, the regeneration becomes not so important as the initial gluon scattering, and $r_{AA}$ becomes higher than unity and increases as a function of centrality. Since the heavy quark production cross section is large at LHC, even at the forward-rapidity, the $r_{AA}$ remains lower than unity. The above energy dependence of $r_{AA}$ can be qualitatively summarized as
$$ r_{AA}=\left\{
\begin{array}{rcl}
>1       &      &\text{SPS}\\
\sim1     &      &\text{RHIC}\\
<1    &      & \text{LHC} .
\end{array} \right.$$

What is the shadowing effect on the ratio $r_{AA}$? Different from the integrated yield which depends strongly on both the cold and hot nuclear matter effects, the averaged transverse momentum is a normalized quantity, the shadowing which changes the parton distribution is minimized in the $r_{AA}$. The small difference between the solid and dashed lines, shown as the hatched band in Fig.\ref{fig23}, is induced by the shadowing effect which is taken from the EKS98 in the calculation. At RHIC energy the band becomes very narrow.

\section{Heavy flavors in electromagnetic and rotational fields}
\label{electro}
It is well known that a strong electromagnetic field and a strong rotational field can be generated in non-central relativistic heavy ion collisions. The maximum magnetic field can reach $eB\sim 5 m_\pi^2$ in semi-central Au+Au collisions at RHIC energy and $\sim 70 m_\pi^2$ in semi-central Pb+Pb collisions at LHC energy~\cite{Deng:2012pc}. As an external field like temperature and baryon chemical potential, the strong electromagnetic field and rotational field will change the QCD symmetry and the QCD phase structure. For instance, in a strong magnetic field the chiral condensate at low temperature is enhanced, called magnetic catalysis, but the critical temperature for the chiral restoration is reduced, called inverse magnetic catalysis~\cite{Shovkovy:2012zn,Bruckmann:2013oba}. In chiral limit, the number imbalance between the left and right hand quarks leads to a chiral current in an external magnetic field, called chiral magnetic effect~\cite{Fukushima:2008xe}. For quarkonia, the strong magnetic field in the early stage results in a non-collective $J/\psi$ flow at high $p_T$~\cite{Guo:2015nsa}. 

The vorticity field can affect the spin polarization of certain hadron production~\cite{Liang:2004ph,Becattini:2007sr}. Recently, the global polarization of $\Lambda$ hyperons in heavy ion collisions was measured by the STAR collaboration~\cite{STAR:2017ckg}. The large averaged polarization indicates that the medium vorticity is at the order of $\omega \approx (9\pm1)\times 10^{21}s^{-1}$, which is the strongest vorticity in nature. Such rotational collective motion in hot medium can induce anomalous transport effects like chiral vortical effect~\cite{Kharzeev:2007tn,Kharzeev:2010gr} and chiral vortical wave~\cite{Jiang:2015cva} which predict a baryon current or a baryonic charge quadrupole along the fluid rotation axis.

The strong electromagnetic field and vorticity field will affect the quarkonium evolution, yield and momentum spectra. In this section, we will summarize the various spectacular phenomena in heavy flavor sector induced by electromagnetic field and vorticity field.
\subsection{Electromagnetic field in heavy ion collisions}
\label{electro2}
We first discuss the framework which is used to calculate the electromagnetic field in heavy ion collisions. We consider the magnetic field at position ${\bf x=(x_\bot},z)$ caused by a particle with initial position ${\bf x}'_\bot$ and electric charge $e$ moving in the z-direction with rapidity $Y_0$. The magnetic field created by such a spectator nucleon at time $t$ can be calculated by either boosting the electric field or using the Lienard-Wiechert potentials,
\begin{eqnarray}
e{\bf B}_s^{\pm}&=&\alpha_{em} \sinh(\pm Y_0){({\bf x_\bot'}-{\bf x_\bot})\times {\bf e}_z\over [({\bf x_\bot'}-{\bf x_\bot})^2+(t\sinh Y-z \cosh Y)^2)]^{3/2}} \nonumber\\
&=&\alpha_{em} \sinh(\pm Y_0){({\bf x_\bot'}-{\bf x_\bot})\times {\bf e}_z\over [({\bf x_\bot'}-{\bf x_\bot})^2+\tau^2\sinh(\eta\mp Y_0)^2)]^{3/2}},
\end{eqnarray}
where the sign $\pm$ means the magnetic field produced by the nucleon moving in $\pm$ z-direction. Assuming that the spectators do not participate in any scattering, they will keep traveling with the beam rapidity $Y_0$. The magnetic field caused by all the spectators can be expressed as a sum of contributions from all the participant nucleons in the two nuclei~\cite{Kharzeev:2007jp}:
\begin{equation}
e{\bf B}_s^{\pm}(\tau,\eta,{\bf x_\bot'})=Z\alpha_{em} \sinh(\pm Y_0)\int d^2{\bf x_\bot'}\rho_\pm({\bf x_\bot'})[1-\theta_\mp({\bf x_\bot'})]{({\bf x_\bot'}-{\bf x_\bot})\times {\bf e}_z\over [({\bf x_\bot'}-{\bf x_\bot})^2+\tau^2\sinh(\eta\mp Y_0)^2)]^{3/2}},
\end{equation}
where $Z$ is the charge number of the nuclei, $\rho({\bf x})$ is the nucleon number density in nuclei, and the step function $\theta_\mp({\bf x'}_\bot)=\theta[R^2-({\bf x'}_\bot\mp{\bf b}/2)^2]$ is used to separate the spectators from the participants. Because the participants lose some rapidity in collisions, their contribution to the magnetic field is 
\begin{eqnarray}
e{\bf B}_p^{\pm}(\tau,\eta,{\bf x_\bot'})&=&Z\alpha_{em} \int d^2{\bf x_\bot'}\int_{-Y_0}^{Y_0}d(\pm Y)f(\pm Y)\sinh(\pm Y)\rho_\pm({\bf x_\bot'})\theta_\mp({\bf x_\bot'}) \nonumber\\
&&\times {({\bf x_\bot'}-{\bf x_\bot})\times {\bf e}_z\over [({\bf x_\bot'}-{\bf x_\bot})^2+\tau^2\sinh(\eta\mp Y)^2)]^{3/2}},
\end{eqnarray}
where $Y$ is the participant rapidity, and $f(Y)$ is the expirical rapidity distribution, 
\begin{equation}
f(\pm Y)={a\over 2\sinh(aY_0)}e^{\pm aY},\ \ \ \ \ \  -Y_0 \leq Y \leq Y_0.
\end{equation}
The experimental data show $a\approx1/2$, which is consistent with the baryon junction stopping mechanism~\cite{Kharzeev:2007jp}. Since the produced particles are globally charge neutral, we expect that, the contribution from the produced particles to the magnetic field is very small and can be neglected. We only take into account the contributions from the spectators and participants, and the total magnetic field is the sum of them, ${\bf B}={\bf B}^{+}_s+{\bf B}_{s}^-+{\bf B}_{p}^++{\bf B}_{p}^-$.

\begin{figure}[!htb]
	{$$\includegraphics[width=0.33\textwidth]{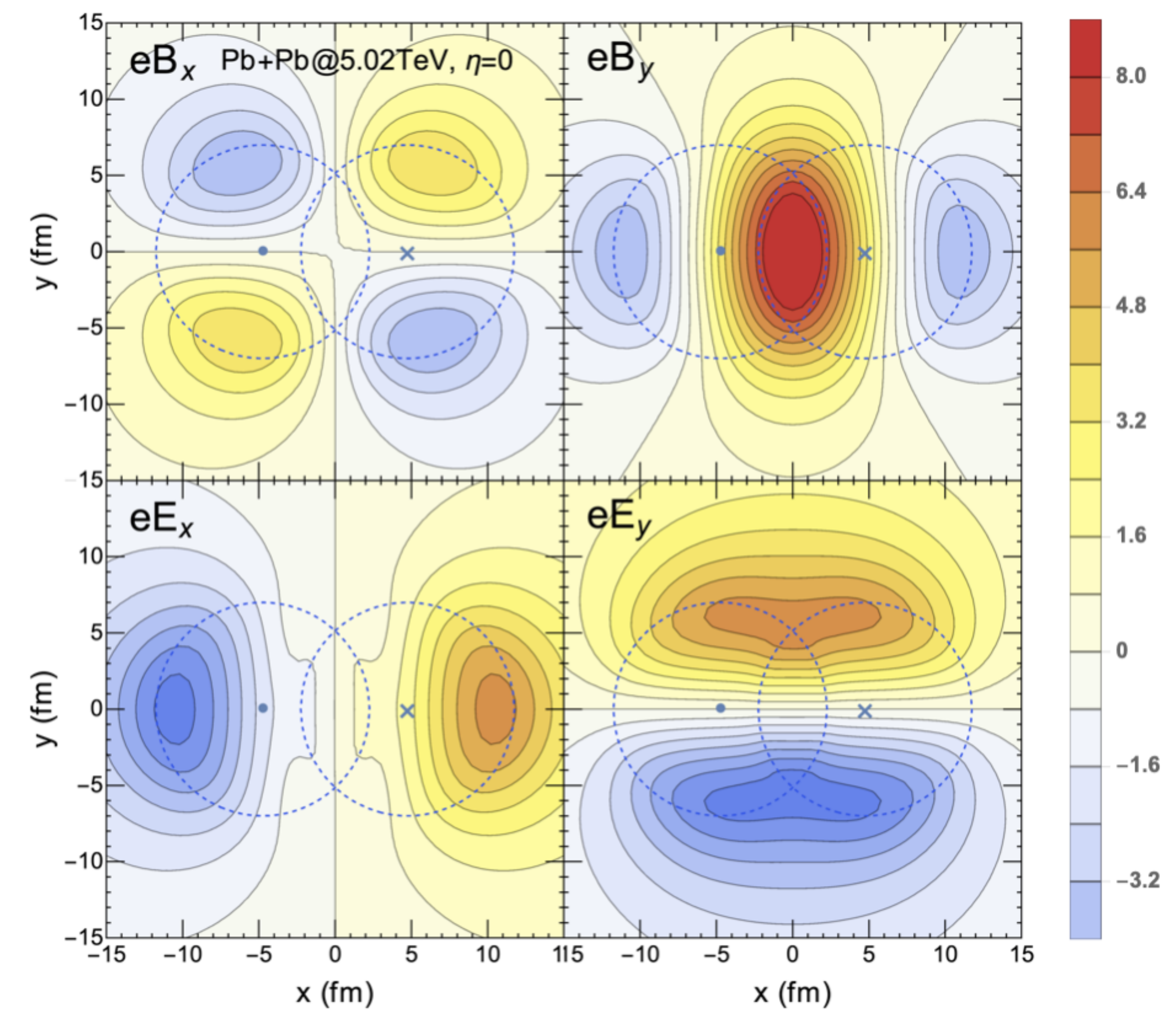} \ ~ \ ~ \includegraphics[width=0.326\textwidth]{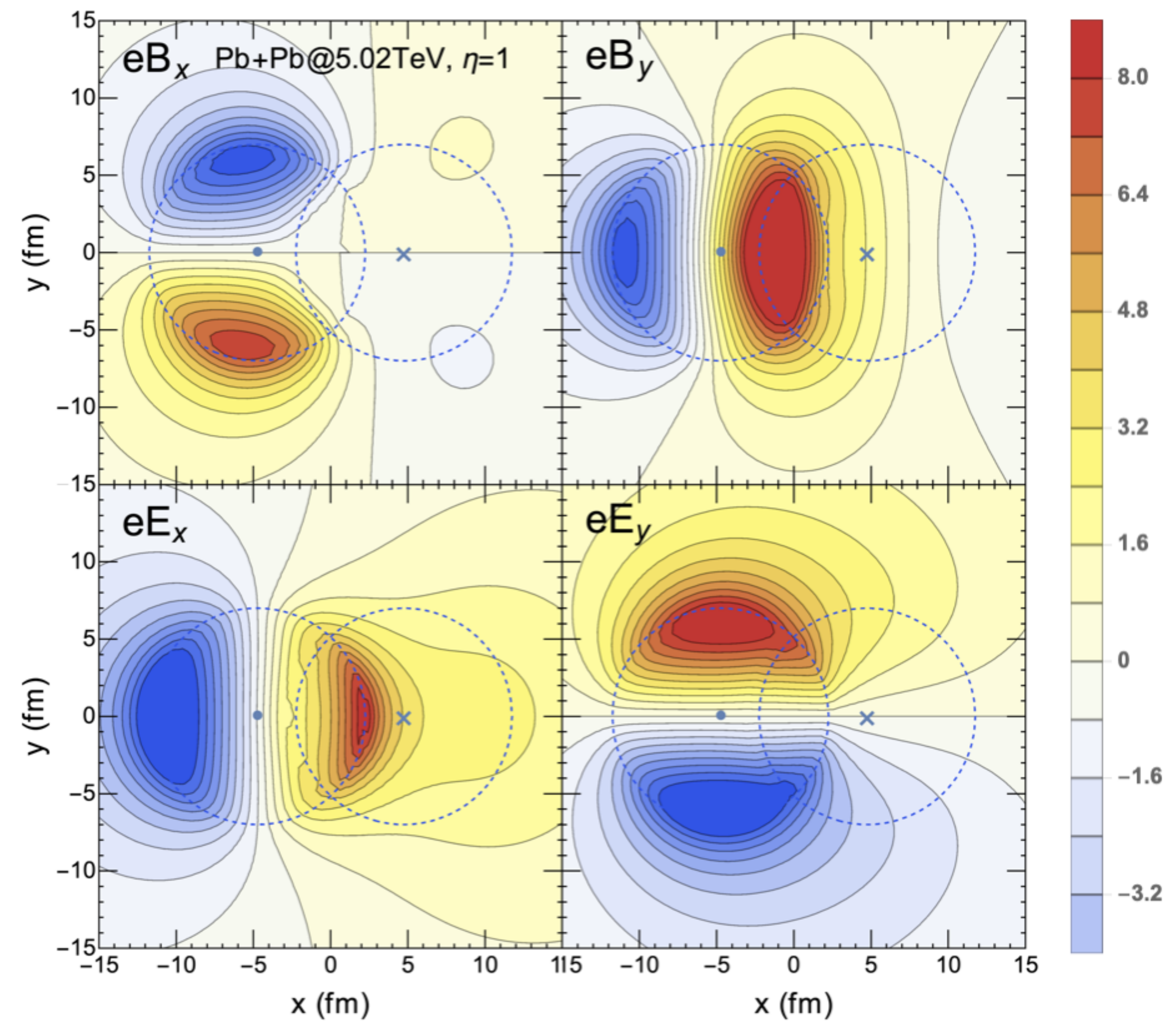}$$
		\caption{The transverse distributions of the initial electromagnetic components $E_x$, $E_y$, $B_x$ and $B_y$ at rapidity $\eta=0$ (left panel) and $\eta=1$ (right panel) in Pb+Pb collisions with impact parameter $b=9.5$ fm. }
		\label{fig24}}
\end{figure}

The corresponding electric field can be obtained from Lorentz transformation, 
\begin{eqnarray}
&&e{\bf E}_x^\pm(\tau,\eta,{\bf x_\bot})=e{\bf B}_y^{\pm}(\tau,\eta,{\bf x_\bot})\coth(\pm Y_0), \nonumber\\
&&e{\bf E}_y^\pm(\tau,\eta,{\bf x_\bot})=e{\bf B}_x^{\pm}(\tau,\eta,{\bf x_\bot})\coth(\mp Y_0).
\end{eqnarray}
The electromagnetic fields are inhomogeneous. The transverse distributions of different components of the initial electromagnetic fields at different rapidity in heavy ion collisions at LHC are shown in Fig.\ref{fig24}. It is easy to understand that, the longitudinal components $B_z$ and $E_z$ are much weaker than the transverse components. The y-component of the magnetic field $B_y$ is the most strong one in both most central rapidity and quite forward rapidity, while the x-component of the electric field $E_x$ is the most strong component in quite forward rapidity.

\begin{figure}[!htb]
	{$$\includegraphics[width=0.338\textwidth]{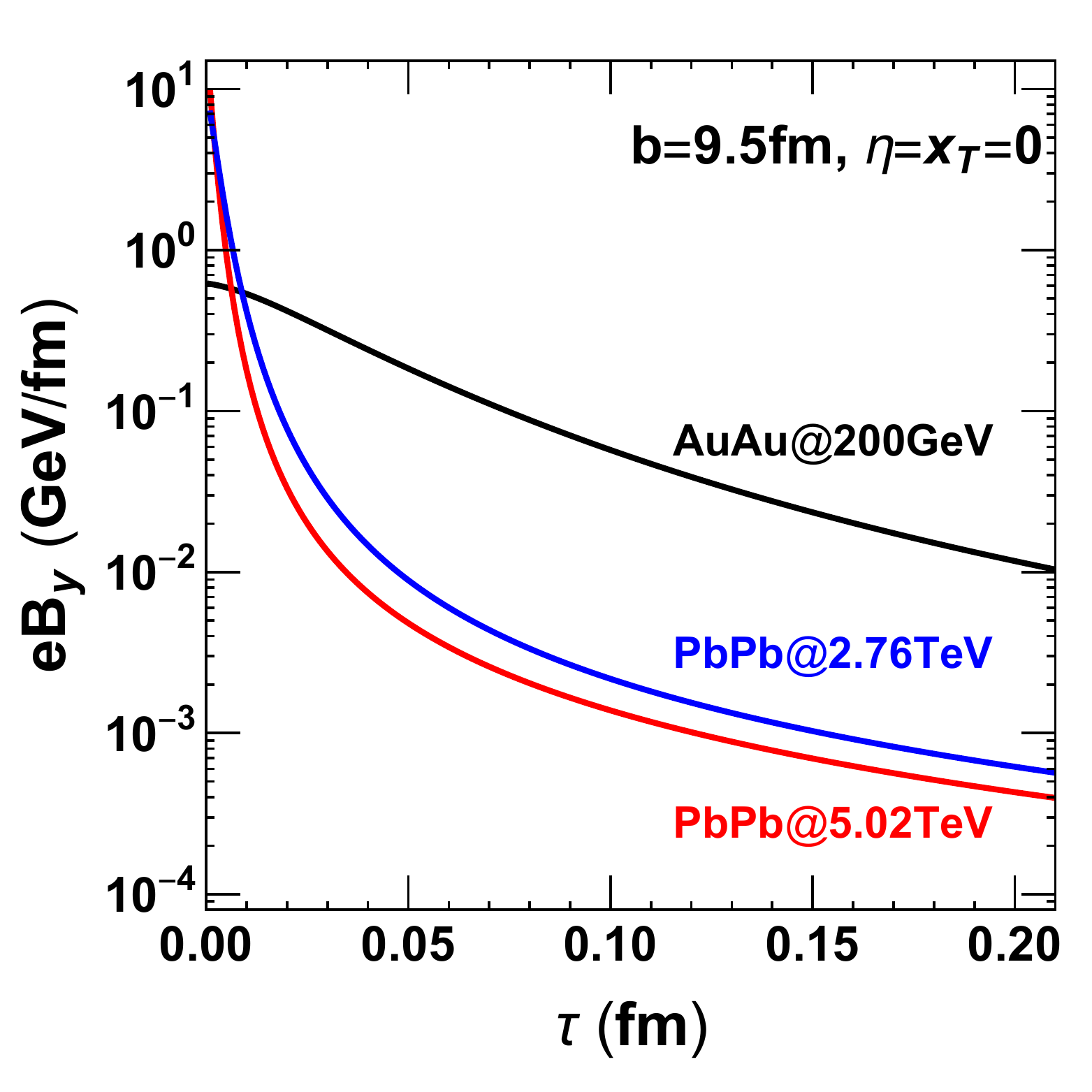} \ ~ \includegraphics[width=0.34\textwidth]{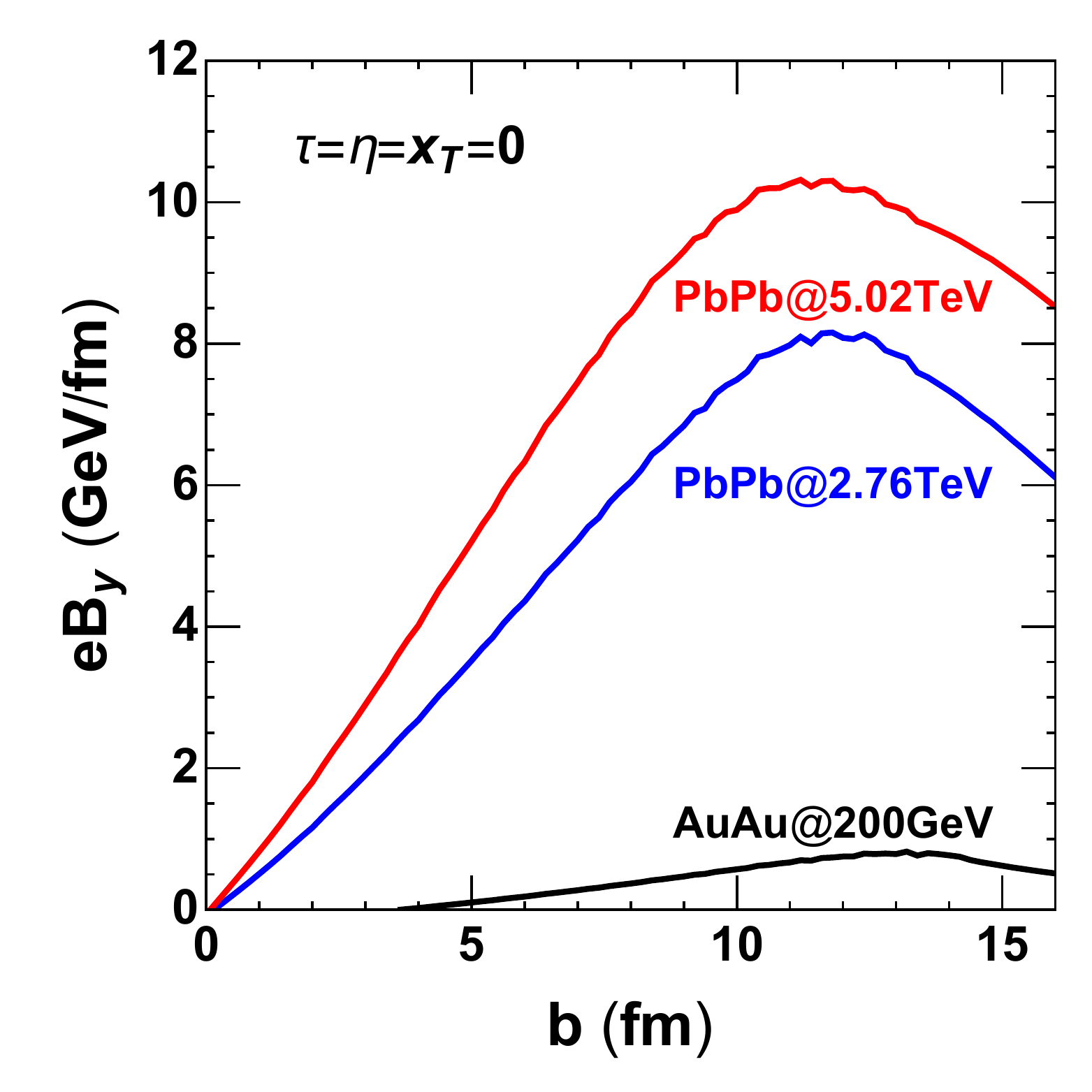}$$
		\caption{ The time dependence of the magnetic component $B_y$ (left panel) and the impact parameter dependence of the initial $B_y$ (right panel) in different colliding systems. }
		\label{fig25}}
\end{figure}

Fig.\ref{fig25} shows the time evolution of the y-component of the magnetic field in heavy ion collisions at RHIC and LHC. The initial magnetic field produced at LHC is much stronger than that at RHIC. However, the lifetime at LHC is much shorter than at RHIC. From the impact parameter dependence of the magnetic field, see the right panel of Fig.\ref{fig25}, the magnetic field is almost proportional to $b$ at small $b$ and reaches the peak value at $b\sim 12$ fm which is almost twice of the nuclear radius $2R_A$.

The classical (external) electromagnetic field decreases very fast in vacuum and may disappears before the QGP formation. However, the charged partons created in collisions will be influenced by the external electromagnetic field, and their feedback will enhance the total electromagnetic field. This electromagnetic response may lengthen the lifetime of the electromagnetic field in QGP. There are lots of works to study the electric conductivity $\sigma_{el}$ of the QGP matter, such as in perturbative QCD~\cite{Arnold:2003zc}, lattice QCD~\cite{Gupta:2003zh,Aarts:2007wj,Buividovich:2010tn,Ding:2010ga,Burnier:2012ts,Brandt:2012jc,Amato:2013naa}, effective models~\cite{Finazzo:2013efa,Sahoo:2018dxn}, and other transport approaches~\cite{Cassing:2013iz,Greif:2014oia,Puglisi:2014sha,Feng:2017tsh}. Aiming to estimate the electromagnetic response of QGP, we need to solve the Maxwell's equations in medium,
\begin{eqnarray}
\nabla\times {\bf B} &=&\epsilon\mu{\partial {\bf E}\over \partial t}+\mu\sigma_{el}({\bf E}+\tilde {\bf v}\times {\bf B})+\mu{\bf j}, \nonumber\\
\nabla \cdot {\bf B}&=&0, \nonumber\\
\nabla \times {\bf E}&=&-{\partial {\bf B}\over \partial t},\nonumber\\
\nabla \cdot {\bf E}&=&\rho,
\end{eqnarray}
where $\epsilon$ and $\mu$ are the permittivity and permeability of the QGP matter, $\tilde {\bf v}$ is the velocity of the medium, and ${\bf j}=\rho {\bf v}$ is the electric current generated by the external source. Under the assumption of vanishing electric charge density, the Maxwell's equations can be rewritten as 
\begin{eqnarray}
&&{\partial {\bf B} \over \partial t}={\bm \nabla}\times(\tilde {\bf v}\times {\bf B})+{1\over \sigma_{el} \mu}\left(\nabla^2{\bf B}-{\partial^2 {\bf B} \over \partial t^2}\right), \nonumber\\
&&{\partial {\bf E} \over \partial t} + {\partial \tilde {\bf v}\over \partial t}\times {\bf B}=\tilde {\bf v} \times ({\bm \nabla} \times {\bf E})+{1\over \sigma_{el} \mu}\left(\nabla^2{\bf E}-\epsilon\mu{\partial^2 {\bf E} \over \partial t^2}\right).
\label{maxell}
\end{eqnarray}
The first terms on the right hand sides are the convection terms, and the second terms are the diffusion terms. The ratio between the two can be defined as the magnetic Reynolds number $R_m$~\cite{Deng:2012pc}. It is easy to see that the Reynolds number $R_m\propto \sigma_{el} \mu$. Due to the large theoretical uncertainty about electric conductivity $\sigma_{el}$ of the QGP matter, the value of $R_m$ is still in a wide range. People usually consider the two limits: $R_m\ll 1$ and $R_m\gg 1$. For $R_m\ll 1$, the convection terms can be neglected and the equations become
\begin{eqnarray}
&&{\partial {\bf B} \over \partial t}={1\over \sigma_{el}}\left(\nabla^2{\bf B}-{\partial^2 {\bf B} \over \partial t^2}\right), \nonumber\\
&&{\partial {\bf E} \over \partial t} + {\partial \tilde {\bf v}\over \partial t}\times {\bf B}={1\over \sigma_{el}\mu}\left(\nabla^2{\bf E}-{\partial^2 {\bf E} \over \partial t^2}\right).
\end{eqnarray}
If the electrical conductivity $\sigma_{el}$ of the QGP is treated as a constant~\cite{Gursoy:2014aka,Tuchin:2013ie}, and the medium is taken as a static fireball, the time dependent electromagnetic field can be analytically solved by the method of Green functions. The method is recently extended to a dynamical medium~\cite{Stewart:2017zsu} with Bjorken expansion. 

For $R_m\gg 1$, the second terms on the right hand sides of Eq.\ref{maxell} can be neglected, and the equations become
\begin{eqnarray}
{\partial {\bf B} \over \partial t}&=& \nabla \times (\tilde {\bf v}\times {\bf B}), \nonumber\\
{\bf E}&=&-\tilde {\bf v}\times {\bf B}.
\end{eqnarray}
Under the simplifications~\cite{Deng:2012pc} of neglecting the influence of the electromagnetic fields on the evolution of the velocity $\tilde {\bf v}$, taking the Bjorken expansion $\tilde v_z=z/t$, and linearizing the ideal hydrodynamic equations to describe the transverse flow velocity $\tilde {\bf v}_\bot$, the time dependent magnetic field can be derived analytically.

\begin{figure}[!htb]
	{$$\includegraphics[width=0.31\textwidth]{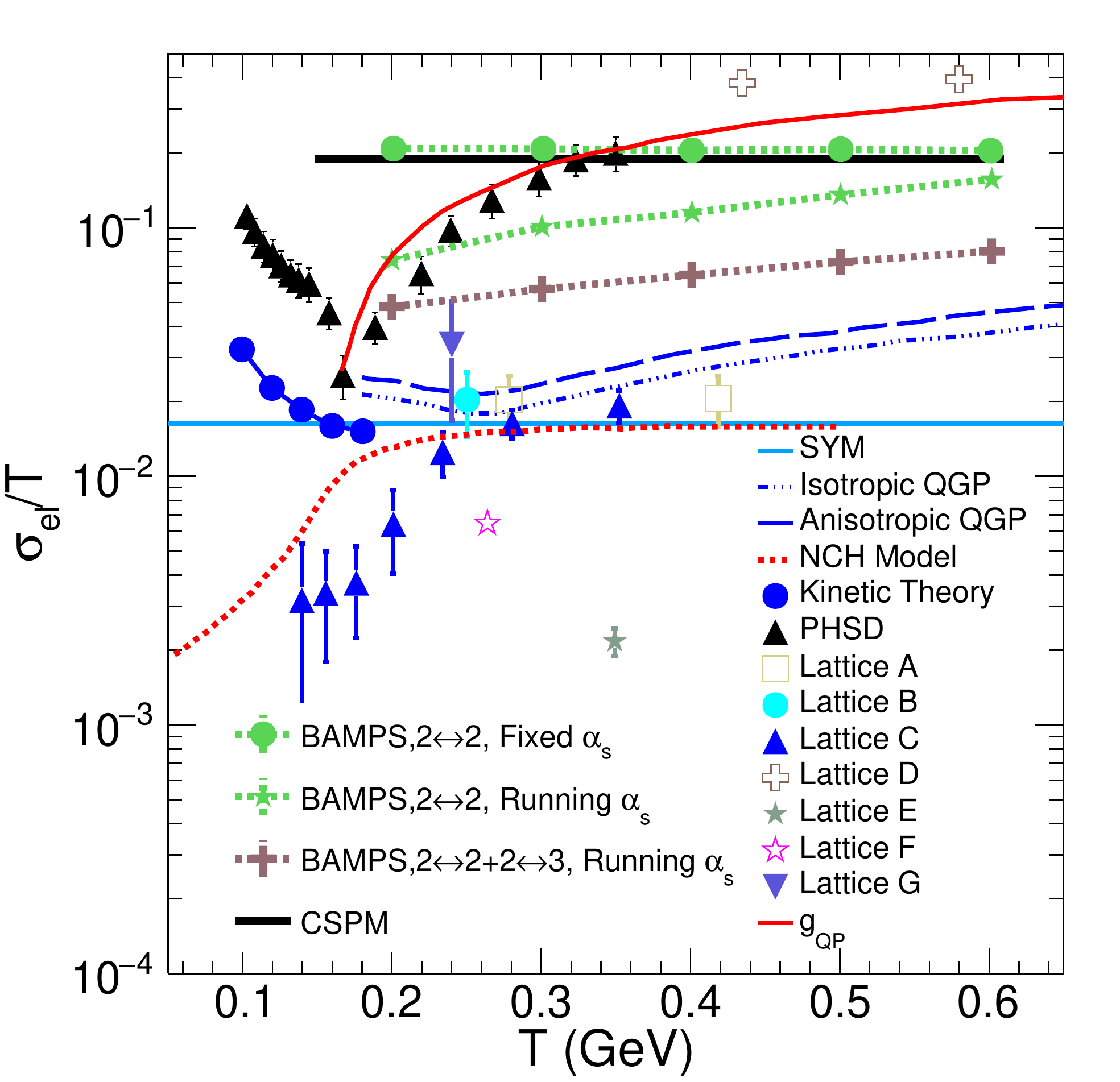} \ ~ \includegraphics[width=5.5cm,height=5.5cm]{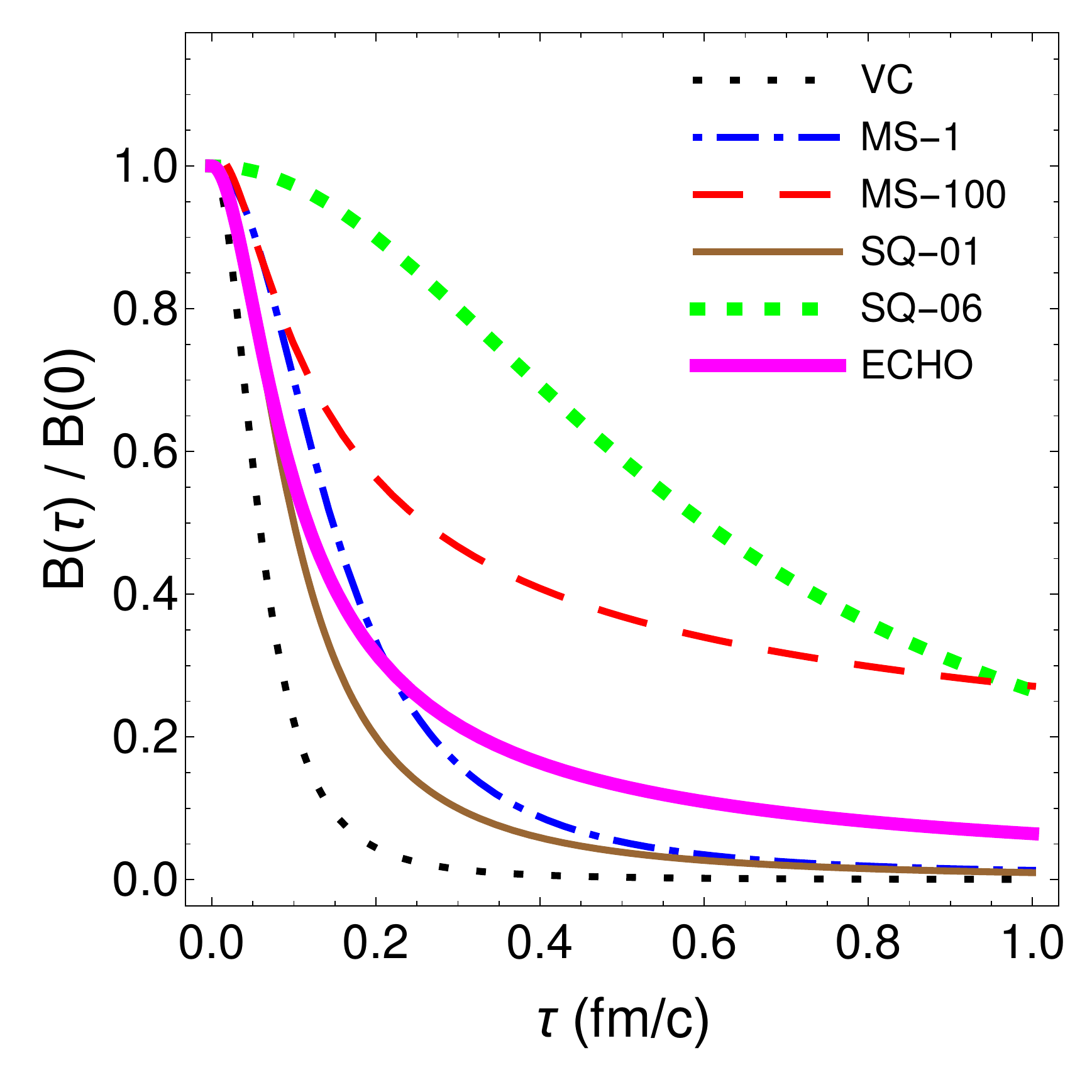}$$
		\caption{ The electric conductivity $\sigma_{el}$ of the QGP matter as a function of temperature (left panel) and the normalized magnetic component $B_y$ as a function of proper time (right panel). Both are theoretical calculations with different models. The figures are taken from Refs.~\cite{Sahoo:2018dxn,Huang:2017tsq}.}
		\label{fig26}}
\end{figure}

The assumption of constant electric conductivity $\sigma_{el}$ can help us to do a most analytic calculation. However, the lattice QCD simulations and other model calculations show that, $\sigma_{el}$ is temperature dependent, as shown in Fig.\ref{fig26}. It is expected to be proportional to the temperature of the QGP medium. That means that, $\sigma_{el}$ is space and time dependent during the QGP expansion. The electric conductivity $\sigma_{el}$ should rapidly increase from zero to its equilibrium value in the early pre-equilibrium stage, and then decrease as the plasma cools. Taking this space-time dependence into consideration, one needs to solve the hydrodynamic equations together with Maxwell equations. That comes to the magnetohydrodynamics. For Bjorken expansion, the one-dimensional relativistic magnetohydrodynamics has an analytical solution~\cite{Roy:2015kma}. It shows that the magnetic field changes the total energy density, but does not change the evolution of the fluid energy density. The (3+1)D relativistic magnetohydrodynamics including the effects of the electromagnetic fields are recently solved numerically~\cite{Inghirami:2016iru}. 

\subsection{Heavy flavor mesons in electromagnetic field}
\label{meson}

Since heavy quarks are produced in the very early stage of heavy ion collisions when the created electromagnetic fields are strong enough, the heavy flavor hadrons formed later should carry the information of the fields. 

The strong electromagnetic fields will affect the strong interaction between heavy quarks~\cite{Kharzeev:2012ph,Miransky:2015ava}. While gluons seem not directly coupled to the electromagnetic fields, they still undergo significant modifications via effective QED-QCD interactions induced by quark loops, investigated by lattice simulations~\cite{Bonati:2014ksa,Bonati:2016kxj} and effective models~\cite{Galilo:2011nh}.

\begin{figure}[!htb]
	{$$\includegraphics[width=0.33\textwidth]{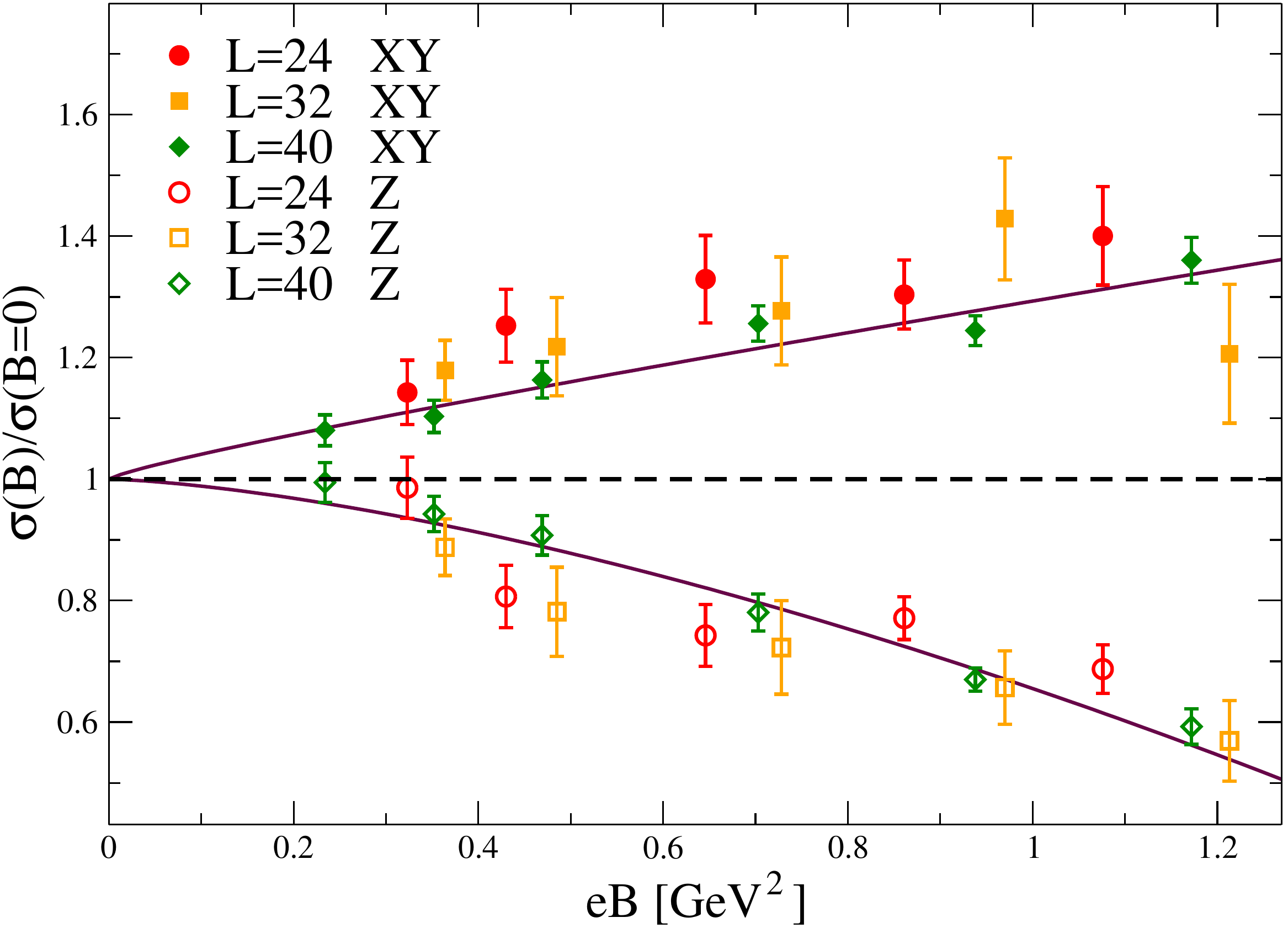} \ ~ \ ~   \includegraphics[width=0.33\textwidth]{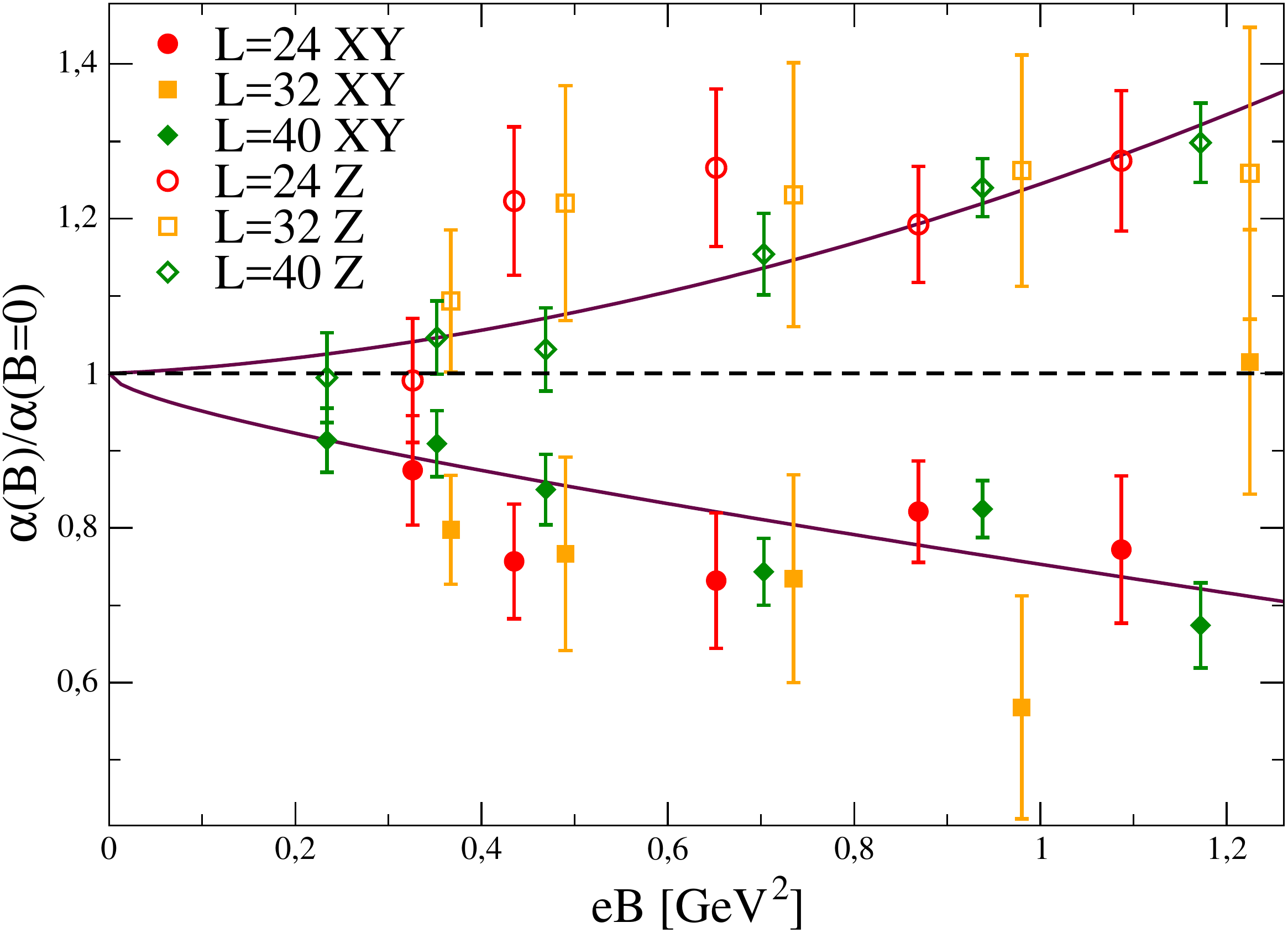}$$
		\caption{ The string tension $\sigma$ (left panel) and Coulomb coupling $\alpha$ (right panel) along the longitudinal and transverse directions as functions of magnetic field. The figures are taken from Ref.~\cite{Bonati:2014ksa}.}
		\label{fig27}}
\end{figure}

For heavy flavor mesons, a very interesting question is how the magnetic field modifies the static $Q\bar Q$ potential. From the lattice simulations and effective model calculations~\cite{Bonati:2014ksa, Dudal:2014jfa,Rougemont:2014efa}, the $Q\bar Q$ potential in an external magnetic field becomes anisotropic, since the magnetic field breaks down the rotational symmetry in vacuum. The string tension $\sigma$ increases (decreases) in the transverse (longitudinal) direction, while the Coulomb coupling $\alpha$ shows an opposite behavior, see Fig.\ref{fig27}. The anisotropic version of the Cornell potential can be expressed as
\begin{equation}
V(r,\theta,B)=-{\alpha(\theta, B)\over r}+\sigma(\theta, B)r+V_0(\theta, B),
\end{equation}
where $\theta$ is the angle between ${\bf r}$ and the magnetic field ${\bf B}$.

The electromagnetic field effect on heavy quark potential is extended from vacuum to heat bath~\cite{Bonati:2016kxj}. It is easy to understand that, the random thermal motion will largely reduce the potential in any direction, and the anisotropy will also be strongly reduced by the thermal motion~\cite{Bonati:2017uvz}.

The static properties of heavy flavor mesons in an external electromagnetic field have been widely discussed with potential models~\cite{Alford:2013jva, Marasinghe:2011bt,Yoshida:2016xgm} and QCD Sum Rules~\cite{Cho:2014loa,Kumar:2018ujk}. The two-body Schr\"odinger equation for a pair of heavy quarks in an external magnetic field is
\begin{equation}
\left[ {({\bf p}_a-q_a{\bf A}_a)^2 \over 2m} +{({\bf p}_b-q_b{\bf A}_b)^2 \over 2m} -{\bm \mu}\cdot {\bf B}+V \right ] \Psi({\bf x_a, x_b})=E\Psi({\bf x_a, x_b}),
\end{equation}
where the vector potential ${\bf A}=({\bf B}\times {\bf x})/2$ is introduced in the equation through minimal coupling, $q_a=-q_b=q$ is the charge of the charm or bottom quark, ${\bm \mu}=q/m({\bf S}_a-{\bf S}_b)$ is the magnetic moment, and the scalar potential $V=V_c+V_s$ include the Cornell potential and spin-spin interaction. Due to the breaking of the translational invariance of the system by the vector potential, the kinetic momentum ${\bf P}_{kin}=\sum_i^2({\bf p}_i-q_i{\bf A}_i)$ is no longer a conserved quantity, but the generalized pseudomomentum operator ${\bf P}_{ps}=\sum_i^2({\bf p}_i+q_i{\bf A}_i)$ commutes with the Hamiltonian of the system,
\begin{equation}
\left[{\bf P}_{ps} , {\bf H} \right] = 0.
\end{equation}
This allows us to factorize the total wave function into a center-of-mass motion and a relative motion,
\begin{equation}
\Psi({\bf R,r}) = e^{i({\bf P}_{ps}-{1\over 2}q{\bf B}\times{\bf r})\cdot {\bf R}}\psi({\bf r}),
\end{equation}
where ${\bf R}=({\bf x_1+x_2})/2$, ${\bf r}={\bf x_1 -x_2}$ are the centre of mass coordinate and relatival coordinate. The relative motion which we are interested in is controlled by the equation,
\begin{equation}
\left[ {{\bf p}^2 \over m}-{\bm \mu}\cdot {\bf B}-{q\over 2m}({\bf P}_{ps}\times {\bf B})\cdot {\bf r}+{q^2\over 4m}({\bf B}\times {\bf r})^2+V \right ] \psi({\bf r})=\left[E-{{\bf P}_{ps}^2 \over 4m}\right ]\psi({\bf r}),
\label{seqb}
\end{equation}
where the second term $-{\bm \mu}\cdot {\bf B}$ is the interaction between the spin magnetic moment and the magnetic field, and the third term corresponds to the Lorentz force. This equation can be solved numerically.

Due to the interaction between the spin magnetic moment and the magnetic field, the total spin is not conserved. This leads to a splitting and coupling of the spin states. The triplet state will split and the state $|S, S_z\rangle=|1, 0\rangle$ will couple with the singlet state $|0, 0\rangle$. For charmonia, there are four states $\eta_c$, $J/\psi^0$, and $J/\psi^{\pm}$, and $\eta_c$ and $J/\psi_0$ are coupled to each other.

The third and fourth term in Eq.\ref{seqb} violate the orbital angular momentum conservation, all vacuum states will couple with each other and the state with nonzero angular momentum quantum number $l$ will split into $2l+1$ new states. For charmonia, $\eta_c$ and $J/\psi$ will couple with $\chi_c$ and $h_c$, and $\chi_c$ will split in both spin space and orbital angular momentum space.

The energy $E$ of the system depends on not only the strength of the magnetic field but also the total kinetic momentum of the system. The particle dispersion relation in magnetic field is very different from the one in ordinary case. This may lead to some controversy in the definition of the mass. However, we can still define $M=2m+E-E_k=2m+E-\frac{\langle {\bf P}_{kin} \rangle^2}{4m}$ as the mass of the bound state and take the total kinetic momentum as a parameter.

\begin{figure}[!htb]
	{$$\includegraphics[width=0.3\textwidth]{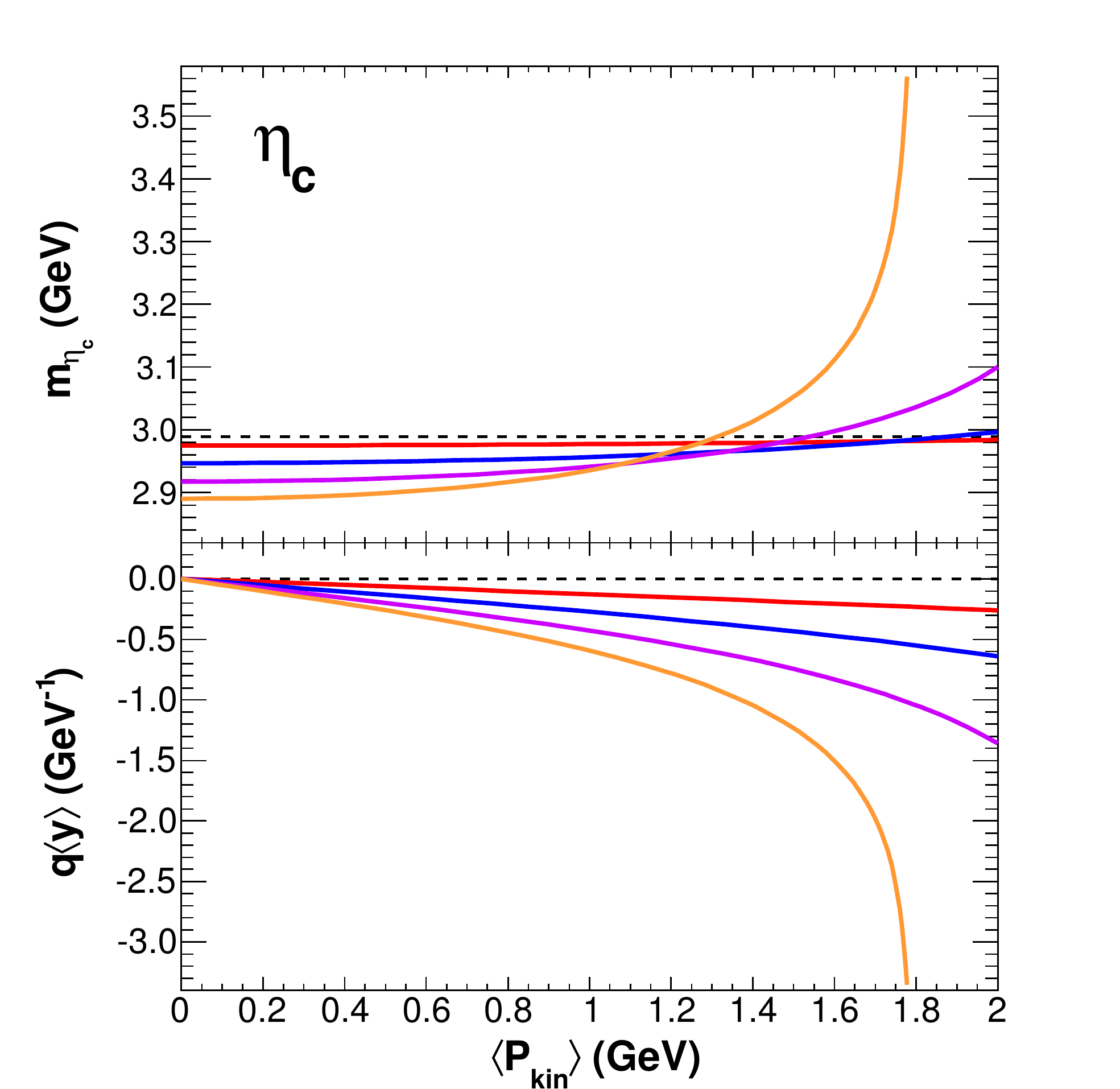}\includegraphics[width=0.3\textwidth]{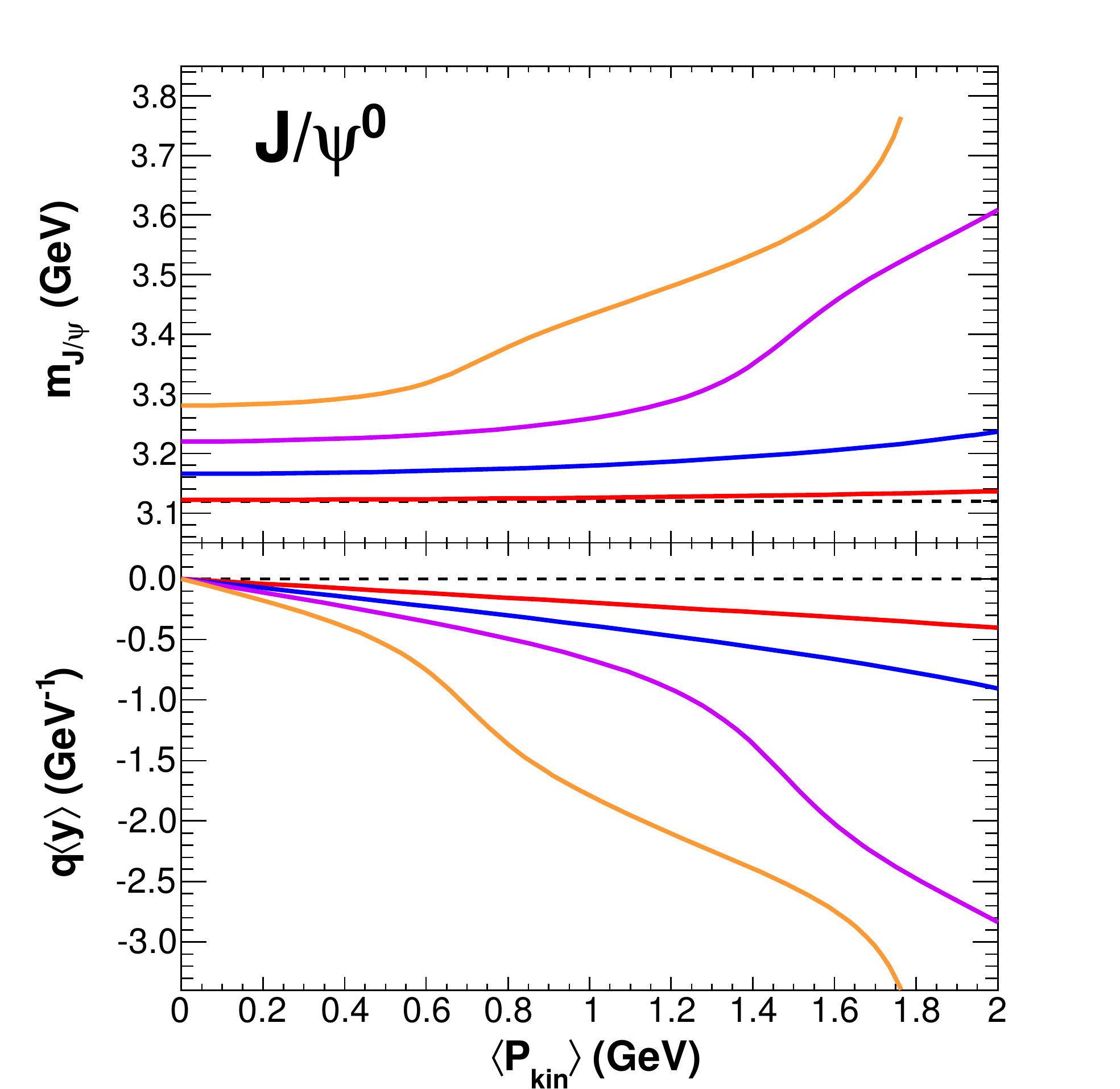}\includegraphics[width=0.3\textwidth]{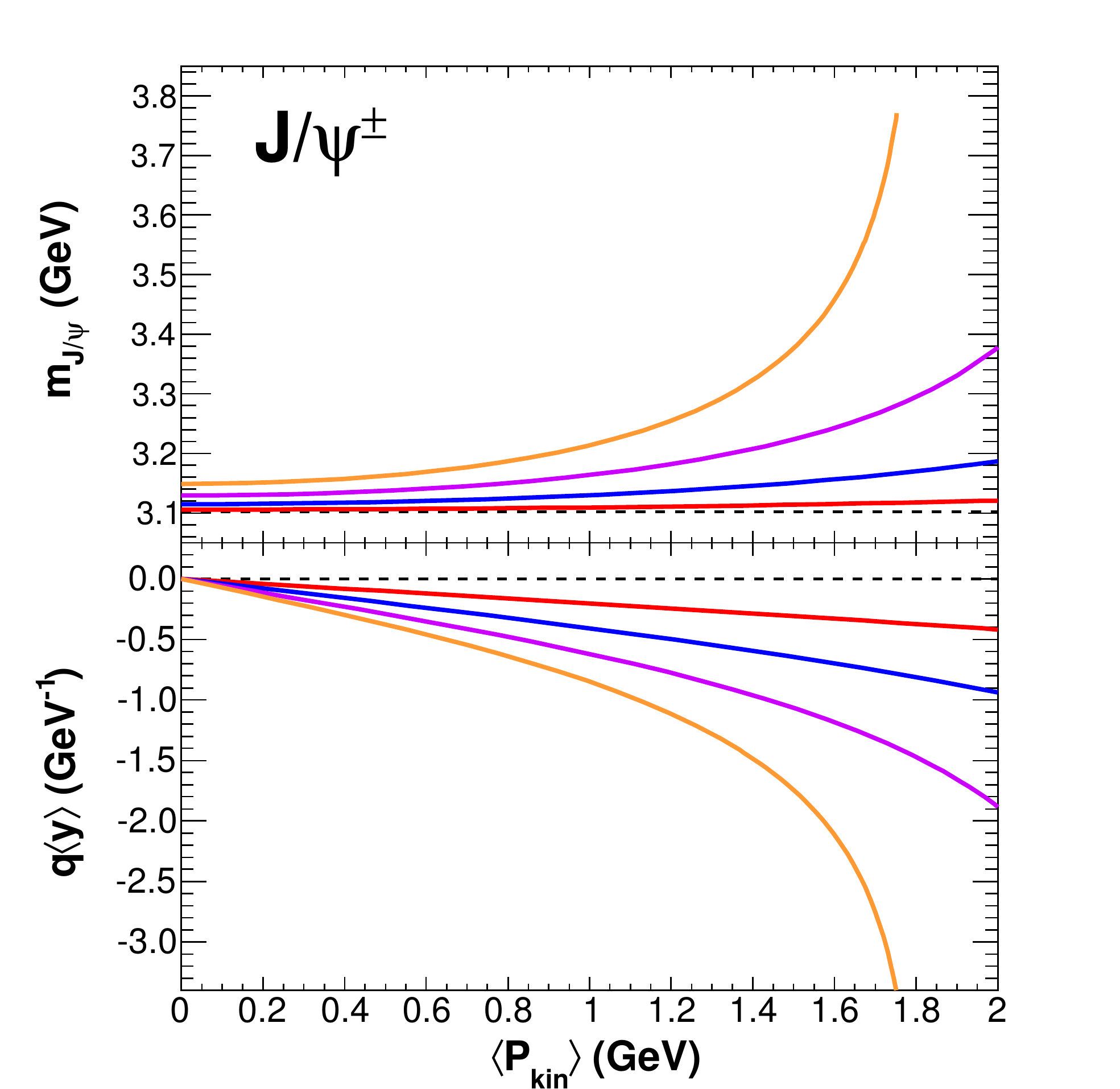}$$
		\caption{ The kinetic momentum dependence of charmonium mass (upper panels) and charmonium electric dipole moment $q\langle y\rangle$ (lower panels) in magnet field $eB$ = 0, 5, 10, 15 and 20 $m_\pi^2$. }
		\label{fig28}}
\end{figure}

The charmonium mass in an external magnetic field as a function of total kinetic momentum is shown in Fig.\ref{fig28}. The behavior of the mass depends strongly on the value of the total kinetic momentum. For small kinetic momentum $\langle P_{kin}\rangle <1$ GeV, the mass of $\eta_c$ decreases while the $J/\psi^0$ mass increases with the magnetic field, called level repulsion caused by the coupling between $\eta_c$ and $J/\psi^0$. The mass difference between $\eta_c$ and $J/\psi^0$ is mainly from the coupling of the spin magnetic moment with the magnetic field $-{\bm \mu}\cdot {\bf B}$. This is also found in QCD Sum Rules~\cite{Cho:2014loa}. For large kinetic momentum $\langle P_{kin} \rangle > 1$ GeV, the Lorentz force term starts to dominate the system and leads to an increasing charmonium mass with the magnetic field. When the anisotropic potential induced by the strong magnetic field is taken into account, the split between $\eta_c$ and $J/\psi^0$ is reduced~\cite{Bonati:2015dka}. Note that, at large enough kinetic momentum $\langle P_{kin}\rangle$, the charmonium states will dissociate under strong magnetic field, even at zero temperature~\cite{Marasinghe:2011bt}. 

The above magnetic field effect on charmonium states can be extended to bottomonium states. Since bottom quark is heavier than charm quark, the magnetic field effect on bottomonia is expected to be smaller compared with charmonia. Similar to charmonia and bottomonia, the mass of open heavy flavor mesons will also mix or split under the magnetic field, discussed with potential models~\cite{Yoshida:2016xgm} and QCD Sum Rules~\cite{Gubler:2015qok,Machado:2013yaa}. The pseudoscalar mesons $D$($B$) couple with vector mesons $D^*$($B^*$), and there exists again level repulsion. 

\subsection{Heavy quark pairs in electromagnetic field}
\label{pair}
Considering the very short lifetime of the electromagnetic fields produced in heavy ion collisions, the formation process from a heavy quark pair $Q\bar Q$ to a heavy flavor quarkonium will be modified by the fields. Since the low momentum quarkonia will be easily dissociated by the later formed fireball, the electromagnetic fields will mainly affect the high momentum quarkonia. 

We first consider the transition among different quarkonium states. The evolution of the wave function for a $c\bar c$ pair produced initially satisfies the time dependent Schr\"odinger equation,
\begin{equation}
i\hbar {\partial \over \partial t}\Psi =H\Psi.
\end{equation}
After separating the relative motion from the center of mass motion and expanding the relative wave function in terms of the charmonium wave functions in vacuum without magnetic field,
\begin{equation}
\Psi({\bf R,r,}t)={1\over 2\pi}e^{i{\bf P}_{ps}\cdot {\bf R}-i{{\bf P}^2\over 4m_c}t}\sum_i C_i(t)e^{-iE_it}\psi_i({\bf r}),
\end{equation}
where the summation is over all the charmonium states, the Schr\"odinger equation for the wave function becomes the equation for the transition amplitude, 
\begin{equation}
{d\over dt}C_i = \sum_j e^{i(E_i-E_j)t}C_j\int d^3{\bf r}\psi_i^*({\bf r})H_B\psi_j({\bf r}),
\end{equation}
where the total relative Hamiltonian is separated into a vacuum part $H_0$ which controls the charmonium states $H_0 \psi_i=E_i\psi_i$ and a magnetic field dependent part 
\begin{equation}
H_B=-{\bm \mu}\cdot {\bf B}-{q\over 2m}({\bf P}_{ps}\times {\bf B})\cdot {\bf r}+{q^2\over 4m}({\bf B}\times {\bf r})^2
\end{equation}
which dominates the transition amplitude. 

The initial relative wave function is very compact and can be described by a Gaussian wave package $\psi({\bf r}) =\exp(-(r-r_0)^2/\sigma_0^2)$, where the width $\sigma_0$ and averaged coordinate $r_0$ can be fixed by fitting the experimentally measured yield ratios between $J/\psi$, $\chi_c$ and $\psi'$ ($|C_{J/\psi}|^2 : |C_{\chi_c}|^2 : |C_{\psi'}|^2 \approx 6:3:1$) in p+p collisions~\cite{Aaij:2012ag,Chatrchyan:2011kc,ATLAS:2014ala} after a time evolution $\tau_f \sim 0.5fm$. 

Since the lifetime of the magnetic field is about $t_B\sim 0.1$ fm/c which is much shorter than the charmonium formation time $\tau_f\sim 0.5$ fm/c, the magnetic field will affect strongly the charmonium fractions $|C_i|^2$ and thus alter the relative yields among different charmonium states.

\begin{figure}[!htb]
	{$$\includegraphics[width=0.36\textwidth]{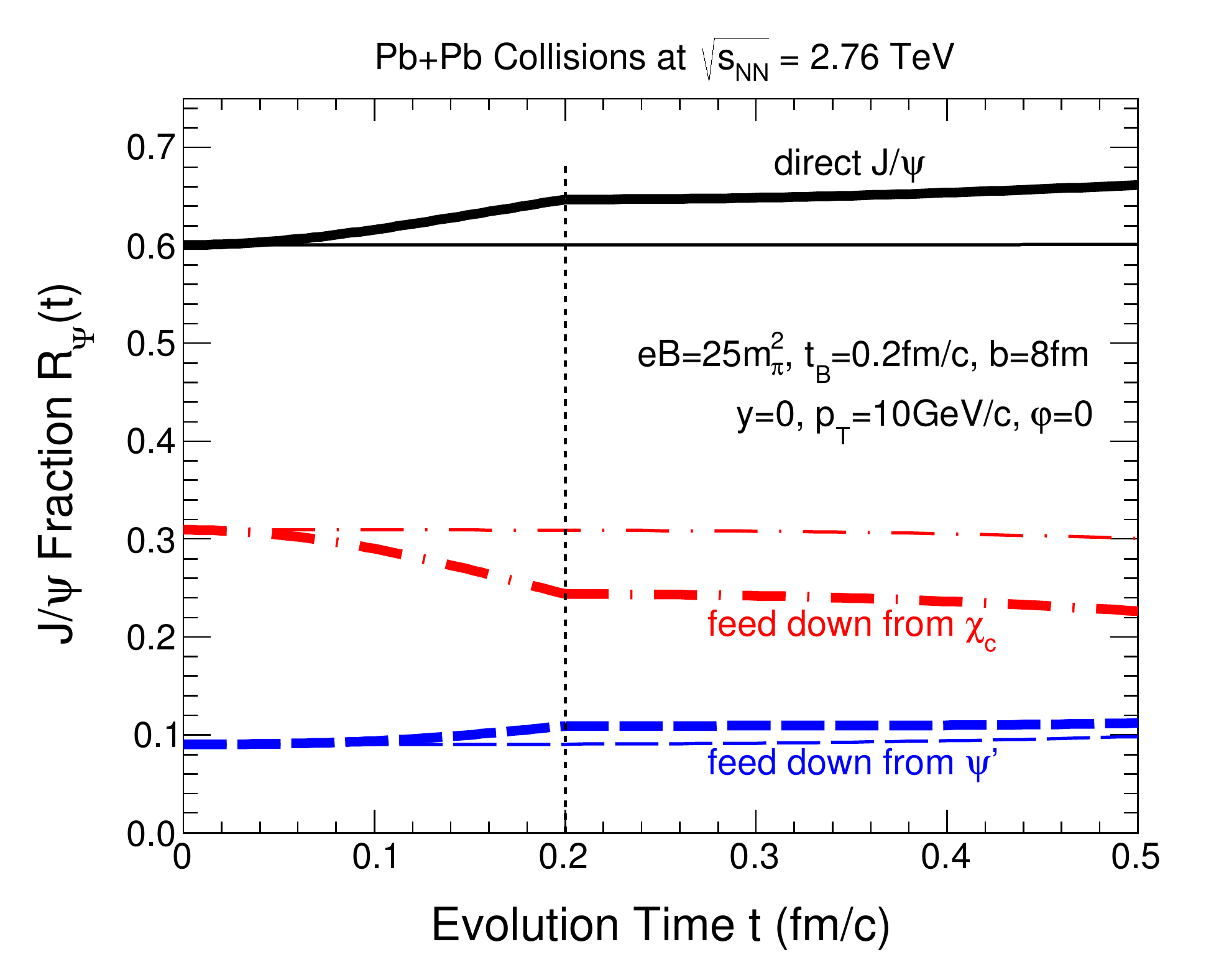}\includegraphics[width=0.36\textwidth]{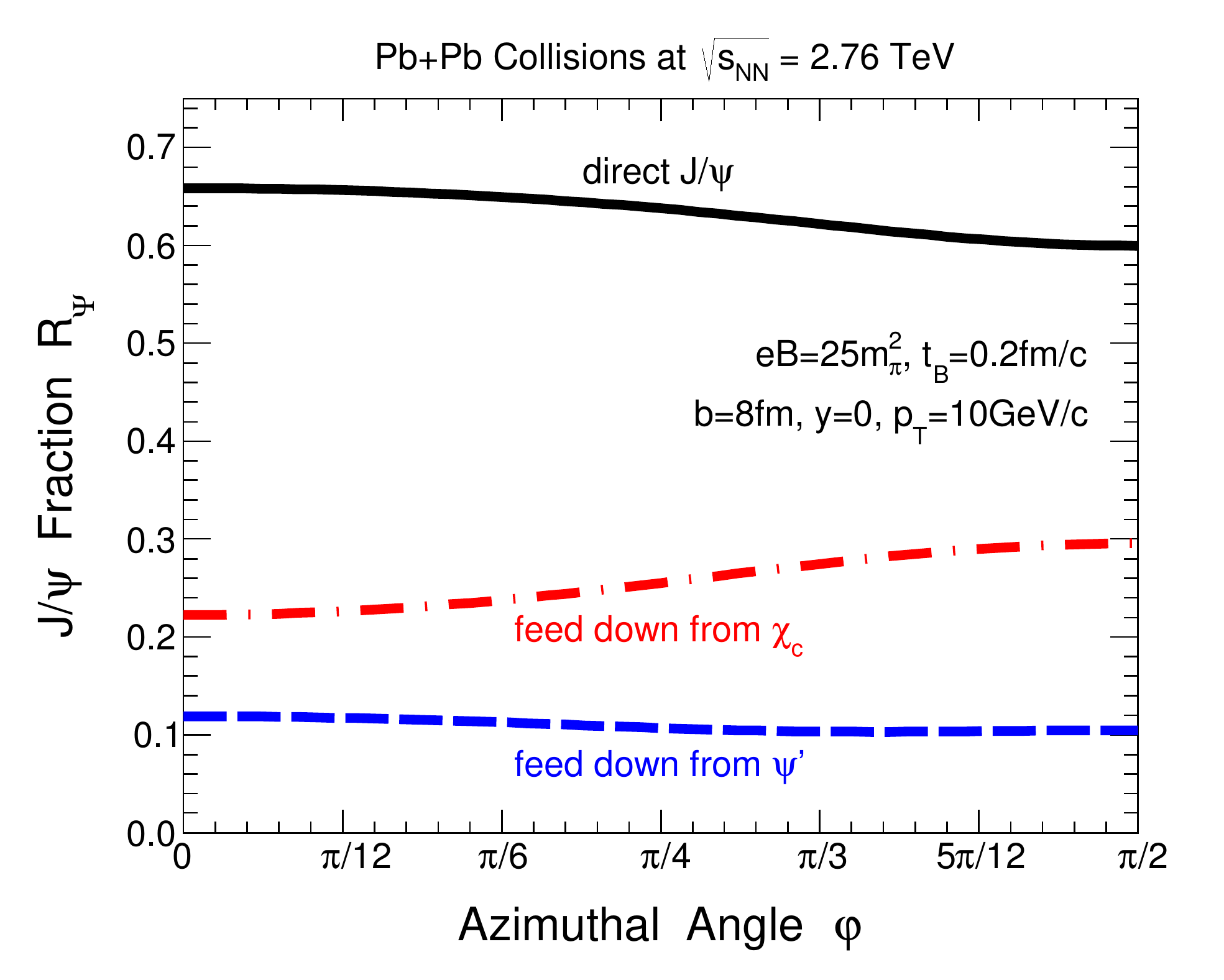}$$
		\caption{ The $J/\psi$ yield fractions in Pb+Pb (thick lines) and p+p (thin lines) collisions as functions of time (left panel) and azimulthal angle (right panel). The figures are taken from Ref.~\cite{Guo:2015nsa}.}
		\label{fig29}}
\end{figure}

The fractions of the final state $J/\psi$ in Pb-Pb collisions at LHC energy is shown in the left panel of Fig.\ref{fig29}. In comparison with p-p collisions where there is no magnetic field, the direct production and the feed down from $\psi'$ are enhanced, while the feed down from $\chi_c$ is suppressed~\cite{Guo:2015nsa}. The strength of the Lorentz force acting on the $c\bar c$ pair is anisotropic. The magnetic field effect is the strongest at $\varphi=0$ ($\varphi=\arctan(p_y/p_x)$) and decreases monotonously with the azimuthal angle. Finally, at $\varphi=\pi/2$, the force disappears and only the weak harmonic potential exists, the fractions approach to their vacuum values, as shown in the right panel of Fig.\ref{fig29}.

The direction dependence of the magnetic field effect on quarkonium formation leads to an anisotropic charmonium production in the transverse plane, and will cause a no-collective flow which is totally different from the collective flow we discussed above. 

\begin{figure}[!htb]
	{$$\includegraphics[width=0.4\textwidth]{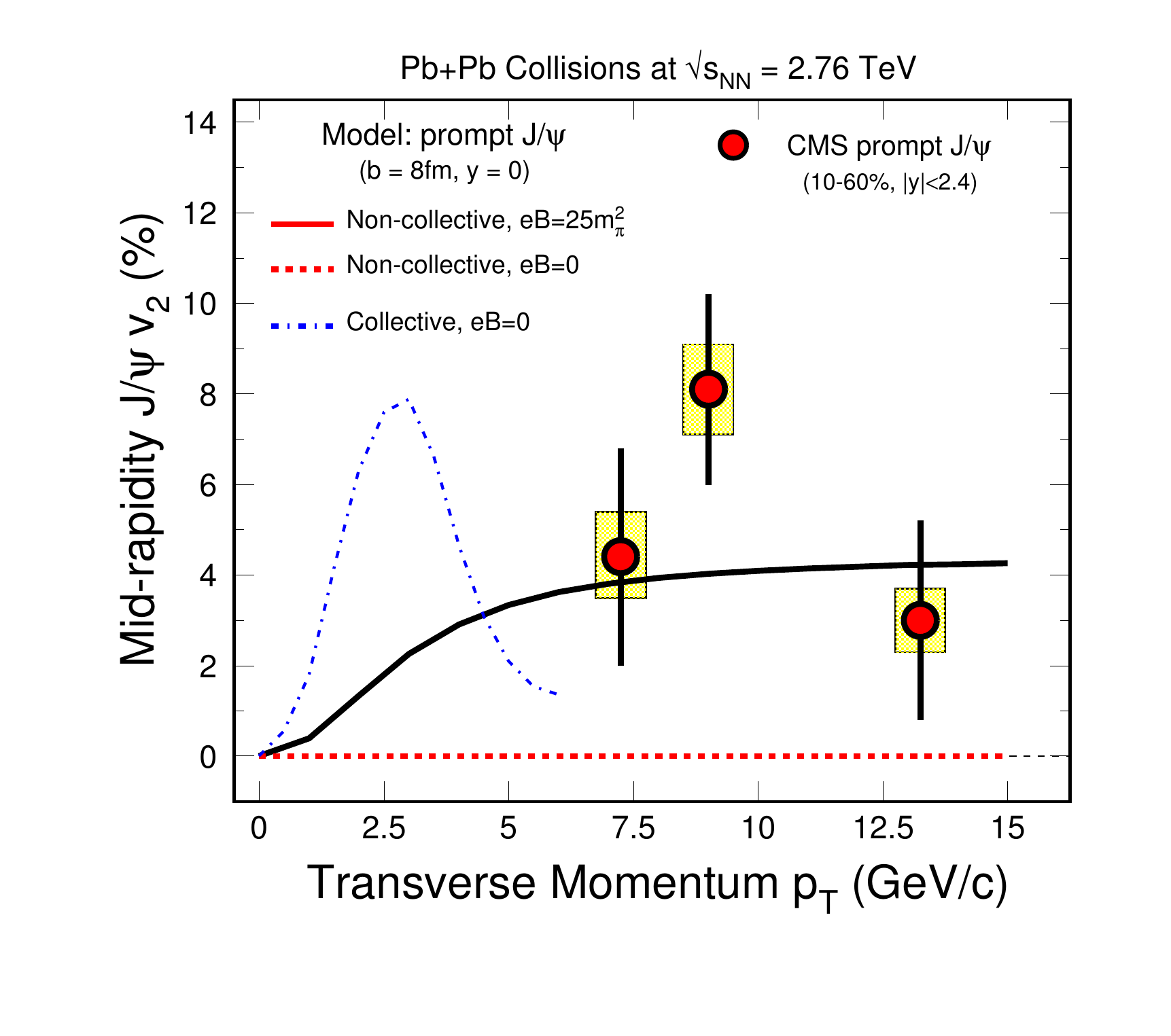}$$
		\caption{The magnetic field induced $J/\psi$ non-collective flow $v_2$ at high transverse momentum. The figure is taken from Ref.~\cite{Guo:2015nsa}.}
		\label{fig30}}
\end{figure}

The elliptic flow $v_2$ of hadrons at low $p_T$ comes from the collective motion of the medium created in heavy ion collisions, and can be well described by hydrodynamic calculations. At the LHC energy, the sizeable $J/\psi$ $v_2$ at low transverse momentum is from the regeneration, due to the heavy quark equilibrium with the medium. The high $p_T$ charmonia generated in the initial stage are not expected to be sensitive to the nature of the hot medium. However, the Lorentz force induced anisotropic production in the transverse plane will result in a non-collective flow. While the low $p_T$ $J/\psi$s will be eaten up by the later formed fireball, the high $p_T$ $J/\psi$s will escape from the fireball and carry the magnetic field effect. This non-collective flow seems to explain reasonably well the CMS data at high $p_T$, see Fig.\ref{fig30}.

\subsection{Heavy quarks in electromagnetic field}
\label{quark}
Recently, a very large directed flow $v_1$ of D mesons is observed in heavy ion collisions at LHC, it is even larger than the flow for light hadrons~\cite{Adam:2019wnk}. Theoretically, this phenomenon is expected to be related to the electromagnetic fields. The heavy quark motion in a medium with electromagnetic fields can be described by a Boltzmann transport equation, 
\begin{equation}
p^\mu\partial_\mu f+mK^\mu{\partial\over \partial p^\mu}f= C[f]
\end{equation}
with $K^\mu=qF^{\mu\nu}u_\nu\ (K^0=0,\ {\bf K}=\gamma q ({\bf E}+{\bf p}/E\times {\bf B})=\gamma{\bf F})$, where ${\bf E}$ and ${\bf B}$ are the electromagnetic fields, ${\bf F}$ is the Lorentz force, and $C[f]$ stands for the interaction between the heavy quark and the medium (collision term). If the momentum transfer is small in each collision between the heavy quark and the medium partons, the Boltzmann equation can be realized by the corresponding Langevin equation,
\begin{eqnarray}
&&{d{\bf x}\over dt}={{\bf p}\over E},\nonumber\\
&&{d{\bf p} \over dt}=-\Gamma(p,T){\bf p}(t)+{\bf \xi}(t)+{\bf F}(t).
\end{eqnarray}

\begin{figure}[!htb]
	{$$\includegraphics[width=0.33\textwidth]{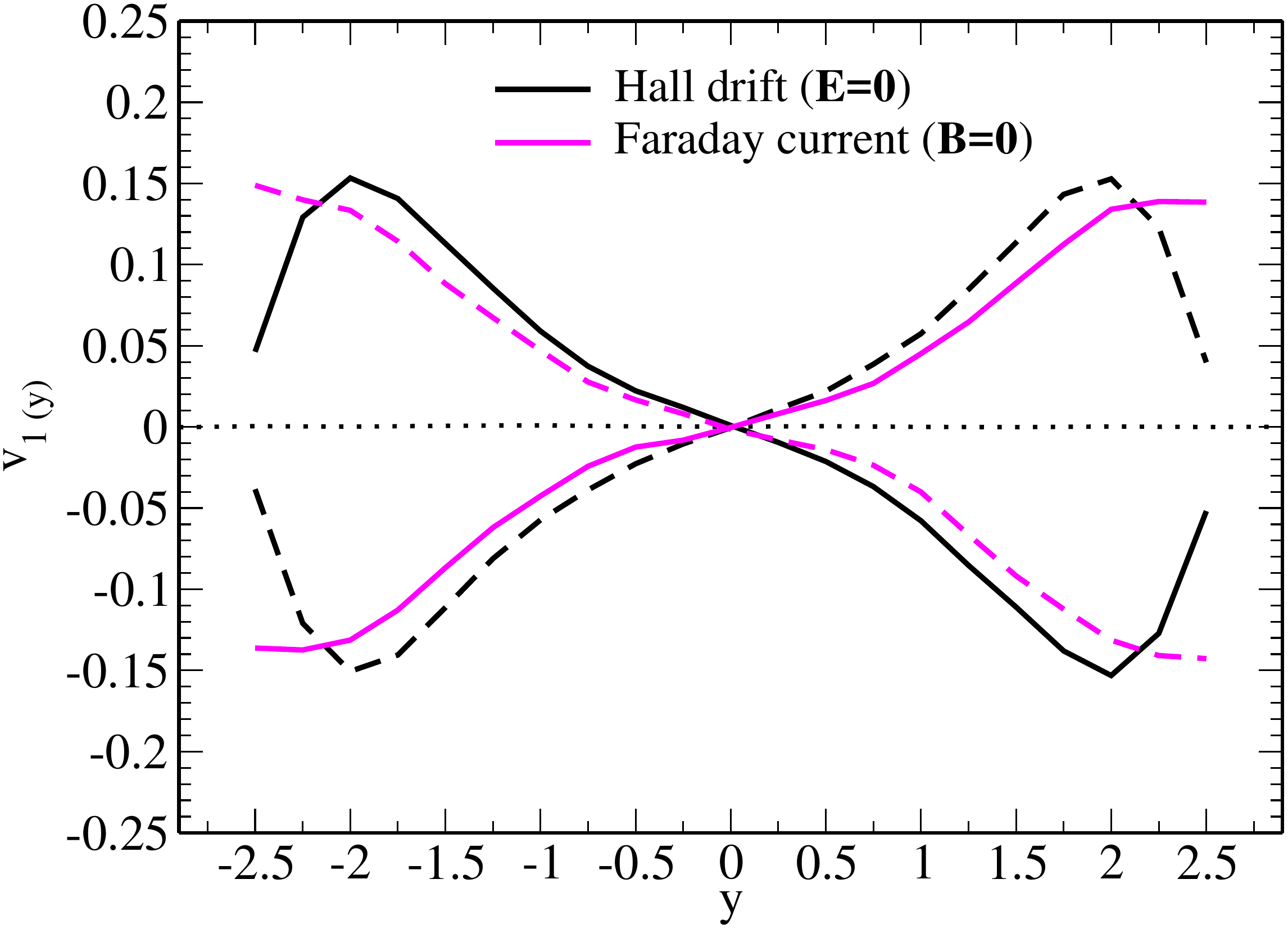} \ ~ \  ~ \includegraphics[width=0.33\textwidth]{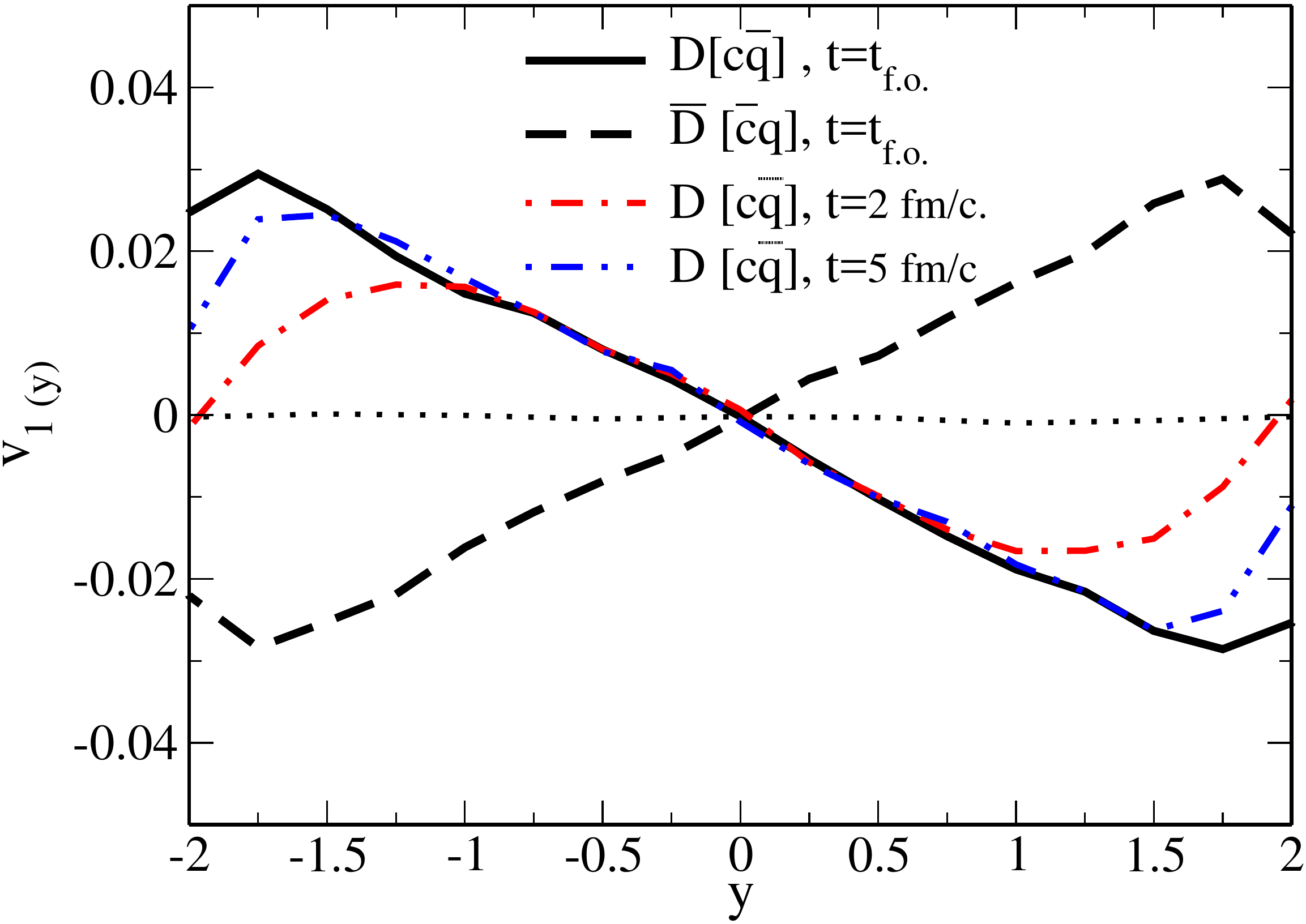}$$
		\caption{ The electric or magnetic force induced directed flow for $D$ mesons (left panel) and the total flow for $D$ and $\bar D$ mesons (right panel) as functions of rapidity in Pb+Pb collisions at $\sqrt{s_{NN}}=2.76$ TeV and $p_T>1$ GeV. The figures are taken from Ref.~\cite{Das:2016cwd}.}
		\label{fig31}}
\end{figure}

Describing the QGP evolution by hydrodynamics with initial time $\tau_0=0.2$ fm/c, choosing the initial spatial and momentum distributions of the charm quarks by the Glauber model and the FONLL scheme, neglecting the charm quark evolution in the pre-equilibrium stage, and calculating the background electromagnetic fields in a static medium with a constant electrical conductivity $\sigma_{el}=0.023$ fm$^{-1}$, the Langevin equation can be numerically solved and the result~\cite{Das:2016cwd} is shown in Fig.\ref{fig31}. 

Without electromagnetic fields, the directed flow $v_1$ of $D$ and $\bar D$ mesons is zero. When the fields are turned on, the electromagnetic interaction between the charged heavy quark and the fields leads to a non-zero flow, and $D$ and $\bar D$ have opposite flow, due to the charge dependent Lorentz force. When we switch off the electric field or the magnetic field, the flow induced by the electric field is opposite to the flow by the magnetic field, since the electric force $q{\bf E}$ is always opposite to the magnetic force $q({\bf v\times B})$. Considering the cancellation between the two, the total flow including both the electric and magnetic forces is much smaller than the flow with only electric or magnetic force. 

Finally, we consider the charm quark thermalization effect. When charm quarks reach kinetic equilibrium with the QGP medium, they totally lose the memory of the information on the initial electromagnetic fields. When considering the medium response to the external electromagnetic fields, the lifetime of the electromagnetic fields would be longer. However, the competition between the Lorentz force and the random kick from partons in the hot medium still exists and controls the magnitude of the flow $v_1$. The numerical results are shown in~\cite{Das:2016cwd}.

The study on electromagnetic field effect on charm quarks can be extended to bottom quarks. On one hand, bottom quarks are created earlier than charm quarks due to the larger mass, the rapid decay of the fields leads to a stronger electromagnetic effect on bottom quarks compared with charm quarks. On the other hand, the charge of a bottom quark is only one half of a charm quark and bottom quarks move much slower. The total electromagnetic effects on bottom quarks depends on the competition of the above two effects. 

In normal Langevin simulations, the heavy quark momentum diffusion coefficients in longitudinal and transverse directions are taken the same value, $\kappa_\parallel= \kappa_\bot$. When the electromagnetic fields are turned on, the medium and in turn the diffusion become anisotropic~\cite{Fukushima:2015wck,Finazzo:2016mhm,Dudal:2018rki}. There will be a stronger diffusion in the longitudinal direction compared to the transversal one. This will lead to an anisotropic drag force on heavy quarks and give rise to a sizable heavy quark elliptic flow.

\subsection{Vector meson photoproduction}
\label{photon}

\begin{figure}[!htb]
	{$$\includegraphics[width=0.37\textwidth]{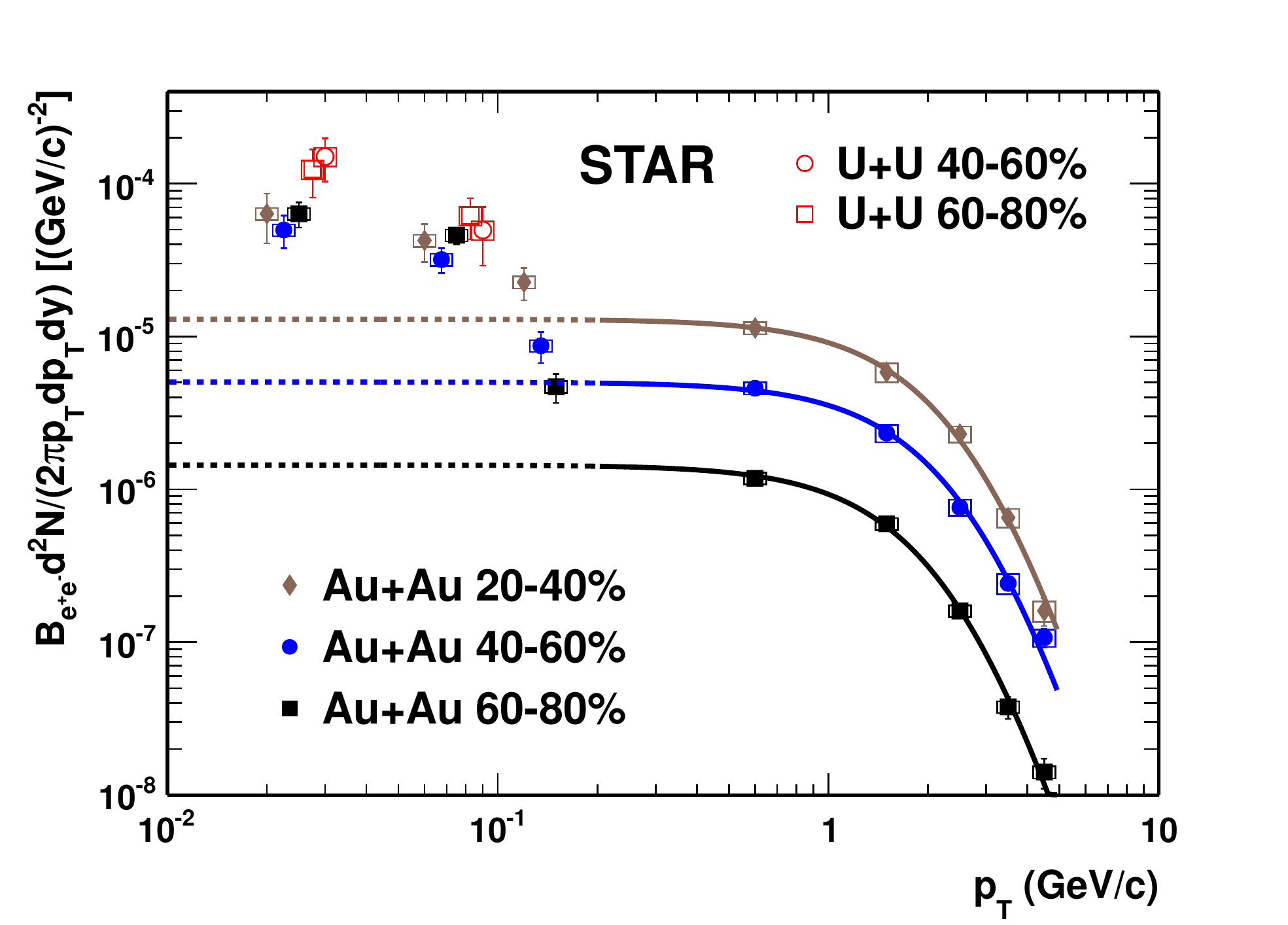}\includegraphics[width=0.25\textwidth]{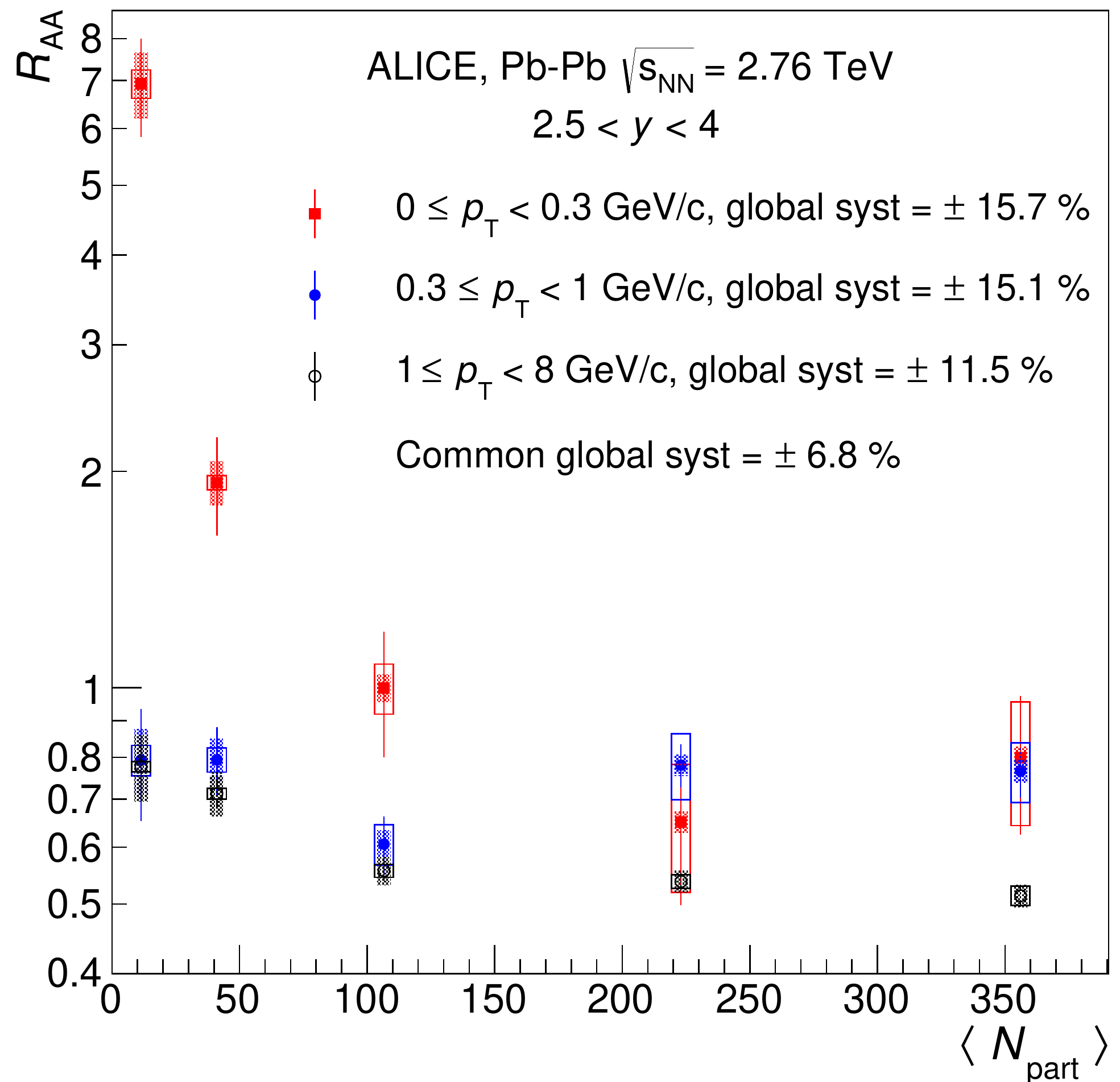}$$
		\caption{ The experimentally measured $J/\psi$ yield as a function of transverse momentum at different centrality (left panel) and the $J/\psi$ nuclear modification factor as a function of number of participants at different transverse momentum bins (right panel) in nuclear collisions at RHIC and LHC energies. The figures are taken from Refs.~\cite{STAR:2019yox,Adam:2015gba}.}
		\label{fig32}}
\end{figure}

The strong electromagnetic fields produced in heavy ion collisions not only affect the production and evolution of heavy flavor hadrons but also induce direct electormagnetic production of vector mesons.
The electromagnetical production has been studied in Ultra-peripheral nuclear collisions (UPC) where the hadronic interaction can be neglected safely~\cite{Bertulani:1987tz,Krauss:1997vr,Bertulani:2005ru,Klein:1999qj,Adam:2015sia,TheALICE:2014dwa}. Virtual photons emitted by one nucleus may fluctuate into $q\bar q$ pairs, scatter off the other nucleus, and emerge as vector mesons (like $J/\psi$ and $\Upsilon$). The magnetic field strength firstly increases and then decreases with impact parameter. With the strong electromagnetic fields, charmonium photoproduction may become more important than the hadroproduction in extremely low $p_T$ region in semi-central nuclear collisions, which has already been observed by recent experiments at RHIC~\cite{STAR:2019yox} and LHC~\cite{Adam:2015gba}, see Fig.\ref{fig32}.

Fermi first proposed that the transverse electromagnetic fields can be approximated as a swarm of equivalent photons, called "equivalent photon approximation"~\cite{Fermi:1924tc}. This idea was extended by Weizs\"acker and Williams~\cite{vonWeizsacker:1934nji} independently and therefore also called "Weizs\"acker-Williams-method". This allows a simple and straightforward calculation of vector meson photoproduction between the target nucleus and electromagnetic fields~\cite{Krauss:1997vr,Klein:1999qj}. With long-range electromagnetic interaction, a hard equivalent photon from one nucleus may penetrate into the other nucleus and interact with quarks or gluons.

The longitudinal component of the electromagnetic field produced by a moving nucleus disappears when the nucleus moves with velocity $v_z\to c$. In this limit the strengths of the electric field and magnetic field are the same, and the two vectors are perpendicular to the velocity ${\bf v}$, $|{\bf E}|=|{\bf B}|$, ${\bf E}\bot {\bf v}$ and ${\bf B} \bot {\bf v}$. The energy flux of the electromagnetic field through a plane perpendicular to the moving direction of the nucleus is described by the Poynting vector,
\begin{equation}
{\bf S}({\bf r},t)={\bf E}({\bf r},t) \times {\bf B}({\bf r},t)\approx |{\bf E}({\bf r},t)|^2{\bf v}
\end{equation}
The last equal holds only in the limit $v_z \to c$. The energy conservation law requires that the energy flux of the fields through a transverse plane is identical to the energy flux of the equivalent photons,
\begin{equation}
\int_{-\infty}^\infty dt\int d{\sigma}_T \cdot {\bf S}({\bf r},t) = \int_0^\infty d\omega \int d{\bf x}_T \omega n(\omega,{\bf x}_T),
\end{equation}
where $n(\omega,{\bf x}_T)$ is the photon number at frequency $\omega$ and certain transverse distance ${\bf x}_T$ from the trajectory of the moving nucleus. The electric field can be calculated from classical Maxwell equations,
\begin{eqnarray}
n(\omega, {\bf x}_T)&=&{1\over \pi \omega}|{\bf E}(\omega, {\bf x}_T)|^2 \nonumber\\
&=&{4Z^2\alpha_{em}\over \omega}\left | \int {d^2{\bf k}_T\over (2\pi)^2}{\bf k}_T {F({\bf k}_T^2+(\omega/\gamma)^2)\over {\bf k}_T^2+(\omega/\gamma)^2} e^{i{\bf x}_T\cdot {\bf k}_T} \right|^2,
\end{eqnarray}
where $\alpha_{em}$ is the electromagnetic coupling constant, $\gamma=\sqrt{s}/(2m_p)$ is the Lorentz boost in the laboratory frame, and the nuclear form factor $F(q)$ can be obtained via the Fourier transformation of the Woods-Saxon distribution.

The difference between coherent and incoherent photoproduction depends on the wavelength of the photons emitted from nuclei. If the wavelength is longer than the size of the nucleus ($2R_{A}$), the photon can not distinguish the nucleons in the nucleus and will interact with the whole nucleus, called coherent photoproduction. If the energy of a photon is very high, it will interact with partons in the nucleus~\cite{Yu:2017pot}. We will concentrate here on the coherent photoproduction.

For coherent photoproduction, the photons interact with the target nucleus coherently to produce vector mesons, the photon density needs to be averaged over the surface of the target nucleus B,
\begin{equation}
n(\omega, {\bf b})=\int_0^{R_A} {rdr \over \pi R_A^2} \int_0^\pi d\phi n(\omega, \sqrt{b^2+2b{\bf x}_T\cos \phi +{\bf x}_T^2}).
\end{equation}

The cross section for the reaction $\gamma A\to V A$ reaction can be derived from a quantum Glauber approach coupled with the parameterized forward scattering cross section~\cite{Klein:1999qj}. With the known cross section and photon density, we obtain the coherent photoproduction as a function of rapidity in AA collisions with impact parameter ${\bf b}$,
\begin{equation}
{dN \over dy}({\bf b})=\omega n(\omega, {\bf b})\sigma_{\gamma A\to VA}(\omega)+(y\to -y).
\end{equation}
This can be used to describe the photoproduction of vector mesons in ultra-peripheral nuclear collisions. For a nuclear collision with general impact parameter ${\bf b}$, however, we need to consider the modifications by the anisotropic QGP~\cite{Shi:2017qep,Chen:2018sir},
\begin{eqnarray}
{dN \over dy}({\bf b})&=&\int d^2{\bf x}_T \omega n(\omega, {\bf b})\sigma_{\gamma A\to VA}(\omega)f_s ({\bf x}_T+{\bf b}/2)[{\mathcal R}_g(x,\mu_F, {\bf x}_T+{\bf b}/2)]^2\times e^{-\int_{\tau_0}^{\tau_f} d\tau \alpha}\nonumber\\
&+&(y\to -y, {\bf b}/2 \to -{\bf b}/2),
\end{eqnarray}
where $f_s$ is the spatial distribution of the photoproduced charmonium state and assumed to be proportional to the thickness function square $T_A^2(x_T)$. ${\mathcal R}_g$ is the gluon shadowing modification factor, and the decay rate $\alpha$ describes the gluon dissociation process in hot medium. The only unknown factor is now the photon-nucleus cross section $\sigma_{\gamma A\to VA}$. It can be obtained from photon-proton cross section with the optical theorem which connects the forward scattering cross section $d\sigma_{\gamma A\to VA}/dt|_{t=0}$ with the quantum Glauber calculation. We write down the photon-nucleus cross section in a differential form~\cite{Klein:1999qj},
\begin{equation}
\sigma_{\gamma A\to VA}={d\sigma_{\gamma A\to VA}\over dt}|_{t=0}\int_{t_{min}}^\infty dt |F(t)|^2.
\end{equation}
For narrow resonance, $t_{min}=(M_V^2/4\omega \gamma)^2$ is the minimum squared momentum transfer needed to produce a vector meson of mass $M_V$ in the laboratory frame. For vector mesons with large width (like $\rho$), it becomes complicated and the cross section is calculated by using a Breit-Wigner resonance. The differential photo-nuclear cross section which comes from vector meson dominance model can be expressed as~\cite{Klein:1999qj}
\begin{eqnarray}
&&{d\sigma_{\gamma A\to VA}\over dt}|_{t=0}=C^2{\alpha_{em} \sigma_{tot}^2(VA) \over 4f_V^2}, \nonumber\\
&&\sigma_{tot}^2(VA) =  2\int d^2{\bf x}_T \left[1-\exp\left (-{1\over 2}\sigma_{tot}(Vp)T_A({\bf x}_T)\right)\right ]
\end{eqnarray}
where $f_V$ is the vector meson-photon coupling, $T_A({\bf x}_T)$ is the nuclear thickness function, the correction factor $C$ is adopted to account for the non-diagonal coupling through higher mass vector mesons~\cite{Hufner:1997jg}. Using again the optical theorem and the connection between the vector meson-proton cross section and the photon-proton cross section, one obtains
\begin{equation}
\sigma_{tot}^2(Vp)={4f_V^2 \over \alpha_{em} C^2}{d\sigma_{\gamma p\to Vp}\over dt} |_{t=0},
\end{equation}
where the cross section $d\sigma_{\gamma p\to Vp}/dt|_{t=0}$ can be calculated in the pQCD framework with Pomeron or vector meson exchange~\cite{Ryskin:1992ui,Jones:2013pga} or directly extracted from the experimental data of $\gamma p \to Vp$~\cite{Klein:2016yzr}.

\begin{figure}[!htb]
	{$$\includegraphics[width=0.35\textwidth]{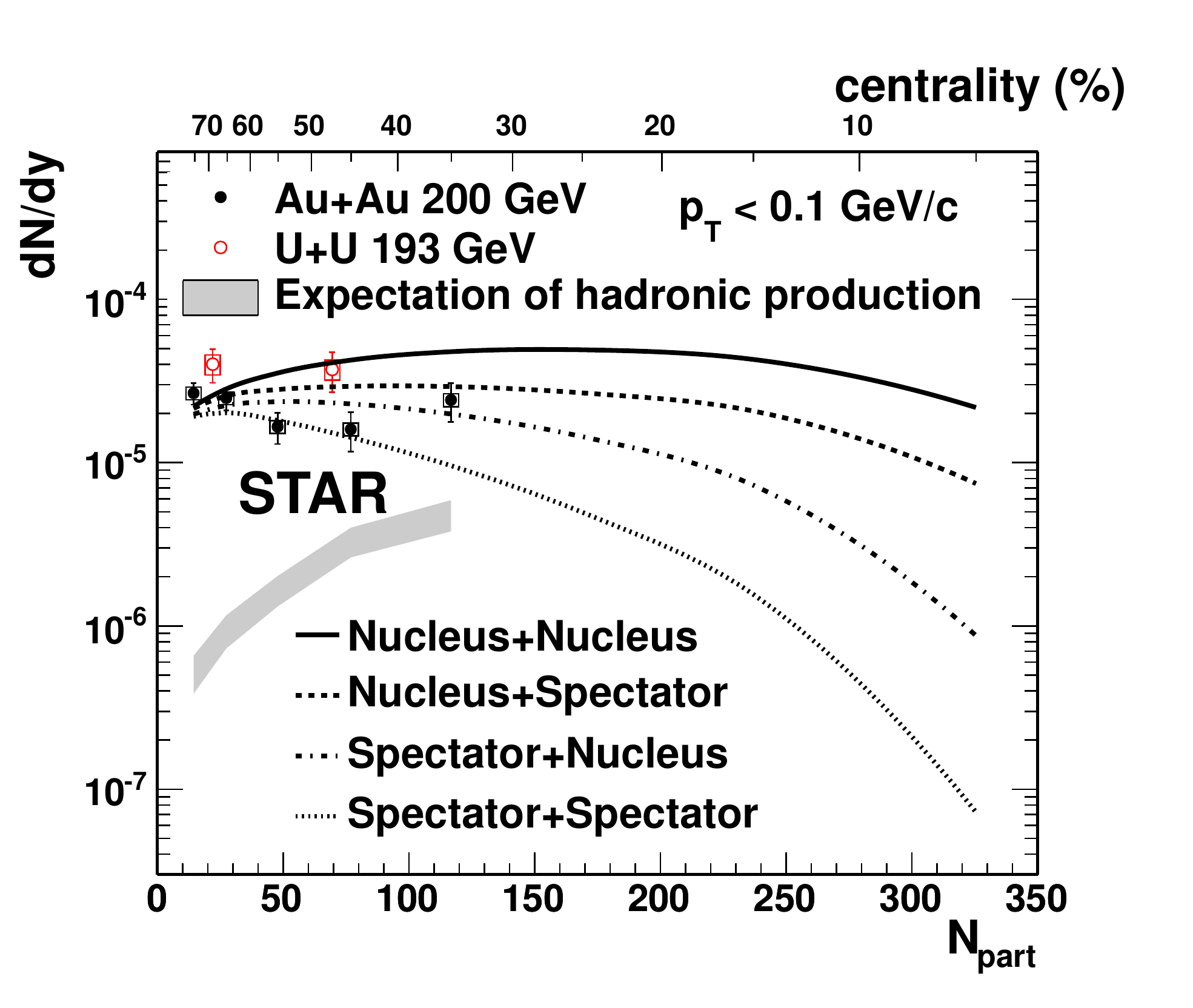}\includegraphics[width=0.33\textwidth]{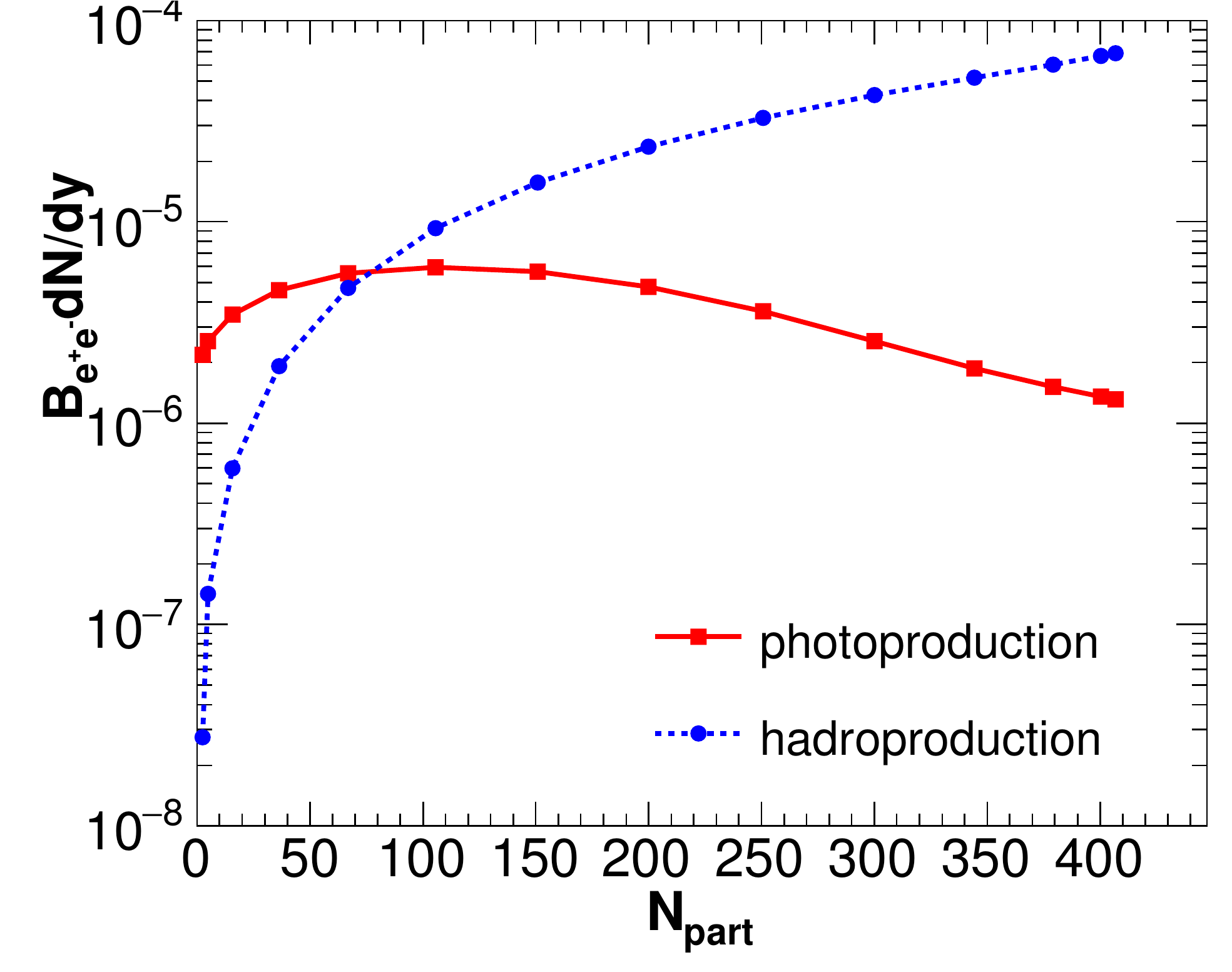}$$
		\caption{ The experimentally measured and theoretically calculated $J/\psi$ yield as a function of number of participants with $p_T < 0.1$ GeV at RHIC energy (left panel) and the model calculation with $p_T < 0.3$ GeV at LHC energy (right panel). The figures are taken from Refs.~\cite{STAR:2019yox,Shi:2017qep}.}
		\label{fig33}}
\end{figure}

With increasing centrality, both the number of binary collisions $N_{coll}$ and the number of participants $N_{part}$ increase monotonously. Since the initial production is proportional to $N_{coll}$ and the regeneration is proportional to $N_{coll}^2$, the number of charmonia through hadroproduction increases significantly with the centrality. On the other hand, the strength of the produced electromagnetic field first increases and then decreases with the centrality, the photoproduction reaches the maximum value in semi-central collisions, as shown in Fig.\ref{fig33}. The very strong enhancement of extremely low $p_T$ $J/\psi$s observed in ultra-peripheral nuclear collisions at LHC can be well explained by the photoproduction~\cite{Shi:2017qep}. 

As studied in~\cite{Jones:2013pga,Klein:2017vua}, the photoproduction of $J/\psi$ and $\Upsilon$ can probe the gluon distribution function at small $x$. The cross section in leading logarithmic approximation using non-relativistic approximation can be expressed as
\begin{equation}
{d\sigma_{\gamma p\to J/\psi p} \over dt}|_{t=0}={\Gamma_{ee} M_{J/\psi}^3\pi^3 \over 48\alpha_{em}}\left[ {\alpha_s(\bar Q^2) \over \bar Q^4}xg(x,\bar Q^2) \right]^2 \left( 1+ {Q^2 \over M_{J/\psi}^2}\right)
\end{equation}
with $\bar Q^2=(Q^2+M_{J/\psi}^2)/4$ and $x=4\bar Q^2/s$, where $Q$ is the photon momentum and $s$ is the central of mass energy. We can see that, the cross section is proportional to the square of the gluon distribution and can be used to extract the gluon density $xg(x,Q^2)$, especially for gluons with small $x$. 

There can be interference between vector mesons emitted from different nucleus.  The sign and degree of the interference depend on the impact parameter, phase of the scattering, meson wavelength, and observation direction. For $J/\psi$ with $p_T<\hbar/b$, it is impossible to distinguish which nucleus emits the photon and which emits the Pomeron. Due to the negative parity of $J/\psi$, the signs of the two amplitudes are opposite, leading to destructive interference~\cite{Klein:1999gv}. The study in Ref.\cite{Zha:2017jch} shows that the interference effect has little effect on the $J/\psi$ yield in peripheral collisions, while it reduces the yield considerably in more central collisions. 

\subsection{Open heavy flavors in rotational field}
\label{rotation1}

There is a nonzero total angular momentum $J\propto b\sqrt{s_{NN}}$ ($b$ is the impact parameter and $\sqrt{s_{NN}}$ the colliding energy) carried by the system of the two colliding nuclei. Although most of this total angular momentum is carried away by the spectators, there is still a sizable fraction that remains in the created QGP and shows a nonzero rotational motion of the fluid.

\begin{figure}[!htb]
	{$$\includegraphics[width=0.38\textwidth]{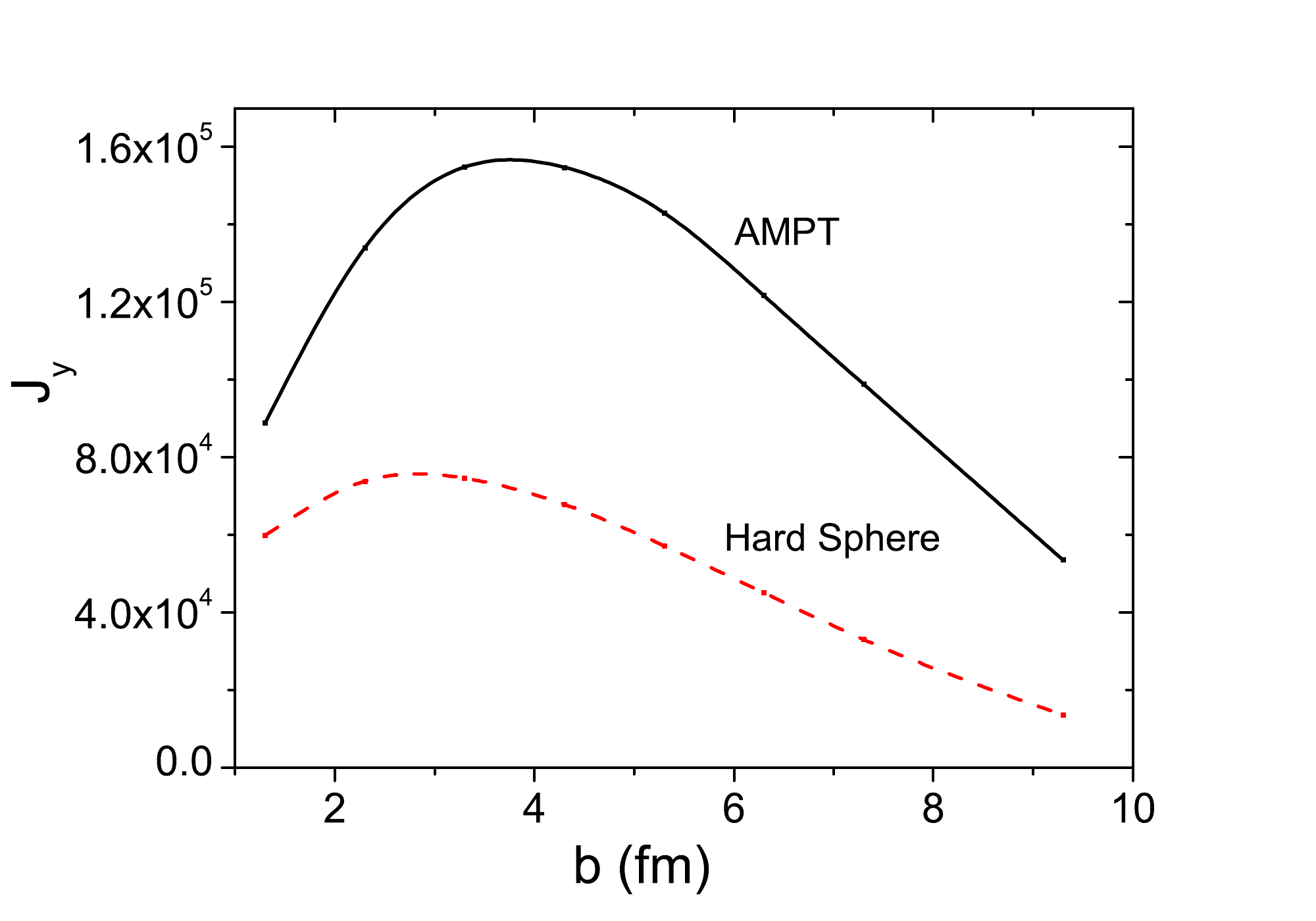}\includegraphics[width=0.38\textwidth]{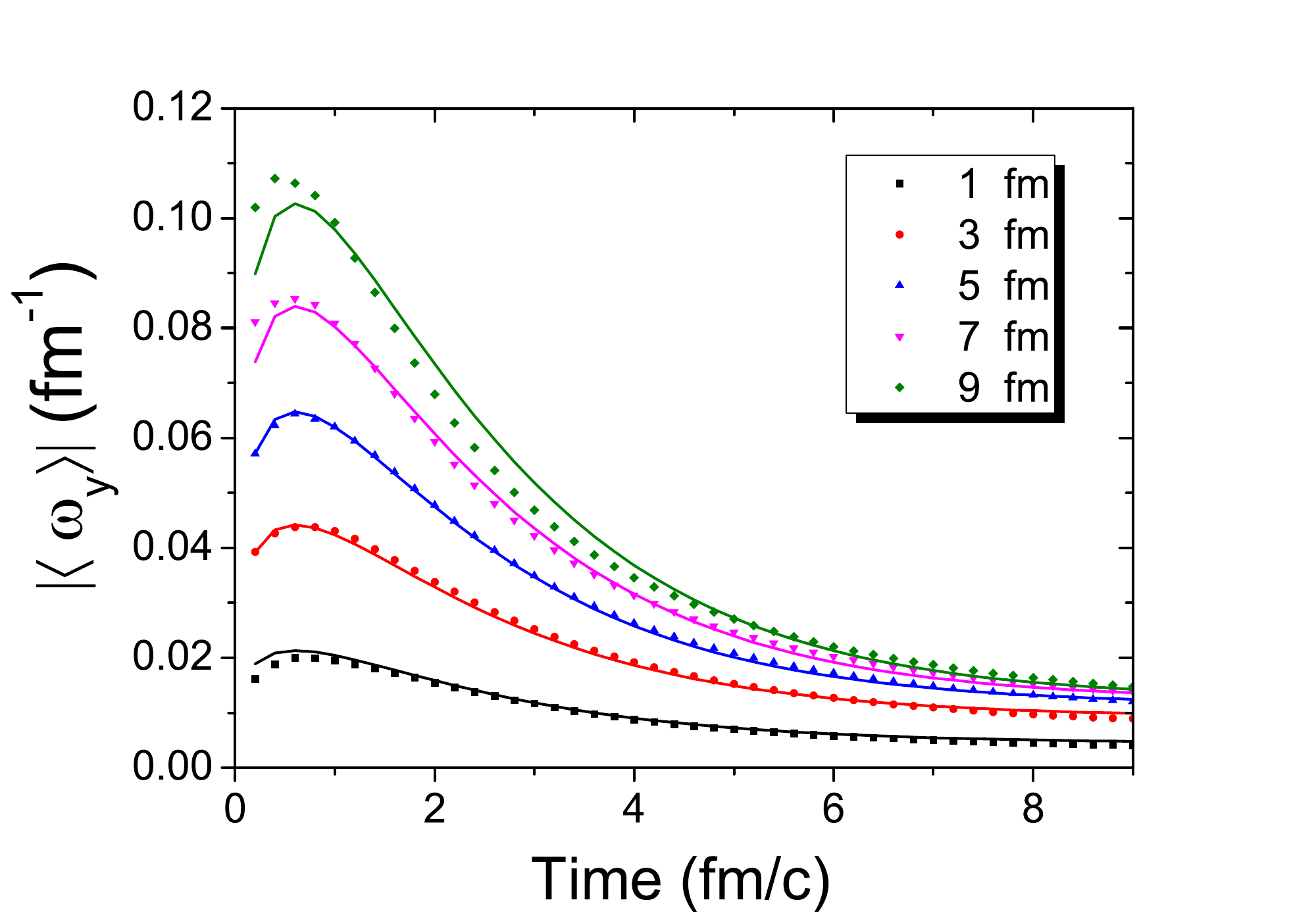}$$
		\caption{ The total angular momentum as a function of impact parameter (left panel) and averaged vorticity as a function of time (right panel) calculated with AMPT model for Au+Au collisions at RHIC energy. The figures are taken from Ref.~\cite{Jiang:2016woz}.}
		\label{fig34}}
\end{figure}

The global rotation of a fluid can be quantified by the total angular momentum, 
\begin{equation}
{\bf J}=\int d^3r {\bf r} \times {\bf p(r)}.
\end{equation}
The fluid vorticity ${\bm \omega}$ can be derived from the local velocity field ${\bf v(r)}$ and related to the momentum and energy densities via ${\bf v(r)=p(r)/\epsilon(r)}$ at each point. In the non-relativistic limit, the vorticity can be expressed as
\begin{equation}
{\bm \omega({\bf r})}= {1\over 2}\nabla \times {\bf v(r)}.
\end{equation}
One can get the relation between the total angular momentum and the averaged vorticity easily. If the system is symmetric around the rotational axis, there is ${\bf J}=1/2\int_V d^3r [\rho^2\epsilon({\bf r})]{\bm \omega}$, where $\rho$ is the distance from the rotational axis, and $[\rho^2\epsilon({\bf r})]$ can be considered as the local fluid "moment of inertia" density~\cite{Jiang:2016woz}. Both ${\bf J}$ and ${\bf \omega}$ are perpendicular to the moving direction of the nuclei. If the impact parameter ${\bf b}$ is along the $x$-axis,  the total angular momentum and the averaged vorticity are along the $y$-axis. It is easy to understand that the most strong angular momentum is in semi-central nuclear collisions, like the behavior of the electromagnetic fields. The AMPT (A Multi-Phase Transport) model simulation~\cite{Jiang:2016woz,Deng:2016gyh} shows the maximum value around $b\sim 4$ fm, see Fig.\ref{fig34}.

For the averaged vorticity $\langle \omega_y\rangle$, it increases with time firstly, reaches the maximum value at around 1 fm/c, and then decreases all the way. The increase is most likely due to the parton scattering during the early stage, and the decrease is due to the expansion of the medium which increases the total moment of inertia at the price of reduced vorticity due to the constraint of constant angular momentum. The averaged vorticity increases from central to peripheral collisions, which is very different from the behavior of the angular momentum. Such difference can be understood as follows: While the vorticity increases with $b$, the fluid moment of inertia in the fireball decreases with $b$, thus the angular momentum shows a non-monotonic behavior due to the two competing trends. 

Hydrodynamics is widely used to describe the vorticity of the hot medium~\cite{Becattini:2015ska,Ivanov:2017dff,Xie:2017upb,Pang:2016igs,Bozek:2010bi}, where the global polarization is related to the fireball's tilted shape in the reaction plane. Such a tilted hydro will also give hadrons a nonzero directed flow $v_1$. Following the strategy in Ref.\cite{Bozek:2010bi}, the initial energy density $\epsilon(\tau_0,x,y,\eta)$ can be constructed within a two-component Glauber model, 
\begin{equation}
\epsilon(\tau_0,x,y,\eta)={\epsilon_0\over N_0} \{[N^+_{part}(x,y)f_+(\eta)+N^-_{part}(x,y)f_-(\eta)](1-\alpha)+\alpha N_{bin}f(\eta) \},
\end{equation}
where the participant nucleons are separated into the forward and backward moving ones, but the binary collisions are assumed to contribute in a symmetric way, $\alpha$ controls the admixture of the participant and binary sources contributed to the total energy density, $N_{bin}=\sigma T(x+b/2,y)T(x-b/2,y)$ with elastic cross section $\sigma$ and thickness function $T(x)$ is the number of binary collisions, and $N^+_{part}$ and $N^-_{part}$ are the participants from the two nuclei,
\begin{eqnarray}
&&N^+_{part}(x,y)=T(x-b/2,y)\left [ 1- e^{-\sigma T(x+b/2,y)} \right ], \nonumber\\
&&N^-_{part}(x,y)=T(x+b/2,y)\left [ 1- e^{-\sigma T(x-b/2,y)} \right ].
\end{eqnarray}
The profile of the initial rapidity distribution is usually taken as a Gaussian,
\begin{equation}
f(\eta)=\exp\left[ -\theta(|\eta|-\eta_p){(|\eta|-\eta_p)^2\over 2\sigma_g^2}\right]
\end{equation}
with a plateau of width $2\eta_p$. The parameters $\eta_p$ and $\sigma_g$ can be fixed by fitting the experiment data of charged particle distribution $dN/d\eta$. The forward and backward distributions are defined as $f_+(\eta)=f(\eta)f_F(\eta)$ and $f_-(\eta)=f(\eta)f_F(-\eta)$. Considering that nucleons from the projectile emit more particles in the forward ($\eta>0$) than in the backward, the profiles which introduce the rapidity-odd component to the initial state can be chosen as 
$$ f_F(\eta)=\left\{
\begin{array}{rcl}
0       &      & {\eta     <      -\eta_m}\\
{\eta +\eta_m \over 2\eta_m}     &      & {-\eta_m\leq\eta \leq \eta_m}\\
1    &      & {\eta_m<\eta}
\end{array} \right.$$
where the parameter $\eta_m$ controls the magnitude of the tilt of the initial energy density. The tilted hydrodynamics reproduces the experimentally observed directed flow $v_1$ of light hadrons~\cite{Bozek:2010bi}.
\begin{figure}[!htb]
	{$$\includegraphics[width=0.48\textwidth]{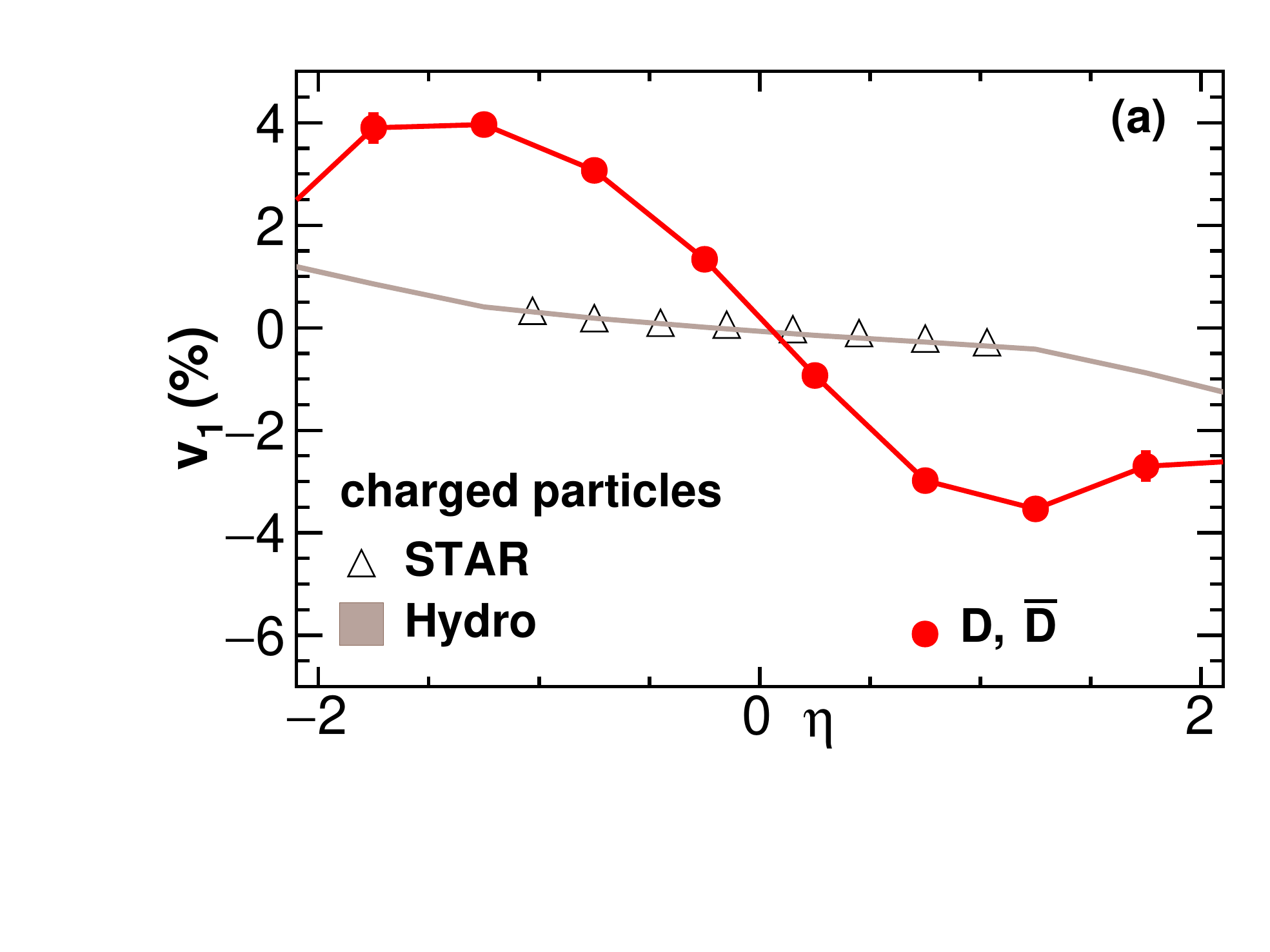}$$
		\caption{ The rapidity dependence of the $D$ meson directed flow $v_1$ in Au+Au collisions at RHIC energy. The figure is taken from Ref.~\cite{Chatterjee:2017ahy}.}
		\label{fig35}}
\end{figure}

Heavy quarks are produced via hard scattering in the initial stage of heavy ion collisions. They witness the entire space-time evolution of the fireball and are ideal probes to the initial stage physics. Considering the rotation, the initially produced heavy quarks should be affected by the initial tilted hydrodynamics more strongly in comparison with light quarks, and the $D$ meson directed flow $v_1$ is expected to be stronger than the light hadrons. 

In the frame of hydrodynamics~\cite{Chatterjee:2017ahy}, by taking the initial charm quark distribution from the $p_T$ spectra obtained in p+p collisions, evolving the charm quarks in the hot medium via the Langevin equation, and hadronizing the charm quarks via Petersen fragmentation when the temperature drops down to 150 MeV, the results show that $D$ and $\bar D$ have the same directed flow $v_1$, and the flow is several times larger than that of light hadrons measured by STAR~\cite{Abelev:2008jga}, as shown in Fig.\ref{fig35}. In comparison with the result of an earlier calculation within the hadron-string-dynamics transport approach~\cite{Bratkovskaya:2004ec}, the $D$ and $\bar D$ directed flow with the tilted hydrodynamics is almost two times larger at the central rapidity.
\subsection{Closed heavy flavors in rotational field}
\label{rotation2}

Finally, we consider the effect of the vorticity field on a heavy quarkonium system. Like the introduction of electromagnetic fields in quantum systems, we introduce a rotation related gauge field $A^\mu$,
\begin{equation}
A^\mu=(A^0, {\bf A})=\left(-{1\over 2}( {\bm \omega}\times {\bf x})^2, {\bm \omega} \times {\bf x}\right)
\end{equation}
in the two-body Schr\"odinger equation through minimal coupling,
\begin{equation}
\left[ {({\bf p}_a-m{\bm \omega}\times {\bf x}_a)^2 \over 2m} +{({\bf p}_b-m{\bm \omega}\times {\bf x}_b)^2 \over 2m} - {1\over 2}m({\bm \omega}\times {\bf x}_a)^2 - {1\over 2}m({\bm \omega}\times {\bf x}_b)^2+V \right ] \Psi({\bf x}_a,{\bf x}_b)=E\Psi({\bf x}_a,{\bf x}_b).
\end{equation}
Taking the transformations in coordinate and momentum spaces, the Hamiltonian can be rewritten as 
\begin{eqnarray}
&&{({\bf p}_a-m{\bm \omega}\times {\bf x}_a)^2 \over 2m} +{({\bf p}_b-m{\bm \omega}\times {\bf x}_b)^2 \over 2m} - {1\over 2}m({\bm \omega}\times {\bf x}_a)^2 - {1\over 2}m({\bm \omega}\times {\bf x}_b)^2 \nonumber\\
&=&{{\bf P}^2\over 4m}-{\bf P}\cdot({\bm \omega}\times {\bf R}) + {{\bf p}^2 \over m}-{\bf p} \cdot ({\bm \omega}\times {\bf r}),
\end{eqnarray}
where ${\bf R}=({\bf x_1+x_2})/2$ and ${\bf r}={\bf x_1 -x_2}$ are the centre of mass coordinate and relatival coordinate, and ${\bf P}$ and ${\bf p}$ the total and relative momenta. If ${\bm \omega}$ is ${\bf r}$ independent, we can separate the center of mass motion from the relative motion, $\Psi({\bf R, r})=\Theta({\bf R})\psi({\bf r})$.
If we take into account the spin of the particles (including the term ${\bm \omega}\cdot {\bf s}$ in the Hamiltonian), the relative part of the wave function is controlled by the equation,
\begin{eqnarray}
\left[ {\bf p}^2 -m{\bm \omega}\cdot {\bf s}-m {\bf p}\cdot ({\bm \omega}\times {\bf r})+ mV(r) \right ] \psi({\bf r})=mE_r\psi({\bf r}), 
\end{eqnarray}
where the term $-m {\bf p}\cdot ({\bm \omega}\times {\bf r})=-m{\bf p}\times{\bm \omega}\cdot {\bf r}$ corresponds to the Coriolis force. Considering the relation $-m{\bf p}\times{\bm \omega}\cdot {\bf r}=-m{\bm \omega}\cdot ({\bf r}\times {\bf p})=-m{\bm \omega}\cdot {\bm l}$ and ${\bm j}={\bm l}+{\bf s}$, the equation can be further simplified as
\begin{eqnarray}
\left[ {\bf p}^2 -m{\bm \omega}\cdot {\bm j}+ mV(r) \right] \psi({\bf r})=mE_r\psi({\bf r}). 
\end{eqnarray}
It is easy to see that, the vorticity leads to an energy shift which is proportional to the total angular momentum of the quarkonium state. 
\section{Conclusion}
\label{conclusion}

Similar to electrons which are used to probe the nucleon structure, heavy quarks, which are produced initially and interact strongly with the hot medium, have long been considered as a sensitive probe of the QGP phase created in high energy heavy ion collisions, especially the initial stage of the new phase with high temperature. However, the situation in heavy ion collisions is very complicated. While we are interested in the hot nuclear matter effect which is the condition for the QCD phase transitions and the creation of the new state of matter, there are also cold nuclear matter effect, non-equilibrium effect and different backgrounds which change the properties of the heavy quarks and heavy flavor hadrons too and then reduce or even remove their sensitivity to the QGP phase. In this case, one needs a comprehensive study on the production and evolution of heavy quarks and heavy flavor hadrons.             

The nuclear modification on heavy flavor hadrons in A+A collisions is relative to the vacuum and p+p collisions. We first discussed the production of heavy quarks and hadrons in vacuum, and payed attention to their properties described in the frame of non-relativistic potential model or relativistic potential model including automatically quark spin interactions. An advantage of the potential models is the easy extension to hot medium by taking the lattice simulated heavy quark potential at finite temperature. Since heavy quarks and also closed heavy flavors are produced in the initial stage of heavy ion collisions, they are affected by both the cold and hot nuclear matter effects before and after the hot medium formation. For the cold nuclear matter effect, we discussed shadowing effect which changes the parton distribution functions and in turn the initial production and regeneration rates of heavy flavor hadrons, Cronin effect which is due to the parton multiple scatterings and leads to a quarkonium transverse momentum broadening, and nuclear absorption which is the background of anomalous quarkonium suppression. For the hot nuclear matter effect on quarkonia, it suppresses, on one hand, the quarkonium yield due to the Debye screening, and on the other hand, enhances the quarkonium regeneration because of the dense charm quarks in the medium.           

After the general discussion on the properties of heavy quark and hadrons in vacuum and at finite temperature, we considered the production and evolution of heavy flavor hadrons under two extreme conditions, the high temperature and strong electromagnetic and rotation fields generated in high energy nuclear collisions at SPS, RHIC and LHC energies. We first reviewed open heavy flavors in hot environment created in heavy ion collisions. The evolution of heavy quarks in the hot medium can be described by transport equations and the interaction between the heavy quark and the medium is reflected in the collisional energy loss and radiative energy loss. During the energy loss process, heavy quarks are continuously thermalized. There are three kinds of mechanisms for heavy quark hadronization after the evolution in the hot medium: fragmentation which contributes mainly to the high momentum hadrons, statistical hadronization with temperature and chemical potential parameters, and coalescence or recombination happening on the boundary of deconfinement phase transition. Different from light hadrons where the coalescence probability (Wigner function) for two or three quarks to combine into a meson or baryon is assumed to be a Gaussian distribution with adjustable width, for heavy flavors the Wigner function can be derived from the meson or baryon wave function controlled by effective models. With this calculable probability one can predict the production of multi-charmed baryons in heavy ion collisions which is significantly enhanced by the coalescence mechanism in comparison with p+p collisions.     

We then turned to quarkonium production and motion in hot QCD phase created in heavy ion collisions. There are two kinds of hot nuclear matter effects on the quarkonium production: the dissociation and the regeneration. The two affect the quarkonium yield in an opposite way, and the degree of both increases with increasing colliding energy. Therefore, the cancellation between the two weakens the sensitivity of the quarkonium yield to the properties of the hot medium. The case is, however, dramatically changed when we focus on the quarkonium transverse momentum distribution. The two hot
nuclear matter effects work in different transverse momentum regions. The dissociation suppresses mainly the initial hard component, and the regeneration enhances the soft component. When the colliding energy increases, the dominant production source changes from the hard process to the soft process. The speed of the change is controlled by the degree of the heavy quark thermalization. If heavy quarks are thermalized fast, the change becomes significant. Therefore, a dominant soft component can be taken as a clear signal of the regeneration, namely the signal of the QGP phase at RHIC and LHC energy. The above idea can be realized through a detailed transport approach for quarkonia in high energy nuclear collisions. The hot medium is described by ideal hydrodynamics, and the quarkonium motion in the medium is governed by transport equations, including the dissociation and regeneration as loss and gain terms and cold nuclear matter effect as initial
condition of the transport. By solving the two groups of coupled equations, it is found that, the nuclear modification factor for averaged transverse momentum square is very sensitive to the hot mediums produced in heavy ion collisions, and it changes from larger than unity at the SPS to around unity at the RHIC and to less than unit at the LHC.      

We finally paid special attention to heavy flavor hadrons under extremely strong electromagnetic and rotational fields which are generated in non-central nuclear collisions. We discussed three kinds of electromagnetic effects on heavy flavors: the heavy quark spin interaction with the electromagnetic fields which leads to a mixing between spin singlet and triplet states of quarkonia and makes a shift of the quarkonium mass due to the quark magnetic moment, the breaking down of the space symmetry which changes the heavy quark potential and results in a sizeable quarkonium elliptic flow at large transverse momentum, and the pure electromagnetic interaction in ultra-peripheral nuclear collisions which causes a significant $J/\psi$ enhancement at extremely low transverse momentum. The question for the electromagnetic field is its lifetime, it may vanish before the QGP formation but can be prolonged if there is a large enough electric conductivity of the QGP phase. From the similarity between the vorticity field and magnetic field, the heavy flavor production and evolution in a rotational field can be described by a transport equation for the heavy flavor plus tilted hydrodynamic equations for the medium.    

There are still unclear issues in studying heavy flavors as a prob of the extreme environment generated in high energy nuclear collisions. While the interaction between heavy quarks and the QGP medium is intensively investigated, we probably ignored the correlation between light and heavy quark hadronizations. For instance, considering the charm quark number conservation during the evolution of the system, the $D_s$ enhancement by the well-known strangeness enhancement will naturally result in a $D^0$ suppression and then a more strong enhancement of the ratio $D_s/D^0$. To extract clear information on the hot QGP phase from the final state heavy flavors, one needs to know the details of the heavy flavors in vacuum and cold nuclear matter. In this case, small systems like p+p and p+A collisions which are the background of the QGP study in A+A collisions become especially important. Similarly, to understand the electromagnetic effect on the hot medium properties carried by heavy flavors in A+A collisions, one needs to calculate precisely the space-time distribution of the initially created electromagnetic fields and the electric conductivity of the medium. Since the study on QCD phase transitions in the future will go to high baryon density region through intermediate energy nuclear collisions at FAIR~\cite{CBM} and NICA~\cite{Kovalenko:2019pgu} and probably to more hot region through ultra high energy nuclear collisions at FCC~\cite{Dainese:2016gch}, one should consider the density effect on charmonium production in the low energy limit, like $\Lambda_c$ enhancement and $\bar D^0/D^0$ enhancement induced by the increasing baryon chemical potential, and the thermal production of charm quarks in the QGP phase in the high energy limit which will lead to a significant enhancement of low momentum heavy flavors.                 

\section*{Acknowledgments}
We thank Baoyi Chen, Yunpeng liu, Zhen Qu, Ralf Rapp, Nu Xu, Zhe Xu, Li Yan and Xianglei Zhu for helpful discussions and fruitful collaborations in the study of heavy flavors and quarkonia. The work is supported by the NSFC grant No. 11890712.



\begin{thebibliography}{20}
\bibitem{Bazavov:2011nk} 
  A.~Bazavov {\it et al.},
  Phys.\ Rev.\ D {\bf 85}, 054503 (2012).
  
\bibitem{Andronic:2015wma} 
A.~Andronic {\it et al.},
Eur.\ Phys.\ J.\ C {\bf 76}, 107 (2016).  
  
\bibitem{Dainese:2016gch} 
  A.~Dainese {\it et al.},
  CERN Yellow Rep.\ , 635 (2017).  
  
\bibitem{Aarts} 
G.~Aarts {\it et al.},
Eur.\ Phys.\ J.\ A {\bf 53}, 93 (2017).

\bibitem{FCC} 
A.~Abada {\it et al.},
Eur.\ Phys.\ J.\ C {\bf 79}, 474 (2019).

\bibitem{Mueller:1985wy} 
  A.~H.~Mueller and J.~w.~Qiu,
  Nucl.\ Phys.\ B {\bf 268}, 427 (1986).
  
\bibitem{Cronin:1974zm} 
  J.~W.~Cronin, H.~J.~Frisch, M.~J.~Shochet, J.~P.~Boymond, R.~Mermod, P.~A.~Piroue and R.~L.~Sumner,
  Phys.\ Rev.\ D {\bf 11}, 3105 (1975).  
  
\bibitem{Gerschel:1988wn} 
  C.~Gerschel and J.~Hufner,
  Phys.\ Lett.\ B {\bf 207}, 253 (1988).    
 
\bibitem{Satz:2005hx} 
  H.~Satz,
  J.\ Phys.\ G {\bf 32}, R25 (2006). 
    
\bibitem{Petreczky:2010yn} 
  P.~Petreczky,
  J.\ Phys.\ G {\bf 37}, 094009 (2010).  
  
\bibitem{Fukushima:2008xe} 
  K.~Fukushima, D.~E.~Kharzeev and H.~J.~Warringa,
  Phys.\ Rev.\ D {\bf 78}, 074033 (2008).  
  
\bibitem{Kharzeev:2007tn} 
  D.~Kharzeev and A.~Zhitnitsky,
  Nucl.\ Phys.\ A {\bf 797}, 67 (2007).
  
\bibitem{Kharzeev:2010gr} 
  D.~E.~Kharzeev and D.~T.~Son,
  Phys.\ Rev.\ Lett.\  {\bf 106}, 062301 (2011).  
  
\bibitem{Collins:1985gm} 
  J.~C.~Collins, D.~E.~Soper and G.~F.~Sterman,
  Nucl.\ Phys.\ B {\bf 263}, 37 (1986).
  
\bibitem{Combridge:1978kx} 
  B.~L.~Combridge,
  Nucl.\ Phys.\ B {\bf 151}, 429 (1979). 

\bibitem{Catani:1990eg} 
  S.~Catani, M.~Ciafaloni and F.~Hautmann,
  Nucl.\ Phys.\ B {\bf 366}, 135 (1991).
  
\bibitem{Baranov:2000gv} 
  S.~P.~Baranov and M.~Smizanska,
  Phys.\ Rev.\ D {\bf 62}, 014012 (2000).
  
\bibitem{Hagler:2000dda} 
  P.~Hagler, R.~Kirschner, A.~Schafer, L.~Szymanowski and O.~Teryaev,
  Phys.\ Rev.\ D {\bf 62}, 071502 (2000).
  
\bibitem{Frixione:1997ma} 
  S.~Frixione, M.~L.~Mangano, P.~Nason and G.~Ridolfi,
  Adv.\ Ser.\ Direct.\ High Energy Phys.\  {\bf 15}, 609 (1998).  
        
\bibitem{Nason:1989zy} 
  P.~Nason, S.~Dawson and R.~K.~Ellis,
  Nucl.\ Phys.\ B {\bf 327}, 49 (1989).
  
\bibitem{Beenakker:1990maa} 
  W.~Beenakker, W.~L.~van Neerven, R.~Meng, G.~A.~Schuler and J.~Smith,
  Nucl.\ Phys.\ B {\bf 351}, 507 (1991).  
  
\bibitem{Mangano:1991jk} 
  M.~L.~Mangano, P.~Nason and G.~Ridolfi,
  Nucl.\ Phys.\ B {\bf 373}, 295 (1992).
  
\bibitem{Cacciari:1993mq} 
  M.~Cacciari and M.~Greco,
  Nucl.\ Phys.\ B {\bf 421}, 530 (1994).
  
\bibitem{Kniehl:1995em} 
  B.~A.~Kniehl, M.~Kramer, G.~Kramer and M.~Spira,
  Phys.\ Lett.\ B {\bf 356}, 539 (1995).
  
\bibitem{Kramer:2001gd} 
  G.~Kramer and H.~Spiesberger,
  Eur.\ Phys.\ J.\ C {\bf 22}, 289 (2001).
  
\bibitem{Kniehl:2005mk} 
  B.~A.~Kniehl, G.~Kramer, I.~Schienbein and H.~Spiesberger,
  Eur.\ Phys.\ J.\ C {\bf 41}, 199 (2005).        
  
\bibitem{Cacciari:2001td} 
  M.~Cacciari, S.~Frixione and P.~Nason,
  JHEP {\bf 0103}, 006 (2001).
  
\bibitem{Sjostrand:2006za} 
  T.~Sjostrand, S.~Mrenna and P.~Z.~Skands,
  JHEP {\bf 0605}, 026 (2006)  
  
\bibitem{Frixione:2002ik} 
  S.~Frixione and B.~R.~Webber,
  JHEP {\bf 0206}, 029 (2002). 
 
\bibitem{Bowler:1981sb} 
  M.~G.~Bowler,
  Z.\ Phys.\ C {\bf 11}, 169 (1981).
  
\bibitem{Andersson:1983ia} 
  B.~Andersson, G.~Gustafson, G.~Ingelman and T.~Sjostrand,
  Phys.\ Rept.\  {\bf 97}, 31 (1983).  
  
\bibitem{Peterson:1982ak} 
  C.~Peterson, D.~Schlatter, I.~Schmitt and P.~M.~Zerwas,
  Phys.\ Rev.\ D {\bf 27}, 105 (1983).  
  
\bibitem{Binnewies:1997xq} 
  J.~Binnewies, B.~A.~Kniehl and G.~Kramer,
  Phys.\ Rev.\ D {\bf 58}, 014014 (1998).
  
\bibitem{Binnewies:1998vm} 
  J.~Binnewies, B.~A.~Kniehl and G.~Kramer,
  Phys.\ Rev.\ D {\bf 58}, 034016 (1998).         
   
\bibitem{Fritzsch:1977ay} 
  H.~Fritzsch,
  Phys.\ Lett.\ B {\bf 67}, 217 (1977).

\bibitem{Amundson:1995em} 
  J.~F.~Amundson, O.~J.~P.~Eboli, E.~M.~Gregores and F.~Halzen,
  Phys.\ Lett.\ B {\bf 372}, 127 (1996).
    
\bibitem{Bedjidian:2004gd} 
  M.~Bedjidian {\it et al.},
  hep-ph/0311048.  
     
\bibitem{Chang:1979nn} 
  C.~H.~Chang,
  Nucl.\ Phys.\ B {\bf 172}, 425 (1980).
    
\bibitem{Baier:1981uk} 
  R.~Baier and R.~Ruckl,
  Phys.\ Lett.\ B {\bf 102}, 364 (1981).
  
\bibitem{Berger:1980ni} 
  E.~L.~Berger and D.~L.~Jones,
  Phys.\ Rev.\ D {\bf 23}, 1521 (1981).    
 
\bibitem{Bodwin:1992qr} 
  G.~T.~Bodwin, E.~Braaten, T.~C.~Yuan and G.~P.~Lepage,
  Phys.\ Rev.\ D {\bf 46}, R3703 (1992).  
  
\bibitem{Bodwin:1994jh} 
  G.~T.~Bodwin, E.~Braaten and G.~P.~Lepage,
  Phys.\ Rev.\ D {\bf 51}, 1125 (1995)
  Erratum: [Phys.\ Rev.\ D {\bf 55}, 5853 (1997)].
  
\bibitem{Chao:2012iv} 
  K.~T.~Chao, Y.~Q.~Ma, H.~S.~Shao, K.~Wang and Y.~J.~Zhang,
  Phys.\ Rev.\ Lett.\  {\bf 108}, 242004 (2012).
  
\bibitem{Butenschoen:2012px} 
  M.~Butenschoen and B.~A.~Kniehl,
  Phys.\ Rev.\ Lett.\  {\bf 108}, 172002 (2012).
  
\bibitem{Gong:2012ug} 
  B.~Gong, L.~P.~Wan, J.~X.~Wang and H.~F.~Zhang,
  Phys.\ Rev.\ Lett.\  {\bf 110}, 042002 (2013).      
    
\bibitem{Gattringer:2010zz} 
  C.~Gattringer and C.~B.~Lang,
  Lect.\ Notes Phys.\  {\bf 788}, 1 (2010).
  
\bibitem{Reinders:1984sr} 
  L.~J.~Reinders, H.~Rubinstein and S.~Yazaki,
  Phys.\ Rept.\  {\bf 127}, 1 (1985).  
  
\bibitem{Shuryak:1981fza} 
  E.~V.~Shuryak,
  Nucl.\ Phys.\ B {\bf 198}, 83 (1982).
  
\bibitem{Bagan:1992tp} 
  E.~Bagan, M.~Chabab, H.~G.~Dosch and S.~Narison,
  Phys.\ Lett.\ B {\bf 287}, 176 (1992). 
  
\bibitem{AliKhan:1999yb} 
  A.~Ali Khan, T.~Bhattacharya, S.~Collins, C.~T.~H.~Davies, R.~Gupta, C.~Morningstar, J.~Shigemitsu and J.~H.~Sloan,
  Phys.\ Rev.\ D {\bf 62}, 054505 (2000).
  
\bibitem{Morita:2009qk} 
  K.~Morita and S.~H.~Lee,
  Phys.\ Rev.\ D {\bf 82}, 054008 (2010).  
  
\bibitem{Brambilla:2004jw} 
  N.~Brambilla, A.~Pineda, J.~Soto and A.~Vairo,
  Rev.\ Mod.\ Phys.\  {\bf 77}, 1423 (2005). 

\bibitem{Matsui:1986dk} 
  T.~Matsui and H.~Satz,
  Phys.\ Lett.\ B {\bf 178}, 416 (1986).
   
\bibitem{Bali:2000gf} 
  G.~S.~Bali,
  Phys.\ Rept.\  {\bf 343}, 1 (2001).
  
\bibitem{Kawanai:2011jt} 
  T.~Kawanai and S.~Sasaki,
  Phys.\ Rev.\ D {\bf 85}, 091503 (2012).    
  
\bibitem{1998FBS}
  Krivec, R., Few-Body Systems, {\bf 25}, 199 (1998).
  
\bibitem{Barnea:1999be} 
  N.~Barnea, W.~Leidemann and G.~Orlandini,
  Phys.\ Rev.\ C {\bf 61}, 054001 (2000).
  
\bibitem{Tanabashi:2018oca} 
  M.~Tanabashi {\it et al.} [Particle Data Group],
  Phys.\ Rev.\ D {\bf 98}, 030001 (2018).
  
\bibitem{Zhao:2017gpq} 
J.~Zhao and P.~Zhuang,
Few Body Syst.\  {\bf 58}, 100 (2017).  
  
\bibitem{Isgur:1979be} 
  N.~Isgur and G.~Karl,
  Phys.\ Rev.\ D {\bf 20}, 1191 (1979).
  
\bibitem{Capstick:1986bm} 
  S.~Capstick and N.~Isgur,
  Phys.\ Rev.\ D {\bf 34}, 2809 (1986).
  
\bibitem{DeRujula:1975qlm} 
  A.~De Rujula, H.~Georgi and S.~L.~Glashow,
  Phys.\ Rev.\ D {\bf 12}, 147 (1975).
  
\bibitem{Salpeter:1951sz}
  E.~E.~Salpeter and H.~A.~Bethe,
  Phys.\ Rev.\  {\bf 84}, 1232 (1951).  
  
\bibitem{Logunov:1963yc} 
  A.~A.~Logunov and A.~N.~Tavkhelidze,
  Nuovo Cim.\  {\bf 29}, 380 (1963).
  
\bibitem{Gross:1969rv} 
  F.~Gross,
  Phys.\ Rev.\  {\bf 186}, 1448 (1969).
  
\bibitem{Sazdjian:1986aw} 
  H.~Sazdjian,
  J.\ Math.\ Phys.\  {\bf 28}, 2618 (1987).
  
\bibitem{Sazdjian:1988be} 
  H.~Sazdjian,
  Annals Phys.\  {\bf 191}, 52 (1989).          
      
\bibitem{Crater:1983ew} 
  H.~w.~Crater and P.~Van Alstine,
  Annals Phys.\  {\bf 148}, 57 (1983).
  
\bibitem{Crater:1987hm} 
  H.~W.~Crater and P.~Van Alstine,
  Phys.\ Rev.\ D {\bf 36}, 3007 (1987).
  
\bibitem{Crater:2010fc} 
  H.~W.~Crater and J.~Schiermeyer,
  Phys.\ Rev.\ D {\bf 82}, 094020 (2010).
 
\bibitem{Liu:2002cn} 
  B.~Liu and H.~Crater,
  Phys.\ Rev.\ C {\bf 67}, 024001 (2003).
  
\bibitem{Shi:2013rga} 
  S.~Shi, X.~Guo and P.~Zhuang,
  Phys.\ Rev.\ D {\bf 88}, 014021 (2013).  

\bibitem{Shi:2019tji} 
S.~Shi, J.~Zhao and P.~Zhuang,
arXiv:1905.10627 [nucl-th]. 
  
\bibitem{Whitney:2011aa} 
  J.~F.~Whitney and H.~W.~Crater,
  Phys.\ Rev.\ D {\bf 89}, 014023 (2014).  

\bibitem{Caswell:1985ui} 
  W.~E.~Caswell and G.~P.~Lepage,
  Phys.\ Lett.\ B {\bf 167}, 437 (1986).
    
\bibitem{Brambilla:1999xf} 
  N.~Brambilla, A.~Pineda, J.~Soto and A.~Vairo,
  Nucl.\ Phys.\ B {\bf 566}, 275 (2000).  
  
\bibitem{Butenschoen:2011yh} 
  M.~Butenschoen and B.~A.~Kniehl,
  Phys.\ Rev.\ D {\bf 84}, 051501 (2011).  
  
\bibitem{Butenschoen:2012qr} 
  M.~Butenschoen and B.~A.~Kniehl,
  Mod.\ Phys.\ Lett.\ A {\bf 28}, 1350027 (2013).  

\bibitem{Han:2014kxa} 
  H.~Han, Y.~Q.~Ma, C.~Meng, H.~S.~Shao, Y.~J.~Zhang and K.~T.~Chao,
  Phys.\ Rev.\ D {\bf 94}, 014028 (2016).
  
\bibitem{Bain:2017wvk} 
  R.~Bain, L.~Dai, A.~Leibovich, Y.~Makris and T.~Mehen,
  Phys.\ Rev.\ Lett.\  {\bf 119}, 032002 (2017).  
  
\bibitem{Kniehl:2002br} 
  B.~A.~Kniehl, A.~A.~Penin, V.~A.~Smirnov and M.~Steinhauser,
  Nucl.\ Phys.\ B {\bf 635}, 357 (2002).
  
\bibitem{Brambilla:2001qk} 
  N.~Brambilla, Y.~Sumino and A.~Vairo,
  Phys.\ Rev.\ D {\bf 65}, 034001 (2002). 
 
\bibitem{LlanesEstrada:2011kc} 
  F.~J.~Llanes-Estrada, O.~I.~Pavlova and R.~Williams,
  Eur.\ Phys.\ J.\ C {\bf 72}, 2019 (2012).  
  
\bibitem{Kniehl:2002yv} 
  B.~A.~Kniehl, A.~A.~Penin, M.~Steinhauser and V.~A.~Smirnov,
  Phys.\ Rev.\ Lett.\  {\bf 90}, 212001 (2003)
  Erratum: [Phys.\ Rev.\ Lett.\  {\bf 91}, 139903 (2003)]. 
  
\bibitem{Donald:2012ga} 
  G.~C.~Donald, C.~T.~H.~Davies, R.~J.~Dowdall, E.~Follana, K.~Hornbostel, J.~Koponen, G.~P.~Lepage and C.~McNeile,
  Phys.\ Rev.\ D {\bf 86}, 094501 (2012). 
  
\bibitem{Basak:2013oya} 
  S.~Basak, S.~Datta, A.~T.~Lytle, M.~Padmanath, P.~Majumdar and N.~Mathur,
  PoS LATTICE {\bf 2013}, 243 (2014).
  
\bibitem{Datta:2003ww} 
  S.~Datta, F.~Karsch, P.~Petreczky and I.~Wetzorke,
  Phys.\ Rev.\ D {\bf 69}, 094507 (2004).
  
\bibitem{Asakawa:2003re} 
  M.~Asakawa and T.~Hatsuda,
  Phys.\ Rev.\ Lett.\  {\bf 92}, 012001 (2004).  
  
\bibitem{Umeda:2002vr} 
  T.~Umeda, K.~Nomura and H.~Matsufuru,
  Eur.\ Phys.\ J.\ C {\bf 39S1}, 9 (2005).
  
\bibitem{Ding:2012sp} 
  H.~T.~Ding, A.~Francis, O.~Kaczmarek, F.~Karsch, H.~Satz and W.~Soeldner,
  Phys.\ Rev.\ D {\bf 86}, 014509 (2012).

\bibitem{Mocsy:2013syh} 
  A.~Mocsy, P.~Petreczky and M.~Strickland,
  Int.\ J.\ Mod.\ Phys.\ A {\bf 28}, 1340012 (2013).
  
\bibitem{Ohno:2011zc} 
  H.~Ohno {\it et al.} [WHOT-QCD Collaboration],
  Phys.\ Rev.\ D {\bf 84}, 094504 (2011).       
  
\bibitem{Bowler:1996ws} 
  K.~C.~Bowler {\it et al.} [UKQCD Collaboration],
  Phys.\ Rev.\ D {\bf 54}, 3619 (1996).
  
\bibitem{Namekawa:2013vu} 
  Y.~Namekawa {\it et al.} [PACS-CS Collaboration],
  Phys.\ Rev.\ D {\bf 87}, 094512 (2013).     
  
\bibitem{Mathur:2018rwu} 
  N.~Mathur and M.~Padmanath,
  Phys.\ Rev.\ D {\bf 99}, 031501 (2019).      
  
\bibitem{Padmanath:2019ybu} 
  M.~Padmanath,
  arXiv:1905.10168 [hep-lat].  
 
\bibitem{Mathur:2018epb} 
N.~Mathur, M.~Padmanath and S.~Mondal,
Phys.\ Rev.\ Lett.\  {\bf 121}, 202002 (2018).  
  
\bibitem{Larsen:2019bwy} 
R.~Larsen, S.~Meinel, S.~Mukherjee and P.~Petreczky,
Phys.\ Rev.\ D {\bf 100}, 074506 (2019).     

\bibitem{Lepage:1992tx} 
G.~P.~Lepage, L.~Magnea, C.~Nakhleh, U.~Magnea and K.~Hornbostel,
Phys.\ Rev.\ D {\bf 46}, 4052 (1992).

\bibitem{Meinel:2010pv} 
  S.~Meinel,
  Phys.\ Rev.\ D {\bf 82}, 114502 (2010).  
  
\bibitem{Dowdall:2011wh} 
  R.~J.~Dowdall {\it et al.} [HPQCD Collaboration],
  Phys.\ Rev.\ D {\bf 85}, 054509 (2012).
    
\bibitem{Mathur:2016hsm} 
  N.~Mathur, M.~Padmanath and R.~Lewis,
  PoS LATTICE {\bf 2016}, 100 (2016).    
  
\bibitem{Aarts:2012ka} 
  G.~Aarts, C.~Allton, S.~Kim, M.~P.~Lombardo, M.~B.~Oktay, S.~M.~Ryan, D.~K.~Sinclair and J.~I.~Skullerud,
  JHEP {\bf 1303}, 084 (2013).  
   
\bibitem{Aarts:2010ek} 
  G.~Aarts, S.~Kim, M.~P.~Lombardo, M.~B.~Oktay, S.~M.~Ryan, D.~K.~Sinclair and J.-I.~Skullerud,
  Phys.\ Rev.\ Lett.\  {\bf 106}, 061602 (2011).
  
\bibitem{Kim:2018yhk} 
  S.~Kim, P.~Petreczky and A.~Rothkopf,
  JHEP {\bf 1811}, 088 (2018).    

\bibitem{Rothkopf:2019ipj} 
A.~Rothkopf,
arXiv:1912.02253 [hep-ph].

\bibitem{Hirai:2007sx} 
M.~Hirai, S.~Kumano and T.-H.~Nagai,
Phys.\ Rev.\ C {\bf 76}, 065207 (2007).  

\bibitem{deFlorian:2003qf} 
D.~de Florian and R.~Sassot,
Phys.\ Rev.\ D {\bf 69}, 074028 (2004).

\bibitem{Eskola:1998df} 
K.~J.~Eskola, V.~J.~Kolhinen and C.~A.~Salgado,
Eur.\ Phys.\ J.\ C {\bf 9}, 61 (1999).

\bibitem{Eskola:2009uj} 
K.~J.~Eskola, H.~Paukkunen and C.~A.~Salgado,
JHEP {\bf 0904}, 065 (2009).

\bibitem{Eskola:2008ca} 
K.~J.~Eskola, H.~Paukkunen and C.~A.~Salgado,
JHEP {\bf 0807}, 102 (2008).

\bibitem{Norton:2003cb} 
P.~R.~Norton,
Rept.\ Prog.\ Phys.\  {\bf 66}, 1253 (2003).  

\bibitem{Klein:2003dj} 
S.~R.~Klein and R.~Vogt,
Phys.\ Rev.\ Lett.\  {\bf 91}, 142301 (2003).

\bibitem{Gerschel:1998zi} 
C.~Gerschel and J.~Hufner,
Ann.\ Rev.\ Nucl.\ Part.\ Sci.\  {\bf 49}, 255 (1999).

\bibitem{Vogt:2001nh} 
R.~Vogt [Hard Probe Collaboration],
Int.\ J.\ Mod.\ Phys.\ E {\bf 12}, 211 (2003).

\bibitem{Lourenco:2008sk} 
C.~Lourenco, R.~Vogt and H.~K.~Woehri,
JHEP {\bf 0902}, 014 (2009).

\bibitem{Brambilla:2010cs} 
N.~Brambilla {\it et al.},
Eur.\ Phys.\ J.\ C {\bf 71}, 1534 (2011).  

\bibitem{Abreu:1999nn} 
M.~C.~Abreu {\it et al.},
Phys.\ Lett.\ B {\bf 466}, 408 (1999).

\bibitem{Kopeliovich:2012be} 
B.~Z.~Kopeliovich, I.~K.~Potashnikova and I.~Schmidt,
EPJ Web Conf.\  {\bf 70}, 00067 (2014).    

\bibitem{Digal:2005ht} 
S.~Digal, O.~Kaczmarek, F.~Karsch and H.~Satz,
Eur.\ Phys.\ J.\ C {\bf 43}, 71 (2005).

\bibitem{Karsch:2003jg} 
F.~Karsch and E.~Laermann,
In *Hwa, R.C. (ed.) et al.: Quark gluon plasma* 1-59
[hep-lat/0305025]. 

\bibitem{Guo:2012hx} 
X.~Guo, S.~Shi and P.~Zhuang,
Phys.\ Lett.\ B {\bf 718}, 143 (2012).  

\bibitem{Digal:2001iu} 
S.~Digal, P.~Petreczky and H.~Satz,
Phys.\ Lett.\ B {\bf 514}, 57 (2001).  

\bibitem{Karsch:1987pv} 
  F.~Karsch, M.~T.~Mehr and H.~Satz,
  Z.\ Phys.\ C {\bf 37}, 617 (1988).  
  
\bibitem{Burnier:2015tda} 
Y.~Burnier, O.~Kaczmarek and A.~Rothkopf,
JHEP {\bf 1512}, 101 (2015).  

\bibitem{Brambilla:2008cx} 
  N.~Brambilla, J.~Ghiglieri, A.~Vairo and P.~Petreczky,
  Phys.\ Rev.\ D {\bf 78}, 014017 (2008).
   
\bibitem{Brambilla:2010vq} 
  N.~Brambilla, M.~A.~Escobedo, J.~Ghiglieri, J.~Soto and A.~Vairo,
  JHEP {\bf 1009}, 038 (2010). 

\bibitem{Laine:2006ns} 
  M.~Laine, O.~Philipsen, P.~Romatschke and M.~Tassler,
  JHEP {\bf 0703}, 054 (2007).

\bibitem{Petreczky:2010tk} 
  P.~Petreczky, C.~Miao and A.~Mocsy,
  Nucl.\ Phys.\ A {\bf 855}, 125 (2011).

\bibitem{Rothkopf:2011db} 
A.~Rothkopf, T.~Hatsuda and S.~Sasaki,
Phys.\ Rev.\ Lett.\  {\bf 108}, 162001 (2012).  
   
\bibitem{Burnier:2014ssa} 
Y.~Burnier, O.~Kaczmarek and A.~Rothkopf,
Phys.\ Rev.\ Lett.\  {\bf 114}, 082001 (2015).

\bibitem{Burnier:2016mxc} 
Y.~Burnier and A.~Rothkopf,
Phys.\ Rev.\ D {\bf 95}, 054511 (2017).    
 
\bibitem{Lafferty:2019jpr} 
D.~Lafferty and A.~Rothkopf,
arXiv:1906.00035 [hep-ph].    
  
\bibitem{Thakur:2013nia} 
L.~Thakur, U.~Kakade and B.~K.~Patra,
Phys.\ Rev.\ D {\bf 89}, 094020 (2014).   

\bibitem{Burnier:2015nsa} 
Y.~Burnier and A.~Rothkopf,
Phys.\ Lett.\ B {\bf 753}, 232 (2016).  

\bibitem{Guo:2019bwa} 
Y.~Guo, L.~Dong, J.~Pan and M.~R.~Moldes,
Phys.\ Rev.\ D {\bf 100}, 036011 (2019).    

\bibitem{Bazavov:2018wmo} 
A.~Bazavov {\it et al.} [TUMQCD Collaboration],
Phys.\ Rev.\ D {\bf 98}, 054511 (2018).  
  
\bibitem{Borsanyi:2015yka} 
  S.~Borsányi, Z.~Fodor, S.~D.~Katz, A.~Pásztor, K.~K.~Szabó and C.~Török,
  JHEP {\bf 1504}, 138 (2015).  
    
\bibitem{Laine:2008cf} 
  M.~Laine,
  Nucl.\ Phys.\ A {\bf 820}, 25C (2009).

\bibitem{Jakovac:2006sf} 
  A.~Jakovac, P.~Petreczky, K.~Petrov and A.~Velytsky,
  Phys.\ Rev.\ D {\bf 75}, 014506 (2007).   
  
\bibitem{Aarts:2011sm} 
  G.~Aarts, C.~Allton, S.~Kim, M.~P.~Lombardo, M.~B.~Oktay, S.~M.~Ryan, D.~K.~Sinclair and J.~I.~Skullerud,
  JHEP {\bf 1111}, 103 (2011).
  
\bibitem{Aarts:2014cda} 
  G.~Aarts, C.~Allton, T.~Harris, S.~Kim, M.~P.~Lombardo, S.~M.~Ryan and J.~I.~Skullerud,
  JHEP {\bf 1407}, 097 (2014).
  
\bibitem{Ikeda:2016czj} 
  A.~Ikeda, M.~Asakawa and M.~Kitazawa,
  Phys.\ Rev.\ D {\bf 95}, 014504 (2017).
 
\bibitem{Kelly:2018hsi} 
  A.~Kelly, A.~Rothkopf and J.~I.~Skullerud,
  Phys.\ Rev.\ D {\bf 97}, 114509 (2018).   
  
\bibitem{Burnier:2013nla} 
  Y.~Burnier and A.~Rothkopf,
  Phys.\ Rev.\ Lett.\  {\bf 111}, 182003 (2013).  
  
\bibitem{Aarts:2013kaa} 
  G.~Aarts, C.~Allton, S.~Kim, M.~P.~Lombardo, S.~M.~Ryan and J.-I.~Skullerud,
  JHEP {\bf 1312}, 064 (2013).    
  
\bibitem{Kim:2014iga} 
  S.~Kim, P.~Petreczky and A.~Rothkopf,
  Phys.\ Rev.\ D {\bf 91}, 054511 (2015).
  
\bibitem{Burnier:2007qm} 
  Y.~Burnier, M.~Laine and M.~Vepsalainen,
  JHEP {\bf 0801}, 043 (2008). 
    
\bibitem{Burnier:2016kqm} 
  Y.~Burnier, O.~Kaczmarek and A.~Rothkopf,
  JHEP {\bf 1610}, 032 (2016). 

\bibitem{Beraudo:2007ky} 
  A.~Beraudo, J.-P.~Blaizot and C.~Ratti,
  Nucl.\ Phys.\ A {\bf 806}, 312 (2008).  
  
\bibitem{Mukherjee:2015mxc} 
  S.~Mukherjee, P.~Petreczky and S.~Sharma,
  Phys.\ Rev.\ D {\bf 93}, 014502 (2016).
  
\bibitem{Bazavov:2014cta} 
  A.~Bazavov, F.~Karsch, Y.~Maezawa, S.~Mukherjee and P.~Petreczky,
  Phys.\ Rev.\ D {\bf 91}, 054503 (2015).    

\bibitem{Peskin:1979va} 
M.~E.~Peskin,
Nucl.\ Phys.\ B {\bf 156}, 365 (1979).

\bibitem{Bhanot:1979vb} 
G.~Bhanot and M.~E.~Peskin,
Nucl.\ Phys.\ B {\bf 156}, 391 (1979).

\bibitem{Polleri:2003kn} 
A.~Polleri, T.~Renk, R.~Schneider and W.~Weise,
Phys.\ Rev.\ C {\bf 70}, 044906 (2004). 

\bibitem{Chen:2018dqg} 
S.~Chen and M.~He,
Phys.\ Lett.\ B {\bf 786}, 260 (2018).  

\bibitem{Grandchamp:2001pf} 
L.~Grandchamp and R.~Rapp,
Phys.\ Lett.\ B {\bf 523}, 60 (2001).

\bibitem{Grandchamp:2002wp} 
L.~Grandchamp and R.~Rapp,
Nucl.\ Phys.\ A {\bf 709}, 415 (2002).  

\bibitem{Brambilla:2011sg} 
N.~Brambilla, M.~A.~Escobedo, J.~Ghiglieri and A.~Vairo,
JHEP {\bf 1112}, 116 (2011).

\bibitem{Brambilla:2013dpa} 
N.~Brambilla, M.~A.~Escobedo, J.~Ghiglieri and A.~Vairo,
JHEP {\bf 1305}, 130 (2013).  

\bibitem{Park:2007zza} 
Y.~Park, K.~I.~Kim, T.~Song, S.~H.~Lee and C.~Y.~Wong,
Phys.\ Rev.\ C {\bf 76}, 044907 (2007).    

\bibitem{Gavai:1994in} 
R.~Gavai, D.~Kharzeev, H.~Satz, G.~A.~Schuler, K.~Sridhar and R.~Vogt,
Int.\ J.\ Mod.\ Phys.\ A {\bf 10}, 3043 (1995).  

\bibitem{GranierdeCassagnac:2008ke} 
R.~Granier de Cassagnac,
J.\ Phys.\ G {\bf 35}, 104023 (2008).

\bibitem{Ferreiro:2008wc} 
E.~G.~Ferreiro, F.~Fleuret, J.~P.~Lansberg and A.~Rakotozafindrabe,
Phys.\ Lett.\ B {\bf 680}, 50 (2009).

\bibitem{Kopeliovich:2010nw} 
B.~Z.~Kopeliovich, I.~K.~Potashnikova, H.~J.~Pirner and I.~Schmidt,
Phys.\ Rev.\ C {\bf 83}, 014912 (2011).    

\bibitem{Yao:2017fuc} 
X.~Yao and B.~Müller,
Phys.\ Rev.\ C {\bf 97}, 014908 (2018)
Erratum: [Phys.\ Rev.\ C {\bf 97}, 049903 (2018)].

\bibitem{Song:2012at} 
T.~Song, K.~C.~Han and C.~M.~Ko,
Phys.\ Rev.\ C {\bf 85}, 054905 (2012). 

\bibitem{Zhang:2007dm} 
B.~W.~Zhang, C.~M.~Ko and W.~Liu,
Phys.\ Rev.\ C {\bf 77}, 024901 (2008).

\bibitem{Zhou:2016wbo} 
K.~Zhou, Z.~Chen, C.~Greiner and P.~Zhuang,
Phys.\ Lett.\ B {\bf 758}, 434 (2016).

\bibitem{Liu:2016zle} 
Y.~Liu and C.~M.~Ko,
J.\ Phys.\ G {\bf 43}, 125108 (2016).  

\bibitem{Thoma:1990fm} 
M.~H.~Thoma and M.~Gyulassy,
Nucl.\ Phys.\ B {\bf 351}, 491 (1991).

\bibitem{Braaten:1991jj} 
E.~Braaten and M.~H.~Thoma,
Phys.\ Rev.\ D {\bf 44}, 1298 (1991).

\bibitem{Braaten:1991we} 
E.~Braaten and M.~H.~Thoma,
Phys.\ Rev.\ D {\bf 44}, R2625 (1991).    

\bibitem{Braaten:1989kk} 
E.~Braaten and R.~D.~Pisarski,
Phys.\ Rev.\ Lett.\  {\bf 64}, 1338 (1990).

\bibitem{Peigne:2008nd} 
S.~Peigne and A.~Peshier,
Phys.\ Rev.\ D {\bf 77}, 114017 (2008).

\bibitem{Peshier:2008zz} 
A.~Peshier,
J.\ Phys.\ G {\bf 35}, 044028 (2008).  

\bibitem{Moore:2004tg} 
  G.~D.~Moore and D.~Teaney,
  Phys.\ Rev.\ C {\bf 71}, 064904 (2005).
 
\bibitem{Berends:1981rb} 
  F.~A.~Berends, R.~Kleiss, P.~De Causmaecker, R.~Gastmans and T.~T.~Wu,
  Phys.\ Lett.\  {\bf 103B}, 124 (1981). 
 
\bibitem{Ellis:1985er} 
  R.~K.~Ellis and J.~C.~Sexton,
  Nucl.\ Phys.\ B {\bf 269}, 445 (1986).
  
\bibitem{Kunszt:1979iy} 
  Z.~Kunszt, E.~Pietarinen and E.~Reya,
  Phys.\ Rev.\ D {\bf 21}, 733 (1980).    
  
\bibitem{Gunion:1981qs} 
  J.~F.~Gunion and G.~Bertsch,
  Phys.\ Rev.\ D {\bf 25}, 746 (1982).
  
\bibitem{Landau:1953um} 
  L.~D.~Landau and I.~Pomeranchuk,
  Dokl.\ Akad.\ Nauk Ser.\ Fiz.\  {\bf 92}, 535 (1953).
  
\bibitem{Migdal:1956tc} 
  A.~B.~Migdal,
  Phys.\ Rev.\  {\bf 103}, 1811 (1956).
  
\bibitem{Gyulassy:2000fs} 
  M.~Gyulassy, P.~Levai and I.~Vitev,
  Phys.\ Rev.\ Lett.\  {\bf 85}, 5535 (2000).
  
\bibitem{Gyulassy:2000er} 
  M.~Gyulassy, P.~Levai and I.~Vitev,
  Nucl.\ Phys.\ B {\bf 594}, 371 (2001).
  
\bibitem{Guo:2000nz} 
  X.~f.~Guo and X.~N.~Wang,
  Phys.\ Rev.\ Lett.\  {\bf 85}, 3591 (2000).
  
\bibitem{Wang:2001ifa} 
  X.~N.~Wang and X.~f.~Guo,
  Nucl.\ Phys.\ A {\bf 696}, 788 (2001).
  
\bibitem{Wiedemann:2000za} 
  U.~A.~Wiedemann,
  Nucl.\ Phys.\ B {\bf 588}, 303 (2000).
  
\bibitem{Salgado:2003gb} 
  C.~A.~Salgado and U.~A.~Wiedemann,
  Phys.\ Rev.\ D {\bf 68}, 014008 (2003).
  
\bibitem{Arnold:2002ja} 
  P.~B.~Arnold, G.~D.~Moore and L.~G.~Yaffe,
  JHEP {\bf 0206}, 030 (2002).
  
\bibitem{Arnold:2002zm} 
  P.~B.~Arnold, G.~D.~Moore and L.~G.~Yaffe,
  JHEP {\bf 0301}, 030 (2003).
          
\bibitem{Mustafa:1997pm} 
  M.~G.~Mustafa, D.~Pal, D.~K.~Srivastava and M.~Thoma,
  Phys.\ Lett.\ B {\bf 428}, 234 (1998).
       
\bibitem{Dokshitzer:2001zm} 
  Y.~L.~Dokshitzer and D.~E.~Kharzeev,
  Phys.\ Lett.\ B {\bf 519}, 199 (2001).
  
\bibitem{Djordjevic:2003zk} 
  M.~Djordjevic and M.~Gyulassy,
  Nucl.\ Phys.\ A {\bf 733}, 265 (2004).      
        
\bibitem{Djordjevic:2008iz} 
  M.~Djordjevic and U.~W.~Heinz,
  Phys.\ Rev.\ Lett.\  {\bf 101}, 022302 (2008).
  
\bibitem{Fochler:2013epa} 
  O.~Fochler, J.~Uphoff, Z.~Xu and C.~Greiner,
  Phys.\ Rev.\ D {\bf 88}, 014018 (2013).
  
\bibitem{Abir:2011jb} 
  R.~Abir, C.~Greiner, M.~Martinez, M.~G.~Mustafa and J.~Uphoff,
  Phys.\ Rev.\ D {\bf 85}, 054012 (2012).    
  
\bibitem{Mustafa:2004dr} 
  M.~G.~Mustafa,
  Phys.\ Rev.\ C {\bf 72}, 014905 (2005).
  
\bibitem{Wicks:2005gt} 
  S.~Wicks, W.~Horowitz, M.~Djordjevic and M.~Gyulassy,
  Nucl.\ Phys.\ A {\bf 784}, 426 (2007).    

\bibitem{Zhang:2003wk} 
  B.~W.~Zhang, E.~Wang and X.~N.~Wang,
  Phys.\ Rev.\ Lett.\  {\bf 93}, 072301 (2004).
   
\bibitem{Armesto:2005iq} 
  N.~Armesto, A.~Dainese, C.~A.~Salgado and U.~A.~Wiedemann,
  Phys.\ Rev.\ D {\bf 71}, 054027 (2005).   
  
\bibitem{Dong:2019byy} 
  X.~Dong, Y.~J.~Lee and R.~Rapp,
  arXiv:1903.07709 [nucl-ex].
 
\bibitem{Sirunyan:2018ktu} 
  A.~M.~Sirunyan {\it et al.} [CMS Collaboration],
  [arXiv:1810.11102 [hep-ex]].   
      
\bibitem{Yan:2007zze} 
  L.~Yan, P.~Zhuang and N.~Xu,
  Int.\ J.\ Mod.\ Phys.\ E {\bf 16}, 2048 (2007). 
  
\bibitem{Xu:2004mz} 
  Z.~Xu and C.~Greiner,
  Phys.\ Rev.\ C {\bf 71}, 064901 (2005).
  
\bibitem{Molnar:2004ph} 
  D.~Molnar,
  J.\ Phys.\ G {\bf 31}, S421 (2005).
  
\bibitem{Zhang:2005ni} 
  B.~Zhang, L.~W.~Chen and C.~M.~Ko,
  Phys.\ Rev.\ C {\bf 72}, 024906 (2005).
  
\bibitem{Rapp:2009my} 
  R.~Rapp and H.~van Hees,
  arXiv:0903.1096 [hep-ph].

\bibitem{CaronHuot:2007gq} 
  S.~Caron-Huot and G.~D.~Moore,
  Phys.\ Rev.\ Lett.\  {\bf 100}, 052301 (2008).
  
\bibitem{Abelev:2006db} 
  B.~I.~Abelev {\it et al.} [STAR Collaboration],
  Phys.\ Rev.\ Lett.\  {\bf 98}, 192301 (2007)
  Erratum: [Phys.\ Rev.\ Lett.\  {\bf 106}, 159902 (2011)].
  
\bibitem{Adare:2006nq} 
  A.~Adare {\it et al.} [PHENIX Collaboration],
  Phys.\ Rev.\ Lett.\  {\bf 98}, 172301 (2007).  
    
\bibitem{ALICE:2012ab} 
  B.~Abelev {\it et al.} [ALICE Collaboration],
  JHEP {\bf 1209}, 112 (2012).
  
\bibitem{Riek:2010fk} 
  F.~Riek and R.~Rapp,
  Phys.\ Rev.\ C {\bf 82}, 035201 (2010).      
        
\bibitem{He:2014cla} 
  M.~He, R.~J.~Fries and R.~Rapp,
  Phys.\ Lett.\ B {\bf 735}, 445 (2014).
  
\bibitem{Qin:2014dqa} 
  S.~x.~Qin and D.~H.~Rischke,
  Phys.\ Lett.\ B {\bf 734}, 157 (2014).    
 
\bibitem{Berrehrah:2013mua} 
  H.~Berrehrah, E.~Bratkovskaya, W.~Cassing, P.~B.~Gossiaux, J.~Aichelin and M.~Bleicher,
  Phys.\ Rev.\ C {\bf 89}, 054901 (2014).
 
\bibitem{Das:2012ck} 
  S.~K.~Das, V.~Chandra and J.~e.~Alam,
  J.\ Phys.\ G {\bf 41}, 015102 (2013).
  
\bibitem{Petreczky:2005nh} 
  P.~Petreczky and D.~Teaney,
  Phys.\ Rev.\ D {\bf 73}, 014508 (2006).  
  
\bibitem{Banerjee:2011ra} 
  D.~Banerjee, S.~Datta, R.~Gavai and P.~Majumdar,
  Phys.\ Rev.\ D {\bf 85}, 014510 (2012).
  
\bibitem{Ding:2015ona} 
  H.~T.~Ding, F.~Karsch and S.~Mukherjee,
  Int.\ J.\ Mod.\ Phys.\ E {\bf 24}, 1530007 (2015).
  
\bibitem{Francis:2015daa} 
  A.~Francis, O.~Kaczmarek, M.~Laine, T.~Neuhaus and H.~Ohno,
  Phys.\ Rev.\ D {\bf 92}, 116003 (2015). 
 
\bibitem{vanHees:2004gq} 
  H.~van Hees and R.~Rapp,
  Phys.\ Rev.\ C {\bf 71}, 034907 (2005).
  
\bibitem{Das:2015ana} 
  S.~K.~Das, F.~Scardina, S.~Plumari and V.~Greco,
  Phys.\ Lett.\ B {\bf 747}, 260 (2015).  
  
\bibitem{Song:2015sfa} 
  T.~Song, H.~Berrehrah, D.~Cabrera, J.~M.~Torres-Rincon, L.~Tolos, W.~Cassing and E.~Bratkovskaya,
  Phys.\ Rev.\ C {\bf 92}, 014910 (2015).
    
\bibitem{Xu:2017obm} 
  Y.~Xu, J.~E.~Bernhard, S.~A.~Bass, M.~Nahrgang and S.~Cao,
  Phys.\ Rev.\ C {\bf 97}, 014907 (2018).
  
\bibitem{Ke:2018tsh} 
  W.~Ke, Y.~Xu and S.~A.~Bass,
  Phys.\ Rev.\ C {\bf 98}, 064901 (2018).     
 
\bibitem{He:2011yi} 
  M.~He, R.~J.~Fries and R.~Rapp,
  Phys.\ Lett.\ B {\bf 701}, 445 (2011). 
 
\bibitem{Tolos:2013kva} 
  L.~Tolos and J.~M.~Torres-Rincon,
  Phys.\ Rev.\ D {\bf 88}, 074019 (2013).
 
\bibitem{Cao:2016gvr} 
  S.~Cao, T.~Luo, G.~Y.~Qin and X.~N.~Wang,
  Phys.\ Rev.\ C {\bf 94}, 014909 (2016).
  
\bibitem{He:2015pra} 
  Y.~He, T.~Luo, X.~N.~Wang and Y.~Zhu,
  Phys.\ Rev.\ C {\bf 91}, 054908 (2015)
  Erratum: [Phys.\ Rev.\ C {\bf 97}, 019902 (2018)].
  
\bibitem{Scardina:2017ipo} 
  F.~Scardina, S.~K.~Das, V.~Minissale, S.~Plumari and V.~Greco,
  Phys.\ Rev.\ C {\bf 96}, 044905 (2017).   
 
\bibitem{Cao:2013ita} 
  S.~Cao, G.~Y.~Qin and S.~A.~Bass,
  Phys.\ Rev.\ C {\bf 88}, 044907 (2013).
  
\bibitem{Cao:2015hia} 
  S.~Cao, G.~Y.~Qin and S.~A.~Bass,
  Phys.\ Rev.\ C {\bf 92}, 024907 (2015).
  
\bibitem{He:2011qa} 
  M.~He, R.~J.~Fries and R.~Rapp,
  Phys.\ Rev.\ C {\bf 86}, 014903 (2012).
  
\bibitem{Cassing:2009vt} 
  W.~Cassing and E.~L.~Bratkovskaya,
  Nucl.\ Phys.\ A {\bf 831}, 215 (2009).

\bibitem{Bratkovskaya:2011wp} 
  E.~L.~Bratkovskaya, W.~Cassing, V.~P.~Konchakovski and O.~Linnyk,
  Nucl.\ Phys.\ A {\bf 856}, 162 (2011).
   
\bibitem{Cao:2018ews} 
  S.~Cao {\it et al.},
  Phys.\ Rev.\ C {\bf 99}, 054907 (2019)  

\bibitem{Rapp:2018qla} 
  R.~Rapp {\it et al.},
  Nucl.\ Phys.\ A {\bf 979}, 21 (2018).
 
\bibitem{Bjorken:1982qr} 
  J.~D.~Bjorken,
  Phys.\ Rev.\ D {\bf 27}, 140 (1983).
  
\bibitem{BraunMunzinger:2003zd} 
  P.~Braun-Munzinger, K.~Redlich and J.~Stachel,
  In *Hwa, R.C. (ed.) et al.: Quark gluon plasma* 491-599.  
  
\bibitem{BraunMunzinger:2000px} 
  P.~Braun-Munzinger and J.~Stachel,
  Phys.\ Lett.\ B {\bf 490}, 196 (2000).
  
\bibitem{Andronic:2003zv} 
  A.~Andronic, P.~Braun-Munzinger, K.~Redlich and J.~Stachel,
  Phys.\ Lett.\ B {\bf 571}, 36 (2003).
  
\bibitem{BraunMunzinger:2007tn} 
  P.~Braun-Munzinger,
  J.\ Phys.\ G {\bf 34}, S471 (2007).

\bibitem{Siemens:1978pb} 
  P.~J.~Siemens and J.~O.~Rasmussen,
  Phys.\ Rev.\ Lett.\  {\bf 42}, 880 (1979).

\bibitem{Schnedermann:1993ws} 
  E.~Schnedermann, J.~Sollfrank and U.~W.~Heinz,
  Phys.\ Rev.\ C {\bf 48}, 2462 (1993).

\bibitem{Cooper:1974mv} 
  F.~Cooper and G.~Frye,
  Phys.\ Rev.\ D {\bf 10}, 186 (1974).

\bibitem{Adam:2018inb} 
  J.~Adam {\it et al.} [STAR Collaboration],
  Phys.\ Rev.\ C {\bf 99}, 034908 (2019).

\bibitem{Adler:2001bp} 
  C.~Adler {\it et al.} [STAR Collaboration],
  Phys.\ Rev.\ Lett.\  {\bf 86}, 4778 (2001)
  Erratum: [Phys.\ Rev.\ Lett.\  {\bf 90}, 119903 (2003)].

\bibitem{Adler:2002uv} 
  C.~Adler {\it et al.} [STAR Collaboration],
  Phys.\ Rev.\ Lett.\  {\bf 89}, 092301 (2002).
  
\bibitem{Sorensen:2003wi} 
  P.~Sorensen [STAR Collaboration],
  J.\ Phys.\ G {\bf 30}, S217 (2004).    

\bibitem{Hwa:2002tu} 
  R.~C.~Hwa and C.~B.~Yang,
  Phys.\ Rev.\ C {\bf 67}, 034902 (2003).  

\bibitem{Molnar:2003ff} 
  D.~Molnar and S.~A.~Voloshin,
  Phys.\ Rev.\ Lett.\  {\bf 91}, 092301 (2003).
  
\bibitem{Lin:2002rw} 
  Z.~w.~Lin and C.~M.~Ko,
  Phys.\ Rev.\ Lett.\  {\bf 89}, 202302 (2002).  
  
\bibitem{Greco:2003vf} 
  V.~Greco, C.~M.~Ko and R.~Rapp,
  Phys.\ Lett.\ B {\bf 595}, 202 (2004).      

\bibitem{Song:2015ykw} 
  T.~Song, H.~Berrehrah, D.~Cabrera, W.~Cassing and E.~Bratkovskaya,
  Phys.\ Rev.\ C {\bf 93}, 034906 (2016).
  
\bibitem{Lee:2007wr} 
  S.~H.~Lee, K.~Ohnishi, S.~Yasui, I.~K.~Yoo and C.~M.~Ko,
  Phys.\ Rev.\ Lett.\  {\bf 100}, 222301 (2008).  

\bibitem{Oh:2009zj} 
  Y.~Oh, C.~M.~Ko, S.~H.~Lee and S.~Yasui,
  Phys.\ Rev.\ C {\bf 79}, 044905 (2009).

\bibitem{Plumari:2017ntm} 
  S.~Plumari, V.~Minissale, S.~K.~Das, G.~Coci and V.~Greco,
  Eur.\ Phys.\ J.\ C {\bf 78}, 348 (2018).  
  
\bibitem{He:2014tga} 
  H.~He, Y.~Liu and P.~Zhuang,
  Phys.\ Lett.\ B {\bf 746}, 59 (2015).
  
\bibitem{Zhao:2016ccp} 
  J.~Zhao, H.~He and P.~Zhuang,
  Phys.\ Lett.\ B {\bf 771}, 349 (2017).
  
\bibitem{Zhao:2017yan} 
  J.~Zhao and B.~Chen,
  Phys.\ Lett.\ B {\bf 776}, 17 (2018).      

\bibitem{Rafelski:1982pu} 
  J.~Rafelski and B.~Muller,
  Phys.\ Rev.\ Lett.\  {\bf 48}, 1066 (1982)
  Erratum: [Phys.\ Rev.\ Lett.\  {\bf 56}, 2334 (1986)].
  
\bibitem{Adamczyk:2017xur} 
  L.~Adamczyk {\it et al.} [STAR Collaboration],
  Phys.\ Rev.\ Lett.\  {\bf 118}, 212301 (2017).
  
\bibitem{Abelev:2013lca} 
  B.~Abelev {\it et al.} [ALICE Collaboration],
  Phys.\ Rev.\ Lett.\  {\bf 111}, 102301 (2013).  
  
\bibitem{Fries:2003kq} 
  R.~J.~Fries, B.~Muller, C.~Nonaka and S.~A.~Bass,
  Phys.\ Rev.\ C {\bf 68}, 044902 (2003).
  
\bibitem{Zimanyi:2005nn} 
J.~Zimanyi, P.~Levai and T.~S.~Biro,
J.\ Phys.\ G {\bf 31}, 711 (2005). 
  
\bibitem{Ravagli:2007xx} 
  L.~Ravagli and R.~Rapp,
  Phys.\ Lett.\ B {\bf 655}, 126 (2007).
  
\bibitem{Ravagli:2008rt} 
  L.~Ravagli, H.~van Hees and R.~Rapp,
  Phys.\ Rev.\ C {\bf 79}, 064902 (2009).
  
\bibitem{Zhao:2018jlw} 
  J.~Zhao, S.~Shi, N.~Xu and P.~Zhuang,
  arXiv:1805.10858 [hep-ph]. 
     
\bibitem{Mattson:2002vu} 
  M.~Mattson {\it et al.} [SELEX Collaboration],
  Phys.\ Rev.\ Lett.\  {\bf 89}, 112001 (2002).

\bibitem{Chistov:2006zj} 
  R.~Chistov {\it et al.} [Belle Collaboration],
  Phys.\ Rev.\ Lett.\  {\bf 97}, 162001 (2006).
  
\bibitem{Aubert:2006qw} 
  B.~Aubert {\it et al.} [BaBar Collaboration],
  Phys.\ Rev.\ D {\bf 74}, 011103 (2006).  
    
\bibitem{Aaij:2013voa} 
  R.~Aaij {\it et al.} [LHCb Collaboration],
  JHEP {\bf 1312}, 090 (2013).
  
\bibitem{Aaij:2017ueg} 
  R.~Aaij {\it et al.} [LHCb Collaboration],
  Phys.\ Rev.\ Lett.\  {\bf 119}, 112001 (2017).
  
\bibitem{Zhao:2017znk} 
  J.~Zhao and P.~Zhuang,
  Phys.\ Lett.\ B {\bf 775}, 84 (2017).  
  
\bibitem{Lourenco:2006vw} 
  C.~Lourenco and H.~K.~Wohri,
  Phys.\ Rept.\  {\bf 433}, 127 (2006).

\bibitem{Zhu:2006er} 
  X.~Zhu, M.~Bleicher, S.~L.~Huang, K.~Schweda, H.~Stoecker, N.~Xu and P.~Zhuang,
  Phys.\ Lett.\ B {\bf 647}, 366 (2007).
  
\bibitem{Zhu:2007ne} 
  X.~Zhu, N.~Xu and P.~Zhuang,
  Phys.\ Rev.\ Lett.\  {\bf 100}, 152301 (2008).
  
\bibitem{Cao:2015cba} 
  S.~Cao, G.~Y.~Qin and S.~A.~Bass,
  Phys.\ Rev.\ C {\bf 92}, 054909 (2015).

\bibitem{Cortese:2005ns} 
  P.~Cortese {\it et al.} [NA50 Collaboration],
  J.\ Phys.\ G {\bf 31}, S809 (2005). 

\bibitem{Blaizot:1996nq} 
  J.~P.~Blaizot and J.~Y.~Ollitrault,
  Phys.\ Rev.\ Lett.\  {\bf 77}, 1703 (1996).

\bibitem{Blaizot:2000ev} 
  J.~P.~Blaizot, M.~Dinh and J.~Y.~Ollitrault,
  Phys.\ Rev.\ Lett.\  {\bf 85}, 4012 (2000).
  
\bibitem{Kharzeev:1994pz} 
  D.~Kharzeev and H.~Satz,
  Phys.\ Lett.\ B {\bf 334}, 155 (1994).  

\bibitem{Gavin:1988hs} 
  S.~Gavin, M.~Gyulassy and A.~Jackson,
  Phys.\ Lett.\ B {\bf 207}, 257 (1988).
  
\bibitem{Vogt:1988fj} 
  R.~Vogt, M.~Prakash, P.~Koch and T.~H.~Hansson,
  Phys.\ Lett.\ B {\bf 207}, 263 (1988).
  
\bibitem{Spieles:1998pa} 
  C.~Spieles, R.~Vogt, L.~Gerland, S.~A.~Bass, M.~Bleicher, H.~Stoecker and W.~Greiner,
  J.\ Phys.\ G {\bf 25}, 2351 (1999).
  
\bibitem{Bratkovskaya:2003ux} 
  E.~L.~Bratkovskaya, W.~Cassing and H.~Stoecker,
  Phys.\ Rev.\ C {\bf 67}, 054905 (2003).
  
\bibitem{Capella:2000zp} 
  A.~Capella, E.~G.~Ferreiro and A.~B.~Kaidalov,
  Phys.\ Rev.\ Lett.\  {\bf 85}, 2080 (2000).  
                
\bibitem{Young:2008he} 
  C.~Young and E.~Shuryak,
  Phys.\ Rev.\ C {\bf 79}, 034907 (2009).

\bibitem{Zhou:2013aea} 
  K.~Zhou, N.~Xu and P.~Zhuang,
  arXiv:1309.7520 [nucl-th].

\bibitem{Thews:2000rj} 
  R.~L.~Thews, M.~Schroedter and J.~Rafelski,
  Phys.\ Rev.\ C {\bf 63}, 054905 (2001).

\bibitem{Thews:2001hy} 
  R.~L.~Thews,
  Nucl.\ Phys.\ A {\bf 702}, 341 (2002).

\bibitem{Grandchamp:2003uw} 
  L.~Grandchamp, R.~Rapp and G.~E.~Brown,
  Phys.\ Rev.\ Lett.\  {\bf 92}, 212301 (2004).

\bibitem{Zhao:2007hh} 
  X.~Zhao and R.~Rapp,
  Phys.\ Lett.\ B {\bf 664}, 253 (2008).  
  
\bibitem{Zhao:2011cv} 
  X.~Zhao and R.~Rapp,
  Nucl.\ Phys.\ A {\bf 859}, 114 (2011).
  
\bibitem{Du:2015wha} 
  X.~Du and R.~Rapp,
  Nucl.\ Phys.\ A {\bf 943}, 147 (2015).
  
\bibitem{Zhu:2004nw} 
  X.~l.~Zhu, P.~f.~Zhuang and N.~Xu,
  Phys.\ Lett.\ B {\bf 607}, 107 (2005).
  
\bibitem{Yan:2006ve} 
  L.~Yan, P.~Zhuang and N.~Xu,
  Phys.\ Rev.\ Lett.\  {\bf 97}, 232301 (2006).      
 
\bibitem{Liu:2009wza} 
  Y.~Liu, Z.~Qu, N.~Xu and P.~Zhuang,
  J.\ Phys.\ G {\bf 37}, 075110 (2010).
  
\bibitem{Zhou:2014kka} 
K.~Zhou, N.~Xu, Z.~Xu and P.~Zhuang,
Phys.\ Rev.\ C {\bf 89}, 054911 (2014).
      
\bibitem{Abelev:2012vra} 
  B.~Abelev {\it et al.} [ALICE Collaboration],
  JHEP {\bf 1207}, 191 (2012). 
  
\bibitem{Sollfrank:1996hd} 
  J.~Sollfrank, P.~Huovinen, M.~Kataja, P.~V.~Ruuskanen, M.~Prakash and R.~Venugopalan,
  Phys.\ Rev.\ C {\bf 55}, 392 (1997).
  
\bibitem{oqs2002} 
H.-P. Breuer and F. Petruccione, The Theory of Open Quantum Systems (Oxford University, New York, 2002).

\bibitem{Lindblad} 
Lindblad, G. Commun. Math. Phys. 48, 119?130 (1976). 
V. Gorini and A. Kossakowski. Journal of Mathematical Physics 17, 821 (1976).

\bibitem{Caldeira:1982iu} 
  A.~O.~Caldeira and A.~J.~Leggett,
  Physica {\bf 121A}, 587 (1983).
    
\bibitem{Feynman:1963fq} 
  R.~P.~Feynman and F.~L.~Vernon, Jr.,
  Annals Phys.\  {\bf 24}, 118 (1963)
  [Annals Phys.\  {\bf 281}, 547 (2000)].

\bibitem{Akamatsu:2012vt} 
  Y.~Akamatsu,
  Phys.\ Rev.\ D {\bf 87}, 045016 (2013).  
  
\bibitem{Akamatsu:2014qsa} 
  Y.~Akamatsu,
  Phys.\ Rev.\ D {\bf 91}, 056002 (2015).      
   
\bibitem{DeBoni:2017ocl} 
  D.~De Boni,
  JHEP {\bf 1708}, 064 (2017).  
 
\bibitem{Rothkopf:2013kya} 
  A.~Rothkopf,
  JHEP {\bf 1404}, 085 (2014).   
   
\bibitem{Kajimoto:2017rel} 
  S.~Kajimoto, Y.~Akamatsu, M.~Asakawa and A.~Rothkopf,
  Phys.\ Rev.\ D {\bf 97}, 014003 (2018).
   
\bibitem{Yao:2018nmy} 
  X.~Yao and T.~Mehen,
  Phys.\ Rev.\ D {\bf 99}, 096028 (2019).  
  
\bibitem{Brambilla:2016wgg} 
  N.~Brambilla, M.~A.~Escobedo, J.~Soto and A.~Vairo,
  Phys.\ Rev.\ D {\bf 96}, 034021 (2017).
 
\bibitem{Brambilla:2017zei} 
  N.~Brambilla, M.~A.~Escobedo, J.~Soto and A.~Vairo,
  Phys.\ Rev.\ D {\bf 97}, 074009 (2018).
  
\bibitem{Katz:2015qja} 
  R.~Katz and P.~B.~Gossiaux,
  Annals Phys.\  {\bf 368}, 267 (2016). 
  
\bibitem{Brambilla:2019tpt} 
N.~Brambilla, M.~A.~Escobedo, A.~Vairo and P.~Vander Griend,
Phys.\ Rev.\ D {\bf 100}, 054025 (2019).

\bibitem{Abelev:2012rv} 
  B.~Abelev {\it et al.} [ALICE Collaboration],
  Phys.\ Rev.\ Lett.\  {\bf 109}, 072301 (2012).
  
\bibitem{Adam:2016rdg} 
  J.~Adam {\it et al.} [ALICE Collaboration],
  Phys.\ Lett.\ B {\bf 766}, 212 (2017).  

\bibitem{ALICE:2013xna} 
  E.~Abbas {\it et al.} [ALICE Collaboration],
  Phys.\ Rev.\ Lett.\  {\bf 111}, 162301 (2013).
  
\bibitem{Acharya:2017tgv} 
  S.~Acharya {\it et al.} [ALICE Collaboration],
  Phys.\ Rev.\ Lett.\  {\bf 119}, 242301 (2017).    
      
\bibitem{Deng:2012pc} 
  W.~T.~Deng and X.~G.~Huang,
  Phys.\ Rev.\ C {\bf 85}, 044907 (2012).
  
\bibitem{Shovkovy:2012zn} 
  I.~A.~Shovkovy,
  Lect.\ Notes Phys.\  {\bf 871}, 13 (2013).
 
\bibitem{Bruckmann:2013oba} 
  F.~Bruckmann, G.~Endrodi and T.~G.~Kovacs,
  JHEP {\bf 1304}, 112 (2013). 
  
\bibitem{Guo:2015nsa} 
  X.~Guo, S.~Shi, N.~Xu, Z.~Xu and P.~Zhuang,
  Phys.\ Lett.\ B {\bf 751}, 215 (2015).  
  
\bibitem{Liang:2004ph} 
  Z.~T.~Liang and X.~N.~Wang,
  Phys.\ Rev.\ Lett.\  {\bf 94}, 102301 (2005)
  Erratum: [Phys.\ Rev.\ Lett.\  {\bf 96}, 039901 (2006)].
  
\bibitem{Becattini:2007sr} 
  F.~Becattini, F.~Piccinini and J.~Rizzo,
  Phys.\ Rev.\ C {\bf 77}, 024906 (2008).  
  
\bibitem{STAR:2017ckg} 
  L.~Adamczyk {\it et al.} [STAR Collaboration],
  Nature {\bf 548}, 62 (2017).
  
\bibitem{Jiang:2015cva} 
  Y.~Jiang, X.~G.~Huang and J.~Liao,
  Phys.\ Rev.\ D {\bf 92}, 071501 (2015).  
 
\bibitem{Kharzeev:2007jp} 
  D.~E.~Kharzeev, L.~D.~McLerran and H.~J.~Warringa,
  Nucl.\ Phys.\ A {\bf 803}, 227 (2008).

\bibitem{Arnold:2003zc} 
  P.~B.~Arnold, G.~D.~Moore and L.~G.~Yaffe,
  JHEP {\bf 0305}, 051 (2003).

\bibitem{Gupta:2003zh} 
  S.~Gupta,
  Phys.\ Lett.\ B {\bf 597}, 57 (2004).
  
\bibitem{Aarts:2007wj} 
  G.~Aarts, C.~Allton, J.~Foley, S.~Hands and S.~Kim,
  Phys.\ Rev.\ Lett.\  {\bf 99}, 022002 (2007).  

\bibitem{Buividovich:2010tn} 
  P.~V.~Buividovich, M.~N.~Chernodub, D.~E.~Kharzeev, T.~Kalaydzhyan, E.~V.~Luschevskaya and M.~I.~Polikarpov,
  Phys.\ Rev.\ Lett.\  {\bf 105}, 132001 (2010).

\bibitem{Ding:2010ga} 
  H.-T.~Ding, A.~Francis, O.~Kaczmarek, F.~Karsch, E.~Laermann and W.~Soeldner,
  Phys.\ Rev.\ D {\bf 83}, 034504 (2011).
  
\bibitem{Burnier:2012ts} 
  Y.~Burnier and M.~Laine,
  Eur.\ Phys.\ J.\ C {\bf 72}, 1902 (2012).  
  
\bibitem{Brandt:2012jc} 
  B.~B.~Brandt, A.~Francis, H.~B.~Meyer and H.~Wittig,
  JHEP {\bf 1303}, 100 (2013).
  
\bibitem{Amato:2013naa} 
  A.~Amato, G.~Aarts, C.~Allton, P.~Giudice, S.~Hands and J.~I.~Skullerud,
  Phys.\ Rev.\ Lett.\  {\bf 111}, 172001 (2013).    
  
\bibitem{Finazzo:2013efa} 
  S.~I.~Finazzo and J.~Noronha,
  Phys.\ Rev.\ D {\bf 89}, 106008 (2014).

\bibitem{Sahoo:2018dxn} 
  P.~Sahoo, S.~K.~Tiwari and R.~Sahoo,
  Phys.\ Rev.\ D {\bf 98}, 054005 (2018).    

\bibitem{Cassing:2013iz} 
  W.~Cassing, O.~Linnyk, T.~Steinert and V.~Ozvenchuk,
  Phys.\ Rev.\ Lett.\  {\bf 110}, 182301 (2013).
  
\bibitem{Greif:2014oia} 
  M.~Greif, I.~Bouras, C.~Greiner and Z.~Xu,
  Phys.\ Rev.\ D {\bf 90}, 094014 (2014).  
  
\bibitem{Puglisi:2014sha} 
  A.~Puglisi, S.~Plumari and V.~Greco,
  Phys.\ Rev.\ D {\bf 90}, 114009 (2014).  
  
\bibitem{Feng:2017tsh} 
  B.~Feng,
  Phys.\ Rev.\ D {\bf 96}, 036009 (2017).   
  
\bibitem{Gursoy:2014aka} 
  U.~Gursoy, D.~Kharzeev and K.~Rajagopal,
  Phys.\ Rev.\ C {\bf 89}, 054905 (2014).  

\bibitem{Tuchin:2013ie} 
  K.~Tuchin,
  Adv.\ High Energy Phys.\  {\bf 2013}, 490495 (2013).
  
\bibitem{Huang:2017tsq} 
  A.~Huang, Y.~Jiang, S.~Shi, J.~Liao and P.~Zhuang,
  Phys.\ Lett.\ B {\bf 777}, 177 (2018).  
  
\bibitem{Stewart:2017zsu} 
  E.~Stewart and K.~Tuchin,
  Phys.\ Rev.\ C {\bf 97}, 044906 (2018).  
  
\bibitem{Roy:2015kma} 
  V.~Roy, S.~Pu, L.~Rezzolla and D.~Rischke,
  Phys.\ Lett.\ B {\bf 750}, 45 (2015). 

\bibitem{Inghirami:2016iru} 
  G.~Inghirami, L.~Del Zanna, A.~Beraudo, M.~H.~Moghaddam, F.~Becattini and M.~Bleicher,
  Eur.\ Phys.\ J.\ C {\bf 76}, 659 (2016).
  
\bibitem{Kharzeev:2012ph} 
  D.~E.~Kharzeev, K.~Landsteiner, A.~Schmitt and H.~U.~Yee,
  Lect.\ Notes Phys.\  {\bf 871}, 1 (2013).

\bibitem{Miransky:2015ava} 
  V.~A.~Miransky and I.~A.~Shovkovy,
  Phys.\ Rept.\  {\bf 576}, 1 (2015).

\bibitem{Bonati:2014ksa} 
  C.~Bonati, M.~D'Elia, M.~Mariti, M.~Mesiti, F.~Negro and F.~Sanfilippo,
  Phys.\ Rev.\ D {\bf 89}, 114502 (2014).

\bibitem{Bonati:2016kxj} 
  C.~Bonati, M.~D'Elia, M.~Mariti, M.~Mesiti, F.~Negro, A.~Rucci and F.~Sanfilippo,
  Phys.\ Rev.\ D {\bf 94}, 094007 (2016).
  
\bibitem{Galilo:2011nh} 
  B.~V.~Galilo and S.~N.~Nedelko,
  Phys.\ Rev.\ D {\bf 84}, 094017 (2011).

\bibitem{Dudal:2014jfa} 
  D.~Dudal and T.~G.~Mertens,
  Phys.\ Rev.\ D {\bf 91}, 086002 (2015).
  
\bibitem{Rougemont:2014efa} 
  R.~Rougemont, R.~Critelli and J.~Noronha,
  Phys.\ Rev.\ D {\bf 91}, 066001 (2015).

\bibitem{Bonati:2017uvz} 
  C.~Bonati, M.~D'Elia, M.~Mariti, M.~Mesiti, F.~Negro, A.~Rucci and F.~Sanfilippo,
  Phys.\ Rev.\ D {\bf 95}, 074515 (2017).

\bibitem{Alford:2013jva} 
  J.~Alford and M.~Strickland,
  Phys.\ Rev.\ D {\bf 88}, 105017 (2013).
 
\bibitem{Marasinghe:2011bt} 
  K.~Marasinghe and K.~Tuchin,
  Phys.\ Rev.\ C {\bf 84}, 044908 (2011).  
 
\bibitem{Yoshida:2016xgm} 
  T.~Yoshida and K.~Suzuki,
  Phys.\ Rev.\ D {\bf 94}, 074043 (2016).
  
\bibitem{Cho:2014loa} 
  S.~Cho, K.~Hattori, S.~H.~Lee, K.~Morita and S.~Ozaki,
  Phys.\ Rev.\ D {\bf 91}, 045025 (2015).  
  
\bibitem{Kumar:2018ujk} 
  R.~Kumar and A.~Kumar,
  Eur.\ Phys.\ J.\ C {\bf 79}, 403 (2019). 
  
\bibitem{Bonati:2015dka} 
  C.~Bonati, M.~D'Elia and A.~Rucci,
  Phys.\ Rev.\ D {\bf 92}, 054014 (2015).  

\bibitem{Gubler:2015qok} 
  P.~Gubler, K.~Hattori, S.~H.~Lee, M.~Oka, S.~Ozaki and K.~Suzuki,
  Phys.\ Rev.\ D {\bf 93}, 054026 (2016).  
  
\bibitem{Machado:2013yaa} 
  C.~S.~Machado, S.~I.~Finazzo, R.~D.~Matheus and J.~Noronha,
  Phys.\ Rev.\ D {\bf 89}, 074027 (2014). 
  
\bibitem{Aaij:2012ag} 
  R.~Aaij {\it et al.} [LHCb Collaboration],
  Eur.\ Phys.\ J.\ C {\bf 72}, 2100 (2012).  
  
\bibitem{Chatrchyan:2011kc} 
  S.~Chatrchyan {\it et al.} [CMS Collaboration],
  JHEP {\bf 1202}, 011 (2012).  
  
\bibitem{ATLAS:2014ala} 
  G.~Aad {\it et al.} [ATLAS Collaboration],
  JHEP {\bf 1407}, 154 (2014).     
 
\bibitem{Adam:2019wnk} 
  J.~Adam {\it et al.} [STAR Collaboration],
  arXiv:1905.02052 [nucl-ex].

\bibitem{Das:2016cwd} 
  S.~K.~Das, S.~Plumari, S.~Chatterjee, J.~Alam, F.~Scardina and V.~Greco,
  Phys.\ Lett.\ B {\bf 768}, 260 (2017).
  
\bibitem{Fukushima:2015wck} 
  K.~Fukushima, K.~Hattori, H.~U.~Yee and Y.~Yin,
  Phys.\ Rev.\ D {\bf 93}, 074028 (2016).  
  
\bibitem{Finazzo:2016mhm} 
  S.~I.~Finazzo, R.~Critelli, R.~Rougemont and J.~Noronha,
  Phys.\ Rev.\ D {\bf 94}, 054020 (2016)
  Erratum: [Phys.\ Rev.\ D {\bf 96}, 019903 (2017)].  
  
\bibitem{Dudal:2018rki} 
  D.~Dudal and T.~G.~Mertens,
  Phys.\ Rev.\ D {\bf 97}, 054035 (2018).

\bibitem{Bertulani:1987tz} 
  C.~A.~Bertulani and G.~Baur,
  Phys.\ Rept.\  {\bf 163}, 299 (1988).
  
\bibitem{Krauss:1997vr} 
  F.~Krauss, M.~Greiner and G.~Soff,
  Prog.\ Part.\ Nucl.\ Phys.\  {\bf 39}, 503 (1997).  
  
\bibitem{Bertulani:2005ru} 
  C.~A.~Bertulani, S.~R.~Klein and J.~Nystrand,
  Ann.\ Rev.\ Nucl.\ Part.\ Sci.\  {\bf 55}, 271 (2005). 
  
\bibitem{Klein:1999qj} 
  S.~Klein and J.~Nystrand,
  Phys.\ Rev.\ C {\bf 60}, 014903 (1999).  

\bibitem{Adam:2015sia} 
  J.~Adam {\it et al.} [ALICE Collaboration],
  Phys.\ Lett.\ B {\bf 751}, 358 (2015).
  
\bibitem{TheALICE:2014dwa} 
  B.~B.~Abelev {\it et al.} [ALICE Collaboration],
  Phys.\ Rev.\ Lett.\  {\bf 113}, 232504 (2014).  
 
\bibitem{STAR:2019yox} 
  K.~Krueger {\it et al.} [STAR Collaboration],
  arXiv:1904.11658 [hep-ex].  
  
\bibitem{Adam:2015gba} 
  J.~Adam {\it et al.} [ALICE Collaboration],
  Phys.\ Rev.\ Lett.\  {\bf 116}, 222301 (2016).
  
\bibitem{Fermi:1924tc} 
  E.~Fermi,
  Z.\ Phys.\  {\bf 29}, 315 (1924).  
 
\bibitem{vonWeizsacker:1934nji} 
  C.~F.~von Weizsacker,
  Z.\ Phys.\  {\bf 88}, 612 (1934).
   
\bibitem{Yu:2017pot} 
  G.~M.~Yu, Y.~B.~Cai, Y.~D.~Li and J.~S.~Wang,
  Phys.\ Rev.\ C {\bf 95}, 014905 (2017)
  Addendum: [Phys.\ Rev.\ C {\bf 95}, 069901 (2017)].
  
\bibitem{Shi:2017qep} 
  W.~Shi, W.~Zha and B.~Chen,
  Phys.\ Lett.\ B {\bf 777}, 399 (2018).
  
\bibitem{Chen:2018sir} 
  B.~Chen, C.~Greiner, W.~Shi, W.~Zha and P.~Zhuang,
  arXiv:1801.01677 [hep-ph].  
 
\bibitem{Hufner:1997jg} 
  J.~Hufner and B.~Z.~Kopeliovich,
  Phys.\ Lett.\ B {\bf 426}, 154 (1998).

\bibitem{Ryskin:1992ui} 
  M.~G.~Ryskin,
  Z.\ Phys.\ C {\bf 57}, 89 (1993).
    
\bibitem{Jones:2013pga} 
  S.~P.~Jones, A.~D.~Martin, M.~G.~Ryskin and T.~Teubner,
  JHEP {\bf 1311}, 085 (2013). 

\bibitem{Klein:2016yzr} 
  S.~R.~Klein, J.~Nystrand, J.~Seger, Y.~Gorbunov and J.~Butterworth,
  Comput.\ Phys.\ Commun.\  {\bf 212}, 258 (2017).
  
\bibitem{Klein:2017vua} 
  S.~R.~Klein,
  Nucl.\ Phys.\ A {\bf 967}, 249 (2017).    

\bibitem{Klein:1999gv} 
  S.~R.~Klein and J.~Nystrand,
  Phys.\ Rev.\ Lett.\  {\bf 84}, 2330 (2000).
    
\bibitem{Zha:2017jch} 
  W.~Zha {\it et al.},
  Phys.\ Rev.\ C {\bf 97}, 044910 (2018). 

\bibitem{Jiang:2016woz} 
  Y.~Jiang, Z.~W.~Lin and J.~Liao,
  Phys.\ Rev.\ C {\bf 94}, 044910 (2016)
  Erratum: [Phys.\ Rev.\ C {\bf 95}, 049904 (2017)].

\bibitem{Deng:2016gyh} 
  W.~T.~Deng and X.~G.~Huang,
  Phys.\ Rev.\ C {\bf 93}, 064907 (2016). 
  
\bibitem{Becattini:2015ska} 
  F.~Becattini {\it et al.},
  Eur.\ Phys.\ J.\ C {\bf 75}, 406 (2015)
  Erratum: [Eur.\ Phys.\ J.\ C {\bf 78}, 354 (2018)].
  
\bibitem{Ivanov:2017dff} 
  Y.~B.~Ivanov and A.~A.~Soldatov,
  Phys.\ Rev.\ C {\bf 95}, 054915 (2017).
  
\bibitem{Xie:2017upb} 
  Y.~Xie, D.~Wang and L.~P.~Csernai,
  Phys.\ Rev.\ C {\bf 95}, 031901 (2017).    

\bibitem{Pang:2016igs} 
  L.~G.~Pang, H.~Petersen, Q.~Wang and X.~N.~Wang,
  Phys.\ Rev.\ Lett.\  {\bf 117}, 192301 (2016).  

\bibitem{Bozek:2010bi} 
  P.~Bozek and I.~Wyskiel,
  Phys.\ Rev.\ C {\bf 81}, 054902 (2010).

\bibitem{Chatterjee:2017ahy} 
  S.~Chatterjee and P.~Bo\.zek,
  Phys.\ Rev.\ Lett.\  {\bf 120}, 192301 (2018).

\bibitem{Abelev:2008jga} 
  B.~I.~Abelev {\it et al.} [STAR Collaboration],
  Phys.\ Rev.\ Lett.\  {\bf 101}, 252301 (2008).

\bibitem{Bratkovskaya:2004ec} 
  E.~L.~Bratkovskaya, W.~Cassing, H.~Stoecker and N.~Xu,
  Phys.\ Rev.\ C {\bf 71}, 044901 (2005).
  
\bibitem{CBM}   
The CBM Physics Book, Lecture Notes in Physics, Vol. 814, ed B Friman and P Senger et al.  
  
\bibitem{Kovalenko:2019pgu} 
  A.~D.~Kovalenko {\it et al.} [Nica Collaboration],
  PoS SPIN {\bf 2018}, 007 (2019).
  
    
\end{thebibliography}
\end{document}